\documentclass[rmp,aps,twocolumn,tight]{revtex4-1}
\usepackage{amsmath}
\usepackage{amssymb}
\usepackage{cancel}
\usepackage{amsfonts}
\usepackage[T1]{fontenc}
\usepackage{bm}
\usepackage{color}
\usepackage{graphicx}
\DeclareSymbolFont{operators}{OT1}{cmr}{m}{n}
\DeclareSymbolFont{letters}{OML}{cmm}{m}{it}
\DeclareSymbolFont{symbols}{OMS}{cmsy}{m}{n}
\DeclareSymbolFont{largesymbols}{OMX}{cmex}{m}{n}

\usepackage{hyperref}
\hypersetup{
    bookmarks=true,         
    unicode=false,          
    pdftoolbar=true,        
    pdfmenubar=true,        
    pdffitwindow=false,     
    pdfstartview={FitH},    
    pdfsubject={Plasma fluid theory},   
    pdfproducer={GNU Make}, 
    pdfkeywords={keyword1} {key2} {key3}, 
    pdfnewwindow=true,      
    colorlinks=true,        
    linkcolor=blue,      
    citecolor=blue,         
    filecolor=blue,      
    urlcolor=blue           
}

\newcommand{\1}{\hat{\bm{1}}}

\makeatother

\begin{document}

\title{Physical constraints for the evolution of life on exoplanets}

\author{Manasvi Lingam}
\email{manasvi.lingam@cfa.harvard.edu}
\affiliation{Institute for Theory and Computation, Harvard University, Cambridge, MA 02138, USA}
\author{Abraham Loeb}
\email{aloeb@cfa.harvard.edu}
\affiliation{Institute for Theory and Computation, Harvard University, Cambridge, MA 02138, USA}
 
\begin{abstract}
Recently, many Earth-sized planets have been discovered around stars other than the Sun that might possess appropriate conditions for life. The development of theoretical methods for assessing the putative habitability of these worlds is of paramount importance, since it serves the dual purpose of identifying and quantifying what types of biosignatures may exist and determining the selection of optimal target stars and planets for subsequent observations. This Colloquium discusses how a multitude of physical factors act in tandem to regulate the propensity of worlds for hosting detectable biospheres. The focus is primarily on planets around low-mass stars, as they are most readily accessible to searches for biosignatures. This Colloquium outlines how factors such as stellar winds, the availability of ultraviolet and visible light, the surface water and land fractions, stellar flares, and associated phenomena place potential constraints on the evolution of life on these planets. 
\end{abstract}

\maketitle

\tableofcontents

\section{Introduction}
\begin{quotation}
\noindent Furthermore, there are infinite worlds both like and unlike this world of ours. For the atoms being infinite in number \dots are borne on far out into space.\\

\noindent -- Epicurus (341-270 BC)\footnote{The translation of the above quote has been adopted from \citet{Bai57}.}
\end{quotation}
Ever since the ancient Greeks, and probably even earlier, many have argued in favor of the existence of worlds (planets and moons) outside our Solar system. However, it was only a few decades ago that the first extrasolar planets (exoplanets) were detected, owing to the observational challenges associated with finding planets because of their extremely low brightness and masses relative to their host stars. One of the notable early discoveries was 51 Pegasi b, the first exoplanet orbiting a solar-type star \citep{MQ95}. Many of the initial discoveries, including 51 Pegasi b, were of exoplanets whose masses were comparable to, or larger than, that of Jupiter. A review of the early developments in exoplanetary science was given by \citet{Per00}. 

From the 1990s until the launch of NASA's \emph{Kepler} spacecraft in 2009 \citep{Bat14,Bor16}, exoplanets were discovered at a steady rate and numbered in the hundreds. In conjunction with the \emph{Kepler} mission, designed with the express purpose of finding exoplanets, a flotilla of ground-based telescopes have propelled the explosive growth of exoplanetary science \citep{WF15}. As of April 2019, the number of confirmed exoplanets is over $4000$ with planetary systems possessing more than one exoplanet exceeding $650$.\footnote{\url{http://exoplanet.eu/catalog/}} Thus, it would be no exaggeration to state that the field of exoplanets represents one of the most rapidly advancing frontiers in astrophysics \citep{Perr18}.

One of the most compelling reasons behind the study of exoplanets entails the prospect of detecting extraterrestrial life beyond Earth.\footnote{Exomoons are also of considerable interest as possible abodes for extraterrestrial life, and are likely to be numerous \citep{HW14}, but we shall restrict ourselves to exoplanets as they have been subjected to more theoretical and observational studies. Very recently, some preliminary evidence has been presented in favor of a Neptune-sized exomoon candidate orbiting the Jupiter-sized planet Kepler-1625b \citep{KT18}.} It is fair to say that the question, ``\emph{Are we alone?}'', is one that has resonated with humanity over millenia. The identification of extraterrestrial life is fraught with immense difficulties and challenges, owing to which it is natural to commence searching for ``life as we know it'' on a planet that shares the basic physical properties of the Earth. Water is conventionally regarded as one of the most essential requirements for life on Earth owing to its unusual physicochemical properties \citep{Ba08}. Hence, most studies tend to adopt a ``follow the water'' approach, i.e. to search for planets where liquid water could exist on their surfaces. 

Before proceeding further, we caution that solvents other than water are conceivable and biochemistries not based on carbon are feasible \citep{Ben04,Ba04,SI08}. One notable example in this category is Titan, which is known to possess methane lakes and seas that contribute to a methane-based hydrological cycle \citep{Hay16,ML16}. We will not tackle this issue herein, despite its undoubted importance, as our understanding of ``life as we do not know it'' is much less developed compared to our knowledge of life on Earth.\footnote{Whenever we shall employ the word ``life'' henceforth, it must be viewed with the proviso understanding that we actually refer to ``life as we know it'', unless explicitly stated otherwise.}

The ``follow the water'' approach, from a theoretical and observational standpoint, has led to the identification of the circumstellar habitable zone (HZ), i.e., the annular region around the host star where a planet is theoretically capable of hosting liquid water on its surface. Aside from having a clement surface temperature, the planet needs to retain an atmosphere since water ice transforms directly into gas phase in vacuum. Mars, whose mass is one-tenth that of the Earth, is believed to have lost the majority of its atmosphere in the distant past \citep{Jak17,DLM18}. 

It is evident that a smaller and less hotter star would emit less radiation, and therefore its HZ would be situated closer to it. However, the HZ depends not only on stellar properties but also a wide range of planetary properties including, but not restricted to, its atmospheric composition (mixture of greenhouse gases), mass and rotation. The HZ has a long and fascinating history dating back to at least the 19th century, which has been discussed by \citet{Gonz05}. The modern formulations of the HZ have matured during the past couple of decades \citep{KWR93,KRK13,KRS14,YBF14,Zs15,KWH16,HKB16,Ram18,SRO19}. Although we focus only on planets within the circumstellar HZ herein, it is worth noting the concept of the HZ has been extended to encompass both exomoons as well as planets around stellar binaries, i.e., the circumplanetary \citep{HW14,DHT17} and circumbinary \citep{KH13,Cun14} habitable zones, respectively. 

There are a couple of important points that must be borne in mind regarding the HZ. First, not all planets within the HZ are guaranteed to actually have liquid water on the surface. Second, even the presence of liquid water is a necessary, but not sufficient, condition for the existence of life. For example, some of the other necessary requirements include sufficient abundances of ``bioessential'' elements such as carbon, hydrogen, oxygen, nitrogen and phosphorus as well as free energy flows \citep{HAS07,MS07}.\footnote{As life is a far-from-equilibrium phenomenon, it is more accurate to state that the conversion of thermodynamic disequilibria facilitates its emergence \citep{BBG17}.} Hence, it is essential to avoid conflating the HZ with the much wider notion of habitability \citep{ML17,TT17}. As per the 2015 NASA Astrobiology Strategy, habitability can be understood in the following terms.\footnote{\url{https://nai.nasa.gov/media/medialibrary/2016/04/NASA_Astrobiology_Strategy_2015_FINAL_041216.pdf}}
\begin{quotation}
Habitability has been defined as the potential of an environment (past  or present) to support life of any kind \dots Habitability is a function of a multitude of environmental parameters whose study is biased by the effects that biology has on these parameters. 
\end{quotation}
Further reviews of this multi-faceted subject can be found in \citet{Dole,LB09,JD10,LiC12,Sea13,Cock16,EA16}. 

Although the above limitations of the HZ definition must be duly recognized, it is equally important to appreciate its strengths. As the HZ depends on a wide range of planetary and stellar parameters, it encompasses a diverse array of potentially habitable environments; an overview of the recent progress in defining and studying HZs was presented by \citet{Ram18}. Perhaps, more importantly, owing to the constraints imposed by observation time and funding, the selection of suitable target stars and planets is of paramount importance \citep{HJ10,KE10,LL18a}. In this respect, the HZ provides a potential methodology for identifying and selecting planets for more detailed follow-up observations. 

In connection with the HZ, a couple of exciting discoveries over the past few years merit a special mention. An exoplanet was discovered in the HZ of Proxima Centauri, the star nearest to the Earth at a distance of $1.3$ pc, namely, $4 \times 10^{16}$ m \citep{AA16}. This exoplanet, named Proxima Centauri b (or Proxima b for short), has a minimum mass of $1.3\,M_\oplus$, where $M_\oplus$ is the mass of the Earth. The second major breakthrough entailed the discovery of seven, roughly Earth-sized, planets orbiting the star TRAPPIST-1 at a distance of $12.1$ pc \citep{GJ16,GT17}. Of these seven planets, at least three of them reside within the HZ, and are therefore capable of hosting liquid water on the surface. The masses of these seven planets fall within $0.3$-$1.16\,M_\oplus$ and the radii range $0.77$-$1.15\,R_\oplus$ \citep{DG18,GD18}, where $R_\oplus$ is the radius of the Earth. Another discovery worth pointing out is the planet LHS 1140b, with a radius and mass of $1.4\,R_\oplus$ and $6.6\,M_\oplus$, respectively, in the HZ of a star at a distance of $12$ pc \citep{DI17}.

Before proceeding further, it is essential to articulate the scope and philosophy of this review. It is not feasible to present a comprehensive account of all aspects of habitability due to length constraints. Instead, only a few select topics that have witnessed notable advances within the past decade are tackled. Virtually all factors that we address have a physical basis and pertain to the properties of the host star. This is because we possess a relatively sound understanding of these aspects, and they are arguably less subject to systemic uncertainties. 

We will, for instance, not address the issue of whether a given planet can sustain a stable and clement climate over Gyr ($10^9$ yrs) timescales, despite its relevance for habitability, because the climate depends on a diverse array of factors such as the orbital eccentricity, axial tilt (obliquity), the existence of a large moon \citep{LJR93}, surface landmasses and oceans, and biological feedback mechanisms to name a few. Discussions of how some of these variables affect the climate have been given by \citet{Kas10,Pie10,CK17}, whereas analyses of the climates of Proxima b and the TRAPPIST-1 system include \citet{TL16,Wolf,BM17,ACL17,TB18,MA18,LMC18}. In the same spirit, we will not tackle geophysical and biogeochemical cycles herein owing to their complexity \citep{SG06}. Furthermore, the biogeochemistry of planets has not been explored in sufficient detail, despite the fact that it can give rise to feedbacks and buffering over short timescales \citep{SB13}, which are comparable to those encountered in this review.

Looking beyond planetary and stellar properties, high-energy astrophysical phenomena within the Galaxy also play a vital role in regulating planetary habitability. This has prompted the analysis and identification of the Galactic Habitable Zone \citep{GBW01,LFG04}, but the limits of this region remain subject to uncertainty depending on the constraints adopted \citep{Pran08,GM18}. Examples of the astrophysical risks for life include high doses of ionizing radiation emanating from quasars, supernovae and gamma ray bursts \citep{Thor95,MT11,PJ14,SAL,BT17,FL18,LGB19}.

We mostly focus on planets in the HZ of the most common stars, namely M-dwarfs. Earlier reviews of the habitability of M-dwarf exoplanets can be found in \citet{TB07,SK07,SBJ16}. Broadly speaking, M-dwarfs are low-mass stars with a mass $M_\star$ in the range $0.075 < M_\star/M_\odot < 0.5$, where $M_\odot$ is the solar mass. M-dwarfs also exhibit significant variation in their radii, effective temperatures, surface magnetic fields and activity \citep{CB00}. One of the most notable features of M-dwarfs is that they appear in two distinct ``flavors'' \citep{CB97,SHC11}. M-dwarfs with $M_\star \gtrsim 0.35\,M_\odot$ are characterized by stellar interiors with an inner radiative zone and an outer convective envelope; in contrast, M-dwarfs with $M_\star \lesssim 0.35\,M_\odot$ are fully convective, i.e., they do not possess the radiative zone. 

Not only are these low-mass stars the most common and long-lived in the Universe \citep{AL97,cha03,TB07}, about $20\%$ of them have Earth-sized planets in their HZs \citep{BDU13,DC15,MPA15}. These planets are also comparatively accessible to detailed observations \citep{Wi10,FA18}, mostly as a consequence of their smaller orbital radii. Finally, the nearest planets in the HZ described earlier - Proxima b, the TRAPPIST-1 system and LHS 1140b - are located around M-dwarfs.

The outline of this Colloquium is as follows. In Sec. \ref{SecSW}, we describe how stellar winds can erode planetary atmospheres and reduce the shielding offered by planetary magnetospheres. The former, in particular, are especially important since they serve as the repositories of most biomarkers, implying that their existence is necessary for detecting non-technological extraterrestrial life outside our Solar system. We discuss how the origin and evolution of life is affected by electromagnetic radiation fluxes from the host star in Sec. \ref{SecEM}. This is followed by analyzing how stellar flares and their associated physical phenomena influence biospheres in Sec. \ref{SecFla}. We conclude with a summary of the central results in Sec. \ref{SecConc}.

\section{Stellar winds}\label{SecSW}
Stellar winds are streams of plasma that originates from the outer regions of stellar coronae \citep{Park58,Pri14}. We will discuss two primary aspects by which stellar winds influence planetary habitability. Additional effects include Ohmic heating in the upper atmosphere \citep{CD14} that may be up to an order of magnitude higher than the heating from extreme ultraviolet (EUV) radiation in the wavelength range of $10$-$120$ nm \citep{CGG18}.

\subsection{Planetary magnetospheres}\label{SSecPMag}
When exoplanets possess an intrinsic magnetic field, the solar wind plasma will be deflected \citep{BT12}. The cavity created by the planetary magnetic field is the magnetosphere. The magnetopause distance ($R_{mp}$), which represents the outer boundary of the magnetosphere, serves as a useful proxy for its size. Let us derive its value here.

Suppose that the planet has a pure dipole magnetic field and that its strength is $B_p$ at the surface of the planet of radius $R_p$. At the magnetopause distance, the corresponding magnetic field $B_{mp}$ is given by
\begin{equation}
    B_{mp} = B_p \left(\frac{R_p}{R_{mp}}\right)^3,
\end{equation}
provided that only the radial dependence has been retained. The distance $R_{mp}$ is computed by demanding that the magnetic pressure is approximately equal to the solar wind pressure $P_{sw}$. The latter has contributions from the kinetic, magnetic and thermal energies of the solar wind, but it is the first component that typically dominates, i.e. $P_{sw} \approx \rho_{sw} v_{sw}^2$, where $\rho_{sw}$ and $v_{sw}$ are the mass density and velocity of the solar wind respectively. Thus, from $B_{mp}^2/\left(2 \mu_0\right) = P_{sw}$, we obtain
\begin{equation}\label{Rmpd}
    R_{mp} = R_p \left(\frac{B_p^2}{2\mu_0 P_{sw}}\right)^{1/6},
\end{equation}
where $\mu_0$ is the vacuum permeability. In actuality, an additional factor of order unity must be introduced for calculating $R_{mp}$ more accurately, due to deviations from an ideal dipole magnetic field \citep{Gom98}; the characteristic value for the Earth is $R_{mp} \approx 10\,R_\oplus$ \citep{KR95}. As $P_{sw}$ is subject to temporal variations depending on the planetary orbit and stellar activity, the value of $R_{mp}$ also varies accordingly. For instance, theoretical models indicate that $P_{sw}$ varies by $1$-$3$ orders of magnitude near Proxima b, which translates to a variation in $R_{mp}$ by a factor of $2$-$5$ \citep{GDC16}. The corresponding average magnetopause distance for Proxima b equals $R_{mp} \sim 2$-$3\,R_\oplus$, which is smaller than the Earth's magnetosphere by a factor of a few. A similar study has also been undertaken for the planets of the TRAPPIST-1 system \citep{GDC17} (see also \citealt{DJL18}).

Before discussing why a smaller magnetosphere may prove to be problematic from the standpoint of habitability, it is instructive to understand, via a toy model, why $R_{mp}$ is smaller for lower mass stars. Denoting the stellar mass-loss rate by $\dot{M}_\star$ and assuming the resultant wind is spherically symmetric, we have
\begin{equation}\label{MaLo}
    \dot{M}_\star = 4\pi a_p^2 \rho_{sw} v_{sw},
\end{equation}
where $a_p$ represents the semi-major axis of the planet. Typically, $v_{sw}$ does not vary significantly beyond the immediate vicinity of the star \citep{Fitz14}, implying that $\rho_{sw} \propto \dot{M}_\star/a_p^2$. Let us consider Proxima b once again. The stellar mass-loss rate of Proxima Centauri has been theoretically predicted to be comparable to that of the Sun \citep{GDC16}, and is consistent with observational constraints \citep{WD02}; see, however, \citet{WL01}. Using this estimate in conjunction with Proxima b's semi-major axis of $0.0485$ AU, we find that $\rho_{sw}$ should be about $400$ times higher near Proxima b than the corresponding solar wind density at the Earth, which is roughly consistent with more detailed numerical calculations \citep{GDC16,DLMC17}. A higher value of $\rho_{sw}$ translates to higher dynamic pressure ($P_{sw} \propto \rho_{sw}$), thereby leading to a compression of the magnetopause distance.

Furthermore, M-dwarfs are typically characterized by very strong magnetic fields \citep{RB09,MD10,Rei12} that are $2$-$3$ orders of magnitude higher than the average magnetic field at the Sun's surface ($\sim 10^{-4}$ T). As a result, the stellar wind pressure also includes a significant component from the star's magnetic field \citep{VJM13}. This serves to raise the value of $P_{sw}$, and thereby decrease $R_{mp}$ in accordance with (\ref{Rmpd}). In general, if $P_{sw}$ is $2$-$3$ orders of magnitude higher for planets in the HZ of low-mass stars, it follows from (\ref{Rmpd}) that the planetary magnetic field $B_p$ will need to be higher by a factor of $\mathcal{O}(10)$ in order to yield a magnetopause distance similar to that of the Earth. Hence, for sufficiently high magnetic fields, it is conceivable that planets could possess Earth-sized magnetospheres \citep{SJV14}.

However, there are grounds for supposing that $B_p$ will be reduced for planets orbiting low-mass stars. As per dynamo theory, the magnetic field strength is expected to scale with the planet's rotation rate ($\Omega_p$) as
\begin{equation}\label{BOmrel}
    B_p \propto \Omega_p^{\alpha},
\end{equation}
with $\alpha \approx 0.5$-$1$ for many of the classical dynamo models \citep{GSM05,LG11}. Other dynamo simulations, in contrast, indicate that $B_p$ is independent of $\Omega_p$ \citep{Ch10}.\footnote{Differentiating between the various dynamo models requires the measurement of exoplanetary magnetic fields. In principle, this could be done through a number of observational avenues based on radio auroral emission, early transit ingress, H$^+_3$ emission and Ly-$\alpha$ absorption profiles of exoplanets \citep{Grie15}.} The majority, although not necessarily all,\footnote{The presence of a thick atmosphere \citep{LWMM15}, semi-liquid interior \citep{Mak15}, or a companion \citep{VH17} could drive the planet into asynchronous rotation.} of the planets in the HZ of low-mass stars are expected to be synchronous rotators, with their rotation periods equal to their orbital periods, owing to the tidal gravitational force of the host star \citep{Dole,BRL11,Bar17}. In this case, the rotation rate will be reduced considerably, and this leads to a corresponding reduction in the magnetic moment (see, however, \citealt{ZBC13}). For example, a planet $0.2$ AU from a star with $M_\star = 0.5\,M_\odot$ will have $\Omega_p \approx 0.03\,\Omega_\oplus$, implying that $B_p \approx 0.17\,B_\oplus$ if $\alpha = 0.5$ in (\ref{BOmrel}). Here, $\Omega_\oplus$ and $B_\oplus$ are the Earth's rotation rate and equatorial magnetic field, respectively.

Thus, planets around low-mass stars will typically have smaller magnetospheres on account of increased stellar wind pressure and possibly weaker magnetic moments. Conventionally, weaker planetary magnetic fields and smaller magnetospheres lead to a lower degree of protection against the stellar wind, thereby resulting in enhanced atmospheric escape; we will return to this topic in Sec. \ref{SSecAE}. However, this paradigm has been challenged by recent theoretical studies, which indicate that the atmospheric escape rate does not always decline with an increase in the magnetic field strength \citep{DLML18,BT18,GMN18,Ling19}. The basic reason can be understood qualitatively as follows. The polar caps are regions close to the magnetic poles with open magnetic field lines, which permit the escape of ions through the polar wind. The latter is dependent on the ambipolar electric field that stems from the difference in the velocities of ions and electrons \citep{Ax68}; electrons typically move faster, thus causing charge separation and inducing an electric field, which accelerates the ions and permits their escape \citep{schunk2009}. The polar wind may become increasingly important for weak dipole magnetic fields with $B_p \sim 10^{-7}$ T for Mars-like exoplanets \citep{SS18}; see also \citet{Ling19}.

The second aspect that we highlight concerns the effect of planetary magnetic fields in regulating the amount of cosmic rays that reach the surface. This is particularly relevant since high-energy radiation can drive the radiolysis (decomposition via radiation) of complex biomolecules \citep{Dart11,HG12}. It is well-known that the planet's magnetospheric shielding (for deflecting charged particles) influences the amount of radiation that reaches the surface \citep{GSG09}, but more recent studies have revealed that the dependence on the magnetic moment $\mathcal{M}_p$ is very sensitive to the atmospheric column density \citep{AHG13}. For example, when one considers planets with an atmospheric column density equal to that of the Earth, changing the magnetic moment from $\mathcal{M}_p = 0$ to $\mathcal{M}_p = 10 \mathcal{M}_\oplus$ (where $\mathcal{M}_\oplus$ is the magnetic moment of the Earth) results in the radiation dose rate dropping by a factor of about $6$. In contrast, if the atmospheric column density is about $10\%$ that of the Earth, the radiation dose rate declines by a factor of $240$ as one moves from $\mathcal{M}_p = 0$ to $\mathcal{M}_p = 10 \mathcal{M}_\oplus$ \citep{GTV16}.

One other aspect that is not explored here concerns the effect of magnetospheric (and atmospheric) shields on the chemistry and abundances of biosignatures \citep{Gren17}. Cosmic rays may react with N$_2$-O$_2$ atmospheres and stimulate the formation of nitrogen oxides (NO$_x$) and drive the depletion of ozone (O$_3$), the latter of which constitutes a widely studied biosignature. It turns out that high-energy particles produced during stellar flares also have similar effects, as discussed in Sec. \ref{SSecSEP}.

\subsection{Atmospheric escape}\label{SSecAE}
Atmospheric escape refers to any mechanism that imparts sufficient energy to particles, enabling them to achieve speeds higher than the threshold to escape the gravitational pull of the planet \citep{Ti15}. Most studies tend to focus on neutral particles, but it is important to appreciate that charged particles can also escape the atmosphere. There are a wide array of mechanisms that enable the escape of ions, and reviews of this subject were given by \citet{LK08} and \citet{BB16}. 

Charged particles are typically accelerated by electric fields in addition to deriving energy from collisions. The electric field ${\bf E}$ can be expressed as
\begin{equation}\label{Ohm}
  {\bf E} = - {\bf v} \times {\bf B} + \frac{{\bf J} \times {\bf B} - \nabla p_e}{n_e e} + \eta {\bf J} + \dots,
\end{equation}
where ${\bf v}$ is the plasma bulk velocity, ${\bf J}$ is the current, ${\bf B}$ represents the magnetic field, $\eta$ denotes the resistivity, $n_e$ is the electron number density and $p_e$ is the electron pressure. The above expression is known as Ohm's Law, and contains additional terms that have been omitted here \citep{Fri14,LMM16,LH17}. Each of the first three terms on the right-hand-side of (\ref{Ohm}) are known to be capable of accelerating charged particles and enabling them to exit the atmosphere. 

Apart from the polar wind encountered in Sec. \ref{SSecPMag}, which relies on ion outflow from polar regions, there are several other ion escape mechanisms. Ion pick up processes enable the acceleration of ions via electric fields embedded within the stellar wind plasma. In addition, there are a multitude of alternative mechanisms involving plasma instabilities and cool ion outflows \citep{Lam13}. Although, in principle, it is feasible to develop simple models for each of the processes and thus estimate the total loss rate, this approach is not practical within the scope of this review.

Instead, it is instructive to consider a toy model for weakly magnetized or unmagnetized planets, i.e. those that have a negligible intrinsic magnetic field. In our Solar system, Venus falls under this category and so does Mars, although the situation with respect to the latter is more complicated owing to the presence of remnant crustal fields \citep{AC98}. As we have seen earlier, weak magnetic fields may be common for planets in the HZ of low-mass stars. Non-thermal escape mechanisms involving ion escape due to interactions with stellar winds could become particularly important for unmagnetized planets that are comparable to the size of the Earth \citep{BB16,DLMC17}.

Our approach partly mirrors the derivation in \citet{ZSR10} and \citet{ZC17}; see also \citet{LL18b}. To begin with, note that the momentum carried by a single proton in the solar wind is $m_p v_{sw}$. In contrast, a chemical species $X$ in the atmosphere requires a maximal momentum gain of $m_X v_{esc}$, where $m_X$ is the mass of the particle and
\begin{equation*}
v_{esc} = \sqrt{2 G M_p/R_p} = 11\,\mathrm{km/s}\,\left(M_p/M_\oplus\right)^{1/2}\left(R_p/R_\oplus\right)^{-1/2}    
\end{equation*}
is the gravitational escape velocity from the planet. Suppose, for instance, that we consider the loss of O$_2^{+}$ as this represents one of the major species lost from CO$_2$-dominated atmospheres; it can be readily verified that $m_X \approx 32\,m_p$. The escape velocity for Earth-sized planets is $\mathcal{O}(10)$ km/s, whereas $v_{sw}$ is typically $\mathcal{O}(100)$ km/s, although $\sim 10^3$ km/s is also possible. Thus, in heuristic terms, the momentum of a proton in the stellar wind can be wholly imparted to a single entity of $X$ to facilitate the escape of the latter. 

Hence, if we determine the rate of protons impinging on the planet, we can also treat this as the escape rate of particles from the planet's atmosphere ($\dot{M}_{p}$). Recall that the stellar mass-loss rate is $\dot{M}_\star$ and the flux at a distance $a_p$ is $\dot{M}_\star/\left(4\pi a_p^2\right)$. The cross-sectional area presented by the planet is $\pi R_p^2$. Thus, the escape rate is
\begin{equation}\label{MLRate}
\dot{M}_p  = \frac{1}{4}\left(\frac{R_p}{a_p}\right)^2 \dot{M}_\star.
\end{equation}
In actuality, not all stellar wind protons will contribute to atmospheric escape, owing to which the factor of $1/4$ could be replaced with an efficiency factor $\varepsilon$ that can be as much as one order of magnitude lower. If we choose $R_p = 0.53\,R_\oplus$ and $a_p = 1.524$ AU for Mars, along with $\dot{M}_\odot \sim 2 \times 10^{-14}\,M_\odot\,\mathrm{yr}^{-1}$ \citep{WD02}, we obtain $\dot{M}_p \sim 0.07$ kg/s, which is only about a factor of two lower than the predicted O$_2^{+}$ loss rate of $\sim 2.6 \times 10^{24}\,\mathrm{s}^{-1}$ \citep{DLM18} that corresponds to $\sim 0.14$ kg/s (using $m_{O_2^+} \approx 32\,m_p$).\footnote{At this rate, losing the mass of Earth's atmosphere ($\sim 5 \times 10^{18}$ kg) would require $\sim 10^{12}$ yrs.} Numerical simulations carried out by \citet{DJL18} indicate that (\ref{MLRate}) accurately reflects the trend for the expected escape rates of the seven TRAPPIST-1 planets. 

Two points regarding (\ref{MLRate}) are worth mentioning. First, the quantity $\dot{M}_\star$ evolves over time, and is much higher when the star is younger \citep{WM05}, implying that the atmospheric escape rates will also be correspondingly higher. Second, it is difficult to estimate $\dot{M}_\star$ as it depends on a variety of stellar parameters such as the mass, radius and rotation rate \citep{CS11}. A simple theoretical prescription that may be reasonably accurate for stars within the range $0.4 < M_\star/M_\odot < 1.1$ is \citep{JG15}:
\begin{equation}\label{MLPres}
    \frac{\dot{M}_\star}{\dot{M}_\odot} = \left(\frac{R_\star}{R_\odot}\right)^2 \left(\frac{\Omega_\star}{\Omega_\odot}\right)^{1.33} \left(\frac{M_\star}{M_\odot}\right)^{-3.36},
\end{equation}
where $M_\star$, $R_\star$ and $\Omega_\star$ are the stellar mass, radius and rotation rate respectively; it must be noted that (\ref{MLPres}) applies only to stars with rotation rate $\Omega_\star < \Omega_c$, where $\Omega_c = 15 \Omega_\odot \left(M_\star/M_\odot\right)^{2.3}$ with $\Omega_\odot$ denoting the solar rotation rate \citep{JG15}. We note that the validity of (\ref{MLPres}) is questionable for most M-dwarfs including Proxima Centauri and TRAPPIST-1.

Next, we may rewrite the mass of the atmosphere ($M_{atm}$) in terms of the surface pressure ($P_s$) via
\begin{equation}\label{Matm}
    M_{atm}  = \frac{4\pi R_p^2 P_s}{g},
\end{equation}
where $g \approx g_\oplus \left(R_p/R_\oplus\right)^{1.7}$ is the planet's surface gravity \citep{VOCS06,ZSJ16}, provided that the planet has a rocky composition and is larger than the size of the Earth. Here, $g_\oplus$ denotes the Earth's surface gravity. With this data, we are free to compute the characteristic timescale ($t_p$) for the depletion of the planet's atmosphere, which is given by $t_p \sim M_{atm}/\dot{M}_p$. The advantage is that $t_p$ is determined purely in terms of basic physical parameters based on (\ref{MLRate}), (\ref{MLPres}) and (\ref{Matm}). It has been implicitly supposed that the rate of atmospheric escape is much higher than the rate of outgassing from the mantle due to geological activity.

If we estimate $t_p$ for Proxima b using the above formulae in conjunction with $\dot{M}_\star \sim \dot{M}_\odot$ and the \emph{fiducial} (but arbitrary) choice of $P_s = 1$ atm, we find that the timescale is $\mathcal{O}\left(10^8\right)$ yrs. This result is in agreement with detailed magnetohydrodynamic numerical simulations of both magnetized and unmagnetized planets that have yielded values ranging between $\mathcal{O}\left(10^7\right)$ to $\mathcal{O}\left(10^9\right)$ yrs, with many converging on $\mathcal{O}\left(10^8\right)$ yrs \citep{DLMC17,AGK17,GG17}. In contrast, for the TRAPPIST-1 system, it has been found that the innermost planet (with $P_s = 1$ atm assumed throughout) may lose its atmosphere over a timescale of $\mathcal{O}\left(10^8\right)$ yrs, while the duration of atmospheric retention for the outermost planet is $\mathcal{O}\left(10^{10}\right)$ yrs \citep{DJL18}. The timescales for the outer TRAPPIST-1 planets are higher than Proxima b because the star TRAPPIST-1 is smaller, less active, and therefore anticipated to have less intense stellar winds. 

Other notable studies of atmospheric escape driven by the stellar wind include \citet{KJO14} and \citet{CMD}. An important point worth appreciating here is that planets in the HZ of solar-type (G-type) stars are not likely to lose the entirety of their atmospheres over Gyr or sub-Gyr timescales through ion escape processes \citep{SEH01,DH17}. We caution the reader that the above timescales only apply to planets with non-massive atmospheres. If the planets have thick atmospheres - as is possible for the TRAPPIST-1 system \citep{GD18} - the corresponding timescales would be increased due to $t_p \sim M_{atm}/\dot{M}_p$. 

Hitherto, we have concerned ourselves with discussing unmagnetized planets when deriving (\ref{MLRate}). When it comes to magnetized planets, finding an equivalent expression is harder and subject to more uncertainty. It was proposed by \citet{BT18} that the analog of (\ref{MLRate}) for magnetized planets is given by $\dot{M}_p^{(mag)} = \mathcal{Q} \dot{M}_p$, where the extra factor $\mathcal{Q}$ is defined as
\begin{equation}
    \mathcal{Q} \sim 7.1 \left(\frac{\chi}{0.1}\right)\left(\frac{R_{mp}/R_p}{10}\right)^2,
\end{equation}
where $\chi$ is a parameter that is proportional to the ratio of the speed associated with magnetic reconnection to the stellar wind speed near the planet. However, it should be cautioned that this expression has not yet been validated by numerical simulations. Thus, setting $\chi \sim 0.1$, we find $\mathcal{Q} > 1$ is achieved when $R_{mp} \gtrsim 3.75\, R_p$. Therefore, for magnetized planets that satisfy this criterion, their escape rates may exceed those of their unmagnetized counterparts. 

The general conclusion that can be drawn (from the above examples) is that planets in the HZ of very low-mass stars are often, but not always, susceptible to losing their atmospheres over sub-Gyr timescales, even reaching a minimum of $\mathcal{O}\left(10^7\right)$ yrs. On Earth, we know that the timescale taken for the origin of life (abiogenesis) was $\leq 0.8$ Gyr \citep{PT18}, and it required $4.5$ Gyr for the emergence of technological intelligence, namely, \emph{Homo sapiens}. Strictly speaking, we have no knowledge whatsoever of what the timescale for abiogenesis is on other planets based on just a single data point from Earth \citep{ST12}. Despite this important caveat, many studies assume that the corresponding timescales are similar on other worlds.  

If we operate under this assumption, it seems plausible that Earth-sized planets in the HZ of M-dwarfs with surface pressures comparable to the Earth might not have enough time for biological evolution to take place. Of course, on account of the near-linear dependence of the depletion timescale on $P_s$ \citep{DLMC17}, planets with much thicker atmospheres than our own will retain them over commensurately longer periods.  In the event that $t_p$ is much less than $t_{HZ}$, i.e. the temporal duration of the HZ, it is the former that will serve as an upper limit on the time over which biological evolution can occur; the expression for $t_{HZ}$ as a function of $M_\star$ is found in \citet{RC13} and \citet{LL18c}.  This is because of the fact that the presence of an atmosphere is necessary for sustaining liquid water on the surface as seen from its phase diagram.\footnote{In the absence of an atmosphere, surficial life is probably ruled out for the most part, but the existence of subsurface biospheres, which are not tackled herein, is still possible.}

Let us suppose that $t_p$ is $\mathcal{O}\left(10^8\right)$ yrs and that $t_p \ll t_{HZ}$. This gives rise to another potential issue. Since the timescale for biological evolution is lower, the biodiversity may also be reduced accordingly. There are some grounds for supposing that the species richness (total number of species) on Earth has grown exponentially over time \citep{Russ83}.\footnote{A more realistic model for the species richness is based on a series of logistic curves spliced together \citep{PuHe00}.} Using this approach, \citet{LL17a} proposed that planets in the HZ of M-dwarfs could attain a peak species richness that is many orders of magnitude lower than the current-day value of $\sim 10^{12}$ on Earth \citep{LL16}, and that M-dwarfs with $M_\star \lesssim 0.2\,M_\odot$ were especially unlikely to host diverse biospheres. Last, we observe that atmospheric erosion will eventually result in lowered column densities, which is problematic because more cosmic rays would reach the surface resulting in enhanced harmful radiation levels, as noted in Sec. \ref{SSecPMag}.

\section{Stellar electromagnetic radiation}\label{SecEM}
\begin{figure}
\includegraphics[width=7.5cm]{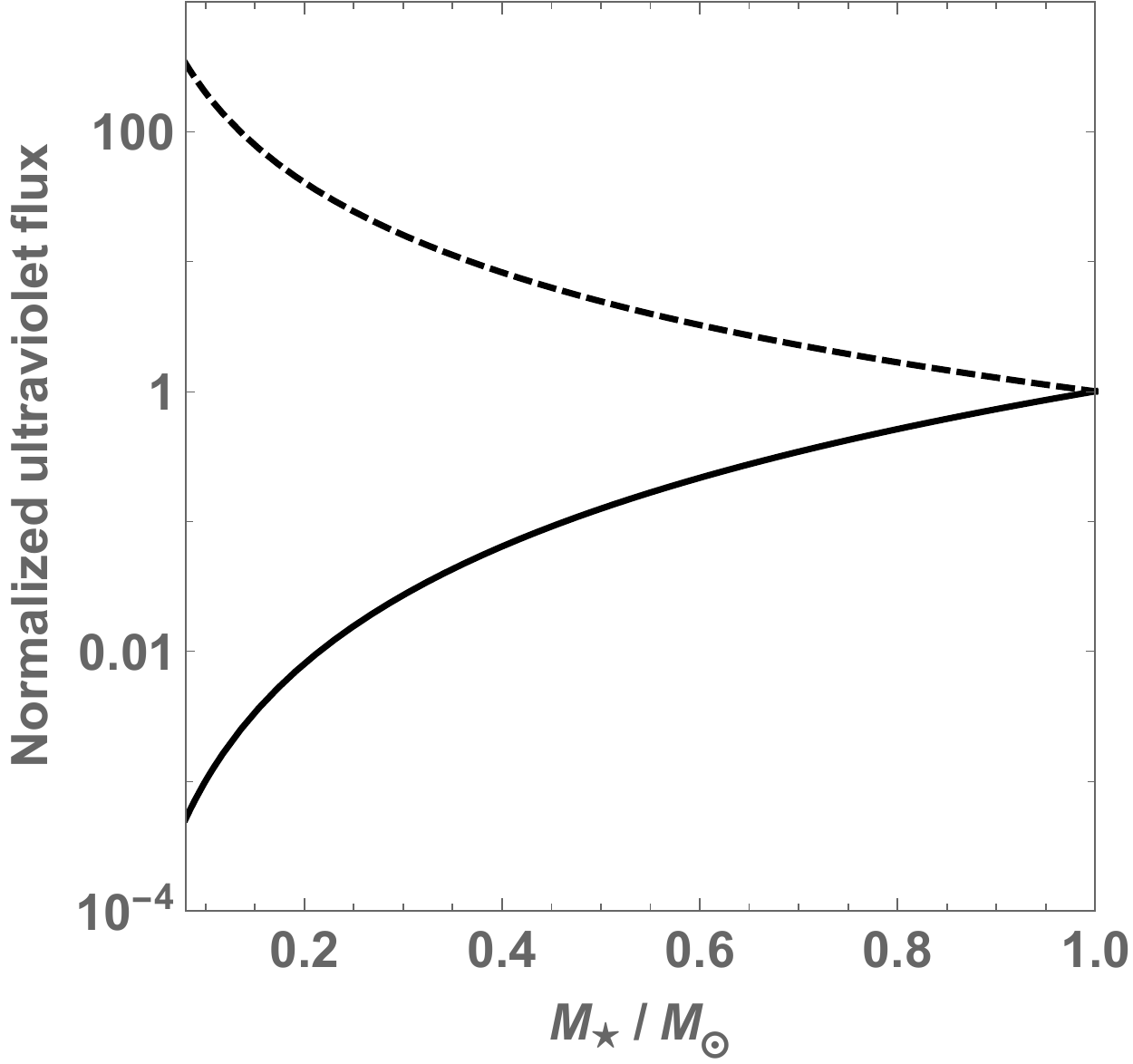} \\
\caption{Schematic representation of the UV fluxes received by Earth-analogs around low-mass stars normalized with respect to the Earth as a function of the stellar mass $M_\star$ (in units of solar mass $M_\odot$). The solid curve denotes the flux of bioactive UV radiation ($\sim 200$-$400$ nm), whereas the dashed curve denotes the flux of Ly$\alpha$ radiation. The former may contribute to the origin of life, driving evolutionary innovations and speciation, DNA damage and inhibiting photosynthesis. The latter is believed to play a key role in the photolysis of molecules such as H$_2$O and CO$_2$, thus enabling water depletion and the build-up of atmospheric O$_2$. Scaling relations have been adopted from \citet{LL18g}.}
\label{FigUVF}
\end{figure}

Of the myriad stellar factors regulating habitability, perhaps the most appreciated amongst them has been the role of electromagnetic radiation emitted by the star. As this represents a vast topic, we will concern ourselves with only a handful of recent developments. An overview of the positive and negative effects on habitability because of electromagnetic radiation has been provided in Table \ref{EMRam}. We have also depicted the two chief UV fluxes of interest in this paper received by Earth-analogs around low-mass stars as a function of $M_\star$ in Figure \ref{FigUVF}. The phrase ``Earth-analogs'' should be used with due caution. We refer here to rocky planets that have the same basic physical parameters as the Earth such as the effective temperature, albedo, atmospheric pressure and radius.

\begin{table*}
\begin{minipage}{155mm}
\caption{Potential positive and negative biological ramifications arising from electromagnetic radiation}
\label{EMRam}
\vspace{0.1 in}
\begin{tabular}{|c|c|c|}
\hline 
Radiation & Consequences & M-dwarf exoplanets\tabularnewline
\hline 
\hline 
XUV & Water worlds to land-water planets via H$_2$O photolysis & High \tabularnewline
\hline 
XUV & Build-up of atmospheric O$_2$ via H$_2$O photolysis for complex life & High\tabularnewline
\hline 
UV-C & Formation of biomolecular building blocks for prebiotic chemistry & Low\tabularnewline
\hline 
UV-Bio & Selection agent for evolutionary innovations and speciation & Low\tabularnewline
\hline 
PAR & Enabling photosynthesis & Low\tabularnewline
\hline 
\hline 
XUV & Complete desiccation of land-water planets via H$_2$O photolysis  & High\tabularnewline
\hline
UV-Bio & DNA damage & Low\tabularnewline
\hline 
UV-Bio & Inhibiting photosynthesis & Low\tabularnewline
\hline 
\end{tabular}
\medskip

{\bf Notes:} XUV ($\sim 0.6$-$120$ nm), UV-C ($\sim 200$-$280$ nm), UV-Bio ($\sim 200$-$400$ nm), PAR ($\sim 400$-$750$ nm). ``High'' and ``Low'' refer to the energy fluxes (in the corresponding wavelength range) received by Earth-analogs around M-dwarfs relative to that of the Earth. The first five rows correspond to the positives, whereas the last three rows represent the negatives. 
\end{minipage}
\end{table*}

\subsection{Evaporation of oceans and buildup of oxygen}\label{SSecEV}
It has been known for a long time that EUV photons are capable of facilitating atmospheric escape \citep{Lam13}. Let us denote the energy flux of EUV photons near the planet by $\mathcal{F}_{EUV}$ and assume that the area over which these photons are intercepted is roughly $\pi R_p^2$. If these photons enable the particles to attain an escape velocity of $v_{esc} = \sqrt{2G M_p/R_p}$, then we have $\frac{1}{2}\dot{M}_p v_{esc}^2 \sim \pi \mathcal{F}_{EUV} R_p^2$. From this relation, we obtain
\begin{equation}\label{EUVEsc}
\dot{M}_p = \frac{\eta_{EUV}}{K_\mathrm{eff}}\frac{\pi R_p^3 \mathcal{F}_{EUV}}{G M_p},  
\end{equation}
where $\eta_{EUV}$ is the heating efficiency factor, while $K_\mathrm{eff}$ is a parameter of order unity that encapsulates the effect of stellar tidal forces \citep{EK07}. In the above expression, we note that the EUV flux can also be replaced by the XUV flux ($0.6$-$120$ nm) that includes the contribution from X-rays \citep{RB16}. This formula has been derived based on the assumption of energy balance, but other regimes are also conceivable \citep{Ow18}; for instance, photon number limitation (as opposed to energy limitation) may regulate the atmospheric escape rate \citep{OA16}.

All stars go through a pre-main-sequence (PMS) phase, during which their luminosity is fuelled mostly by gravitational contraction as opposed to the nuclear fusion of hydrogen \citep{MO07}. The importance of this phase, particularly for low-mass stars, stems from the fact that the luminosity during the PMS phase can be as much as two orders of magnitude higher compared to the point where the star enters the main-sequence \citep{BC02}. Moreover, for very low-mass stars (with $M_\star \lesssim 0.1\,M_\odot$), the PMS phase lasts for a few Gyr, whereas it is $\lesssim 0.1$ Gyr for solar-type stars \citep{RK14}. Thus, the collective effect of the PMS phase is that planets around low-mass stars that would otherwise be situated within the HZ are not located in it as a result of the enhanced luminosity. In other words, these planets exceed the runaway greenhouse threshold, thereby leading to significant losses of water \citep{RK14,LB15,BS17,TGJ18}.

The basic expectation is that the water vapor in the atmosphere injected via the greenhouse effect would undergo photolysis to yield hydrogen and oxygen. As the former has a much lower mass, it is more susceptible to atmospheric escape. This can be verified by converting $\dot{M}_p$ in (\ref{EUVEsc}) from kg/yr to moles/yr, as it yields an inverse dependence on the mass of the chemical species. Hence, the hydrogen would be lost to space, leaving behind massive O$_2$ atmospheres. In reality, the situation is more complex and would depend on factors such as the EUV or XUV flux ($\mathcal{F}_{XUV}$), the planet's mass ($M_p$) and water inventory \citep{Tian}. The rate of atmospheric O$_2$ buildup ($\dot{P}_{O_2}$) was analytically estimated by \citet{LB15}, and is expressible as
\begin{equation}\label{AbO2rat}
\dot{P}_{O_2} \sim 0.138\,\mathrm{bars/Myr}\,\left(\frac{\mathcal{F}_{XUV}}{\mathcal{F}_\oplus}\right)\left(\frac{R}{R_\oplus}\right)^{-1}\left(\frac{\eta_{XUV}}{0.30}\right)
\end{equation}
where $\eta_{XUV}$ denotes the XUV absorption efficiency, $\mathrm{Myr} \equiv 10^6$ yr and $\mathcal{F}_\oplus \approx 4.6 \times 10^{-3}\,\mathrm{J\,m^{-2}\,s^{-1}}$ is the XUV flux incident on Earth. This result can be rewritten in terms of $R_p$ by applying the mass-radius relationship $M_p \propto R_p^{3.7}$ for rocky planets \citep{ZSJ16}. However, when $\mathcal{F}_{XUV}$ exceeds a certain threshold, O$_2$ is ``dragged'' along with hydrogen (due to the emergence of strong hydrodynamic flows) and is therefore subject to escape. The corresponding threshold is given by
\begin{equation}
    \mathcal{F}_c \sim 0.18\,\mathrm{J\,m^{-2}\,s^{-1}}\,\left(\frac{M_p}{M_\oplus}\right)^2\left(\frac{R_p}{R_\oplus}\right)^{-3} \left(\frac{\eta_{XUV}}{0.30}\right)^{-1}.
\end{equation}
Hence, in the regime $\mathcal{F}_{XUV} \geq \mathcal{F}_c$, the rate of atmospheric O$_2$ buildup undergoes saturation and does not depend on $\mathcal{F}_{XUV}$ \citep{LB15}. The corresponding value of $\dot{P}_{O_2}$ for $\mathcal{F}_{XUV} \geq \mathcal{F}_c$ becomes
\begin{equation}\label{AbO2ratsat}
    \dot{P}_{O_2} \sim 5.35\,\mathrm{bar/Myr}\, \left(\frac{M_p}{M_\oplus}\right)^2\left(\frac{R_p}{R_\oplus}\right)^{-4}.
\end{equation}
As noted above, $M_p$ can be eliminated by utilizing the mass-radius relationship for terrestrial planets. In comparison, the rate of O$_2$ contributed to Earth's atmosphere indirectly by photosynthesis (entailing the burial of organic matter) is $\sim 0.06\,\mathrm{bars/Myr}$ \citep{Holl02}. If (\ref{AbO2rat}) or (\ref{AbO2ratsat}) exceeds the atmospheric O$_2$ sinks for Earth-like planets, oxygen can build up in the atmosphere and produce an ozone layer \citep{CK17}.

A number of studies have been conducted that pertain to the production of abiotic O$_2$ and O$_3$ in the atmosphere due to the photolysis of H$_2$O and CO$_2$ by XUV radiation \citep{SM07,WP13,WP14,TF14,HS15,NEM15,GHR15}. Hence, the detection of O$_2$ and O$_3$ at high levels in exoplanetary atmospheres opens up the possibility of a ``false positive'', corresponding to the apparent detection of a biosignature despite it not being produced by biological activity. Several methods have been proposed for distinguishing abiotic O$_2$ from that generated by biological sources \citep{Mea17,MR18}.

As the net effect of UV photolysis of water leads to the depletion of oceans, this has important ramifications for habitability since liquid water is one of the requirements for life. Before discussing this point further, some of the studies pertaining to water loss from Proxima b and the TRAPPIST-1 planets are worth mentioning. In the case of Proxima b, it has been predicted that $< 1 M_{oc,\oplus}$ has been lost over its history \citep{RB16}, where $M_{oc,\oplus} \approx 1.4 \times 10^{21}$ kg is the mass of Earth's oceans today. This could, however, still leave behind an O$_2$ atmosphere of $\sim 100$ bar. In the case of the TRAPPIST-1 planets, the innermost two planets (TRAPPIST-1b and TRAPPIST-1c) may have lost as much as $15 M_{oc,\oplus}$ \citep{BS17} but this value could drop below $1 M_{oc,\oplus}$ for the outer planets \citep{BE17}.

One of the Earth's unique features is that the fraction of the surface covered by land ($0.3$) is comparable to that covered by water ($0.7$). However, in the majority of cases, we can expect one of two possible scenarios to arise, especially around low-mass stars. First, the initial water inventory, which depends on the physical mechanisms involved in the delivery of H$_2$O during terrestrial planet formation \citep{OI18}, may vary widely \citep{RSM07,MC15,RI17}. Hence, there will be many worlds with water inventories much higher than the Earth - some of the planets in the TRAPPIST-1 system are an excellent example, as their water inventory could exceed that of the Earth by $2$ orders of magnitude \citep{GD18,UD18,UHD18,DMG18}. These worlds are expected to comprise only oceans on the surface, thereby lacking landmasses altogether, and both observational \citep{Rog15,WL15,ChKi17,JM18,LHD18} and theoretical \citep{AB17,Sim17} studies indicate they are potentially quite common. The habitability of these ``water worlds'' \citep{Kuc03,LS04} has been the subject of many studies over the past few years \citep{NSR17,KF18,RL18}.

On the other hand, as we have seen previously, XUV radiation during the long PMS phase of low-mass stars is very effective at depleting several oceans worth of water. Moreover, the deep-water cycle transports water from the oceans to the underlying mantle \citep{Hir06}; the water reservoir in the latter may exceed the former by up to an order of magnitude on Earth \citep{Kor08,NZM17}. In many cases, worlds with a moderate initial water inventory could end up being desiccated and eventually transformed into desert planets \citep{AA11,ZS13}. In contrast, for certain water-rich worlds, XUV radiation may actually play a positive role by removing excess water and ensuring that landmasses are exposed on the surface. As per these arguments, we predict the following two features for planets in the HZ of low-mass stars: (a) worlds with Earth-like water inventories are rare, and (b) the water inventory is describable by a bimodal distribution (water- or desert-worlds). Both of these points appear to be consistent with numerical simulations carried out by \citet{TiId}; see also \citet{ZDR18} for a related analysis of solar-type stars.

This aspect poses major issues for habitability for the following reasons. If the worlds are covered almost entirely by land, water availability becomes an issue and the resultant arid environments are expected to have low biomass densities \citep{HS81}. On the other hand, if the surface consists exclusively of oceans, the availability of bioessential elements like phosphorus - widely accepted as the ultimate limiting nutrient for the biological productivity of oceans \citep{Tyrr99} -  becomes problematic because it is primarily delivered via rivers and the atmosphere through continental weathering \citep{PM07}, implying that the absence of continents will lower the influx of dissolved phosphorus by a few orders of magnitude provided that the oceans are alkaline and water-rock interactions exist at the seafloor \citep{LL17c,LL18d}. 

A recent analysis of the surface water fraction ($f_w$) proposed that $0.3 < f_w < 0.9$ is optimal for the buildup of O$_2$ in the atmosphere \citep{LL18e}, which represented a major evolutionary event. As the fraction of Earth-sized worlds in the HZ with water inventories that lie within this range is presumably small ($\lesssim 1\%$) - see \citet{TiId} and \citet{LL18e} - it seems plausible that Earth-like worlds with an admixture of landmasses and oceans are not very common.

\subsection{Origin of life}\label{SSecOOL}
The issue of how and where life originated remains one of science's most profound and enduring mysteries. The reader may consult \citet{McCo13,Lu16,Suth16} for recent overviews of this subject. Currently, life uses deoxyribonucleic acid (DNA) for the storage and transmission of genetic information, while proteins, among several other functions, play the role of catalysts. In the 1980s, it was discovered that ribonucleic acid (RNA) was capable of carrying information as well as facilitating catalysis. This gave impetus to the idea, which was first proposed in the 1960s, that life on Earth began with self-replicating RNA molecules. The hypothesis came to be subsequently known as the RNA world \citep{Gil86}, and has witnessed rapid advancements within the last decade \citep{Joy02,Org04,NKB13,HL15}. 

Before proceeding further, it is crucial to recognize that the RNA world is not the only hypothesis for the origin of life (abiogenesis). A number of models have proposed that self-sustaining metabolic networks arose prior to the evolution of replicators like RNA \citep{Wac07,RBD14,GS18}. The geochemical environment in which life originated has also subject to numerous investigations \citep{SAB13,KM18}. A number of hypotheses have advocated that life arose in land-based environments such as tidal pools \citep{Rom33,Lat04},\footnote{The biological ramifications of tidal modulations arising from the presence of two planetary companions were discussed by \citet{Bal14} and \citet{LL18f}.} intermountain valleys \citep{BKC12}, hot springs and geysers \citep{MB12}. In contrast, one of the most prominent hypotheses suggests that life originated in alkaline hydrothermal vents situated at the ocean floor \citep{MB08,SHW16}.

Many of the surficial geochemical environments customarily feature a wide assortment of minerals, which are believed to have played a crucial role in the emergence and early evolution of life \citep{CSH12,Haz17}. However, it should be recognized that most of the land-based environments in this category are not anticipated to be widely prevalent on water worlds, which were introduced in Sec. \ref{SSecEV}. On the other hand, life's origin in environments at the ocean floor (e.g. hydrothermal vents) remains a distinct possibility unless the water content is high enough to drive the formation of high-pressure ices, thereby cutting off direct water-rock interactions \citep{NH16}.

Broadly speaking, there are three broad classes of macromolecules (apart from carbohydrates) that are necessary for ``life as we know it''. The first is nucleic acids (DNA and RNA), which serve as repositories for genetic information. The second is proteins, as they catalyze biochemical reactions as well as playing a vital role in maintaining cell structure and enabling cell signaling. The third is lipids, which form an integral part of cell membranes and also facilitate energy storage. The building blocks of nucleic acids are nucleotides - for example, RNA is synthesized from the polymerization of ribonucleotides - whereas proteins are polymers of amino acids.

Ever since the 1960s and 1970s, it has been appreciated that UV radiation enabled the synthesis of prebiotic compounds such as amino acids \citep{SK71}. However, it is only within the last decade that the putative significance of UV radiation has been appreciated in greater detail owing to a number of recent developments outlined next \citep{Suth17}.
\begin{itemize}
    \item Studies have demonstrated that UV light accords a selective advantage to RNA-like molecules \citep{GIM10,SSG16}, implying that it may play a crucial role in enabling their polymerization \citep{DCG12}.
    \item Several laboratory experiments display a tendency to form complex organic mixtures (``tar'') that essentially represent a dead end insofar as abiogenesis is concerned. This issue is known as the ``asphalt problem'', but it might be bypassed in certain geochemical environments with UV radiation playing a potentially important role \citep{BKC12}.
    \item The synthesis of biomolecular building blocks is very challenging, especially without regular manual intervention and under conditions that ostensibly resemble that of early Earth. Remarkably, several notable breakthroughs have been achieved recently that rely on UV light. The organic compounds synthesized include: (i) RNA components, e.g. ribonucleotides \citep{PGS09,IP17}, (ii) simple sugar-related molecules \citep{RS12,RS13,TFM18}, (iii) precursors of nucleic acids, amino acids, lipids and carbohydrates from interlinked pathways involving plausible feedstock molecules such as hydrogen cyanide \citep{Pat15,XRR18}, and (iv) iron-sulfur clusters, which play a vital role in metabolism \citep{BVS17}.
    \item RNA nucleotides are stable to irradiation by UV photons, which has been used to argue that they originated in the high-UV environments on early Earth \citep{RT13,Beck16,RS16}. 
\end{itemize}
Thus, taken collectively, there is compelling, but not necessarily definitive, evidence that UV light played an important role in the origin of life on the planetary surface.

This brings us to the question of which planets in the HZ receive higher fluxes of \emph{quiescent} (i.e., background) UV radiation. Several studies indicate planets in the HZ of M-dwarfs receive bioactive UV fluxes (with $200 < \lambda < 400$ nm) at the surface that are around $100$-$1000$ times lower compared to early Earth \citep{RSK15,RWS17}; see also \citet{BLM07}. It must also be appreciated that the UV flux reaching the surface depends on a number of planetary characteristics such as the atmospheric composition and column density. As the bioactive UV flux depends on stellar properties, the region around the host star where the planet receives enough UV photons for prebiotic chemistry (the UV zone) will not always overlap with the classical HZ. 

This issue was investigated in \citet{GZZ10}, where it was concluded that the UV zone lies inward of the HZ when $T_\mathrm{eff} < 4600$ K. A more recent analysis, based on the latest prebiotic pathways for synthesizing RNA, protein and lipid precursors \citep{Pat15,XRR18}, found that stars with $T_\mathrm{eff} < 4400$ K were not likely to drive the formation of these building blocks if the potential role of stellar flares (see Sec. \ref{SecFla}) is neglected \citep{RXT18}. In addition to the stellar temperature, which serves as a rough proxy for the stellar mass, metallicity, i.e., abundance of elements other than H or He, can also influence the location of the UV zone \citep{OK16} and its overlap with the HZ.

While these studies deal with the spatial overlap between the UV zone and the HZ, it must be appreciated that the rates of these prebiotic reactions may also depend on the bioactive UV fluxes \citep{RWS17}. Hence, planets in the HZ of M-dwarfs could require much longer timescales for abiogenesis. This fact was utilized by \citet{LL18g} to suggest that the timescale for abiogenesis ($t_A$) is
\begin{equation}
    t_A \sim t_{A,\oplus}\left(\frac{M_\star}{M_\odot}\right)^{-\kappa},
\end{equation}
where $\kappa \approx 3$ for $M_\star \lesssim M_\odot$, $\kappa \approx 1$ for $M_\star \gtrsim M_\odot$ and $t_{A,\oplus}$ is the timescale for the origin of life on Earth, which has a strict upper bound of $0.8$ Gyr. Coupled to the fact that Earth-analogs around M-dwarfs are expected to lose their atmospheres more rapidly, it was proposed that stars with $M_\star \lesssim 0.4 M_\odot$ are relatively unlikely to host biospheres \citep{LL18c}.

\subsection{Evolution of complex life}
We will briefly discuss how stellar radiation, particularly in the UV and visible ranges, influences the trajectories of biological evolution. 

\subsubsection{Biological damage due to UV radiation}\label{SSSecUVD}
It is well-known that UV radiation, particularly in the range $180 < \lambda < 300$ nm, is capable of suppressing photosynthesis and causing damage to DNA and other biomolecules \citep{VGCD,Sag73,TS94,CSD05}. Yet, on the other hand, it must be recognized that UV radiation might potentially function as a selection agent, a driver of evolutionary innovations and speciation \citep{Sag73,Roth99,EG05}.

As this topic has been the subject of many investigations, we will concern ourselves with describing only a couple of recent examples. Before embarking on that discussion, we note that a number of environments effectively shield organisms from the damaging effects of UV radiation such as a layer of soil, ocean \citep{CM98}, hazes in the atmosphere \citep{ADG16}, and other screening compounds \citep{CK99}. Another possibility is that complex evolutionary adaptations, such as those evinced by a wide array of microbes \citep{GG11,GS13,PBO17,JLB17}, could protect organisms from high doses of UV radiation.

The biologically effective irradiance (BEI) was computed for Earth-analogs around different stars in \citet{RSK15}. The BEI can be envisioned as the product of the surface UV flux and the DNA action spectrum, where the latter quantifies the extent of DNA damage at different wavelengths. Two broad conclusions can be immediately drawn. First, the BEI for very low-mass stars (with $M_\star \lesssim 0.1 M_\odot$) is always about $\gtrsim 100$ times smaller than the Earth at the \emph{same} epoch. Second, the presence of an ozone layer, not surprisingly, plays a major role in determining the BEI. For example, it was estimated that the BEI at $3.9$ Gyr ago on Earth was about $600$ times the present-day value. 

A similar study was undertaken by \citet{OMJ17}, who concluded that two other parameters which influence the BEI are the stellar activity and the atmospheric pressure along with the presence (or absence) of an ozone layer. It was found that planets around more active stars, with partially eroded atmospheres, are likely to receive high UV doses at the surface. The issue of rarefied atmospheres is particularly relevant, since we have seen earlier that stellar winds and UV radiation can drive atmospheric escape and erosion. However, even in the worst-case scenario, it was found that a combination of evolutionary adaptations and ecological niches are theoretically capable of sustaining life \citep{OMJK19}. 

\subsubsection{Photosynthesis}\label{SSSecPhot}
On Earth, most life is dependent, either directly or indirectly, upon photosynthesis. The radiation typically employed by photoautotrophs (photosynthetic organisms) ranges between wavelengths of $400$ and $700$ nm \citep{HB11}, and goes by the name of photosynthetically active radiation (PAR). If we approximate the stellar spectrum as black body radiation, Wien's displacement law implies
\begin{equation}
    \lambda_{max} T_\mathrm{eff} = 2.9 \times 10^{-3}\,\mathrm{m \cdot K},
\end{equation}
where $\lambda_{max}$ is the wavelength at which the black body spectrum peaks. Choosing $T_\mathrm{eff} \approx 5778$ K for the Sun yields $\lambda_{max} \approx 500$ nm, which sits almost in the middle of PAR range. In qualitative terms, this makes sense because photoautotrophs may have evolved to optimize the number of photons that they collect for synthesizing organic compounds \citep{KS07a}. 

Estimating the PAR range for other stars, especially M-dwarfs, is very difficult since it depends on the spectral properties of the host star, the atmospheric composition and the pigments that enable photosynthesis \citep{KS07a,BSZ14}. One of the most striking features of vegetation on Earth is the ``red edge'', i.e. an increase in the reflectance by roughly an order of magnitude at $\sim 700$ nm. The importance of the red edge stems from the fact that it represents a viable biosignature that could be detectable with future telescopes \citep{STS05,Kal17}. Owing to the aforementioned issues, it is not easy to predict where the analog of red edge would exist for M-dwarfs. It does, however, seem plausible that the edge would be shifted to near-infrared wavelengths around $1.1$ $\mu$m \citep{KS07b}; see, however, \citet{TMT17}. Other methods for detecting signatures of photosynthesis include distinctive features in circularly and linearly polarized spectra \citep{SHG09,BKH16} and angle-dependent reflectivity \citep{DW10}.

If we restrict our attention to the same PAR as on Earth, we are confronted with a potential problem. As low-mass stars tend to radiate primarily in the infrared, the availability of photons in the PAR range will be reduced. This raises the question of whether photosynthesis on planets around M-dwarfs is feasible. As most of the photosynthesis on Earth leads to the production of O$_2$ \citep{FHJ16}, many studies have concentrated on this particular process. The simplified reaction presented below illustrates the formation of organic matter and oxygen via oxygenic photosynthesis.
\begin{equation}\label{OP}
 \mathrm{CO_2} + 2 \mathrm{H_2O} + h\nu \rightarrow \mathrm{CH_2O} + \mathrm{H_2O} + \mathrm{O_2}.   
\end{equation}
It was argued by \citet{GW17} that oxygenic photosynthesis in the PAR is feasible on planets around low-mass stars, primarily because the side facing the star (assuming synchronous rotation) will receive continuous illumination that compensates for the moderate photon flux. A detailed analysis of the prospects for photosynthesis on Proxima b was undertaken by \citet{RLR18} - the rate of synthesis of organic compounds via photosynthesis was estimated to be $\sim 17\%$ that of the Earth despite the photon flux in the PAR range being only $3\%$ of the Earth. However, it was also argued therein that oxygenic photosynthesis might not evolve on planets around M-dwarfs as this process accords minimal competitive advantage. \citet{LCP18} investigated whether enough photons would be available to support an Earth-like biosphere on M-dwarf exoplanets, and found that many of them are incapable of doing so; see also \citet{LL19}. In particular, if the maximum wavelength of PAR is specified to be $750$ nm, none of the TRAPPIST-1 planets in the HZ appear to have the capacity for sustaining Earth-like biospheres. Relatively sparse biospheres due to the limited PAR fluxes may also have the additional disadvantage of producing weak O$_2$ signals that are not easily detectable. 

\subsubsection{Oxygen and complex life}
There are sufficient grounds to conclude that the buildup of O$_2$ in the atmosphere may have constituted a rate-limiting step insofar the development of complex life is concerned on other worlds \citep{Kno85,OMP16,Jud17}. Yet, it should also be recognized that similar environmental conditions and evolutionary trajectories for complex life need not prevail on all worlds. The development of the ozone layer and the expansion of aerobic metabolism, which releases about an order of magnitude more energy compared to its anaerobic counterparts \citep{KB08}, are two examples of the profound changes triggered by the rise of O$_2$. The oxygen levels in Earth's atmosphere as a function of geological time were subject to fluctuations, but there is compelling evidence that the O$_2$ levels increased from a very small value to roughly $1\%$ of the present atmospheric level (PAL) around $\sim 2.4$ Ga \citep{GCBS17,KN17}. Many different mechanisms have been advanced to explain this rise in oxygen, known as the Great Oxygenation Event (GOE), such as changes in volcanism and the oxidation state of continents \citep{Kast13,LRP14}. 

One of the chief hypotheses for the GOE that we focus on concerns the production of O$_2$ in the atmosphere through UV photolysis \citep{CZM01,CaKa17}. Organic matter is produced from oxygenic photosynthesis via (\ref{OP}), and this is subsequently decomposed by other microbes to release methane (CH$_4$). Methane undergoes UV photolysis with H escaping to space and the carbon combining with O$_2$ to form CO$_2$. The net reaction is expressible as follows:
\begin{equation}\label{BioUV}
    2 \mathrm{H_2O} + ``\mathrm{biosphere}" + h\nu \rightarrow \mathrm{O}_2  + 4 \mathrm{H}\, (\uparrow \mathrm{space}).
\end{equation}
The key point is that this reaction does not represent the abiotic photolysis of water since reactions mediated by the biosphere are necessary. In addition, as we have seen in Sec. \ref{SSecEV}, there are several abiotic pathways by which O$_2$ is produced through UV photolysis on planets, particularly those around low-mass stars. These mechanisms could enable the build-up of O$_2$ at rates much faster than those driven by biological factors.

It is therefore conceivable that the time required for O$_2$ levels to reach a certain value (e.g. $1\%$ PAL) via (\ref{BioUV}) is much shorter for planets that receive high UV fluxes at the appropriate wavelengths. An important feature of M-dwarfs is that their ratio of far-UV ($117$-$175$ nm) to near-UV ($175$-$320$ nm) fluxes are $\sim 1000$ times higher than the corresponding ratio for the Sun \citep{TF14}, because of enhanced UV emission from the chromosphere and transition regions \citep{LFA13}. The Ly$\alpha$ flux ($\Phi_{L}$), which is responsible for a large fraction of water and methane photolysis, for Earth-analogs can be approximated by a power law of the form $\Phi_L \propto M_\star^{\zeta}$, where $\zeta \approx -2.3$ for $M_\star \lesssim M_\odot$ and $\zeta \approx 3.3$ for $M_\star \gtrsim M_\odot$ \citep{LL18g}. Note that the flux for stars more massive than the Sun is higher as a larger fraction of the black body radiation is emitted in the UV regime.

Hence, if the oxygenation time ($t_{O_2}$) scales inversely with $\Phi_L$, the evolution of complex aerobic life from microbial anaerobic organisms may require a shorter amount of time on Earth-analogs orbiting stars that have a higher mass than the Sun \citep{Liv99}, although these stars are less abundant with respect to solar-mass stars. 

\section{Stellar flares}\label{SecFla}
Stellar flares are explosive phenomena on the stellar surface that lead to the release of energy in various forms such as electromagnetic radiation, plasma and energetic particles \citep{Benz17}. The central engine behind this energy release is believed to be magnetic reconnection \citep{PF02}, which entails changes in magnetic topology through the breaking and reconnection of field lines, thus resulting in the rapid conversion of magnetic energy into other forms of energy \citep{Bis00}. Classical magnetic reconnection models, which were originally developed in the 1950s, result in energy release over long timescales that are not supported by observations, but subsequent developments in this area have achieved much progress in terms of addressing this issue \citep{ShiMa11,Pri14,CLHB16}.

The largest flares documented in modern history have energies of $\sim 10^{25}$ J, with the largest one on record being the Carrington event from 1859 \citep{Carr59} that released a total energy of approximately $5 \times 10^{25}$ J \citep{CD13}. However, it is important to appreciate that flares with energies $\geq 10^{26}$ J, known as superflares, are also feasible on theoretical and observational grounds. The \emph{Kepler} mission has recorded a wealth of observational data regarding the statistics of superflares \citep{Mae12,Shi13,Dav16,NSW17,NMH19}. For solar-type stars with similar rotation rates as the Sun, it has been found that 
\begin{equation} \label{Freq}
    \frac{d N}{d E} \propto E^{-\alpha},
\end{equation}
with $\alpha \approx 1.5$-$2$, where $N(E)$ represents the occurrence rate of superflares and $E$ denotes the energy of these superflares \citep{Mae12,Mae15,GZS19}. The power-law exponent is close to that documented for normal flares on the Sun \citep{Han11}. The maximum amount of energy that may be released, in theory, during a flare event is expressible as \citep{KS13}:
\begin{equation} \label{FlareEn}
E \sim 10^{28}\,\mathrm{J}\, \left(\frac{\epsilon}{0.1}\right) \left(\frac{B_A}{0.1\,\mathrm{T}}\right)^2 \left(\frac{f_A}{0.3}\right)^{3/2}\left(\frac{R_\star}{R_\odot}\right)^3.
\end{equation}
Note that $R_\star$ and $R_\odot$ are the stellar and solar radii respectively, whereas $\epsilon$ represents the ``efficiency'' of converting magnetic energy into flare energy. In this formula, $B_A$ denotes the magnetic field strength in the active region(s), and $f_A$ is the fraction of the star's surface that is covered by the active region(s). The active regions constitute the sites of solar flares and other stellar activity, and sunspots serve as indicators of active regions. 

In the case of the Sun, it has been theorized that flares of $\sim 10^{27}$ J can occur over a timescale of $\sim 2000$ yrs \citep{Shi13}, but this is hard to verify on account of relatively sparse direct evidence. The studies by \citet{MNMN} and \citet{MMN13}, which discovered rapid variations in the concentrations of radionuclides in tree rings might be indicative of superflare activity. The extreme upper limit on the maximum flare energy released from Sun-like stars appears to be $\sim 10^{30}$ J \citep{Sch12,LL17b}, although it remains highly uncertain as to whether such superflares could actually arise in reality.

In our subsequent discussion, we will mostly concentrate on planets in the HZ of M-dwarfs as these stars are known to be very active and typically, but not always, produce flares and superflares at rates higher than G-type stars like the Sun \citep{SK07,Mae12}. Both TRAPPIST-1 and Proxima Centauri have been documented to produce flares regularly - the number of flares with energies $\gtrsim 10^{26}$ J is predicted to be $\sim 10^{-2}$ per day based on flare statistics \citep{DKS16,VK17,HT18}; see also \citet{MW18}. Recall that these planets are situated much closer to their host stars, may have weak magnetic fields and could be subjected to rapid atmospheric erosion from stellar winds. 

\subsection{Electromagnetic radiation}\label{SSecFER}
It has long been appreciated that flares lead to enhanced UV fluxes at the surfaces of planets within the HZ of M-dwarfs \citep{HDJ99,SS04}. In the majority of instances, this has been viewed as a negative strike against the habitability of M-dwarf exoplanets, despite the fact that many UV shielding mechanisms are feasible, as described in Sec. \ref{SSSecUVD}. Nonetheless, we shall highlight a couple of recent studies that illustrate the detrimental effects before discussing the potential positives later. 

In an important study, \citet{SWM10} simulated the fluxes of UV radiation that were delivered to the surface of an Earth-analog at a distance of $0.16$ AU orbiting the star AD Leonis ($M_\star \approx 0.4 M_\odot$ and age $< 0.3$ Gyr) during a flare with a total energy of $\sim 10^{27}$ J. It was found that the amount of UV flux increased by a factor of around $50$ at the peak of the flare compared to the quiescent phase prior to the onset of the flare. Although this was seemingly a significant increase, most of the UV enhancement was in the UV-A ($315$-$400$ nm) range, which has a much weaker effect on damaging DNA (by a factor of $\sim 100$) compared to shorter UV wavelengths. However, an important aspect worth appreciating is that this result presumed the existence of ozone. Taking ozone depletion into account (see Sec. \ref{SSecSEP}), the same flare was estimated to produce higher UV-A and UV-B ($280$-$315$ nm) fluxes at the peak of the flare.

\citet{EV18} analyzed how a hypothetical Earth-analog at $1$ AU around the solar-analog Kepler-96 (with $M_\star \approx M_\odot$) would respond to a superflare with energy $E \sim 1.8 \times 10^{28}$ J. This led to an enhancement in the UV flux by nearly two orders of magnitude, and the BEI (see Sec. \ref{SSSecUVD}) was found to exceed even that of radiation-resistant extremophile \emph{Deinococcus radiodurans} in the absence of ozone. However, the presence of an ozone layer or an ocean depth of $\sim 12$ m would suffice to protect microbes akin to \emph{D. radiodurans} from the UV radiation emitted during superflares. 

Moving on to the positive effects, we have noted in Sec. \ref{SSecOOL} that planets around M-dwarfs may be characterized by a paucity of bioactive UV radiation. However, it is possible, in principle, for prebiotic reactions to take place during the flaring phase because of the enhanced UV fluxes, while remaining inactive during the quiescent phase \citep{BLM07,RWS17}. This hypothesis requires further experimental tests before it can be confirmed or invalidated. Based on the available flare statistics, \citet{RXT18} proposed that $\sim 20\%$ of M-dwarfs are sufficiently active so as to provide sufficient UV fluxes for synthesizing the precursors of biomolecules. 

In the same spirit, we saw in Sec. \ref{SSSecPhot} that Earth-like biospheres based on photosynthesis may not be sustainable on low-mass stars. However, an important point worth recognizing is that flares can also deliver photons in the PAR range. Hence, when the effects of stellar flares are included, theoretical calculations seem to indicate that the PAR flux may be raised by approximately one order of magnitude \citep{MB18} (see, however, \citealt{LL19}), although the averaged photosynthetic efficiency of planets in the HZ of M-dwarfs is still anticipated to be lower than that of the Earth \citep{Sch19}. Finally, the high fluxes of UV radiation incident on the surface during flaring events are expected to result in intermittently enhanced mutation rates, thereby serving as agents of ecological and evolutionary change \citep{SSW04}.

\subsection{Coronal mass ejections}
Coronal mass ejections (CMEs) are large loads of plasma and magnetic fields expelled from the host star. They are generally, but not always, linked with flares \citep{WH12}. The majority of CMEs carry a mass of $\lesssim 10^{13}$ kg and move at velocities of order $100$-$1000$ km/s. One of the important reasons for studying CMEs is that they can facilitate the acceleration of energetic particles through shock waves \citep{KKP17}, which is addressed in Sec. \ref{SSecSEP}. We will, instead, focus on their effects on planetary magnetospheres and atmospheric erosion.

To begin with, it is instructive to consider the parameters of a sizable CME. A CME with parameters commensurate with the famous Carrington event is predicted to have a density that is $\sim 50$ times higher than the current solar wind density and a velocity that is $\sim 4$ times greater \citep{ng14}. Thus, based on the preceding discussion in Sec. \ref{SSecPMag}, we find that the corresponding ram pressure would be $\sim 800$ times higher than that of the present-day solar wind. Substituting this result into (\ref{Rmpd}), we find that the magnetopause distance for an Earth-analog would be compressed to roughly one-third of its steady-state value for the current solar wind. 

From (\ref{MaLo}), we see that the mass-loss rate should be enhanced by a factor of $\sim 200$, and utilizing (\ref{MLRate}) reveals that the atmospheric escape rate should increase by a factor of $\sim 200$ with respect to the steady-state as well. This prediction is in good agreement with numerical simulations that yield an enhancement of $\sim 110$ \citep{DH17}. Observational evidence from the ongoing Mars Atmosphere and Volatile Evolution (MAVEN) mission has also revealed that CMEs (smaller than Carrington-type events) can result in the enhancement of Martian atmospheric ion escape rates by roughly an order of magnitude \citep{dong15a,jak15b}.

Although the above discussion pertains to our Solar system, we observe that these two implications also apply to extrasolar systems. For instance, owing to a combination of the weak planetary magnetic fields, close-in distances and frequency of CME impacts \citep{KOK16}, the effects of CMEs are expected to significantly compress the magnetospheres of M-dwarf exoplanets \citep{KRL07} and enhance the escape rates leading to cumulative atmospheric losses of $\sim 10$-$100$ bars \citep{LLK07}. At the same time, we wish to underscore the fact that the mass-loss rates from active stars, especially M-dwarfs, due to CMEs is not tightly constrained. This is because of the fact that most studies rely on extrapolations based on putative correlations between the X-ray fluences of flares and the kinetic energies and masses of CMEs, although the extent to which such extrapolations are valid remains unknown \citep{OLH17}. The presence of large-scale magnetic fields may, for instance, impose upper limits on the kinetic energy of CMEs \citep{AGD18}.

Bearing these caveats in mind, we note that the mass-loss rates due to CMEs from magnetically active stars may be $1$-$3$ orders of magnitude higher than the current mass carried away by the steady-state solar wind today \citep{DCY13,Cran17}. Hence, from (\ref{MLRate}), it also follows that the atmospheric escape rates will be enhanced by the same degree for active (usually young) stars. In other words, the timescale for the total depletion of the atmosphere computed in Sec. \ref{SSecAE} could represent an upper bound. The magnetospheres will also be compressed by a factor of a few owing to the enhanced value of the ram pressure in (\ref{Rmpd}).

Thus, when viewed cumulatively, it seems plausible that CMEs exacerbate the issues discussed in the context of stellar winds in Sec. \ref{SecSW}.

\subsection{Stellar proton events}\label{SSecSEP}
The majority of large stellar flares are accompanied by the production of stellar/solar energetic particles (SEPs), and these phenomena are known as stellar/solar proton events (SPEs). The mechanisms behind their production are complex, but can be divided into two broad categories: ``impulsive'' events involving magnetic reconnection and ``gradual'' events involving fast shock waves propelled by CMEs \citep{Rea13}. Of the two, it is the latter that produce higher fluences of SEPs \citep{DG16}, and are therefore expected to be more prominent. The kinetic energies of SEPs (mostly protons or electrons) are usually in the keV and MeV ranges, but the maximum values can reach several GeV \citep{Mew06}. As direct measurements of extrasolar SEP fluences are not currently feasible, most models rely on extrapolations from empirical scalings \citep{TMS16,YFL17}. It must, however, be appreciated that these predictions may break down at high SEP fluences \citep{Hud15,Uso17}.

It has been duly recognized, since nearly half a century ago, that SEPs lead to the formation of hydrogen and nitrogen oxides \citep{Cru79,LPF05}. The latter, in particular, have been shown to facilitate the depletion of ozone \citep{Sol99} through catalytic reactions, of which one of them is
\begin{equation}
    \mathrm{NO} + \mathrm{O}_3 \rightarrow \mathrm{NO}_2 + \mathrm{O}_2,
\end{equation}
where NO and NO$_2$ are nitric oxide and nitrogen dioxide, respectively \citep{CIR75}. Hence, high-fluence SPEs can lead to the destruction of the ozone layer \citep{TSM18}, which in turn results in high doses of UV radiation reaching the surface. In principle, this might be offset by the presence of high concentrations of N$_2$O (which absorb UV radiation) \citep{RKZ13,RKS15}. As noted previously, high-UV radiation fluxes do not preclude the existence of life under a layer of water or in subsurface environments, but they may impose constraints on surface habitability. 

\citet{SWM10} analyzed how a putative SPE associated with the AD Leonis flare, discussed earlier in Sec. \ref{SSecFER}, affected the ozone layer. It was found that the ozone depletion reached a maximum of $94\%$ provided that the Earth-analog was unmagnetized. Very recently, a superflare with energy $\sim 10^{26.5}$ J was detected from Proxima Centauri \citep{HT18}. In the presence of an ozone layer, the effects on habitability were not significant. However, when the effects of SPEs were taken into account, it was found that $90\%$ of the ozone layer could be lost within five years. In this scenario, the amount of UV-C radiation ($< 280$ nm) that reached the surface was almost two orders of magnitude higher than the critical flux that would kill $90\%$ of a given population of the radiation-resistant microbe \emph{D. radiodurans}. 

A similar study was undertaken to assess the impact of repeated SPEs on a hypothetical Earth-analog around the star GJ1243, and it was found that ozone depletion of $94\%$ could occur in as little as $10$ yrs \citep{TSM18}. The net result of all these studies appears to be that complete ozone depletion might take place within $\mathcal{O}\left(10^5\right)$ yrs, although it must be recognized that this timescale depends on the flaring frequency, planetary magnetic field and atmospheric composition; see also \citet{TVG16} for a discussion of how SPEs impact M-dwarf exoplanets. Consequently, it would result in a quiescent UV-C flux that may lie above the tolerance threshold of \emph{D. radiodurans}; see, however, \citet{OMJK19}. This issue will probably be exacerbated by subsequent flares that increase the UV-C flux by $2$-$3$ orders of magnitude. 

By examining the available empirical evidence for the correlations between flare energy, the fluence of SEPs, and the degree of ozone depletion, \citet{LL17b} presented a phenomenological relation between the ozone depletion ($\mathcal{D}_{O_3}$) and the flare energy:
\begin{equation} \label{OzEn}
   \mathcal{D}_{O_3} \sim  2.8\%\,\left(\frac{E}{10^{25}\,\mathrm{J}}\right)^{9/25} \left(\frac{a_p}{1\,\mathrm{AU}}\right)^{-18/25}.
\end{equation}
Hence, it can be verified that a flare with $E \sim 2 \times 10^{29}$ J would lead to complete ozone depletion ($\mathcal{D}_{O_3} \approx 100\%$) for a terrestrial planet at $1$ AU, whereas the corresponding value at $0.1$ AU is $E \sim 2 \times 10^{27}$ J. If we make use of the relations $a_p \propto L_\star^{1/2} \propto M_\star^{7/4}$ \citep{BV92} for Earth-analogs, we obtain $\mathcal{D}_{O_3} \propto E^{0.36} M_\star^{-1.26}$. 

Setting aside the issue of ozone depletion, another problematic issue arising from SEPs is that their impact with atmospheres induces the formation of secondary particle cascades that increase the radiation dose received at the surface \citep{MT11,AT14}. This topic was recently studied by \citet{Atri} for a wide range of planetary parameters. In general, it was found that the radiation dose at the surface increased when: (a) the atmospheric column density was lowered, (b) the flare energy was increased, (c) the planetary magnetic moment was reduced, and (d) the star-planet distance was lowered. In the case of exoplanets around M-dwarfs, it is conceivable that one or more of (a)-(d) are applicable. While this does not rule out the prospects for life, the critical radiation doses for macroscopic multicellular organisms on Earth (e.g. mammals) are exceeded under certain circumstances. 

However, there are potential benefits associated with SPEs as well. To begin with, the abiotic synthesis of the building blocks of life requires suitable energy sources and pathways \citep{MU59,CS92,DeWe10}. One of the most important feedstock molecules for prebiotic synthesis is hydrogen cyanide \citep{Suth16}. Detailed numerical simulations carried out by \citet{Aira16} demonstrate that hydrogen cyanide (HCN) can be synthesized at concentrations of tens of parts per million by volume in the lower atmosphere during SPEs. The production of the aforementioned nitrogen oxides via SEPs can also be associated with a positive component: these compounds are attractive electron acceptors that may facilitate the emergence of metabolic pathways and life at alkaline hydrothermal vents \citep{WCG17} or shallow ponds and lakes \citep{RTR19}.

Looking beyond feedstock molecules like HCN and formaldehyde (CH$_2$O), laboratory experiments have demonstrated that a wide array of biomolecular building blocks such as amino acids and nucleobases, which are essential components of nucleotides (i.e., monomers of nucleic acids), are synthesized with relatively high efficiency when gasesous mixtures are irradiated by high-energy protons \citep{KKT98,Miy02,Dart11}. The averaged energy flux ($\Phi_\mathrm{SEP}$) delivered to the surface of a planet with a $1$-bar atmosphere via SEPs is \citep{LDF18}:
\begin{equation} \label{PhiMD}
    \Phi_\mathrm{SEP} \sim 50\,\mathrm{J\,m^{-2}\,yr^{-1}}\,\left(\frac{\dot{N}}{1\,\mathrm{day}^{-1}}\right) \left(\frac{a_p}{1\,\mathrm{AU}}\right)^{-2},
\end{equation}
where $\dot{N}$ denotes the number of ``large'' SPEs expected to impact the planet per day, which could reach a maximum of order unity for young Sun-like stars as well as M-dwarfs \citep{KOK16}. It is evident from (\ref{PhiMD}) that M-dwarf exoplanets will receive higher SEP fluxes due to the smaller values of $a_p$ \citep{FDA19}. In turn, this could result in the enhanced synthesis of prebiotic compounds \citep{NOSD}, thereby partially offsetting the deficiency of UV light delineated in Sec. \ref{SSecOOL}. By drawing upon laboratory experiments, which are admittedly idealized, \citet{LDF18} suggested that the production rates of organics were given by
\begin{equation} \label{AmAcR}
    \dot{\mathcal{M}}_A \sim 10^7\,\mathrm{kg}/\mathrm{yr}\,\left(\frac{\Phi_\mathrm{SEP}}{100\,\mathrm{J\,m^{-2}\,yr^{-1}}}\right) \left(\frac{R_p}{R_\oplus}\right)^2,
\end{equation}
\begin{equation} \label{NucR}
    \dot{\mathcal{M}}_N \sim 10^4\,\mathrm{kg}/\mathrm{yr}\,\left(\frac{\Phi_\mathrm{SEP}}{100\,\mathrm{J\,m^{-2}\,yr^{-1}}}\right) \left(\frac{R_p}{R_\oplus}\right)^2,
\end{equation}
where $\dot{\mathcal{M}}_A$ and $\dot{\mathcal{M}}_N$ denoted the rate of production of amino acids and nucleobases, respectively. On early Earth, it was found that $\dot{\mathcal{M}}_A$ was comparable to the production rate of amino acids from electrical discharges (lightning), and $3$-$4$ orders of magnitude higher than the delivery rate of these compounds through meteorites. 

From a geological standpoint, the primordial Earth possibly required a high concentration of greenhouse gases to prevent its oceans from freezing due to the lower luminosity of the Sun ($70\%$ of the present-day value) during this epoch \citep{SM72}. This has led to many proposals to explain how temperatures above freezing were achieved \citep{Feu12}. Nitrous oxide (N$_2$O) is efficiently formed during SPEs \citep{Aira16},\footnote{Moreover, it may also build up on planets with low-UV fluxes as the latter would otherwise destroy N$_2$O \citep{RKZ13}.} and has a greenhouse potential that is $\sim 300$ times higher than CO$_2$ over a $100$-yr period \citep{VML17}. The latter feature lends credence to the idea that it may have contributed to the greenhouse warming of early Earth \citep{CGF10} and possibly other habitable planets. Signatures of N$_2$O and biosignature gases like O$_2$ could be ``highlighted'' during episodes of magnetic activity, thereby making them more detectable \citep{AJM17}.

\section{Discussion}\label{SecConc}
M-dwarfs like Proxima Centauri and TRAPPIST-1 are $\lesssim 10$ times more abundant than the Sun \citep{cha03,RLG08} and have stellar lifetimes that are $\sim 100$-$1000$ times greater \citep{AL97,TB07,LBS16}. Furthermore, exoplanets around these stars are easier to detect and their atmospheres can be analyzed via transit spectroscopy \citep{Wi10,FA18}, thus enabling the ready detection of biomarkers. Hence, in light of these facts, we are confronted with the following question: how is exoplanetary habitability influenced by the properties of the host star? One possible answer is that physical mechanisms could act in concert to suppress the likelihood of life's emergence on M-dwarf exoplanets relative to their counterparts around solar-type stars.

In this Colloquium, we have encountered a number of potential reasons as to why the habitability of M-dwarf exoplanets might be reduced. These include atmospheric erosion due to intense stellar winds, coronal mass ejections and UV radiation, paucity of photons for the origin of life and photosynthesis, and an inhospitable surface environment due to radiation from stellar flares and solar proton events. Yet, it is equally vital to appreciate that numerous nonlinear feedback mechanisms are at play, and that none of the aforementioned factors rule out the prospects for life altogether.

For instance, exoplanets around M-dwarfs could have started out with massive hydrogen-helium atmospheres and lost them via atmospheric erosion involving UV radiation, stellar winds or active phases of supermassive black holes to yield potentially habitable worlds \citep{LBL15,CFL18}. It is also possible that planets started with high water inventories that were depleted through UV photolysis and hydrogen escape to yield worlds with a mixture of land and oceans on the surface. In addition, planets may have formed outside the HZ and escaped the brunt of the long and intense pre-main-sequence phase of M-dwarfs, before eventually migrating inwards into the HZ at a later stage; see \citet{TRP17} and \citet{OLS17}. Finally, M-dwarf planetary systems might possess inherent advantages such as the enhanced transport of life between planets via lithopanspermia by several orders of magnitude compared to the Earth-Mars system \citep{SL16,LL17d,KBL17}. However, in each of the above instances, either a high degree of fine tuning might be required or the feasibility of the proposed mechanisms remains indeterminate. 

The question posed earlier is important from a practical standpoint because of the fact that resources such as the observation time, technological capabilities of current facilities and funding are all limited. Hence, the selection of the most optimal target stars and planets when searching for biosignatures will be of the utmost importance in the future, when we are confronted with a diverse array of planetary systems around different stars \citep{KE10,HJ10,LL18a}. In this respect, the importance of theoretical modeling as a tool for identifying which targets merit the highest consideration becomes manifest. Moreover, theoretical models can also help us figure out what kind of hypothetical biospheres may exist and what to potentially expect when searching for signatures of life.

The ultimate answer to this question will, of course, necessitate detailed observations that are free of preconceived biases. With the upcoming launch of the James Webb Space Telescope, the ongoing development of ground-based extremely large telescopes, and mission concepts for future space telescopes, the characterization of terrestrial planets and searching for biosignatures is expected to become feasible in the upcoming decades \citep{RLM14,SDB15,BI16,MKR17,KDG18,FA18,Mad19}. These developments might not only help up answer the age-old question of whether we are alone but also assess the rarity or commonality of life in the Universe.

\acknowledgments
We thank the reviewers for their perspicacious and detailed reports which were very valuable for improving the paper. This work was supported in part by the Breakthrough Prize Foundation, Harvard University's Faculty of Arts and Sciences, and the Institute for Theory and Computation (ITC) at Harvard University.


\begin{thebibliography}{395}%
\makeatletter
\providecommand \@ifxundefined [1]{%
 \@ifx{#1\undefined}
}%
\providecommand \@ifnum [1]{%
 \ifnum #1\expandafter \@firstoftwo
 \else \expandafter \@secondoftwo
 \fi
}%
\providecommand \@ifx [1]{%
 \ifx #1\expandafter \@firstoftwo
 \else \expandafter \@secondoftwo
 \fi
}%
\providecommand \natexlab [1]{#1}%
\providecommand \enquote  [1]{``#1''}%
\providecommand \bibnamefont  [1]{#1}%
\providecommand \bibfnamefont [1]{#1}%
\providecommand \citenamefont [1]{#1}%
\providecommand \href@noop [0]{\@secondoftwo}%
\providecommand \href [0]{\begingroup \@sanitize@url \@href}%
\providecommand \@href[1]{\@@startlink{#1}\@@href}%
\providecommand \@@href[1]{\endgroup#1\@@endlink}%
\providecommand \@sanitize@url [0]{\catcode `\\12\catcode `\$12\catcode
  `\&12\catcode `\#12\catcode `\^12\catcode `\_12\catcode `\%12\relax}%
\providecommand \@@startlink[1]{}%
\providecommand \@@endlink[0]{}%
\providecommand \url  [0]{\begingroup\@sanitize@url \@url }%
\providecommand \@url [1]{\endgroup\@href {#1}{\urlprefix }}%
\providecommand \urlprefix  [0]{URL }%
\providecommand \Eprint [0]{\href }%
\providecommand \doibase [0]{http://dx.doi.org/}%
\providecommand \selectlanguage [0]{\@gobble}%
\providecommand \bibinfo  [0]{\@secondoftwo}%
\providecommand \bibfield  [0]{\@secondoftwo}%
\providecommand \translation [1]{[#1]}%
\providecommand \BibitemOpen [0]{}%
\providecommand \bibitemStop [0]{}%
\providecommand \bibitemNoStop [0]{.\EOS\space}%
\providecommand \EOS [0]{\spacefactor3000\relax}%
\providecommand \BibitemShut  [1]{\csname bibitem#1\endcsname}%
\let\auto@bib@innerbib\@empty
\bibitem [{\citenamefont {{Bailey}}(1957)}]{Bai57}%
  \BibitemOpen
  \bibfield  {author} {\bibinfo {author} {\bibfnamefont {C.}~\bibnamefont
  {{Bailey}}},\ }in\ \href@noop {} {\emph {\bibinfo {booktitle} {The Stoic and
  Epicurean Philosophers}}},\ \bibinfo {editor} {edited by\ \bibinfo {editor}
  {\bibfnamefont {W.~J.}\ \bibnamefont {{Oates}}}}\ (\bibinfo  {publisher} {The
  Modern Library},\ \bibinfo {year} {1957})\ pp.\ \bibinfo {pages}
  {3--68}\BibitemShut {NoStop}%
\bibitem [{\citenamefont {{Mayor}}\ and\ \citenamefont
  {{Queloz}}(1995)}]{MQ95}%
  \BibitemOpen
  \bibfield  {author} {\bibinfo {author} {\bibfnamefont {M.}~\bibnamefont
  {{Mayor}}}\ and\ \bibinfo {author} {\bibfnamefont {D.}~\bibnamefont
  {{Queloz}}},\ }\href {\doibase 10.1038/378355a0} {\bibfield  {journal}
  {\bibinfo  {journal} {Nature}\ }\textbf {\bibinfo {volume} {378}},\ \bibinfo
  {pages} {355} (\bibinfo {year} {1995})}\BibitemShut {NoStop}%
\bibitem [{\citenamefont {{Perryman}}(2000)}]{Per00}%
  \BibitemOpen
  \bibfield  {author} {\bibinfo {author} {\bibfnamefont {M.~A.~C.}\
  \bibnamefont {{Perryman}}},\ }\href {\doibase 10.1088/0034-4885/63/8/202}
  {\bibfield  {journal} {\bibinfo  {journal} {Rep. Prog. Phys.}\ }\textbf
  {\bibinfo {volume} {63}},\ \bibinfo {pages} {1209} (\bibinfo {year}
  {2000})}\BibitemShut {NoStop}%
\bibitem [{\citenamefont {{Batalha}}(2014)}]{Bat14}%
  \BibitemOpen
  \bibfield  {author} {\bibinfo {author} {\bibfnamefont {N.~M.}\ \bibnamefont
  {{Batalha}}},\ }\href {\doibase 10.1073/pnas.1304196111} {\bibfield
  {journal} {\bibinfo  {journal} {Proc. Natl. Acad. Sci. USA}\ }\textbf
  {\bibinfo {volume} {111}},\ \bibinfo {pages} {12647} (\bibinfo {year}
  {2014})}\BibitemShut {NoStop}%
\bibitem [{\citenamefont {{Borucki}}(2016)}]{Bor16}%
  \BibitemOpen
  \bibfield  {author} {\bibinfo {author} {\bibfnamefont {W.~J.}\ \bibnamefont
  {{Borucki}}},\ }\href {\doibase 10.1088/0034-4885/79/3/036901} {\bibfield
  {journal} {\bibinfo  {journal} {Rep. Prog. Phys.}\ }\textbf {\bibinfo
  {volume} {79}},\ \bibinfo {eid} {036901} (\bibinfo {year}
  {2016})}\BibitemShut {NoStop}%
\bibitem [{\citenamefont {{Winn}}\ and\ \citenamefont
  {{Fabrycky}}(2015)}]{WF15}%
  \BibitemOpen
  \bibfield  {author} {\bibinfo {author} {\bibfnamefont {J.~N.}\ \bibnamefont
  {{Winn}}}\ and\ \bibinfo {author} {\bibfnamefont {D.~C.}\ \bibnamefont
  {{Fabrycky}}},\ }\href {\doibase 10.1146/annurev-astro-082214-122246}
  {\bibfield  {journal} {\bibinfo  {journal} {Annu. Rev. Astron. Astrophys.}\
  }\textbf {\bibinfo {volume} {53}},\ \bibinfo {pages} {409} (\bibinfo {year}
  {2015})}\BibitemShut {NoStop}%
\bibitem [{\citenamefont {{Perryman}}(2018)}]{Perr18}%
  \BibitemOpen
  \bibfield  {author} {\bibinfo {author} {\bibfnamefont {M.}~\bibnamefont
  {{Perryman}}},\ }\href@noop {} {\emph {\bibinfo {title} {{The Exoplanet
  Handbook}}}},\ \bibinfo {edition} {2nd}\ ed.\ (\bibinfo  {publisher}
  {Cambridge University Press},\ \bibinfo {year} {2018})\BibitemShut {NoStop}%
\bibitem [{\citenamefont {{Heller}}\ \emph {et~al.}(2014)\citenamefont
  {{Heller}}, \citenamefont {{Williams}}, \citenamefont {{Kipping}},
  \citenamefont {{Limbach}}, \citenamefont {{Turner}}, \citenamefont
  {{Greenberg}}, \citenamefont {{Sasaki}}, \citenamefont {{Bolmont}},
  \citenamefont {{Grasset}}, \citenamefont {{Lewis}}, \citenamefont
  {{Barnes}},\ and\ \citenamefont {{Zuluaga}}}]{HW14}%
  \BibitemOpen
  \bibfield  {author} {\bibinfo {author} {\bibfnamefont {R.}~\bibnamefont
  {{Heller}}}, \bibinfo {author} {\bibfnamefont {D.}~\bibnamefont
  {{Williams}}}, \bibinfo {author} {\bibfnamefont {D.}~\bibnamefont
  {{Kipping}}}, \bibinfo {author} {\bibfnamefont {M.~A.}\ \bibnamefont
  {{Limbach}}}, \bibinfo {author} {\bibfnamefont {E.}~\bibnamefont {{Turner}}},
  \bibinfo {author} {\bibfnamefont {R.}~\bibnamefont {{Greenberg}}}, \bibinfo
  {author} {\bibfnamefont {T.}~\bibnamefont {{Sasaki}}}, \bibinfo {author}
  {\bibfnamefont {{\'E}.}~\bibnamefont {{Bolmont}}}, \bibinfo {author}
  {\bibfnamefont {O.}~\bibnamefont {{Grasset}}}, \bibinfo {author}
  {\bibfnamefont {K.}~\bibnamefont {{Lewis}}}, \bibinfo {author} {\bibfnamefont
  {R.}~\bibnamefont {{Barnes}}}, \ and\ \bibinfo {author} {\bibfnamefont
  {J.~I.}\ \bibnamefont {{Zuluaga}}},\ }\href {\doibase 10.1089/ast.2014.1147}
  {\bibfield  {journal} {\bibinfo  {journal} {Astrobiology}\ }\textbf {\bibinfo
  {volume} {14}},\ \bibinfo {pages} {798} (\bibinfo {year} {2014})}\BibitemShut
  {NoStop}%
\bibitem [{\citenamefont {{Teachey}}\ and\ \citenamefont
  {{Kipping}}(2018)}]{KT18}%
  \BibitemOpen
  \bibfield  {author} {\bibinfo {author} {\bibfnamefont {A.}~\bibnamefont
  {{Teachey}}}\ and\ \bibinfo {author} {\bibfnamefont {D.~M.}\ \bibnamefont
  {{Kipping}}},\ }\href {\doibase 10.1126/sciadv.aav1784} {\bibfield  {journal}
  {\bibinfo  {journal} {Sci. Adv.}\ }\textbf {\bibinfo {volume} {4}},\ \bibinfo
  {pages} {eaav1784} (\bibinfo {year} {2018})}\BibitemShut {NoStop}%
\bibitem [{\citenamefont {{Ball}}(2008)}]{Ba08}%
  \BibitemOpen
  \bibfield  {author} {\bibinfo {author} {\bibfnamefont {P.}~\bibnamefont
  {{Ball}}},\ }\href {\doibase 10.1021/cr068037a} {\bibfield  {journal}
  {\bibinfo  {journal} {Chem. Rev.}\ }\textbf {\bibinfo {volume} {108}},\
  \bibinfo {pages} {74} (\bibinfo {year} {2008})}\BibitemShut {NoStop}%
\bibitem [{\citenamefont {{Benner}}\ \emph {et~al.}(2004)\citenamefont
  {{Benner}}, \citenamefont {{Ricardo}},\ and\ \citenamefont
  {{Carrigan}}}]{Ben04}%
  \BibitemOpen
  \bibfield  {author} {\bibinfo {author} {\bibfnamefont {S.~A.}\ \bibnamefont
  {{Benner}}}, \bibinfo {author} {\bibfnamefont {A.}~\bibnamefont {{Ricardo}}},
  \ and\ \bibinfo {author} {\bibfnamefont {M.~A.}\ \bibnamefont {{Carrigan}}},\
  }\href {\doibase 10.1016/j.cbpa.2004.10.003} {\bibfield  {journal} {\bibinfo
  {journal} {Curr. Opin. Chem. Biol.}\ }\textbf {\bibinfo {volume} {8}},\
  \bibinfo {pages} {672} (\bibinfo {year} {2004})}\BibitemShut {NoStop}%
\bibitem [{\citenamefont {{Bains}}(2004)}]{Ba04}%
  \BibitemOpen
  \bibfield  {author} {\bibinfo {author} {\bibfnamefont {W.}~\bibnamefont
  {{Bains}}},\ }\href {\doibase 10.1089/153110704323175124} {\bibfield
  {journal} {\bibinfo  {journal} {Astrobiology}\ }\textbf {\bibinfo {volume}
  {4}},\ \bibinfo {pages} {137} (\bibinfo {year} {2004})}\BibitemShut {NoStop}%
\bibitem [{\citenamefont {{Schulze-Makuch}}\ and\ \citenamefont
  {{Irwin}}(2008)}]{SI08}%
  \BibitemOpen
  \bibfield  {author} {\bibinfo {author} {\bibfnamefont {D.}~\bibnamefont
  {{Schulze-Makuch}}}\ and\ \bibinfo {author} {\bibfnamefont {L.~N.}\
  \bibnamefont {{Irwin}}},\ }\href {\doibase 10.1007/978-3-540-76817-3} {\emph
  {\bibinfo {title} {{Life in the Universe: Expectations and Constraints}}}},\
  \bibinfo {edition} {2nd}\ ed.\ (\bibinfo  {publisher} {Springer-Verlag},\
  \bibinfo {year} {2008})\BibitemShut {NoStop}%
\bibitem [{\citenamefont {{Hayes}}(2016)}]{Hay16}%
  \BibitemOpen
  \bibfield  {author} {\bibinfo {author} {\bibfnamefont {A.~G.}\ \bibnamefont
  {{Hayes}}},\ }\href {\doibase 10.1146/annurev-earth-060115-012247} {\bibfield
   {journal} {\bibinfo  {journal} {Annu. Rev. Earth Planet. Sci.}\ }\textbf
  {\bibinfo {volume} {44}},\ \bibinfo {pages} {57} (\bibinfo {year}
  {2016})}\BibitemShut {NoStop}%
\bibitem [{\citenamefont {{Mitchell}}\ and\ \citenamefont
  {{Lora}}(2016)}]{ML16}%
  \BibitemOpen
  \bibfield  {author} {\bibinfo {author} {\bibfnamefont {J.~L.}\ \bibnamefont
  {{Mitchell}}}\ and\ \bibinfo {author} {\bibfnamefont {J.~M.}\ \bibnamefont
  {{Lora}}},\ }\href {\doibase 10.1146/annurev-earth-060115-012428} {\bibfield
  {journal} {\bibinfo  {journal} {Annu. Rev. Earth Planet. Sci.}\ }\textbf
  {\bibinfo {volume} {44}},\ \bibinfo {pages} {353} (\bibinfo {year}
  {2016})}\BibitemShut {NoStop}%
\bibitem [{\citenamefont {{Jakosky}}\ \emph {et~al.}(2017)\citenamefont
  {{Jakosky}}, \citenamefont {{Slipski}}, \citenamefont {{Benna}},
  \citenamefont {{Mahaffy}}, \citenamefont {{Elrod}}, \citenamefont {{Yelle}},
  \citenamefont {{Stone}},\ and\ \citenamefont {{Alsaeed}}}]{Jak17}%
  \BibitemOpen
  \bibfield  {author} {\bibinfo {author} {\bibfnamefont {B.~M.}\ \bibnamefont
  {{Jakosky}}}, \bibinfo {author} {\bibfnamefont {M.}~\bibnamefont
  {{Slipski}}}, \bibinfo {author} {\bibfnamefont {M.}~\bibnamefont {{Benna}}},
  \bibinfo {author} {\bibfnamefont {P.}~\bibnamefont {{Mahaffy}}}, \bibinfo
  {author} {\bibfnamefont {M.}~\bibnamefont {{Elrod}}}, \bibinfo {author}
  {\bibfnamefont {R.}~\bibnamefont {{Yelle}}}, \bibinfo {author} {\bibfnamefont
  {S.}~\bibnamefont {{Stone}}}, \ and\ \bibinfo {author} {\bibfnamefont
  {N.}~\bibnamefont {{Alsaeed}}},\ }\href {\doibase 10.1126/science.aai7721}
  {\bibfield  {journal} {\bibinfo  {journal} {Science}\ }\textbf {\bibinfo
  {volume} {355}},\ \bibinfo {pages} {1408} (\bibinfo {year}
  {2017})}\BibitemShut {NoStop}%
\bibitem [{\citenamefont {{Dong}}\ \emph
  {et~al.}(2018{\natexlab{a}})\citenamefont {{Dong}}, \citenamefont {{Lee}},
  \citenamefont {{Ma}}, \citenamefont {{Lingam}}, \citenamefont {{Bougher}},
  \citenamefont {{Luhmann}}, \citenamefont {{Curry}}, \citenamefont {{Toth}},
  \citenamefont {{Nagy}}, \citenamefont {{Tenishev}}, \citenamefont {{Fang}},
  \citenamefont {{Mitchell}}, \citenamefont {{Brain}},\ and\ \citenamefont
  {{Jakosky}}}]{DLM18}%
  \BibitemOpen
  \bibfield  {author} {\bibinfo {author} {\bibfnamefont {C.}~\bibnamefont
  {{Dong}}}, \bibinfo {author} {\bibfnamefont {Y.}~\bibnamefont {{Lee}}},
  \bibinfo {author} {\bibfnamefont {Y.}~\bibnamefont {{Ma}}}, \bibinfo {author}
  {\bibfnamefont {M.}~\bibnamefont {{Lingam}}}, \bibinfo {author}
  {\bibfnamefont {S.}~\bibnamefont {{Bougher}}}, \bibinfo {author}
  {\bibfnamefont {J.}~\bibnamefont {{Luhmann}}}, \bibinfo {author}
  {\bibfnamefont {S.}~\bibnamefont {{Curry}}}, \bibinfo {author} {\bibfnamefont
  {G.}~\bibnamefont {{Toth}}}, \bibinfo {author} {\bibfnamefont
  {A.}~\bibnamefont {{Nagy}}}, \bibinfo {author} {\bibfnamefont
  {V.}~\bibnamefont {{Tenishev}}}, \bibinfo {author} {\bibfnamefont
  {X.}~\bibnamefont {{Fang}}}, \bibinfo {author} {\bibfnamefont
  {D.}~\bibnamefont {{Mitchell}}}, \bibinfo {author} {\bibfnamefont
  {D.}~\bibnamefont {{Brain}}}, \ and\ \bibinfo {author} {\bibfnamefont
  {B.}~\bibnamefont {{Jakosky}}},\ }\href {\doibase 10.3847/2041-8213/aac489}
  {\bibfield  {journal} {\bibinfo  {journal} {Astrophys. J. Lett.}\ }\textbf
  {\bibinfo {volume} {859}},\ \bibinfo {eid} {L14} (\bibinfo {year}
  {2018}{\natexlab{a}})}\BibitemShut {NoStop}%
\bibitem [{\citenamefont {{Gonzalez}}(2005)}]{Gonz05}%
  \BibitemOpen
  \bibfield  {author} {\bibinfo {author} {\bibfnamefont {G.}~\bibnamefont
  {{Gonzalez}}},\ }\href {\doibase 10.1007/s11084-005-5010-8} {\bibfield
  {journal} {\bibinfo  {journal} {Orig. Life Evol. Biosph.}\ }\textbf {\bibinfo
  {volume} {35}},\ \bibinfo {pages} {555} (\bibinfo {year} {2005})}\BibitemShut
  {NoStop}%
\bibitem [{\citenamefont {{Kasting}}\ \emph {et~al.}(1993)\citenamefont
  {{Kasting}}, \citenamefont {{Whitmire}},\ and\ \citenamefont
  {{Reynolds}}}]{KWR93}%
  \BibitemOpen
  \bibfield  {author} {\bibinfo {author} {\bibfnamefont {J.~F.}\ \bibnamefont
  {{Kasting}}}, \bibinfo {author} {\bibfnamefont {D.~P.}\ \bibnamefont
  {{Whitmire}}}, \ and\ \bibinfo {author} {\bibfnamefont {R.~T.}\ \bibnamefont
  {{Reynolds}}},\ }\href {\doibase 10.1006/icar.1993.1010} {\bibfield
  {journal} {\bibinfo  {journal} {Icarus}\ }\textbf {\bibinfo {volume} {101}},\
  \bibinfo {pages} {108} (\bibinfo {year} {1993})}\BibitemShut {NoStop}%
\bibitem [{\citenamefont {{Kopparapu}}\ \emph {et~al.}(2013)\citenamefont
  {{Kopparapu}}, \citenamefont {{Ramirez}}, \citenamefont {{Kasting}},
  \citenamefont {{Eymet}}, \citenamefont {{Robinson}}, \citenamefont
  {{Mahadevan}}, \citenamefont {{Terrien}}, \citenamefont {{Domagal-Goldman}},
  \citenamefont {{Meadows}},\ and\ \citenamefont {{Deshpande}}}]{KRK13}%
  \BibitemOpen
  \bibfield  {author} {\bibinfo {author} {\bibfnamefont {R.~K.}\ \bibnamefont
  {{Kopparapu}}}, \bibinfo {author} {\bibfnamefont {R.}~\bibnamefont
  {{Ramirez}}}, \bibinfo {author} {\bibfnamefont {J.~F.}\ \bibnamefont
  {{Kasting}}}, \bibinfo {author} {\bibfnamefont {V.}~\bibnamefont {{Eymet}}},
  \bibinfo {author} {\bibfnamefont {T.~D.}\ \bibnamefont {{Robinson}}},
  \bibinfo {author} {\bibfnamefont {S.}~\bibnamefont {{Mahadevan}}}, \bibinfo
  {author} {\bibfnamefont {R.~C.}\ \bibnamefont {{Terrien}}}, \bibinfo {author}
  {\bibfnamefont {S.}~\bibnamefont {{Domagal-Goldman}}}, \bibinfo {author}
  {\bibfnamefont {V.}~\bibnamefont {{Meadows}}}, \ and\ \bibinfo {author}
  {\bibfnamefont {R.}~\bibnamefont {{Deshpande}}},\ }\href {\doibase
  10.1088/0004-637X/765/2/131} {\bibfield  {journal} {\bibinfo  {journal}
  {Astrophys. J.}\ }\textbf {\bibinfo {volume} {765}},\ \bibinfo {eid} {131}
  (\bibinfo {year} {2013})}\BibitemShut {NoStop}%
\bibitem [{\citenamefont {{Kopparapu}}\ \emph {et~al.}(2014)\citenamefont
  {{Kopparapu}}, \citenamefont {{Ramirez}}, \citenamefont {{SchottelKotte}},
  \citenamefont {{Kasting}}, \citenamefont {{Domagal-Goldman}},\ and\
  \citenamefont {{Eymet}}}]{KRS14}%
  \BibitemOpen
  \bibfield  {author} {\bibinfo {author} {\bibfnamefont {R.~K.}\ \bibnamefont
  {{Kopparapu}}}, \bibinfo {author} {\bibfnamefont {R.~M.}\ \bibnamefont
  {{Ramirez}}}, \bibinfo {author} {\bibfnamefont {J.}~\bibnamefont
  {{SchottelKotte}}}, \bibinfo {author} {\bibfnamefont {J.~F.}\ \bibnamefont
  {{Kasting}}}, \bibinfo {author} {\bibfnamefont {S.}~\bibnamefont
  {{Domagal-Goldman}}}, \ and\ \bibinfo {author} {\bibfnamefont
  {V.}~\bibnamefont {{Eymet}}},\ }\href {\doibase 10.1088/2041-8205/787/2/L29}
  {\bibfield  {journal} {\bibinfo  {journal} {Astrophys. J. Lett.}\ }\textbf
  {\bibinfo {volume} {787}},\ \bibinfo {eid} {L29} (\bibinfo {year}
  {2014})}\BibitemShut {NoStop}%
\bibitem [{\citenamefont {{Yang}}\ \emph {et~al.}(2014)\citenamefont {{Yang}},
  \citenamefont {{Bou{\'e}}}, \citenamefont {{Fabrycky}},\ and\ \citenamefont
  {{Abbot}}}]{YBF14}%
  \BibitemOpen
  \bibfield  {author} {\bibinfo {author} {\bibfnamefont {J.}~\bibnamefont
  {{Yang}}}, \bibinfo {author} {\bibfnamefont {G.}~\bibnamefont {{Bou{\'e}}}},
  \bibinfo {author} {\bibfnamefont {D.~C.}\ \bibnamefont {{Fabrycky}}}, \ and\
  \bibinfo {author} {\bibfnamefont {D.~S.}\ \bibnamefont {{Abbot}}},\ }\href
  {\doibase 10.1088/2041-8205/787/1/L2} {\bibfield  {journal} {\bibinfo
  {journal} {Astrophys. J. Lett.}\ }\textbf {\bibinfo {volume} {787}},\
  \bibinfo {eid} {L2} (\bibinfo {year} {2014})}\BibitemShut {NoStop}%
\bibitem [{\citenamefont {{Zsom}}(2015)}]{Zs15}%
  \BibitemOpen
  \bibfield  {author} {\bibinfo {author} {\bibfnamefont {A.}~\bibnamefont
  {{Zsom}}},\ }\href {\doibase 10.1088/0004-637X/813/1/9} {\bibfield  {journal}
  {\bibinfo  {journal} {Astrophys. J.}\ }\textbf {\bibinfo {volume} {813}},\
  \bibinfo {eid} {9} (\bibinfo {year} {2015})}\BibitemShut {NoStop}%
\bibitem [{\citenamefont {{Kopparapu}}\ \emph {et~al.}(2016)\citenamefont
  {{Kopparapu}}, \citenamefont {{Wolf}}, \citenamefont {{Haqq-Misra}},
  \citenamefont {{Yang}}, \citenamefont {{Kasting}}, \citenamefont {{Meadows}},
  \citenamefont {{Terrien}},\ and\ \citenamefont {{Mahadevan}}}]{KWH16}%
  \BibitemOpen
  \bibfield  {author} {\bibinfo {author} {\bibfnamefont {R.~k.}\ \bibnamefont
  {{Kopparapu}}}, \bibinfo {author} {\bibfnamefont {E.~T.}\ \bibnamefont
  {{Wolf}}}, \bibinfo {author} {\bibfnamefont {J.}~\bibnamefont
  {{Haqq-Misra}}}, \bibinfo {author} {\bibfnamefont {J.}~\bibnamefont
  {{Yang}}}, \bibinfo {author} {\bibfnamefont {J.~F.}\ \bibnamefont
  {{Kasting}}}, \bibinfo {author} {\bibfnamefont {V.}~\bibnamefont
  {{Meadows}}}, \bibinfo {author} {\bibfnamefont {R.}~\bibnamefont
  {{Terrien}}}, \ and\ \bibinfo {author} {\bibfnamefont {S.}~\bibnamefont
  {{Mahadevan}}},\ }\href {\doibase 10.3847/0004-637X/819/1/84} {\bibfield
  {journal} {\bibinfo  {journal} {Astrophys. J.}\ }\textbf {\bibinfo {volume}
  {819}},\ \bibinfo {eid} {84} (\bibinfo {year} {2016})}\BibitemShut {NoStop}%
\bibitem [{\citenamefont {{Haqq-Misra}}\ \emph {et~al.}(2016)\citenamefont
  {{Haqq-Misra}}, \citenamefont {{Kopparapu}}, \citenamefont {{Batalha}},
  \citenamefont {{Harman}},\ and\ \citenamefont {{Kasting}}}]{HKB16}%
  \BibitemOpen
  \bibfield  {author} {\bibinfo {author} {\bibfnamefont {J.}~\bibnamefont
  {{Haqq-Misra}}}, \bibinfo {author} {\bibfnamefont {R.~K.}\ \bibnamefont
  {{Kopparapu}}}, \bibinfo {author} {\bibfnamefont {N.~E.}\ \bibnamefont
  {{Batalha}}}, \bibinfo {author} {\bibfnamefont {C.~E.}\ \bibnamefont
  {{Harman}}}, \ and\ \bibinfo {author} {\bibfnamefont {J.~F.}\ \bibnamefont
  {{Kasting}}},\ }\href {\doibase 10.3847/0004-637X/827/2/120} {\bibfield
  {journal} {\bibinfo  {journal} {Astrophys. J.}\ }\textbf {\bibinfo {volume}
  {827}},\ \bibinfo {eid} {120} (\bibinfo {year} {2016})}\BibitemShut {NoStop}%
\bibitem [{\citenamefont {{Ramirez}}(2018)}]{Ram18}%
  \BibitemOpen
  \bibfield  {author} {\bibinfo {author} {\bibfnamefont {R.~M.}\ \bibnamefont
  {{Ramirez}}},\ }\href {\doibase 10.3390/geosciences8080280} {\bibfield
  {journal} {\bibinfo  {journal} {Geosciences}\ }\textbf {\bibinfo {volume}
  {8}},\ \bibinfo {eid} {280} (\bibinfo {year} {2018})}\BibitemShut {NoStop}%
\bibitem [{\citenamefont {{Schwieterman}}\ \emph {et~al.}(2018)\citenamefont
  {{Schwieterman}}, \citenamefont {{Reinhard}}, \citenamefont {{Olson}},
  \citenamefont {{Harman}},\ and\ \citenamefont {{Lyons}}}]{SRO19}%
  \BibitemOpen
  \bibfield  {author} {\bibinfo {author} {\bibfnamefont {E.~W.}\ \bibnamefont
  {{Schwieterman}}}, \bibinfo {author} {\bibfnamefont {C.~T.}\ \bibnamefont
  {{Reinhard}}}, \bibinfo {author} {\bibfnamefont {S.~L.}\ \bibnamefont
  {{Olson}}}, \bibinfo {author} {\bibfnamefont {C.~E.}\ \bibnamefont
  {{Harman}}}, \ and\ \bibinfo {author} {\bibfnamefont {T.~W.}\ \bibnamefont
  {{Lyons}}},\ }\href@noop {} {\bibfield  {journal} {\bibinfo  {journal}
  {Astrophys. J.}\ }\textbf {\bibinfo {volume} {878}},\ \bibinfo {eid} {19}
  (\bibinfo {year} {2018})}\BibitemShut {NoStop}%
\bibitem [{\citenamefont {{Dobos}}\ \emph {et~al.}(2017)\citenamefont
  {{Dobos}}, \citenamefont {{Heller}},\ and\ \citenamefont {{Turner}}}]{DHT17}%
  \BibitemOpen
  \bibfield  {author} {\bibinfo {author} {\bibfnamefont {V.}~\bibnamefont
  {{Dobos}}}, \bibinfo {author} {\bibfnamefont {R.}~\bibnamefont {{Heller}}}, \
  and\ \bibinfo {author} {\bibfnamefont {E.~L.}\ \bibnamefont {{Turner}}},\
  }\href {\doibase 10.1051/0004-6361/201730541} {\bibfield  {journal} {\bibinfo
   {journal} {Astron. Astrophys.}\ }\textbf {\bibinfo {volume} {601}},\
  \bibinfo {eid} {A91} (\bibinfo {year} {2017})}\BibitemShut {NoStop}%
\bibitem [{\citenamefont {{Kane}}\ and\ \citenamefont {{Hinkel}}(2013)}]{KH13}%
  \BibitemOpen
  \bibfield  {author} {\bibinfo {author} {\bibfnamefont {S.~R.}\ \bibnamefont
  {{Kane}}}\ and\ \bibinfo {author} {\bibfnamefont {N.~R.}\ \bibnamefont
  {{Hinkel}}},\ }\href {\doibase 10.1088/0004-637X/762/1/7} {\bibfield
  {journal} {\bibinfo  {journal} {Astrophys. J.}\ }\textbf {\bibinfo {volume}
  {762}},\ \bibinfo {eid} {7} (\bibinfo {year} {2013})}\BibitemShut {NoStop}%
\bibitem [{\citenamefont {{Cuntz}}(2014)}]{Cun14}%
  \BibitemOpen
  \bibfield  {author} {\bibinfo {author} {\bibfnamefont {M.}~\bibnamefont
  {{Cuntz}}},\ }\href {\doibase 10.1088/0004-637X/780/1/14} {\bibfield
  {journal} {\bibinfo  {journal} {Astrophys. J.}\ }\textbf {\bibinfo {volume}
  {780}},\ \bibinfo {eid} {14} (\bibinfo {year} {2014})}\BibitemShut {NoStop}%
\bibitem [{\citenamefont {{Hoehler}}\ \emph {et~al.}(2007)\citenamefont
  {{Hoehler}}, \citenamefont {{Amend}},\ and\ \citenamefont {{Shock}}}]{HAS07}%
  \BibitemOpen
  \bibfield  {author} {\bibinfo {author} {\bibfnamefont {T.~M.}\ \bibnamefont
  {{Hoehler}}}, \bibinfo {author} {\bibfnamefont {J.~P.}\ \bibnamefont
  {{Amend}}}, \ and\ \bibinfo {author} {\bibfnamefont {E.~L.}\ \bibnamefont
  {{Shock}}},\ }\href {\doibase 10.1089/ast.2007.0207} {\bibfield  {journal}
  {\bibinfo  {journal} {Astrobiology}\ }\textbf {\bibinfo {volume} {7}},\
  \bibinfo {pages} {819} (\bibinfo {year} {2007})}\BibitemShut {NoStop}%
\bibitem [{\citenamefont {{Morowitz}}\ and\ \citenamefont
  {{Smith}}(2007)}]{MS07}%
  \BibitemOpen
  \bibfield  {author} {\bibinfo {author} {\bibfnamefont {H.}~\bibnamefont
  {{Morowitz}}}\ and\ \bibinfo {author} {\bibfnamefont {E.}~\bibnamefont
  {{Smith}}},\ }\href {\doibase 10.1002/cplx.20191} {\bibfield  {journal}
  {\bibinfo  {journal} {Complexity}\ }\textbf {\bibinfo {volume} {13}},\
  \bibinfo {pages} {51} (\bibinfo {year} {2007})}\BibitemShut {NoStop}%
\bibitem [{\citenamefont {{Branscomb}}\ \emph {et~al.}(2017)\citenamefont
  {{Branscomb}}, \citenamefont {{Biancalani}}, \citenamefont {{Goldenfeld}},\
  and\ \citenamefont {{Russell}}}]{BBG17}%
  \BibitemOpen
  \bibfield  {author} {\bibinfo {author} {\bibfnamefont {E.}~\bibnamefont
  {{Branscomb}}}, \bibinfo {author} {\bibfnamefont {T.}~\bibnamefont
  {{Biancalani}}}, \bibinfo {author} {\bibfnamefont {N.}~\bibnamefont
  {{Goldenfeld}}}, \ and\ \bibinfo {author} {\bibfnamefont {M.}~\bibnamefont
  {{Russell}}},\ }\href {\doibase 10.1016/j.physrep.2017.02.001} {\bibfield
  {journal} {\bibinfo  {journal} {Phys. Rep.}\ }\textbf {\bibinfo {volume}
  {677}},\ \bibinfo {pages} {1} (\bibinfo {year} {2017})}\BibitemShut {NoStop}%
\bibitem [{\citenamefont {{Moore}}\ \emph {et~al.}(2017)\citenamefont
  {{Moore}}, \citenamefont {{Lenardic}}, \citenamefont {{Jellinek}},
  \citenamefont {{Johnson}}, \citenamefont {{Goldblatt}},\ and\ \citenamefont
  {{Lorenz}}}]{ML17}%
  \BibitemOpen
  \bibfield  {author} {\bibinfo {author} {\bibfnamefont {W.~B.}\ \bibnamefont
  {{Moore}}}, \bibinfo {author} {\bibfnamefont {A.}~\bibnamefont {{Lenardic}}},
  \bibinfo {author} {\bibfnamefont {A.~M.}\ \bibnamefont {{Jellinek}}},
  \bibinfo {author} {\bibfnamefont {C.~L.}\ \bibnamefont {{Johnson}}}, \bibinfo
  {author} {\bibfnamefont {C.}~\bibnamefont {{Goldblatt}}}, \ and\ \bibinfo
  {author} {\bibfnamefont {R.~D.}\ \bibnamefont {{Lorenz}}},\ }\href {\doibase
  10.1038/s41550-017-0043} {\bibfield  {journal} {\bibinfo  {journal} {Nat.
  Astron.}\ }\textbf {\bibinfo {volume} {1}},\ \bibinfo {eid} {0043} (\bibinfo
  {year} {2017})}\BibitemShut {NoStop}%
\bibitem [{\citenamefont {{Tasker}}\ \emph {et~al.}(2017)\citenamefont
  {{Tasker}}, \citenamefont {{Tan}}, \citenamefont {{Heng}}, \citenamefont
  {{Kane}}, \citenamefont {{Spiegel}}, \citenamefont {{Brasser}}, \citenamefont
  {{Casey}}, \citenamefont {{Desch}}, \citenamefont {{Dorn}}, \citenamefont
  {{Hernlund}}, \citenamefont {{Houser}}, \citenamefont {{Laneuville}},
  \citenamefont {{Lasbleis}}, \citenamefont {{Libert}}, \citenamefont
  {{Noack}}, \citenamefont {{Unterborn}},\ and\ \citenamefont
  {{Wicks}}}]{TT17}%
  \BibitemOpen
  \bibfield  {author} {\bibinfo {author} {\bibfnamefont {E.}~\bibnamefont
  {{Tasker}}}, \bibinfo {author} {\bibfnamefont {J.}~\bibnamefont {{Tan}}},
  \bibinfo {author} {\bibfnamefont {K.}~\bibnamefont {{Heng}}}, \bibinfo
  {author} {\bibfnamefont {S.}~\bibnamefont {{Kane}}}, \bibinfo {author}
  {\bibfnamefont {D.}~\bibnamefont {{Spiegel}}}, \bibinfo {author}
  {\bibfnamefont {R.}~\bibnamefont {{Brasser}}}, \bibinfo {author}
  {\bibfnamefont {A.}~\bibnamefont {{Casey}}}, \bibinfo {author} {\bibfnamefont
  {S.}~\bibnamefont {{Desch}}}, \bibinfo {author} {\bibfnamefont
  {C.}~\bibnamefont {{Dorn}}}, \bibinfo {author} {\bibfnamefont
  {J.}~\bibnamefont {{Hernlund}}}, \bibinfo {author} {\bibfnamefont
  {C.}~\bibnamefont {{Houser}}}, \bibinfo {author} {\bibfnamefont
  {M.}~\bibnamefont {{Laneuville}}}, \bibinfo {author} {\bibfnamefont
  {M.}~\bibnamefont {{Lasbleis}}}, \bibinfo {author} {\bibfnamefont {A.-S.}\
  \bibnamefont {{Libert}}}, \bibinfo {author} {\bibfnamefont {L.}~\bibnamefont
  {{Noack}}}, \bibinfo {author} {\bibfnamefont {C.}~\bibnamefont
  {{Unterborn}}}, \ and\ \bibinfo {author} {\bibfnamefont {J.}~\bibnamefont
  {{Wicks}}},\ }\href {\doibase 10.1038/s41550-017-0042} {\bibfield  {journal}
  {\bibinfo  {journal} {Nat. Astron.}\ }\textbf {\bibinfo {volume} {1}},\
  \bibinfo {eid} {0042} (\bibinfo {year} {2017})}\BibitemShut {NoStop}%
\bibitem [{\citenamefont {{Dole}}(1964)}]{Dole}%
  \BibitemOpen
  \bibfield  {author} {\bibinfo {author} {\bibfnamefont {S.~H.}\ \bibnamefont
  {{Dole}}},\ }\href@noop {} {\emph {\bibinfo {title} {{Habitable planets for
  man}}}}\ (\bibinfo  {publisher} {Blaisdell Pub.~Co.},\ \bibinfo {year}
  {1964})\BibitemShut {NoStop}%
\bibitem [{\citenamefont {{Lammer}}\ \emph {et~al.}(2009)\citenamefont
  {{Lammer}}, \citenamefont {{Bredeh{\"o}ft}}, \citenamefont {{Coustenis}},
  \citenamefont {{Khodachenko}}, \citenamefont {{Kaltenegger}}, \citenamefont
  {{Grasset}}, \citenamefont {{Prieur}}, \citenamefont {{Raulin}},
  \citenamefont {{Ehrenfreund}}, \citenamefont {{Yamauchi}}, \citenamefont
  {{Wahlund}}, \citenamefont {{Grie{\ss}meier}}, \citenamefont {{Stangl}},
  \citenamefont {{Cockell}}, \citenamefont {{Kulikov}}, \citenamefont
  {{Grenfell}},\ and\ \citenamefont {{Rauer}}}]{LB09}%
  \BibitemOpen
  \bibfield  {author} {\bibinfo {author} {\bibfnamefont {H.}~\bibnamefont
  {{Lammer}}}, \bibinfo {author} {\bibfnamefont {J.~H.}\ \bibnamefont
  {{Bredeh{\"o}ft}}}, \bibinfo {author} {\bibfnamefont {A.}~\bibnamefont
  {{Coustenis}}}, \bibinfo {author} {\bibfnamefont {M.~L.}\ \bibnamefont
  {{Khodachenko}}}, \bibinfo {author} {\bibfnamefont {L.}~\bibnamefont
  {{Kaltenegger}}}, \bibinfo {author} {\bibfnamefont {O.}~\bibnamefont
  {{Grasset}}}, \bibinfo {author} {\bibfnamefont {D.}~\bibnamefont {{Prieur}}},
  \bibinfo {author} {\bibfnamefont {F.}~\bibnamefont {{Raulin}}}, \bibinfo
  {author} {\bibfnamefont {P.}~\bibnamefont {{Ehrenfreund}}}, \bibinfo {author}
  {\bibfnamefont {M.}~\bibnamefont {{Yamauchi}}}, \bibinfo {author}
  {\bibfnamefont {J.-E.}\ \bibnamefont {{Wahlund}}}, \bibinfo {author}
  {\bibfnamefont {J.-M.}\ \bibnamefont {{Grie{\ss}meier}}}, \bibinfo {author}
  {\bibfnamefont {G.}~\bibnamefont {{Stangl}}}, \bibinfo {author}
  {\bibfnamefont {C.~S.}\ \bibnamefont {{Cockell}}}, \bibinfo {author}
  {\bibfnamefont {Y.~N.}\ \bibnamefont {{Kulikov}}}, \bibinfo {author}
  {\bibfnamefont {J.~L.}\ \bibnamefont {{Grenfell}}}, \ and\ \bibinfo {author}
  {\bibfnamefont {H.}~\bibnamefont {{Rauer}}},\ }\href {\doibase
  10.1007/s00159-009-0019-z} {\bibfield  {journal} {\bibinfo  {journal}
  {Astron. Astrophys. Rev.}\ }\textbf {\bibinfo {volume} {17}},\ \bibinfo
  {pages} {181} (\bibinfo {year} {2009})}\BibitemShut {NoStop}%
\bibitem [{\citenamefont {{Javaux}}\ and\ \citenamefont
  {{Dehant}}(2010)}]{JD10}%
  \BibitemOpen
  \bibfield  {author} {\bibinfo {author} {\bibfnamefont {E.~J.}\ \bibnamefont
  {{Javaux}}}\ and\ \bibinfo {author} {\bibfnamefont {V.}~\bibnamefont
  {{Dehant}}},\ }\href {\doibase 10.1007/s00159-010-0030-4} {\bibfield
  {journal} {\bibinfo  {journal} {Astron. Astrophys. Rev.}\ }\textbf {\bibinfo
  {volume} {18}},\ \bibinfo {pages} {383} (\bibinfo {year} {2010})}\BibitemShut
  {NoStop}%
\bibitem [{\citenamefont {{Lineweaver}}\ and\ \citenamefont
  {{Chopra}}(2012)}]{LiC12}%
  \BibitemOpen
  \bibfield  {author} {\bibinfo {author} {\bibfnamefont {C.~H.}\ \bibnamefont
  {{Lineweaver}}}\ and\ \bibinfo {author} {\bibfnamefont {A.}~\bibnamefont
  {{Chopra}}},\ }\href {\doibase 10.1146/annurev-earth-042711-105531}
  {\bibfield  {journal} {\bibinfo  {journal} {Annu. Rev. Earth Planet. Sci.}\
  }\textbf {\bibinfo {volume} {40}},\ \bibinfo {pages} {597} (\bibinfo {year}
  {2012})}\BibitemShut {NoStop}%
\bibitem [{\citenamefont {{Seager}}(2013)}]{Sea13}%
  \BibitemOpen
  \bibfield  {author} {\bibinfo {author} {\bibfnamefont {S.}~\bibnamefont
  {{Seager}}},\ }\href {\doibase 10.1126/science.1232226} {\bibfield  {journal}
  {\bibinfo  {journal} {Science}\ }\textbf {\bibinfo {volume} {340}},\ \bibinfo
  {pages} {577} (\bibinfo {year} {2013})}\BibitemShut {NoStop}%
\bibitem [{\citenamefont {{Cockell}}\ \emph {et~al.}(2016)\citenamefont
  {{Cockell}}, \citenamefont {{Bush}}, \citenamefont {{Bryce}}, \citenamefont
  {{Direito}}, \citenamefont {{Fox-Powell}}, \citenamefont {{Harrison}},
  \citenamefont {{Lammer}}, \citenamefont {{Landenmark}}, \citenamefont
  {{Martin-Torres}}, \citenamefont {{Nicholson}}, \citenamefont {{Noack}},
  \citenamefont {{O'Malley-James}}, \citenamefont {{Payler}}, \citenamefont
  {{Rushby}}, \citenamefont {{Samuels}}, \citenamefont {{Schwendner}},
  \citenamefont {{Wadsworth}},\ and\ \citenamefont {{Zorzano}}}]{Cock16}%
  \BibitemOpen
  \bibfield  {author} {\bibinfo {author} {\bibfnamefont {C.~S.}\ \bibnamefont
  {{Cockell}}}, \bibinfo {author} {\bibfnamefont {T.}~\bibnamefont {{Bush}}},
  \bibinfo {author} {\bibfnamefont {C.}~\bibnamefont {{Bryce}}}, \bibinfo
  {author} {\bibfnamefont {S.}~\bibnamefont {{Direito}}}, \bibinfo {author}
  {\bibfnamefont {M.}~\bibnamefont {{Fox-Powell}}}, \bibinfo {author}
  {\bibfnamefont {J.~P.}\ \bibnamefont {{Harrison}}}, \bibinfo {author}
  {\bibfnamefont {H.}~\bibnamefont {{Lammer}}}, \bibinfo {author}
  {\bibfnamefont {H.}~\bibnamefont {{Landenmark}}}, \bibinfo {author}
  {\bibfnamefont {J.}~\bibnamefont {{Martin-Torres}}}, \bibinfo {author}
  {\bibfnamefont {N.}~\bibnamefont {{Nicholson}}}, \bibinfo {author}
  {\bibfnamefont {L.}~\bibnamefont {{Noack}}}, \bibinfo {author} {\bibfnamefont
  {J.}~\bibnamefont {{O'Malley-James}}}, \bibinfo {author} {\bibfnamefont
  {S.~J.}\ \bibnamefont {{Payler}}}, \bibinfo {author} {\bibfnamefont
  {A.}~\bibnamefont {{Rushby}}}, \bibinfo {author} {\bibfnamefont
  {T.}~\bibnamefont {{Samuels}}}, \bibinfo {author} {\bibfnamefont
  {P.}~\bibnamefont {{Schwendner}}}, \bibinfo {author} {\bibfnamefont
  {J.}~\bibnamefont {{Wadsworth}}}, \ and\ \bibinfo {author} {\bibfnamefont
  {M.~P.}\ \bibnamefont {{Zorzano}}},\ }\href {\doibase 10.1089/ast.2015.1295}
  {\bibfield  {journal} {\bibinfo  {journal} {Astrobiology}\ }\textbf {\bibinfo
  {volume} {16}},\ \bibinfo {pages} {89} (\bibinfo {year} {2016})}\BibitemShut
  {NoStop}%
\bibitem [{\citenamefont {{Ehlmann}}\ \emph {et~al.}(2016)\citenamefont
  {{Ehlmann}}, \citenamefont {{Anderson}}, \citenamefont {{Andrews-Hanna}},
  \citenamefont {{Catling}}, \citenamefont {{Christensen}}, \citenamefont
  {{Cohen}}, \citenamefont {{Dressing}}, \citenamefont {{Edwards}},
  \citenamefont {{Elkins-Tanton}}, \citenamefont {{Farley}}, \citenamefont
  {{Fassett}}, \citenamefont {{Fischer}}, \citenamefont {{Fraeman}},
  \citenamefont {{Golombek}}, \citenamefont {{Hamilton}}, \citenamefont
  {{Hayes}}, \citenamefont {{Herd}}, \citenamefont {{Horgan}}, \citenamefont
  {{Hu}}, \citenamefont {{Jakosky}}, \citenamefont {{Johnson}}, \citenamefont
  {{Kasting}}, \citenamefont {{Kerber}}, \citenamefont {{Kinch}}, \citenamefont
  {{Kite}}, \citenamefont {{Knutson}}, \citenamefont {{Lunine}}, \citenamefont
  {{Mahaffy}}, \citenamefont {{Mangold}}, \citenamefont {{McCubbin}},
  \citenamefont {{Mustard}}, \citenamefont {{Niles}}, \citenamefont
  {{Quantin-Nataf}}, \citenamefont {{Rice}}, \citenamefont {{Stack}},
  \citenamefont {{Stevenson}}, \citenamefont {{Stewart}}, \citenamefont
  {{Toplis}}, \citenamefont {{Usui}}, \citenamefont {{Weiss}}, \citenamefont
  {{Werner}}, \citenamefont {{Wordsworth}}, \citenamefont {{Wray}},
  \citenamefont {{Yingst}}, \citenamefont {{Yung}},\ and\ \citenamefont
  {{Zahnle}}}]{EA16}%
  \BibitemOpen
  \bibfield  {author} {\bibinfo {author} {\bibfnamefont {B.~L.}\ \bibnamefont
  {{Ehlmann}}}, \bibinfo {author} {\bibfnamefont {F.~S.}\ \bibnamefont
  {{Anderson}}}, \bibinfo {author} {\bibfnamefont {J.}~\bibnamefont
  {{Andrews-Hanna}}}, \bibinfo {author} {\bibfnamefont {D.~C.}\ \bibnamefont
  {{Catling}}}, \bibinfo {author} {\bibfnamefont {P.~R.}\ \bibnamefont
  {{Christensen}}}, \bibinfo {author} {\bibfnamefont {B.~A.}\ \bibnamefont
  {{Cohen}}}, \bibinfo {author} {\bibfnamefont {C.~D.}\ \bibnamefont
  {{Dressing}}}, \bibinfo {author} {\bibfnamefont {C.~S.}\ \bibnamefont
  {{Edwards}}}, \bibinfo {author} {\bibfnamefont {L.~T.}\ \bibnamefont
  {{Elkins-Tanton}}}, \bibinfo {author} {\bibfnamefont {K.~A.}\ \bibnamefont
  {{Farley}}}, \bibinfo {author} {\bibfnamefont {C.~I.}\ \bibnamefont
  {{Fassett}}}, \bibinfo {author} {\bibfnamefont {W.~W.}\ \bibnamefont
  {{Fischer}}}, \bibinfo {author} {\bibfnamefont {A.~A.}\ \bibnamefont
  {{Fraeman}}}, \bibinfo {author} {\bibfnamefont {M.~P.}\ \bibnamefont
  {{Golombek}}}, \bibinfo {author} {\bibfnamefont {V.~E.}\ \bibnamefont
  {{Hamilton}}}, \bibinfo {author} {\bibfnamefont {A.~G.}\ \bibnamefont
  {{Hayes}}}, \bibinfo {author} {\bibfnamefont {C.~D.~K.}\ \bibnamefont
  {{Herd}}}, \bibinfo {author} {\bibfnamefont {B.}~\bibnamefont {{Horgan}}},
  \bibinfo {author} {\bibfnamefont {R.}~\bibnamefont {{Hu}}}, \bibinfo {author}
  {\bibfnamefont {B.~M.}\ \bibnamefont {{Jakosky}}}, \bibinfo {author}
  {\bibfnamefont {J.~R.}\ \bibnamefont {{Johnson}}}, \bibinfo {author}
  {\bibfnamefont {J.~F.}\ \bibnamefont {{Kasting}}}, \bibinfo {author}
  {\bibfnamefont {L.}~\bibnamefont {{Kerber}}}, \bibinfo {author}
  {\bibfnamefont {K.~M.}\ \bibnamefont {{Kinch}}}, \bibinfo {author}
  {\bibfnamefont {E.~S.}\ \bibnamefont {{Kite}}}, \bibinfo {author}
  {\bibfnamefont {H.~A.}\ \bibnamefont {{Knutson}}}, \bibinfo {author}
  {\bibfnamefont {J.~I.}\ \bibnamefont {{Lunine}}}, \bibinfo {author}
  {\bibfnamefont {P.~R.}\ \bibnamefont {{Mahaffy}}}, \bibinfo {author}
  {\bibfnamefont {N.}~\bibnamefont {{Mangold}}}, \bibinfo {author}
  {\bibfnamefont {F.~M.}\ \bibnamefont {{McCubbin}}}, \bibinfo {author}
  {\bibfnamefont {J.~F.}\ \bibnamefont {{Mustard}}}, \bibinfo {author}
  {\bibfnamefont {P.~B.}\ \bibnamefont {{Niles}}}, \bibinfo {author}
  {\bibfnamefont {C.}~\bibnamefont {{Quantin-Nataf}}}, \bibinfo {author}
  {\bibfnamefont {M.~S.}\ \bibnamefont {{Rice}}}, \bibinfo {author}
  {\bibfnamefont {K.~M.}\ \bibnamefont {{Stack}}}, \bibinfo {author}
  {\bibfnamefont {D.~J.}\ \bibnamefont {{Stevenson}}}, \bibinfo {author}
  {\bibfnamefont {S.~T.}\ \bibnamefont {{Stewart}}}, \bibinfo {author}
  {\bibfnamefont {M.~J.}\ \bibnamefont {{Toplis}}}, \bibinfo {author}
  {\bibfnamefont {T.}~\bibnamefont {{Usui}}}, \bibinfo {author} {\bibfnamefont
  {B.~P.}\ \bibnamefont {{Weiss}}}, \bibinfo {author} {\bibfnamefont {S.~C.}\
  \bibnamefont {{Werner}}}, \bibinfo {author} {\bibfnamefont {R.~D.}\
  \bibnamefont {{Wordsworth}}}, \bibinfo {author} {\bibfnamefont {J.~J.}\
  \bibnamefont {{Wray}}}, \bibinfo {author} {\bibfnamefont {R.~A.}\
  \bibnamefont {{Yingst}}}, \bibinfo {author} {\bibfnamefont {Y.~L.}\
  \bibnamefont {{Yung}}}, \ and\ \bibinfo {author} {\bibfnamefont {K.~J.}\
  \bibnamefont {{Zahnle}}},\ }\href {\doibase 10.1002/2016JE005134} {\bibfield
  {journal} {\bibinfo  {journal} {J. Geophys. Res. E}\ }\textbf {\bibinfo
  {volume} {121}},\ \bibinfo {pages} {1927} (\bibinfo {year}
  {2016})}\BibitemShut {NoStop}%
\bibitem [{\citenamefont {{Horner}}\ and\ \citenamefont
  {{Jones}}(2010)}]{HJ10}%
  \BibitemOpen
  \bibfield  {author} {\bibinfo {author} {\bibfnamefont {J.}~\bibnamefont
  {{Horner}}}\ and\ \bibinfo {author} {\bibfnamefont {B.~W.}\ \bibnamefont
  {{Jones}}},\ }\href {\doibase 10.1017/S1473550410000261} {\bibfield
  {journal} {\bibinfo  {journal} {Int. J. Astrobiol.}\ }\textbf {\bibinfo
  {volume} {9}},\ \bibinfo {pages} {273} (\bibinfo {year} {2010})}\BibitemShut
  {NoStop}%
\bibitem [{\citenamefont {{Kaltenegger}}\ \emph {et~al.}(2010)\citenamefont
  {{Kaltenegger}}, \citenamefont {{Eiroa}}, \citenamefont {{Ribas}},
  \citenamefont {{Paresce}}, \citenamefont {{Leitzinger}}, \citenamefont
  {{Odert}}, \citenamefont {{Hanslmeier}}, \citenamefont {{Fridlund}},
  \citenamefont {{Lammer}}, \citenamefont {{Beichman}}, \citenamefont
  {{Danchi}}, \citenamefont {{Henning}}, \citenamefont {{Herbst}},
  \citenamefont {{L{\'e}ger}}, \citenamefont {{Liseau}}, \citenamefont
  {{Lunine}}, \citenamefont {{Penny}}, \citenamefont {{Quirrenbach}},
  \citenamefont {{R{\"o}ttgering}}, \citenamefont {{Selsis}}, \citenamefont
  {{Schneider}}, \citenamefont {{Stam}}, \citenamefont {{Tinetti}},\ and\
  \citenamefont {{White}}}]{KE10}%
  \BibitemOpen
  \bibfield  {author} {\bibinfo {author} {\bibfnamefont {L.}~\bibnamefont
  {{Kaltenegger}}}, \bibinfo {author} {\bibfnamefont {C.}~\bibnamefont
  {{Eiroa}}}, \bibinfo {author} {\bibfnamefont {I.}~\bibnamefont {{Ribas}}},
  \bibinfo {author} {\bibfnamefont {F.}~\bibnamefont {{Paresce}}}, \bibinfo
  {author} {\bibfnamefont {M.}~\bibnamefont {{Leitzinger}}}, \bibinfo {author}
  {\bibfnamefont {P.}~\bibnamefont {{Odert}}}, \bibinfo {author} {\bibfnamefont
  {A.}~\bibnamefont {{Hanslmeier}}}, \bibinfo {author} {\bibfnamefont
  {M.}~\bibnamefont {{Fridlund}}}, \bibinfo {author} {\bibfnamefont
  {H.}~\bibnamefont {{Lammer}}}, \bibinfo {author} {\bibfnamefont
  {C.}~\bibnamefont {{Beichman}}}, \bibinfo {author} {\bibfnamefont
  {W.}~\bibnamefont {{Danchi}}}, \bibinfo {author} {\bibfnamefont
  {T.}~\bibnamefont {{Henning}}}, \bibinfo {author} {\bibfnamefont
  {T.}~\bibnamefont {{Herbst}}}, \bibinfo {author} {\bibfnamefont
  {A.}~\bibnamefont {{L{\'e}ger}}}, \bibinfo {author} {\bibfnamefont
  {R.}~\bibnamefont {{Liseau}}}, \bibinfo {author} {\bibfnamefont
  {J.}~\bibnamefont {{Lunine}}}, \bibinfo {author} {\bibfnamefont
  {A.}~\bibnamefont {{Penny}}}, \bibinfo {author} {\bibfnamefont
  {A.}~\bibnamefont {{Quirrenbach}}}, \bibinfo {author} {\bibfnamefont
  {H.}~\bibnamefont {{R{\"o}ttgering}}}, \bibinfo {author} {\bibfnamefont
  {F.}~\bibnamefont {{Selsis}}}, \bibinfo {author} {\bibfnamefont
  {J.}~\bibnamefont {{Schneider}}}, \bibinfo {author} {\bibfnamefont
  {D.}~\bibnamefont {{Stam}}}, \bibinfo {author} {\bibfnamefont
  {G.}~\bibnamefont {{Tinetti}}}, \ and\ \bibinfo {author} {\bibfnamefont
  {G.~J.}\ \bibnamefont {{White}}},\ }\href {\doibase 10.1089/ast.2009.0367}
  {\bibfield  {journal} {\bibinfo  {journal} {Astrobiology}\ }\textbf {\bibinfo
  {volume} {10}},\ \bibinfo {pages} {103} (\bibinfo {year} {2010})}\BibitemShut
  {NoStop}%
\bibitem [{\citenamefont {{Lingam}}\ and\ \citenamefont
  {{Loeb}}(2018{\natexlab{a}})}]{LL18a}%
  \BibitemOpen
  \bibfield  {author} {\bibinfo {author} {\bibfnamefont {M.}~\bibnamefont
  {{Lingam}}}\ and\ \bibinfo {author} {\bibfnamefont {A.}~\bibnamefont
  {{Loeb}}},\ }\href {\doibase 10.3847/2041-8213/aabd86} {\bibfield  {journal}
  {\bibinfo  {journal} {Astrophys. J. Lett.}\ }\textbf {\bibinfo {volume}
  {857}},\ \bibinfo {eid} {L17} (\bibinfo {year}
  {2018}{\natexlab{a}})}\BibitemShut {NoStop}%
\bibitem [{\citenamefont {{Anglada-Escud{\'e}}}\ \emph
  {et~al.}(2016)\citenamefont {{Anglada-Escud{\'e}}}, \citenamefont {{Amado}},
  \citenamefont {{Barnes}}, \citenamefont {{Berdi{\~n}as}}, \citenamefont
  {{Butler}}, \citenamefont {{Coleman}}, \citenamefont {{de La Cueva}},
  \citenamefont {{Dreizler}}, \citenamefont {{Endl}}, \citenamefont
  {{Giesers}}, \citenamefont {{Jeffers}}, \citenamefont {{Jenkins}},
  \citenamefont {{Jones}}, \citenamefont {{Kiraga}}, \citenamefont
  {{K{\"u}rster}}, \citenamefont {{L{\'o}pez-Gonz{\'a}lez}}, \citenamefont
  {{Marvin}}, \citenamefont {{Morales}}, \citenamefont {{Morin}}, \citenamefont
  {{Nelson}}, \citenamefont {{Ortiz}}, \citenamefont {{Ofir}}, \citenamefont
  {{Paardekooper}}, \citenamefont {{Reiners}}, \citenamefont
  {{Rodr{\'{\i}}guez}}, \citenamefont {{Rodr{\'{\i}}guez-L{\'o}pez}},
  \citenamefont {{Sarmiento}}, \citenamefont {{Strachan}}, \citenamefont
  {{Tsapras}}, \citenamefont {{Tuomi}},\ and\ \citenamefont
  {{Zechmeister}}}]{AA16}%
  \BibitemOpen
  \bibfield  {author} {\bibinfo {author} {\bibfnamefont {G.}~\bibnamefont
  {{Anglada-Escud{\'e}}}}, \bibinfo {author} {\bibfnamefont {P.~J.}\
  \bibnamefont {{Amado}}}, \bibinfo {author} {\bibfnamefont {J.}~\bibnamefont
  {{Barnes}}}, \bibinfo {author} {\bibfnamefont {Z.~M.}\ \bibnamefont
  {{Berdi{\~n}as}}}, \bibinfo {author} {\bibfnamefont {R.~P.}\ \bibnamefont
  {{Butler}}}, \bibinfo {author} {\bibfnamefont {G.~A.~L.}\ \bibnamefont
  {{Coleman}}}, \bibinfo {author} {\bibfnamefont {I.}~\bibnamefont {{de La
  Cueva}}}, \bibinfo {author} {\bibfnamefont {S.}~\bibnamefont {{Dreizler}}},
  \bibinfo {author} {\bibfnamefont {M.}~\bibnamefont {{Endl}}}, \bibinfo
  {author} {\bibfnamefont {B.}~\bibnamefont {{Giesers}}}, \bibinfo {author}
  {\bibfnamefont {S.~V.}\ \bibnamefont {{Jeffers}}}, \bibinfo {author}
  {\bibfnamefont {J.~S.}\ \bibnamefont {{Jenkins}}}, \bibinfo {author}
  {\bibfnamefont {H.~R.~A.}\ \bibnamefont {{Jones}}}, \bibinfo {author}
  {\bibfnamefont {M.}~\bibnamefont {{Kiraga}}}, \bibinfo {author}
  {\bibfnamefont {M.}~\bibnamefont {{K{\"u}rster}}}, \bibinfo {author}
  {\bibfnamefont {M.~J.}\ \bibnamefont {{L{\'o}pez-Gonz{\'a}lez}}}, \bibinfo
  {author} {\bibfnamefont {C.~J.}\ \bibnamefont {{Marvin}}}, \bibinfo {author}
  {\bibfnamefont {N.}~\bibnamefont {{Morales}}}, \bibinfo {author}
  {\bibfnamefont {J.}~\bibnamefont {{Morin}}}, \bibinfo {author} {\bibfnamefont
  {R.~P.}\ \bibnamefont {{Nelson}}}, \bibinfo {author} {\bibfnamefont {J.~L.}\
  \bibnamefont {{Ortiz}}}, \bibinfo {author} {\bibfnamefont {A.}~\bibnamefont
  {{Ofir}}}, \bibinfo {author} {\bibfnamefont {S.-J.}\ \bibnamefont
  {{Paardekooper}}}, \bibinfo {author} {\bibfnamefont {A.}~\bibnamefont
  {{Reiners}}}, \bibinfo {author} {\bibfnamefont {E.}~\bibnamefont
  {{Rodr{\'{\i}}guez}}}, \bibinfo {author} {\bibfnamefont {C.}~\bibnamefont
  {{Rodr{\'{\i}}guez-L{\'o}pez}}}, \bibinfo {author} {\bibfnamefont {L.~F.}\
  \bibnamefont {{Sarmiento}}}, \bibinfo {author} {\bibfnamefont {J.~P.}\
  \bibnamefont {{Strachan}}}, \bibinfo {author} {\bibfnamefont
  {Y.}~\bibnamefont {{Tsapras}}}, \bibinfo {author} {\bibfnamefont
  {M.}~\bibnamefont {{Tuomi}}}, \ and\ \bibinfo {author} {\bibfnamefont
  {M.}~\bibnamefont {{Zechmeister}}},\ }\href {\doibase 10.1038/nature19106}
  {\bibfield  {journal} {\bibinfo  {journal} {Nature}\ }\textbf {\bibinfo
  {volume} {536}},\ \bibinfo {pages} {437} (\bibinfo {year}
  {2016})}\BibitemShut {NoStop}%
\bibitem [{\citenamefont {{Gillon}}\ \emph {et~al.}(2016)\citenamefont
  {{Gillon}}, \citenamefont {{Jehin}}, \citenamefont {{Lederer}}, \citenamefont
  {{Delrez}}, \citenamefont {{de Wit}}, \citenamefont {{Burdanov}},
  \citenamefont {{Van Grootel}}, \citenamefont {{Burgasser}}, \citenamefont
  {{Triaud}}, \citenamefont {{Opitom}}, \citenamefont {{Demory}}, \citenamefont
  {{Sahu}}, \citenamefont {{Bardalez Gagliuffi}}, \citenamefont {{Magain}},\
  and\ \citenamefont {{Queloz}}}]{GJ16}%
  \BibitemOpen
  \bibfield  {author} {\bibinfo {author} {\bibfnamefont {M.}~\bibnamefont
  {{Gillon}}}, \bibinfo {author} {\bibfnamefont {E.}~\bibnamefont {{Jehin}}},
  \bibinfo {author} {\bibfnamefont {S.~M.}\ \bibnamefont {{Lederer}}}, \bibinfo
  {author} {\bibfnamefont {L.}~\bibnamefont {{Delrez}}}, \bibinfo {author}
  {\bibfnamefont {J.}~\bibnamefont {{de Wit}}}, \bibinfo {author}
  {\bibfnamefont {A.}~\bibnamefont {{Burdanov}}}, \bibinfo {author}
  {\bibfnamefont {V.}~\bibnamefont {{Van Grootel}}}, \bibinfo {author}
  {\bibfnamefont {A.~J.}\ \bibnamefont {{Burgasser}}}, \bibinfo {author}
  {\bibfnamefont {A.~H.~M.~J.}\ \bibnamefont {{Triaud}}}, \bibinfo {author}
  {\bibfnamefont {C.}~\bibnamefont {{Opitom}}}, \bibinfo {author}
  {\bibfnamefont {B.-O.}\ \bibnamefont {{Demory}}}, \bibinfo {author}
  {\bibfnamefont {D.~K.}\ \bibnamefont {{Sahu}}}, \bibinfo {author}
  {\bibfnamefont {D.}~\bibnamefont {{Bardalez Gagliuffi}}}, \bibinfo {author}
  {\bibfnamefont {P.}~\bibnamefont {{Magain}}}, \ and\ \bibinfo {author}
  {\bibfnamefont {D.}~\bibnamefont {{Queloz}}},\ }\href {\doibase
  10.1038/nature17448} {\bibfield  {journal} {\bibinfo  {journal} {Nature}\
  }\textbf {\bibinfo {volume} {533}},\ \bibinfo {pages} {221} (\bibinfo {year}
  {2016})}\BibitemShut {NoStop}%
\bibitem [{\citenamefont {{Gillon}}\ \emph {et~al.}(2017)\citenamefont
  {{Gillon}}, \citenamefont {{Triaud}}, \citenamefont {{Demory}}, \citenamefont
  {{Jehin}}, \citenamefont {{Agol}}, \citenamefont {{Deck}}, \citenamefont
  {{Lederer}}, \citenamefont {{de Wit}}, \citenamefont {{Burdanov}},
  \citenamefont {{Ingalls}}, \citenamefont {{Bolmont}}, \citenamefont
  {{Leconte}}, \citenamefont {{Raymond}}, \citenamefont {{Selsis}},
  \citenamefont {{Turbet}}, \citenamefont {{Barkaoui}}, \citenamefont
  {{Burgasser}}, \citenamefont {{Burleigh}}, \citenamefont {{Carey}},
  \citenamefont {{Chaushev}}, \citenamefont {{Copperwheat}}, \citenamefont
  {{Delrez}}, \citenamefont {{Fernandes}}, \citenamefont {{Holdsworth}},
  \citenamefont {{Kotze}}, \citenamefont {{Van Grootel}}, \citenamefont
  {{Almleaky}}, \citenamefont {{Benkhaldoun}}, \citenamefont {{Magain}},\ and\
  \citenamefont {{Queloz}}}]{GT17}%
  \BibitemOpen
  \bibfield  {author} {\bibinfo {author} {\bibfnamefont {M.}~\bibnamefont
  {{Gillon}}}, \bibinfo {author} {\bibfnamefont {A.~H.~M.~J.}\ \bibnamefont
  {{Triaud}}}, \bibinfo {author} {\bibfnamefont {B.-O.}\ \bibnamefont
  {{Demory}}}, \bibinfo {author} {\bibfnamefont {E.}~\bibnamefont {{Jehin}}},
  \bibinfo {author} {\bibfnamefont {E.}~\bibnamefont {{Agol}}}, \bibinfo
  {author} {\bibfnamefont {K.~M.}\ \bibnamefont {{Deck}}}, \bibinfo {author}
  {\bibfnamefont {S.~M.}\ \bibnamefont {{Lederer}}}, \bibinfo {author}
  {\bibfnamefont {J.}~\bibnamefont {{de Wit}}}, \bibinfo {author}
  {\bibfnamefont {A.}~\bibnamefont {{Burdanov}}}, \bibinfo {author}
  {\bibfnamefont {J.~G.}\ \bibnamefont {{Ingalls}}}, \bibinfo {author}
  {\bibfnamefont {E.}~\bibnamefont {{Bolmont}}}, \bibinfo {author}
  {\bibfnamefont {J.}~\bibnamefont {{Leconte}}}, \bibinfo {author}
  {\bibfnamefont {S.~N.}\ \bibnamefont {{Raymond}}}, \bibinfo {author}
  {\bibfnamefont {F.}~\bibnamefont {{Selsis}}}, \bibinfo {author}
  {\bibfnamefont {M.}~\bibnamefont {{Turbet}}}, \bibinfo {author}
  {\bibfnamefont {K.}~\bibnamefont {{Barkaoui}}}, \bibinfo {author}
  {\bibfnamefont {A.}~\bibnamefont {{Burgasser}}}, \bibinfo {author}
  {\bibfnamefont {M.~R.}\ \bibnamefont {{Burleigh}}}, \bibinfo {author}
  {\bibfnamefont {S.~J.}\ \bibnamefont {{Carey}}}, \bibinfo {author}
  {\bibfnamefont {A.}~\bibnamefont {{Chaushev}}}, \bibinfo {author}
  {\bibfnamefont {C.~M.}\ \bibnamefont {{Copperwheat}}}, \bibinfo {author}
  {\bibfnamefont {L.}~\bibnamefont {{Delrez}}}, \bibinfo {author}
  {\bibfnamefont {C.~S.}\ \bibnamefont {{Fernandes}}}, \bibinfo {author}
  {\bibfnamefont {D.~L.}\ \bibnamefont {{Holdsworth}}}, \bibinfo {author}
  {\bibfnamefont {E.~J.}\ \bibnamefont {{Kotze}}}, \bibinfo {author}
  {\bibfnamefont {V.}~\bibnamefont {{Van Grootel}}}, \bibinfo {author}
  {\bibfnamefont {Y.}~\bibnamefont {{Almleaky}}}, \bibinfo {author}
  {\bibfnamefont {Z.}~\bibnamefont {{Benkhaldoun}}}, \bibinfo {author}
  {\bibfnamefont {P.}~\bibnamefont {{Magain}}}, \ and\ \bibinfo {author}
  {\bibfnamefont {D.}~\bibnamefont {{Queloz}}},\ }\href {\doibase
  10.1038/nature21360} {\bibfield  {journal} {\bibinfo  {journal} {Nature}\
  }\textbf {\bibinfo {volume} {542}},\ \bibinfo {pages} {456} (\bibinfo {year}
  {2017})}\BibitemShut {NoStop}%
\bibitem [{\citenamefont {{Delrez}}\ \emph {et~al.}(2018)\citenamefont
  {{Delrez}}, \citenamefont {{Gillon}}, \citenamefont {{Triaud}}, \citenamefont
  {{Demory}}, \citenamefont {{de Wit}}, \citenamefont {{Ingalls}},
  \citenamefont {{Agol}}, \citenamefont {{Bolmont}}, \citenamefont
  {{Burdanov}}, \citenamefont {{Burgasser}}, \citenamefont {{Carey}},
  \citenamefont {{Jehin}}, \citenamefont {{Leconte}}, \citenamefont
  {{Lederer}}, \citenamefont {{Queloz}}, \citenamefont {{Selsis}},\ and\
  \citenamefont {{Van Grootel}}}]{DG18}%
  \BibitemOpen
  \bibfield  {author} {\bibinfo {author} {\bibfnamefont {L.}~\bibnamefont
  {{Delrez}}}, \bibinfo {author} {\bibfnamefont {M.}~\bibnamefont {{Gillon}}},
  \bibinfo {author} {\bibfnamefont {A.~H.~M.~J.}\ \bibnamefont {{Triaud}}},
  \bibinfo {author} {\bibfnamefont {B.-O.}\ \bibnamefont {{Demory}}}, \bibinfo
  {author} {\bibfnamefont {J.}~\bibnamefont {{de Wit}}}, \bibinfo {author}
  {\bibfnamefont {J.~G.}\ \bibnamefont {{Ingalls}}}, \bibinfo {author}
  {\bibfnamefont {E.}~\bibnamefont {{Agol}}}, \bibinfo {author} {\bibfnamefont
  {E.}~\bibnamefont {{Bolmont}}}, \bibinfo {author} {\bibfnamefont
  {A.}~\bibnamefont {{Burdanov}}}, \bibinfo {author} {\bibfnamefont {A.~J.}\
  \bibnamefont {{Burgasser}}}, \bibinfo {author} {\bibfnamefont {S.~J.}\
  \bibnamefont {{Carey}}}, \bibinfo {author} {\bibfnamefont {E.}~\bibnamefont
  {{Jehin}}}, \bibinfo {author} {\bibfnamefont {J.}~\bibnamefont {{Leconte}}},
  \bibinfo {author} {\bibfnamefont {S.}~\bibnamefont {{Lederer}}}, \bibinfo
  {author} {\bibfnamefont {D.}~\bibnamefont {{Queloz}}}, \bibinfo {author}
  {\bibfnamefont {F.}~\bibnamefont {{Selsis}}}, \ and\ \bibinfo {author}
  {\bibfnamefont {V.}~\bibnamefont {{Van Grootel}}},\ }\href {\doibase
  10.1093/mnras/sty051} {\bibfield  {journal} {\bibinfo  {journal} {Mon. Not.
  R. Astron. Soc.}\ }\textbf {\bibinfo {volume} {475}},\ \bibinfo {pages}
  {3577} (\bibinfo {year} {2018})}\BibitemShut {NoStop}%
\bibitem [{\citenamefont {{Grimm}}\ \emph {et~al.}(2018)\citenamefont
  {{Grimm}}, \citenamefont {{Demory}}, \citenamefont {{Gillon}}, \citenamefont
  {{Dorn}}, \citenamefont {{Agol}}, \citenamefont {{Burdanov}}, \citenamefont
  {{Delrez}}, \citenamefont {{Sestovic}}, \citenamefont {{Triaud}},
  \citenamefont {{Turbet}}, \citenamefont {{Bolmont}}, \citenamefont
  {{Caldas}}, \citenamefont {{Wit}}, \citenamefont {{Jehin}}, \citenamefont
  {{Leconte}}, \citenamefont {{Raymond}}, \citenamefont {{Grootel}},
  \citenamefont {{Burgasser}}, \citenamefont {{Carey}}, \citenamefont
  {{Fabrycky}}, \citenamefont {{Heng}}, \citenamefont {{Hernandez}},
  \citenamefont {{Ingalls}}, \citenamefont {{Lederer}}, \citenamefont
  {{Selsis}},\ and\ \citenamefont {{Queloz}}}]{GD18}%
  \BibitemOpen
  \bibfield  {author} {\bibinfo {author} {\bibfnamefont {S.~L.}\ \bibnamefont
  {{Grimm}}}, \bibinfo {author} {\bibfnamefont {B.-O.}\ \bibnamefont
  {{Demory}}}, \bibinfo {author} {\bibfnamefont {M.}~\bibnamefont {{Gillon}}},
  \bibinfo {author} {\bibfnamefont {C.}~\bibnamefont {{Dorn}}}, \bibinfo
  {author} {\bibfnamefont {E.}~\bibnamefont {{Agol}}}, \bibinfo {author}
  {\bibfnamefont {A.}~\bibnamefont {{Burdanov}}}, \bibinfo {author}
  {\bibfnamefont {L.}~\bibnamefont {{Delrez}}}, \bibinfo {author}
  {\bibfnamefont {M.}~\bibnamefont {{Sestovic}}}, \bibinfo {author}
  {\bibfnamefont {A.~H.~M.~J.}\ \bibnamefont {{Triaud}}}, \bibinfo {author}
  {\bibfnamefont {M.}~\bibnamefont {{Turbet}}}, \bibinfo {author}
  {\bibfnamefont {{\'E}.}~\bibnamefont {{Bolmont}}}, \bibinfo {author}
  {\bibfnamefont {A.}~\bibnamefont {{Caldas}}}, \bibinfo {author}
  {\bibfnamefont {J.~d.}\ \bibnamefont {{Wit}}}, \bibinfo {author}
  {\bibfnamefont {E.}~\bibnamefont {{Jehin}}}, \bibinfo {author} {\bibfnamefont
  {J.}~\bibnamefont {{Leconte}}}, \bibinfo {author} {\bibfnamefont {S.~N.}\
  \bibnamefont {{Raymond}}}, \bibinfo {author} {\bibfnamefont {V.~V.}\
  \bibnamefont {{Grootel}}}, \bibinfo {author} {\bibfnamefont {A.~J.}\
  \bibnamefont {{Burgasser}}}, \bibinfo {author} {\bibfnamefont
  {S.}~\bibnamefont {{Carey}}}, \bibinfo {author} {\bibfnamefont
  {D.}~\bibnamefont {{Fabrycky}}}, \bibinfo {author} {\bibfnamefont
  {K.}~\bibnamefont {{Heng}}}, \bibinfo {author} {\bibfnamefont {D.~M.}\
  \bibnamefont {{Hernandez}}}, \bibinfo {author} {\bibfnamefont {J.~G.}\
  \bibnamefont {{Ingalls}}}, \bibinfo {author} {\bibfnamefont {S.}~\bibnamefont
  {{Lederer}}}, \bibinfo {author} {\bibfnamefont {F.}~\bibnamefont {{Selsis}}},
  \ and\ \bibinfo {author} {\bibfnamefont {D.}~\bibnamefont {{Queloz}}},\
  }\href {\doibase 10.1051/0004-6361/201732233} {\bibfield  {journal} {\bibinfo
   {journal} {Astron. Astrophys.}\ }\textbf {\bibinfo {volume} {613}},\
  \bibinfo {eid} {A68} (\bibinfo {year} {2018})}\BibitemShut {NoStop}%
\bibitem [{\citenamefont {{Dittmann}}\ \emph {et~al.}(2017)\citenamefont
  {{Dittmann}}, \citenamefont {{Irwin}}, \citenamefont {{Charbonneau}},
  \citenamefont {{Bonfils}}, \citenamefont {{Astudillo-Defru}}, \citenamefont
  {{Haywood}}, \citenamefont {{Berta-Thompson}}, \citenamefont {{Newton}},
  \citenamefont {{Rodriguez}}, \citenamefont {{Winters}}, \citenamefont
  {{Tan}}, \citenamefont {{Almenara}}, \citenamefont {{Bouchy}}, \citenamefont
  {{Delfosse}}, \citenamefont {{Forveille}}, \citenamefont {{Lovis}},
  \citenamefont {{Murgas}}, \citenamefont {{Pepe}}, \citenamefont {{Santos}},
  \citenamefont {{Udry}}, \citenamefont {{W{\"u}nsche}}, \citenamefont
  {{Esquerdo}}, \citenamefont {{Latham}},\ and\ \citenamefont
  {{Dressing}}}]{DI17}%
  \BibitemOpen
  \bibfield  {author} {\bibinfo {author} {\bibfnamefont {J.~A.}\ \bibnamefont
  {{Dittmann}}}, \bibinfo {author} {\bibfnamefont {J.~M.}\ \bibnamefont
  {{Irwin}}}, \bibinfo {author} {\bibfnamefont {D.}~\bibnamefont
  {{Charbonneau}}}, \bibinfo {author} {\bibfnamefont {X.}~\bibnamefont
  {{Bonfils}}}, \bibinfo {author} {\bibfnamefont {N.}~\bibnamefont
  {{Astudillo-Defru}}}, \bibinfo {author} {\bibfnamefont {R.~D.}\ \bibnamefont
  {{Haywood}}}, \bibinfo {author} {\bibfnamefont {Z.~K.}\ \bibnamefont
  {{Berta-Thompson}}}, \bibinfo {author} {\bibfnamefont {E.~R.}\ \bibnamefont
  {{Newton}}}, \bibinfo {author} {\bibfnamefont {J.~E.}\ \bibnamefont
  {{Rodriguez}}}, \bibinfo {author} {\bibfnamefont {J.~G.}\ \bibnamefont
  {{Winters}}}, \bibinfo {author} {\bibfnamefont {T.-G.}\ \bibnamefont
  {{Tan}}}, \bibinfo {author} {\bibfnamefont {J.-M.}\ \bibnamefont
  {{Almenara}}}, \bibinfo {author} {\bibfnamefont {F.}~\bibnamefont
  {{Bouchy}}}, \bibinfo {author} {\bibfnamefont {X.}~\bibnamefont
  {{Delfosse}}}, \bibinfo {author} {\bibfnamefont {T.}~\bibnamefont
  {{Forveille}}}, \bibinfo {author} {\bibfnamefont {C.}~\bibnamefont
  {{Lovis}}}, \bibinfo {author} {\bibfnamefont {F.}~\bibnamefont {{Murgas}}},
  \bibinfo {author} {\bibfnamefont {F.}~\bibnamefont {{Pepe}}}, \bibinfo
  {author} {\bibfnamefont {N.~C.}\ \bibnamefont {{Santos}}}, \bibinfo {author}
  {\bibfnamefont {S.}~\bibnamefont {{Udry}}}, \bibinfo {author} {\bibfnamefont
  {A.}~\bibnamefont {{W{\"u}nsche}}}, \bibinfo {author} {\bibfnamefont {G.~A.}\
  \bibnamefont {{Esquerdo}}}, \bibinfo {author} {\bibfnamefont {D.~W.}\
  \bibnamefont {{Latham}}}, \ and\ \bibinfo {author} {\bibfnamefont {C.~D.}\
  \bibnamefont {{Dressing}}},\ }\href {\doibase 10.1038/nature22055} {\bibfield
   {journal} {\bibinfo  {journal} {Nature}\ }\textbf {\bibinfo {volume}
  {544}},\ \bibinfo {pages} {333} (\bibinfo {year} {2017})}\BibitemShut
  {NoStop}%
\bibitem [{\citenamefont {{Laskar}}\ \emph {et~al.}(1993)\citenamefont
  {{Laskar}}, \citenamefont {{Joutel}},\ and\ \citenamefont
  {{Robutel}}}]{LJR93}%
  \BibitemOpen
  \bibfield  {author} {\bibinfo {author} {\bibfnamefont {J.}~\bibnamefont
  {{Laskar}}}, \bibinfo {author} {\bibfnamefont {F.}~\bibnamefont {{Joutel}}},
  \ and\ \bibinfo {author} {\bibfnamefont {P.}~\bibnamefont {{Robutel}}},\
  }\href {\doibase 10.1038/361615a0} {\bibfield  {journal} {\bibinfo  {journal}
  {Nature}\ }\textbf {\bibinfo {volume} {361}},\ \bibinfo {pages} {615}
  (\bibinfo {year} {1993})}\BibitemShut {NoStop}%
\bibitem [{\citenamefont {{Kasting}}(2010)}]{Kas10}%
  \BibitemOpen
  \bibfield  {author} {\bibinfo {author} {\bibfnamefont {J.}~\bibnamefont
  {{Kasting}}},\ }\href@noop {} {\emph {\bibinfo {title} {{How to Find a
  Habitable Planet}}}}\ (\bibinfo  {publisher} {Princeton University Press},\
  \bibinfo {year} {2010})\BibitemShut {NoStop}%
\bibitem [{\citenamefont {{Pierrehumbert}}(2010)}]{Pie10}%
  \BibitemOpen
  \bibfield  {author} {\bibinfo {author} {\bibfnamefont {R.~T.}\ \bibnamefont
  {{Pierrehumbert}}},\ }\href@noop {} {\emph {\bibinfo {title} {{Principles of
  Planetary Climate}}}}\ (\bibinfo  {publisher} {Cambridge University Press},\
  \bibinfo {year} {2010})\BibitemShut {NoStop}%
\bibitem [{\citenamefont {{Catling}}\ and\ \citenamefont
  {{Kasting}}(2017{\natexlab{a}})}]{CK17}%
  \BibitemOpen
  \bibfield  {author} {\bibinfo {author} {\bibfnamefont {D.~C.}\ \bibnamefont
  {{Catling}}}\ and\ \bibinfo {author} {\bibfnamefont {J.~F.}\ \bibnamefont
  {{Kasting}}},\ }\href@noop {} {\emph {\bibinfo {title} {{Atmospheric
  Evolution on Inhabited and Lifeless Worlds}}}}\ (\bibinfo  {publisher}
  {Cambridge University Press},\ \bibinfo {year} {2017})\BibitemShut {NoStop}%
\bibitem [{\citenamefont {{Turbet}}\ \emph {et~al.}(2016)\citenamefont
  {{Turbet}}, \citenamefont {{Leconte}}, \citenamefont {{Selsis}},
  \citenamefont {{Bolmont}}, \citenamefont {{Forget}}, \citenamefont {{Ribas}},
  \citenamefont {{Raymond}},\ and\ \citenamefont
  {{Anglada-Escud{\'e}}}}]{TL16}%
  \BibitemOpen
  \bibfield  {author} {\bibinfo {author} {\bibfnamefont {M.}~\bibnamefont
  {{Turbet}}}, \bibinfo {author} {\bibfnamefont {J.}~\bibnamefont {{Leconte}}},
  \bibinfo {author} {\bibfnamefont {F.}~\bibnamefont {{Selsis}}}, \bibinfo
  {author} {\bibfnamefont {E.}~\bibnamefont {{Bolmont}}}, \bibinfo {author}
  {\bibfnamefont {F.}~\bibnamefont {{Forget}}}, \bibinfo {author}
  {\bibfnamefont {I.}~\bibnamefont {{Ribas}}}, \bibinfo {author} {\bibfnamefont
  {S.~N.}\ \bibnamefont {{Raymond}}}, \ and\ \bibinfo {author} {\bibfnamefont
  {G.}~\bibnamefont {{Anglada-Escud{\'e}}}},\ }\href {\doibase
  10.1051/0004-6361/201629577} {\bibfield  {journal} {\bibinfo  {journal}
  {Astron. Astrophys.}\ }\textbf {\bibinfo {volume} {596}},\ \bibinfo {eid}
  {A112} (\bibinfo {year} {2016})}\BibitemShut {NoStop}%
\bibitem [{\citenamefont {{Wolf}}(2017)}]{Wolf}%
  \BibitemOpen
  \bibfield  {author} {\bibinfo {author} {\bibfnamefont {E.~T.}\ \bibnamefont
  {{Wolf}}},\ }\href {\doibase 10.3847/2041-8213/aa693a} {\bibfield  {journal}
  {\bibinfo  {journal} {Astrophys. J. Lett.}\ }\textbf {\bibinfo {volume}
  {839}},\ \bibinfo {eid} {L1} (\bibinfo {year} {2017})}\BibitemShut {NoStop}%
\bibitem [{\citenamefont {{Boutle}}\ \emph {et~al.}(2017)\citenamefont
  {{Boutle}}, \citenamefont {{Mayne}}, \citenamefont {{Drummond}},
  \citenamefont {{Manners}}, \citenamefont {{Goyal}}, \citenamefont {{Hugo
  Lambert}}, \citenamefont {{Acreman}},\ and\ \citenamefont
  {{Earnshaw}}}]{BM17}%
  \BibitemOpen
  \bibfield  {author} {\bibinfo {author} {\bibfnamefont {I.~A.}\ \bibnamefont
  {{Boutle}}}, \bibinfo {author} {\bibfnamefont {N.~J.}\ \bibnamefont
  {{Mayne}}}, \bibinfo {author} {\bibfnamefont {B.}~\bibnamefont {{Drummond}}},
  \bibinfo {author} {\bibfnamefont {J.}~\bibnamefont {{Manners}}}, \bibinfo
  {author} {\bibfnamefont {J.}~\bibnamefont {{Goyal}}}, \bibinfo {author}
  {\bibfnamefont {F.}~\bibnamefont {{Hugo Lambert}}}, \bibinfo {author}
  {\bibfnamefont {D.~M.}\ \bibnamefont {{Acreman}}}, \ and\ \bibinfo {author}
  {\bibfnamefont {P.~D.}\ \bibnamefont {{Earnshaw}}},\ }\href {\doibase
  10.1051/0004-6361/201630020} {\bibfield  {journal} {\bibinfo  {journal}
  {Astron. Astrophys.}\ }\textbf {\bibinfo {volume} {601}},\ \bibinfo {eid}
  {A120} (\bibinfo {year} {2017})}\BibitemShut {NoStop}%
\bibitem [{\citenamefont {{Alberti}}\ \emph {et~al.}(2017)\citenamefont
  {{Alberti}}, \citenamefont {{Carbone}}, \citenamefont {{Lepreti}},\ and\
  \citenamefont {{Vecchio}}}]{ACL17}%
  \BibitemOpen
  \bibfield  {author} {\bibinfo {author} {\bibfnamefont {T.}~\bibnamefont
  {{Alberti}}}, \bibinfo {author} {\bibfnamefont {V.}~\bibnamefont
  {{Carbone}}}, \bibinfo {author} {\bibfnamefont {F.}~\bibnamefont
  {{Lepreti}}}, \ and\ \bibinfo {author} {\bibfnamefont {A.}~\bibnamefont
  {{Vecchio}}},\ }\href {\doibase 10.3847/1538-4357/aa78a2} {\bibfield
  {journal} {\bibinfo  {journal} {Astrophys. J.}\ }\textbf {\bibinfo {volume}
  {844}},\ \bibinfo {eid} {19} (\bibinfo {year} {2017})}\BibitemShut {NoStop}%
\bibitem [{\citenamefont {{Turbet}}\ \emph {et~al.}(2018)\citenamefont
  {{Turbet}}, \citenamefont {{Bolmont}}, \citenamefont {{Leconte}},
  \citenamefont {{Forget}}, \citenamefont {{Selsis}}, \citenamefont {{Tobie}},
  \citenamefont {{Caldas}}, \citenamefont {{Naar}},\ and\ \citenamefont
  {{Gillon}}}]{TB18}%
  \BibitemOpen
  \bibfield  {author} {\bibinfo {author} {\bibfnamefont {M.}~\bibnamefont
  {{Turbet}}}, \bibinfo {author} {\bibfnamefont {E.}~\bibnamefont {{Bolmont}}},
  \bibinfo {author} {\bibfnamefont {J.}~\bibnamefont {{Leconte}}}, \bibinfo
  {author} {\bibfnamefont {F.}~\bibnamefont {{Forget}}}, \bibinfo {author}
  {\bibfnamefont {F.}~\bibnamefont {{Selsis}}}, \bibinfo {author}
  {\bibfnamefont {G.}~\bibnamefont {{Tobie}}}, \bibinfo {author} {\bibfnamefont
  {A.}~\bibnamefont {{Caldas}}}, \bibinfo {author} {\bibfnamefont
  {J.}~\bibnamefont {{Naar}}}, \ and\ \bibinfo {author} {\bibfnamefont
  {M.}~\bibnamefont {{Gillon}}},\ }\href {\doibase 10.1051/0004-6361/201731620}
  {\bibfield  {journal} {\bibinfo  {journal} {Astron. Astrophys.}\ }\textbf
  {\bibinfo {volume} {612}},\ \bibinfo {eid} {A86} (\bibinfo {year}
  {2018})}\BibitemShut {NoStop}%
\bibitem [{\citenamefont {{Meadows}}\ \emph
  {et~al.}(2018{\natexlab{a}})\citenamefont {{Meadows}}, \citenamefont
  {{Arney}}, \citenamefont {{Schwieterman}}, \citenamefont {{Lustig-Yaeger}},
  \citenamefont {{Lincowski}}, \citenamefont {{Robinson}}, \citenamefont
  {{Domagal-Goldman}}, \citenamefont {{Deitrick}}, \citenamefont {{Barnes}},
  \citenamefont {{Fleming}}, \citenamefont {{Luger}}, \citenamefont
  {{Driscoll}}, \citenamefont {{Quinn}},\ and\ \citenamefont {{Crisp}}}]{MA18}%
  \BibitemOpen
  \bibfield  {author} {\bibinfo {author} {\bibfnamefont {V.~S.}\ \bibnamefont
  {{Meadows}}}, \bibinfo {author} {\bibfnamefont {G.~N.}\ \bibnamefont
  {{Arney}}}, \bibinfo {author} {\bibfnamefont {E.~W.}\ \bibnamefont
  {{Schwieterman}}}, \bibinfo {author} {\bibfnamefont {J.}~\bibnamefont
  {{Lustig-Yaeger}}}, \bibinfo {author} {\bibfnamefont {A.~P.}\ \bibnamefont
  {{Lincowski}}}, \bibinfo {author} {\bibfnamefont {T.}~\bibnamefont
  {{Robinson}}}, \bibinfo {author} {\bibfnamefont {S.~D.}\ \bibnamefont
  {{Domagal-Goldman}}}, \bibinfo {author} {\bibfnamefont {R.}~\bibnamefont
  {{Deitrick}}}, \bibinfo {author} {\bibfnamefont {R.~K.}\ \bibnamefont
  {{Barnes}}}, \bibinfo {author} {\bibfnamefont {D.~P.}\ \bibnamefont
  {{Fleming}}}, \bibinfo {author} {\bibfnamefont {R.}~\bibnamefont {{Luger}}},
  \bibinfo {author} {\bibfnamefont {P.~E.}\ \bibnamefont {{Driscoll}}},
  \bibinfo {author} {\bibfnamefont {T.~R.}\ \bibnamefont {{Quinn}}}, \ and\
  \bibinfo {author} {\bibfnamefont {D.}~\bibnamefont {{Crisp}}},\ }\href
  {\doibase 10.1089/ast.2016.1589} {\bibfield  {journal} {\bibinfo  {journal}
  {Astrobiology}\ }\textbf {\bibinfo {volume} {18}},\ \bibinfo {pages} {133}
  (\bibinfo {year} {2018}{\natexlab{a}})}\BibitemShut {NoStop}%
\bibitem [{\citenamefont {{Lincowski}}\ \emph {et~al.}(2018)\citenamefont
  {{Lincowski}}, \citenamefont {{Meadows}}, \citenamefont {{Crisp}},
  \citenamefont {{Robinson}}, \citenamefont {{Luger}}, \citenamefont
  {{Lustig-Yaeger}},\ and\ \citenamefont {{Arney}}}]{LMC18}%
  \BibitemOpen
  \bibfield  {author} {\bibinfo {author} {\bibfnamefont {A.~P.}\ \bibnamefont
  {{Lincowski}}}, \bibinfo {author} {\bibfnamefont {V.~S.}\ \bibnamefont
  {{Meadows}}}, \bibinfo {author} {\bibfnamefont {D.}~\bibnamefont {{Crisp}}},
  \bibinfo {author} {\bibfnamefont {T.~D.}\ \bibnamefont {{Robinson}}},
  \bibinfo {author} {\bibfnamefont {R.}~\bibnamefont {{Luger}}}, \bibinfo
  {author} {\bibfnamefont {J.}~\bibnamefont {{Lustig-Yaeger}}}, \ and\ \bibinfo
  {author} {\bibfnamefont {G.~N.}\ \bibnamefont {{Arney}}},\ }\href {\doibase
  10.3847/1538-4357/aae36a} {\bibfield  {journal} {\bibinfo  {journal}
  {Astrophys. J.}\ }\textbf {\bibinfo {volume} {867}},\ \bibinfo {eid} {76}
  (\bibinfo {year} {2018})}\BibitemShut {NoStop}%
\bibitem [{\citenamefont {{Sarmiento}}\ and\ \citenamefont
  {{Gruber}}(2006)}]{SG06}%
  \BibitemOpen
  \bibfield  {author} {\bibinfo {author} {\bibfnamefont {J.~L.}\ \bibnamefont
  {{Sarmiento}}}\ and\ \bibinfo {author} {\bibfnamefont {N.}~\bibnamefont
  {{Gruber}}},\ }\href@noop {} {\emph {\bibinfo {title} {{Ocean Biogeochemical
  Dynamics}}}}\ (\bibinfo  {publisher} {Princeton University Press},\ \bibinfo
  {year} {2006})\BibitemShut {NoStop}%
\bibitem [{\citenamefont {{Schlesinger}}\ and\ \citenamefont
  {{Bernhardt}}(2013)}]{SB13}%
  \BibitemOpen
  \bibfield  {author} {\bibinfo {author} {\bibfnamefont {W.~H.}\ \bibnamefont
  {{Schlesinger}}}\ and\ \bibinfo {author} {\bibfnamefont {E.~S.}\ \bibnamefont
  {{Bernhardt}}},\ }\href@noop {} {\emph {\bibinfo {title} {{Biogeochemistry:
  An Analysis of Global Change}}}},\ \bibinfo {edition} {3rd}\ ed.\ (\bibinfo
  {publisher} {Academic Press},\ \bibinfo {year} {2013})\BibitemShut {NoStop}%
\bibitem [{\citenamefont {{Gonzalez}}\ \emph {et~al.}(2001)\citenamefont
  {{Gonzalez}}, \citenamefont {{Brownlee}},\ and\ \citenamefont
  {{Ward}}}]{GBW01}%
  \BibitemOpen
  \bibfield  {author} {\bibinfo {author} {\bibfnamefont {G.}~\bibnamefont
  {{Gonzalez}}}, \bibinfo {author} {\bibfnamefont {D.}~\bibnamefont
  {{Brownlee}}}, \ and\ \bibinfo {author} {\bibfnamefont {P.}~\bibnamefont
  {{Ward}}},\ }\href {\doibase 10.1006/icar.2001.6617} {\bibfield  {journal}
  {\bibinfo  {journal} {Icarus}\ }\textbf {\bibinfo {volume} {152}},\ \bibinfo
  {pages} {185} (\bibinfo {year} {2001})}\BibitemShut {NoStop}%
\bibitem [{\citenamefont {{Lineweaver}}\ \emph {et~al.}(2004)\citenamefont
  {{Lineweaver}}, \citenamefont {{Fenner}},\ and\ \citenamefont
  {{Gibson}}}]{LFG04}%
  \BibitemOpen
  \bibfield  {author} {\bibinfo {author} {\bibfnamefont {C.~H.}\ \bibnamefont
  {{Lineweaver}}}, \bibinfo {author} {\bibfnamefont {Y.}~\bibnamefont
  {{Fenner}}}, \ and\ \bibinfo {author} {\bibfnamefont {B.~K.}\ \bibnamefont
  {{Gibson}}},\ }\href {\doibase 10.1126/science.1092322} {\bibfield  {journal}
  {\bibinfo  {journal} {Science}\ }\textbf {\bibinfo {volume} {303}},\ \bibinfo
  {pages} {59} (\bibinfo {year} {2004})}\BibitemShut {NoStop}%
\bibitem [{\citenamefont {{Prantzos}}(2008)}]{Pran08}%
  \BibitemOpen
  \bibfield  {author} {\bibinfo {author} {\bibfnamefont {N.}~\bibnamefont
  {{Prantzos}}},\ }\href {\doibase 10.1007/s11214-007-9236-9} {\bibfield
  {journal} {\bibinfo  {journal} {Space Sci. Rev.}\ }\textbf {\bibinfo {volume}
  {135}},\ \bibinfo {pages} {313} (\bibinfo {year} {2008})}\BibitemShut
  {NoStop}%
\bibitem [{\citenamefont {{Gowanlock}}\ and\ \citenamefont
  {{Morrison}}(2018)}]{GM18}%
  \BibitemOpen
  \bibfield  {author} {\bibinfo {author} {\bibfnamefont {M.~G.}\ \bibnamefont
  {{Gowanlock}}}\ and\ \bibinfo {author} {\bibfnamefont {I.~S.}\ \bibnamefont
  {{Morrison}}},\ }\enquote {\bibinfo {title} {{The Habitability of our
  Evolving Galaxy}},}\ in\ \href {\doibase 10.1016/B978-0-12-811940-2.00007-1}
  {\emph {\bibinfo {booktitle} {Habitability of the Universe Before Earth}}},\
  \bibinfo {series} {Astrobiology: Exploring Life on Earth and Beyond},
  Vol.~\bibinfo {volume} {1},\ \bibinfo {editor} {edited by\ \bibinfo {editor}
  {\bibfnamefont {P.~H.}\ \bibnamefont {{Rampelotto}}}, \bibinfo {editor}
  {\bibfnamefont {J.}~\bibnamefont {{Seckbach}}}, \ and\ \bibinfo {editor}
  {\bibfnamefont {R.}~\bibnamefont {{Gordon}}}}\ (\bibinfo  {publisher}
  {Academic Press},\ \bibinfo {year} {2018})\ pp.\ \bibinfo {pages}
  {149--171}\BibitemShut {NoStop}%
\bibitem [{\citenamefont {{Thorsett}}(1995)}]{Thor95}%
  \BibitemOpen
  \bibfield  {author} {\bibinfo {author} {\bibfnamefont {S.~E.}\ \bibnamefont
  {{Thorsett}}},\ }\href {\doibase 10.1086/187858} {\bibfield  {journal}
  {\bibinfo  {journal} {Astrophys. J. Lett.}\ }\textbf {\bibinfo {volume}
  {444}},\ \bibinfo {pages} {L53} (\bibinfo {year} {1995})}\BibitemShut
  {NoStop}%
\bibitem [{\citenamefont {{Melott}}\ and\ \citenamefont
  {{Thomas}}(2011)}]{MT11}%
  \BibitemOpen
  \bibfield  {author} {\bibinfo {author} {\bibfnamefont {A.~L.}\ \bibnamefont
  {{Melott}}}\ and\ \bibinfo {author} {\bibfnamefont {B.~C.}\ \bibnamefont
  {{Thomas}}},\ }\href {\doibase 10.1089/ast.2010.0603} {\bibfield  {journal}
  {\bibinfo  {journal} {Astrobiology}\ }\textbf {\bibinfo {volume} {11}},\
  \bibinfo {pages} {343} (\bibinfo {year} {2011})}\BibitemShut {NoStop}%
\bibitem [{\citenamefont {{Piran}}\ and\ \citenamefont
  {{Jimenez}}(2014)}]{PJ14}%
  \BibitemOpen
  \bibfield  {author} {\bibinfo {author} {\bibfnamefont {T.}~\bibnamefont
  {{Piran}}}\ and\ \bibinfo {author} {\bibfnamefont {R.}~\bibnamefont
  {{Jimenez}}},\ }\href {\doibase 10.1103/PhysRevLett.113.231102} {\bibfield
  {journal} {\bibinfo  {journal} {Phys. Rev. Lett.}\ }\textbf {\bibinfo
  {volume} {113}},\ \bibinfo {eid} {231102} (\bibinfo {year}
  {2014})}\BibitemShut {NoStop}%
\bibitem [{\citenamefont {{Sloan}}\ \emph {et~al.}(2017)\citenamefont
  {{Sloan}}, \citenamefont {{Alves Batista}},\ and\ \citenamefont
  {{Loeb}}}]{SAL}%
  \BibitemOpen
  \bibfield  {author} {\bibinfo {author} {\bibfnamefont {D.}~\bibnamefont
  {{Sloan}}}, \bibinfo {author} {\bibfnamefont {R.}~\bibnamefont {{Alves
  Batista}}}, \ and\ \bibinfo {author} {\bibfnamefont {A.}~\bibnamefont
  {{Loeb}}},\ }\href {\doibase 10.1038/s41598-017-05796-x} {\bibfield
  {journal} {\bibinfo  {journal} {Sci. Rep.}\ }\textbf {\bibinfo {volume}
  {7}},\ \bibinfo {eid} {5419} (\bibinfo {year} {2017})}\BibitemShut {NoStop}%
\bibitem [{\citenamefont {{Balbi}}\ and\ \citenamefont
  {{Tombesi}}(2017)}]{BT17}%
  \BibitemOpen
  \bibfield  {author} {\bibinfo {author} {\bibfnamefont {A.}~\bibnamefont
  {{Balbi}}}\ and\ \bibinfo {author} {\bibfnamefont {F.}~\bibnamefont
  {{Tombesi}}},\ }\href {\doibase 10.1038/s41598-017-16110-0} {\bibfield
  {journal} {\bibinfo  {journal} {Sci. Rep.}\ }\textbf {\bibinfo {volume}
  {7}},\ \bibinfo {eid} {16626} (\bibinfo {year} {2017})}\BibitemShut {NoStop}%
\bibitem [{\citenamefont {{Forbes}}\ and\ \citenamefont {{Loeb}}(2018)}]{FL18}%
  \BibitemOpen
  \bibfield  {author} {\bibinfo {author} {\bibfnamefont {J.~C.}\ \bibnamefont
  {{Forbes}}}\ and\ \bibinfo {author} {\bibfnamefont {A.}~\bibnamefont
  {{Loeb}}},\ }\href {\doibase 10.1093/mnras/sty1433} {\bibfield  {journal}
  {\bibinfo  {journal} {Mon. Not. R. Astron. Soc.}\ }\textbf {\bibinfo {volume}
  {479}},\ \bibinfo {pages} {171} (\bibinfo {year} {2018})}\BibitemShut
  {NoStop}%
\bibitem [{\citenamefont {{Lingam}}\ \emph {et~al.}(2019)\citenamefont
  {{Lingam}}, \citenamefont {{Ginsburg}},\ and\ \citenamefont
  {{Bialy}}}]{LGB19}%
  \BibitemOpen
  \bibfield  {author} {\bibinfo {author} {\bibfnamefont {M.}~\bibnamefont
  {{Lingam}}}, \bibinfo {author} {\bibfnamefont {I.}~\bibnamefont
  {{Ginsburg}}}, \ and\ \bibinfo {author} {\bibfnamefont {S.}~\bibnamefont
  {{Bialy}}},\ }\href {\doibase 10.3847/1538-4357/ab1b2f} {\bibfield  {journal}
  {\bibinfo  {journal} {Astrophys. J.}\ }\textbf {\bibinfo {volume} {877}},\
  \bibinfo {eid} {62} (\bibinfo {year} {2019})}\BibitemShut {NoStop}%
\bibitem [{\citenamefont {{Tarter}}\ \emph {et~al.}(2007)\citenamefont
  {{Tarter}}, \citenamefont {{Backus}}, \citenamefont {{Mancinelli}},
  \citenamefont {{Aurnou}}, \citenamefont {{Backman}}, \citenamefont {{Basri}},
  \citenamefont {{Boss}}, \citenamefont {{Clarke}}, \citenamefont {{Deming}},
  \citenamefont {{Doyle}}, \citenamefont {{Feigelson}}, \citenamefont
  {{Freund}}, \citenamefont {{Grinspoon}}, \citenamefont {{Haberle}},
  \citenamefont {{Hauck}}, \citenamefont {{Heath}}, \citenamefont {{Henry}},
  \citenamefont {{Hollingsworth}}, \citenamefont {{Joshi}}, \citenamefont
  {{Kilston}}, \citenamefont {{Liu}}, \citenamefont {{Meikle}}, \citenamefont
  {{Reid}}, \citenamefont {{Rothschild}}, \citenamefont {{Scalo}},
  \citenamefont {{Segura}}, \citenamefont {{Tang}}, \citenamefont {{Tiedje}},
  \citenamefont {{Turnbull}}, \citenamefont {{Walkowicz}}, \citenamefont
  {{Weber}},\ and\ \citenamefont {{Young}}}]{TB07}%
  \BibitemOpen
  \bibfield  {author} {\bibinfo {author} {\bibfnamefont {J.~C.}\ \bibnamefont
  {{Tarter}}}, \bibinfo {author} {\bibfnamefont {P.~R.}\ \bibnamefont
  {{Backus}}}, \bibinfo {author} {\bibfnamefont {R.~L.}\ \bibnamefont
  {{Mancinelli}}}, \bibinfo {author} {\bibfnamefont {J.~M.}\ \bibnamefont
  {{Aurnou}}}, \bibinfo {author} {\bibfnamefont {D.~E.}\ \bibnamefont
  {{Backman}}}, \bibinfo {author} {\bibfnamefont {G.~S.}\ \bibnamefont
  {{Basri}}}, \bibinfo {author} {\bibfnamefont {A.~P.}\ \bibnamefont {{Boss}}},
  \bibinfo {author} {\bibfnamefont {A.}~\bibnamefont {{Clarke}}}, \bibinfo
  {author} {\bibfnamefont {D.}~\bibnamefont {{Deming}}}, \bibinfo {author}
  {\bibfnamefont {L.~R.}\ \bibnamefont {{Doyle}}}, \bibinfo {author}
  {\bibfnamefont {E.~D.}\ \bibnamefont {{Feigelson}}}, \bibinfo {author}
  {\bibfnamefont {F.}~\bibnamefont {{Freund}}}, \bibinfo {author}
  {\bibfnamefont {D.~H.}\ \bibnamefont {{Grinspoon}}}, \bibinfo {author}
  {\bibfnamefont {R.~M.}\ \bibnamefont {{Haberle}}}, \bibinfo {author}
  {\bibfnamefont {S.~A.}\ \bibnamefont {{Hauck}}, \bibfnamefont {II}}, \bibinfo
  {author} {\bibfnamefont {M.~J.}\ \bibnamefont {{Heath}}}, \bibinfo {author}
  {\bibfnamefont {T.~J.}\ \bibnamefont {{Henry}}}, \bibinfo {author}
  {\bibfnamefont {J.~L.}\ \bibnamefont {{Hollingsworth}}}, \bibinfo {author}
  {\bibfnamefont {M.~M.}\ \bibnamefont {{Joshi}}}, \bibinfo {author}
  {\bibfnamefont {S.}~\bibnamefont {{Kilston}}}, \bibinfo {author}
  {\bibfnamefont {M.~C.}\ \bibnamefont {{Liu}}}, \bibinfo {author}
  {\bibfnamefont {E.}~\bibnamefont {{Meikle}}}, \bibinfo {author}
  {\bibfnamefont {I.~N.}\ \bibnamefont {{Reid}}}, \bibinfo {author}
  {\bibfnamefont {L.~J.}\ \bibnamefont {{Rothschild}}}, \bibinfo {author}
  {\bibfnamefont {J.}~\bibnamefont {{Scalo}}}, \bibinfo {author} {\bibfnamefont
  {A.}~\bibnamefont {{Segura}}}, \bibinfo {author} {\bibfnamefont {C.~M.}\
  \bibnamefont {{Tang}}}, \bibinfo {author} {\bibfnamefont {J.~M.}\
  \bibnamefont {{Tiedje}}}, \bibinfo {author} {\bibfnamefont {M.~C.}\
  \bibnamefont {{Turnbull}}}, \bibinfo {author} {\bibfnamefont {L.~M.}\
  \bibnamefont {{Walkowicz}}}, \bibinfo {author} {\bibfnamefont {A.~L.}\
  \bibnamefont {{Weber}}}, \ and\ \bibinfo {author} {\bibfnamefont {R.~E.}\
  \bibnamefont {{Young}}},\ }\href {\doibase 10.1089/ast.2006.0124} {\bibfield
  {journal} {\bibinfo  {journal} {Astrobiology}\ }\textbf {\bibinfo {volume}
  {7}},\ \bibinfo {pages} {30} (\bibinfo {year} {2007})}\BibitemShut {NoStop}%
\bibitem [{\citenamefont {{Scalo}}\ \emph {et~al.}(2007)\citenamefont
  {{Scalo}}, \citenamefont {{Kaltenegger}}, \citenamefont {{Segura}},
  \citenamefont {{Fridlund}}, \citenamefont {{Ribas}}, \citenamefont
  {{Kulikov}}, \citenamefont {{Grenfell}}, \citenamefont {{Rauer}},
  \citenamefont {{Odert}}, \citenamefont {{Leitzinger}}, \citenamefont
  {{Selsis}}, \citenamefont {{Khodachenko}}, \citenamefont {{Eiroa}},
  \citenamefont {{Kasting}},\ and\ \citenamefont {{Lammer}}}]{SK07}%
  \BibitemOpen
  \bibfield  {author} {\bibinfo {author} {\bibfnamefont {J.}~\bibnamefont
  {{Scalo}}}, \bibinfo {author} {\bibfnamefont {L.}~\bibnamefont
  {{Kaltenegger}}}, \bibinfo {author} {\bibfnamefont {A.~G.}\ \bibnamefont
  {{Segura}}}, \bibinfo {author} {\bibfnamefont {M.}~\bibnamefont
  {{Fridlund}}}, \bibinfo {author} {\bibfnamefont {I.}~\bibnamefont {{Ribas}}},
  \bibinfo {author} {\bibfnamefont {Y.~N.}\ \bibnamefont {{Kulikov}}}, \bibinfo
  {author} {\bibfnamefont {J.~L.}\ \bibnamefont {{Grenfell}}}, \bibinfo
  {author} {\bibfnamefont {H.}~\bibnamefont {{Rauer}}}, \bibinfo {author}
  {\bibfnamefont {P.}~\bibnamefont {{Odert}}}, \bibinfo {author} {\bibfnamefont
  {M.}~\bibnamefont {{Leitzinger}}}, \bibinfo {author} {\bibfnamefont
  {F.}~\bibnamefont {{Selsis}}}, \bibinfo {author} {\bibfnamefont {M.~L.}\
  \bibnamefont {{Khodachenko}}}, \bibinfo {author} {\bibfnamefont
  {C.}~\bibnamefont {{Eiroa}}}, \bibinfo {author} {\bibfnamefont
  {J.}~\bibnamefont {{Kasting}}}, \ and\ \bibinfo {author} {\bibfnamefont
  {H.}~\bibnamefont {{Lammer}}},\ }\href {\doibase 10.1089/ast.2006.0125}
  {\bibfield  {journal} {\bibinfo  {journal} {Astrobiology}\ }\textbf {\bibinfo
  {volume} {7}},\ \bibinfo {pages} {85} (\bibinfo {year} {2007})}\BibitemShut
  {NoStop}%
\bibitem [{\citenamefont {{Shields}}\ \emph {et~al.}(2016)\citenamefont
  {{Shields}}, \citenamefont {{Ballard}},\ and\ \citenamefont
  {{Johnson}}}]{SBJ16}%
  \BibitemOpen
  \bibfield  {author} {\bibinfo {author} {\bibfnamefont {A.~L.}\ \bibnamefont
  {{Shields}}}, \bibinfo {author} {\bibfnamefont {S.}~\bibnamefont
  {{Ballard}}}, \ and\ \bibinfo {author} {\bibfnamefont {J.~A.}\ \bibnamefont
  {{Johnson}}},\ }\href {\doibase 10.1016/j.physrep.2016.10.003} {\bibfield
  {journal} {\bibinfo  {journal} {Phys. Rep.}\ }\textbf {\bibinfo {volume}
  {663}},\ \bibinfo {pages} {1} (\bibinfo {year} {2016})}\BibitemShut {NoStop}%
\bibitem [{\citenamefont {{Chabrier}}\ and\ \citenamefont
  {{Baraffe}}(2000)}]{CB00}%
  \BibitemOpen
  \bibfield  {author} {\bibinfo {author} {\bibfnamefont {G.}~\bibnamefont
  {{Chabrier}}}\ and\ \bibinfo {author} {\bibfnamefont {I.}~\bibnamefont
  {{Baraffe}}},\ }\href {\doibase 10.1146/annurev.astro.38.1.337} {\bibfield
  {journal} {\bibinfo  {journal} {Annu. Rev. Astron. Astrophys.}\ }\textbf
  {\bibinfo {volume} {38}},\ \bibinfo {pages} {337} (\bibinfo {year}
  {2000})}\BibitemShut {NoStop}%
\bibitem [{\citenamefont {{Chabrier}}\ and\ \citenamefont
  {{Baraffe}}(1997)}]{CB97}%
  \BibitemOpen
  \bibfield  {author} {\bibinfo {author} {\bibfnamefont {G.}~\bibnamefont
  {{Chabrier}}}\ and\ \bibinfo {author} {\bibfnamefont {I.}~\bibnamefont
  {{Baraffe}}},\ }\href@noop {} {\bibfield  {journal} {\bibinfo  {journal}
  {Astron. Astrophys.}\ }\textbf {\bibinfo {volume} {327}},\ \bibinfo {pages}
  {1039} (\bibinfo {year} {1997})}\BibitemShut {NoStop}%
\bibitem [{\citenamefont {{Stassun}}\ \emph {et~al.}(2011)\citenamefont
  {{Stassun}}, \citenamefont {{Hebb}}, \citenamefont {{Covey}}, \citenamefont
  {{West}}, \citenamefont {{Irwin}}, \citenamefont {{Jackson}}, \citenamefont
  {{Jardine}}, \citenamefont {{Morin}}, \citenamefont {{Mullan}},\ and\
  \citenamefont {{Reid}}}]{SHC11}%
  \BibitemOpen
  \bibfield  {author} {\bibinfo {author} {\bibfnamefont {K.~G.}\ \bibnamefont
  {{Stassun}}}, \bibinfo {author} {\bibfnamefont {L.}~\bibnamefont {{Hebb}}},
  \bibinfo {author} {\bibfnamefont {K.}~\bibnamefont {{Covey}}}, \bibinfo
  {author} {\bibfnamefont {A.~A.}\ \bibnamefont {{West}}}, \bibinfo {author}
  {\bibfnamefont {J.}~\bibnamefont {{Irwin}}}, \bibinfo {author} {\bibfnamefont
  {R.}~\bibnamefont {{Jackson}}}, \bibinfo {author} {\bibfnamefont
  {M.}~\bibnamefont {{Jardine}}}, \bibinfo {author} {\bibfnamefont
  {J.}~\bibnamefont {{Morin}}}, \bibinfo {author} {\bibfnamefont
  {D.}~\bibnamefont {{Mullan}}}, \ and\ \bibinfo {author} {\bibfnamefont
  {I.~N.}\ \bibnamefont {{Reid}}},\ }in\ \href@noop {} {\emph {\bibinfo
  {booktitle} {16th Cambridge Workshop on Cool Stars, Stellar Systems, and the
  Sun}}},\ \bibinfo {series} {Astronomical Society of the Pacific Conference
  Series}, Vol.\ \bibinfo {volume} {448},\ \bibinfo {editor} {edited by\
  \bibinfo {editor} {\bibfnamefont {C.}~\bibnamefont {{Johns-Krull}}}, \bibinfo
  {editor} {\bibfnamefont {M.~K.}\ \bibnamefont {{Browning}}}, \ and\ \bibinfo
  {editor} {\bibfnamefont {A.~A.}\ \bibnamefont {{West}}}}\ (\bibinfo {year}
  {2011})\ pp.\ \bibinfo {pages} {505--516}\BibitemShut {NoStop}%
\bibitem [{\citenamefont {{Adams}}\ and\ \citenamefont
  {{Laughlin}}(1997)}]{AL97}%
  \BibitemOpen
  \bibfield  {author} {\bibinfo {author} {\bibfnamefont {F.~C.}\ \bibnamefont
  {{Adams}}}\ and\ \bibinfo {author} {\bibfnamefont {G.}~\bibnamefont
  {{Laughlin}}},\ }\href {\doibase 10.1103/RevModPhys.69.337} {\bibfield
  {journal} {\bibinfo  {journal} {Rev. Mod. Phys.}\ }\textbf {\bibinfo {volume}
  {69}},\ \bibinfo {pages} {337} (\bibinfo {year} {1997})}\BibitemShut
  {NoStop}%
\bibitem [{\citenamefont {{Chabrier}}(2003)}]{cha03}%
  \BibitemOpen
  \bibfield  {author} {\bibinfo {author} {\bibfnamefont {G.}~\bibnamefont
  {{Chabrier}}},\ }\href {\doibase 10.1086/376392} {\bibfield  {journal}
  {\bibinfo  {journal} {Publ. Astron. Soc. Pac.}\ }\textbf {\bibinfo {volume}
  {115}},\ \bibinfo {pages} {763} (\bibinfo {year} {2003})}\BibitemShut
  {NoStop}%
\bibitem [{\citenamefont {{Bonfils}}\ \emph {et~al.}(2013)\citenamefont
  {{Bonfils}}, \citenamefont {{Delfosse}}, \citenamefont {{Udry}},
  \citenamefont {{Forveille}}, \citenamefont {{Mayor}}, \citenamefont
  {{Perrier}}, \citenamefont {{Bouchy}}, \citenamefont {{Gillon}},
  \citenamefont {{Lovis}}, \citenamefont {{Pepe}}, \citenamefont {{Queloz}},
  \citenamefont {{Santos}}, \citenamefont {{S{\'e}gransan}},\ and\
  \citenamefont {{Bertaux}}}]{BDU13}%
  \BibitemOpen
  \bibfield  {author} {\bibinfo {author} {\bibfnamefont {X.}~\bibnamefont
  {{Bonfils}}}, \bibinfo {author} {\bibfnamefont {X.}~\bibnamefont
  {{Delfosse}}}, \bibinfo {author} {\bibfnamefont {S.}~\bibnamefont {{Udry}}},
  \bibinfo {author} {\bibfnamefont {T.}~\bibnamefont {{Forveille}}}, \bibinfo
  {author} {\bibfnamefont {M.}~\bibnamefont {{Mayor}}}, \bibinfo {author}
  {\bibfnamefont {C.}~\bibnamefont {{Perrier}}}, \bibinfo {author}
  {\bibfnamefont {F.}~\bibnamefont {{Bouchy}}}, \bibinfo {author}
  {\bibfnamefont {M.}~\bibnamefont {{Gillon}}}, \bibinfo {author}
  {\bibfnamefont {C.}~\bibnamefont {{Lovis}}}, \bibinfo {author} {\bibfnamefont
  {F.}~\bibnamefont {{Pepe}}}, \bibinfo {author} {\bibfnamefont
  {D.}~\bibnamefont {{Queloz}}}, \bibinfo {author} {\bibfnamefont {N.~C.}\
  \bibnamefont {{Santos}}}, \bibinfo {author} {\bibfnamefont {D.}~\bibnamefont
  {{S{\'e}gransan}}}, \ and\ \bibinfo {author} {\bibfnamefont {J.-L.}\
  \bibnamefont {{Bertaux}}},\ }\href {\doibase 10.1051/0004-6361/201014704}
  {\bibfield  {journal} {\bibinfo  {journal} {Astron. Astrophys.}\ }\textbf
  {\bibinfo {volume} {549}},\ \bibinfo {eid} {A109} (\bibinfo {year}
  {2013})}\BibitemShut {NoStop}%
\bibitem [{\citenamefont {{Dressing}}\ and\ \citenamefont
  {{Charbonneau}}(2015)}]{DC15}%
  \BibitemOpen
  \bibfield  {author} {\bibinfo {author} {\bibfnamefont {C.~D.}\ \bibnamefont
  {{Dressing}}}\ and\ \bibinfo {author} {\bibfnamefont {D.}~\bibnamefont
  {{Charbonneau}}},\ }\href {\doibase 10.1088/0004-637X/807/1/45} {\bibfield
  {journal} {\bibinfo  {journal} {Astrophys. J.}\ }\textbf {\bibinfo {volume}
  {807}},\ \bibinfo {eid} {45} (\bibinfo {year} {2015})}\BibitemShut {NoStop}%
\bibitem [{\citenamefont {{Mulders}}\ \emph
  {et~al.}(2015{\natexlab{a}})\citenamefont {{Mulders}}, \citenamefont
  {{Pascucci}},\ and\ \citenamefont {{Apai}}}]{MPA15}%
  \BibitemOpen
  \bibfield  {author} {\bibinfo {author} {\bibfnamefont {G.~D.}\ \bibnamefont
  {{Mulders}}}, \bibinfo {author} {\bibfnamefont {I.}~\bibnamefont
  {{Pascucci}}}, \ and\ \bibinfo {author} {\bibfnamefont {D.}~\bibnamefont
  {{Apai}}},\ }\href {\doibase 10.1088/0004-637X/798/2/112} {\bibfield
  {journal} {\bibinfo  {journal} {Astrophys. J.}\ }\textbf {\bibinfo {volume}
  {798}},\ \bibinfo {eid} {112} (\bibinfo {year}
  {2015}{\natexlab{a}})}\BibitemShut {NoStop}%
\bibitem [{\citenamefont {{Winn}}(2010)}]{Wi10}%
  \BibitemOpen
  \bibfield  {author} {\bibinfo {author} {\bibfnamefont {J.~N.}\ \bibnamefont
  {{Winn}}},\ }\enquote {\bibinfo {title} {{Exoplanet Transits and
  Occultations}},}\ in\ \href@noop {} {\emph {\bibinfo {booktitle}
  {Exoplanets}}},\ \bibinfo {editor} {edited by\ \bibinfo {editor}
  {\bibfnamefont {S.}~\bibnamefont {{Seager}}}}\ (\bibinfo  {publisher}
  {University of Arizona Press},\ \bibinfo {year} {2010})\ pp.\ \bibinfo
  {pages} {55--77}\BibitemShut {NoStop}%
\bibitem [{\citenamefont {{Fujii}}\ \emph {et~al.}(2018)\citenamefont
  {{Fujii}}, \citenamefont {{Angerhausen}}, \citenamefont {{Deitrick}},
  \citenamefont {{Domagal-Goldman}}, \citenamefont {{Grenfell}}, \citenamefont
  {{Hori}}, \citenamefont {{Kane}}, \citenamefont {{Pall{\'e}}}, \citenamefont
  {{Rauer}}, \citenamefont {{Siegler}}, \citenamefont {{Stapelfeldt}},\ and\
  \citenamefont {{Stevenson}}}]{FA18}%
  \BibitemOpen
  \bibfield  {author} {\bibinfo {author} {\bibfnamefont {Y.}~\bibnamefont
  {{Fujii}}}, \bibinfo {author} {\bibfnamefont {D.}~\bibnamefont
  {{Angerhausen}}}, \bibinfo {author} {\bibfnamefont {R.}~\bibnamefont
  {{Deitrick}}}, \bibinfo {author} {\bibfnamefont {S.}~\bibnamefont
  {{Domagal-Goldman}}}, \bibinfo {author} {\bibfnamefont {J.~L.}\ \bibnamefont
  {{Grenfell}}}, \bibinfo {author} {\bibfnamefont {Y.}~\bibnamefont {{Hori}}},
  \bibinfo {author} {\bibfnamefont {S.~R.}\ \bibnamefont {{Kane}}}, \bibinfo
  {author} {\bibfnamefont {E.}~\bibnamefont {{Pall{\'e}}}}, \bibinfo {author}
  {\bibfnamefont {H.}~\bibnamefont {{Rauer}}}, \bibinfo {author} {\bibfnamefont
  {N.}~\bibnamefont {{Siegler}}}, \bibinfo {author} {\bibfnamefont
  {K.}~\bibnamefont {{Stapelfeldt}}}, \ and\ \bibinfo {author} {\bibfnamefont
  {K.~B.}\ \bibnamefont {{Stevenson}}},\ }\href {\doibase
  10.1089/ast.2017.1733} {\bibfield  {journal} {\bibinfo  {journal}
  {Astrobiology}\ }\textbf {\bibinfo {volume} {18}},\ \bibinfo {pages} {739}
  (\bibinfo {year} {2018})}\BibitemShut {NoStop}%
\bibitem [{\citenamefont {{Parker}}(1958)}]{Park58}%
  \BibitemOpen
  \bibfield  {author} {\bibinfo {author} {\bibfnamefont {E.~N.}\ \bibnamefont
  {{Parker}}},\ }\href {\doibase 10.1086/146579} {\bibfield  {journal}
  {\bibinfo  {journal} {Astrophys. J.}\ }\textbf {\bibinfo {volume} {128}},\
  \bibinfo {pages} {664} (\bibinfo {year} {1958})}\BibitemShut {NoStop}%
\bibitem [{\citenamefont {{Priest}}(2014)}]{Pri14}%
  \BibitemOpen
  \bibfield  {author} {\bibinfo {author} {\bibfnamefont {E.}~\bibnamefont
  {{Priest}}},\ }\href@noop {} {\emph {\bibinfo {title} {{Magnetohydrodynamics
  of the Sun}}}}\ (\bibinfo  {publisher} {Cambridge University Press},\
  \bibinfo {year} {2014})\BibitemShut {NoStop}%
\bibitem [{\citenamefont {{Cohen}}\ \emph {et~al.}(2014)\citenamefont
  {{Cohen}}, \citenamefont {{Drake}}, \citenamefont {{Glocer}}, \citenamefont
  {{Garraffo}}, \citenamefont {{Poppenhaeger}}, \citenamefont {{Bell}},
  \citenamefont {{Ridley}},\ and\ \citenamefont {{Gombosi}}}]{CD14}%
  \BibitemOpen
  \bibfield  {author} {\bibinfo {author} {\bibfnamefont {O.}~\bibnamefont
  {{Cohen}}}, \bibinfo {author} {\bibfnamefont {J.~J.}\ \bibnamefont
  {{Drake}}}, \bibinfo {author} {\bibfnamefont {A.}~\bibnamefont {{Glocer}}},
  \bibinfo {author} {\bibfnamefont {C.}~\bibnamefont {{Garraffo}}}, \bibinfo
  {author} {\bibfnamefont {K.}~\bibnamefont {{Poppenhaeger}}}, \bibinfo
  {author} {\bibfnamefont {J.~M.}\ \bibnamefont {{Bell}}}, \bibinfo {author}
  {\bibfnamefont {A.~J.}\ \bibnamefont {{Ridley}}}, \ and\ \bibinfo {author}
  {\bibfnamefont {T.~I.}\ \bibnamefont {{Gombosi}}},\ }\href {\doibase
  10.1088/0004-637X/790/1/57} {\bibfield  {journal} {\bibinfo  {journal}
  {Astrophys. J.}\ }\textbf {\bibinfo {volume} {790}},\ \bibinfo {eid} {57}
  (\bibinfo {year} {2014})}\BibitemShut {NoStop}%
\bibitem [{\citenamefont {{Cohen}}\ \emph {et~al.}(2018)\citenamefont
  {{Cohen}}, \citenamefont {{Glocer}}, \citenamefont {{Garraffo}},
  \citenamefont {{Drake}},\ and\ \citenamefont {{Bell}}}]{CGG18}%
  \BibitemOpen
  \bibfield  {author} {\bibinfo {author} {\bibfnamefont {O.}~\bibnamefont
  {{Cohen}}}, \bibinfo {author} {\bibfnamefont {A.}~\bibnamefont {{Glocer}}},
  \bibinfo {author} {\bibfnamefont {C.}~\bibnamefont {{Garraffo}}}, \bibinfo
  {author} {\bibfnamefont {J.~J.}\ \bibnamefont {{Drake}}}, \ and\ \bibinfo
  {author} {\bibfnamefont {J.~M.}\ \bibnamefont {{Bell}}},\ }\href {\doibase
  10.3847/2041-8213/aab5b5} {\bibfield  {journal} {\bibinfo  {journal}
  {Astrophys. J. Lett.}\ }\textbf {\bibinfo {volume} {856}},\ \bibinfo {eid}
  {L11} (\bibinfo {year} {2018})}\BibitemShut {NoStop}%
\bibitem [{\citenamefont {{Baumjohann}}\ and\ \citenamefont
  {{Treumann}}(2012)}]{BT12}%
  \BibitemOpen
  \bibfield  {author} {\bibinfo {author} {\bibfnamefont {W.}~\bibnamefont
  {{Baumjohann}}}\ and\ \bibinfo {author} {\bibfnamefont {R.~A.}\ \bibnamefont
  {{Treumann}}},\ }\href {\doibase 10.1142/p850} {\emph {\bibinfo {title}
  {{Basic Space Plasma Physics}}}}\ (\bibinfo  {publisher} {Imperial College
  Press},\ \bibinfo {year} {2012})\BibitemShut {NoStop}%
\bibitem [{\citenamefont {{Gombosi}}(1998)}]{Gom98}%
  \BibitemOpen
  \bibfield  {author} {\bibinfo {author} {\bibfnamefont {T.~I.}\ \bibnamefont
  {{Gombosi}}},\ }\href@noop {} {\emph {\bibinfo {title} {{Physics of the Space
  Environment}}}},\ Cambridge Atmospheric and Space Science Series\ (\bibinfo
  {publisher} {Cambridge University Press},\ \bibinfo {year}
  {1998})\BibitemShut {NoStop}%
\bibitem [{\citenamefont {{Kivelson}}\ and\ \citenamefont
  {{Russell}}(1995)}]{KR95}%
  \BibitemOpen
  \bibfield  {author} {\bibinfo {author} {\bibfnamefont {M.~G.}\ \bibnamefont
  {{Kivelson}}}\ and\ \bibinfo {author} {\bibfnamefont {C.~T.}\ \bibnamefont
  {{Russell}}},\ }\href@noop {} {\emph {\bibinfo {title} {{Introduction to
  Space Physics}}}}\ (\bibinfo  {publisher} {Cambridge University Press},\
  \bibinfo {year} {1995})\BibitemShut {NoStop}%
\bibitem [{\citenamefont {{Garraffo}}\ \emph {et~al.}(2016)\citenamefont
  {{Garraffo}}, \citenamefont {{Drake}},\ and\ \citenamefont
  {{Cohen}}}]{GDC16}%
  \BibitemOpen
  \bibfield  {author} {\bibinfo {author} {\bibfnamefont {C.}~\bibnamefont
  {{Garraffo}}}, \bibinfo {author} {\bibfnamefont {J.~J.}\ \bibnamefont
  {{Drake}}}, \ and\ \bibinfo {author} {\bibfnamefont {O.}~\bibnamefont
  {{Cohen}}},\ }\href {\doibase 10.3847/2041-8205/833/1/L4} {\bibfield
  {journal} {\bibinfo  {journal} {Astrophys. J. Lett.}\ }\textbf {\bibinfo
  {volume} {833}},\ \bibinfo {eid} {L4} (\bibinfo {year} {2016})}\BibitemShut
  {NoStop}%
\bibitem [{\citenamefont {{Garraffo}}\ \emph {et~al.}(2017)\citenamefont
  {{Garraffo}}, \citenamefont {{Drake}}, \citenamefont {{Cohen}}, \citenamefont
  {{Alvarado-G{\'o}mez}},\ and\ \citenamefont {{Moschou}}}]{GDC17}%
  \BibitemOpen
  \bibfield  {author} {\bibinfo {author} {\bibfnamefont {C.}~\bibnamefont
  {{Garraffo}}}, \bibinfo {author} {\bibfnamefont {J.~J.}\ \bibnamefont
  {{Drake}}}, \bibinfo {author} {\bibfnamefont {O.}~\bibnamefont {{Cohen}}},
  \bibinfo {author} {\bibfnamefont {J.~D.}\ \bibnamefont
  {{Alvarado-G{\'o}mez}}}, \ and\ \bibinfo {author} {\bibfnamefont {S.~P.}\
  \bibnamefont {{Moschou}}},\ }\href {\doibase 10.3847/2041-8213/aa79ed}
  {\bibfield  {journal} {\bibinfo  {journal} {Astrophys. J. Lett.}\ }\textbf
  {\bibinfo {volume} {843}},\ \bibinfo {eid} {L33} (\bibinfo {year}
  {2017})}\BibitemShut {NoStop}%
\bibitem [{\citenamefont {{Dong}}\ \emph
  {et~al.}(2018{\natexlab{b}})\citenamefont {{Dong}}, \citenamefont {{Jin}},
  \citenamefont {{Lingam}}, \citenamefont {{Airapetian}}, \citenamefont
  {{Ma}},\ and\ \citenamefont {{van der Holst}}}]{DJL18}%
  \BibitemOpen
  \bibfield  {author} {\bibinfo {author} {\bibfnamefont {C.}~\bibnamefont
  {{Dong}}}, \bibinfo {author} {\bibfnamefont {M.}~\bibnamefont {{Jin}}},
  \bibinfo {author} {\bibfnamefont {M.}~\bibnamefont {{Lingam}}}, \bibinfo
  {author} {\bibfnamefont {V.~S.}\ \bibnamefont {{Airapetian}}}, \bibinfo
  {author} {\bibfnamefont {Y.}~\bibnamefont {{Ma}}}, \ and\ \bibinfo {author}
  {\bibfnamefont {B.}~\bibnamefont {{van der Holst}}},\ }\href {\doibase
  10.1073/pnas.1708010115} {\bibfield  {journal} {\bibinfo  {journal} {Proc.
  Natl. Acad. Sci. USA}\ }\textbf {\bibinfo {volume} {115}},\ \bibinfo {pages}
  {260} (\bibinfo {year} {2018}{\natexlab{b}})}\BibitemShut {NoStop}%
\bibitem [{\citenamefont {{Fitzpatrick}}(2014)}]{Fitz14}%
  \BibitemOpen
  \bibfield  {author} {\bibinfo {author} {\bibfnamefont {R.}~\bibnamefont
  {{Fitzpatrick}}},\ }\href@noop {} {\emph {\bibinfo {title} {{Plasma Physics:
  An Introduction}}}}\ (\bibinfo  {publisher} {CRC Press},\ \bibinfo {year}
  {2014})\BibitemShut {NoStop}%
\bibitem [{\citenamefont {{Wargelin}}\ and\ \citenamefont
  {{Drake}}(2002)}]{WD02}%
  \BibitemOpen
  \bibfield  {author} {\bibinfo {author} {\bibfnamefont {B.~J.}\ \bibnamefont
  {{Wargelin}}}\ and\ \bibinfo {author} {\bibfnamefont {J.~J.}\ \bibnamefont
  {{Drake}}},\ }\href {\doibase 10.1086/342270} {\bibfield  {journal} {\bibinfo
   {journal} {Astrophys. J.}\ }\textbf {\bibinfo {volume} {578}},\ \bibinfo
  {pages} {503} (\bibinfo {year} {2002})}\BibitemShut {NoStop}%
\bibitem [{\citenamefont {{Wood}}\ \emph {et~al.}(2001)\citenamefont {{Wood}},
  \citenamefont {{Linsky}}, \citenamefont {{M{\"u}ller}},\ and\ \citenamefont
  {{Zank}}}]{WL01}%
  \BibitemOpen
  \bibfield  {author} {\bibinfo {author} {\bibfnamefont {B.~E.}\ \bibnamefont
  {{Wood}}}, \bibinfo {author} {\bibfnamefont {J.~L.}\ \bibnamefont
  {{Linsky}}}, \bibinfo {author} {\bibfnamefont {H.-R.}\ \bibnamefont
  {{M{\"u}ller}}}, \ and\ \bibinfo {author} {\bibfnamefont {G.~P.}\
  \bibnamefont {{Zank}}},\ }\href {\doibase 10.1086/318888} {\bibfield
  {journal} {\bibinfo  {journal} {Astrophys. J. Lett.}\ }\textbf {\bibinfo
  {volume} {547}},\ \bibinfo {pages} {L49} (\bibinfo {year}
  {2001})}\BibitemShut {NoStop}%
\bibitem [{\citenamefont {{Dong}}\ \emph
  {et~al.}(2017{\natexlab{a}})\citenamefont {{Dong}}, \citenamefont {{Lingam}},
  \citenamefont {{Ma}},\ and\ \citenamefont {{Cohen}}}]{DLMC17}%
  \BibitemOpen
  \bibfield  {author} {\bibinfo {author} {\bibfnamefont {C.}~\bibnamefont
  {{Dong}}}, \bibinfo {author} {\bibfnamefont {M.}~\bibnamefont {{Lingam}}},
  \bibinfo {author} {\bibfnamefont {Y.}~\bibnamefont {{Ma}}}, \ and\ \bibinfo
  {author} {\bibfnamefont {O.}~\bibnamefont {{Cohen}}},\ }\href {\doibase
  10.3847/2041-8213/aa6438} {\bibfield  {journal} {\bibinfo  {journal}
  {Astrophys. J. Lett.}\ }\textbf {\bibinfo {volume} {837}},\ \bibinfo {eid}
  {L26} (\bibinfo {year} {2017}{\natexlab{a}})}\BibitemShut {NoStop}%
\bibitem [{\citenamefont {{Reiners}}\ and\ \citenamefont
  {{Basri}}(2009)}]{RB09}%
  \BibitemOpen
  \bibfield  {author} {\bibinfo {author} {\bibfnamefont {A.}~\bibnamefont
  {{Reiners}}}\ and\ \bibinfo {author} {\bibfnamefont {G.}~\bibnamefont
  {{Basri}}},\ }\href {\doibase 10.1051/0004-6361:200811450} {\bibfield
  {journal} {\bibinfo  {journal} {Astron. Astrophys.}\ }\textbf {\bibinfo
  {volume} {496}},\ \bibinfo {pages} {787} (\bibinfo {year}
  {2009})}\BibitemShut {NoStop}%
\bibitem [{\citenamefont {{Morin}}\ \emph {et~al.}(2010)\citenamefont
  {{Morin}}, \citenamefont {{Donati}}, \citenamefont {{Petit}}, \citenamefont
  {{Delfosse}}, \citenamefont {{Forveille}},\ and\ \citenamefont
  {{Jardine}}}]{MD10}%
  \BibitemOpen
  \bibfield  {author} {\bibinfo {author} {\bibfnamefont {J.}~\bibnamefont
  {{Morin}}}, \bibinfo {author} {\bibfnamefont {J.-F.}\ \bibnamefont
  {{Donati}}}, \bibinfo {author} {\bibfnamefont {P.}~\bibnamefont {{Petit}}},
  \bibinfo {author} {\bibfnamefont {X.}~\bibnamefont {{Delfosse}}}, \bibinfo
  {author} {\bibfnamefont {T.}~\bibnamefont {{Forveille}}}, \ and\ \bibinfo
  {author} {\bibfnamefont {M.~M.}\ \bibnamefont {{Jardine}}},\ }\href {\doibase
  10.1111/j.1365-2966.2010.17101.x} {\bibfield  {journal} {\bibinfo  {journal}
  {Mon. Not. R. Astron. Soc.}\ }\textbf {\bibinfo {volume} {407}},\ \bibinfo
  {pages} {2269} (\bibinfo {year} {2010})}\BibitemShut {NoStop}%
\bibitem [{\citenamefont {{Reiners}}(2012)}]{Rei12}%
  \BibitemOpen
  \bibfield  {author} {\bibinfo {author} {\bibfnamefont {A.}~\bibnamefont
  {{Reiners}}},\ }\href {\doibase 10.12942/lrsp-2012-1} {\bibfield  {journal}
  {\bibinfo  {journal} {Living Rev. Sol. Phys.}\ }\textbf {\bibinfo {volume}
  {9}},\ \bibinfo {eid} {1} (\bibinfo {year} {2012})}\BibitemShut {NoStop}%
\bibitem [{\citenamefont {{Vidotto}}\ \emph {et~al.}(2013)\citenamefont
  {{Vidotto}}, \citenamefont {{Jardine}}, \citenamefont {{Morin}},
  \citenamefont {{Donati}}, \citenamefont {{Lang}},\ and\ \citenamefont
  {{Russell}}}]{VJM13}%
  \BibitemOpen
  \bibfield  {author} {\bibinfo {author} {\bibfnamefont {A.~A.}\ \bibnamefont
  {{Vidotto}}}, \bibinfo {author} {\bibfnamefont {M.}~\bibnamefont
  {{Jardine}}}, \bibinfo {author} {\bibfnamefont {J.}~\bibnamefont {{Morin}}},
  \bibinfo {author} {\bibfnamefont {J.-F.}\ \bibnamefont {{Donati}}}, \bibinfo
  {author} {\bibfnamefont {P.}~\bibnamefont {{Lang}}}, \ and\ \bibinfo {author}
  {\bibfnamefont {A.~J.~B.}\ \bibnamefont {{Russell}}},\ }\href {\doibase
  10.1051/0004-6361/201321504} {\bibfield  {journal} {\bibinfo  {journal}
  {Astron. Astrophys.}\ }\textbf {\bibinfo {volume} {557}},\ \bibinfo {eid}
  {A67} (\bibinfo {year} {2013})}\BibitemShut {NoStop}%
\bibitem [{\citenamefont {{See}}\ \emph {et~al.}(2014)\citenamefont {{See}},
  \citenamefont {{Jardine}}, \citenamefont {{Vidotto}}, \citenamefont
  {{Petit}}, \citenamefont {{Marsden}}, \citenamefont {{Jeffers}},\ and\
  \citenamefont {{do Nascimento}}}]{SJV14}%
  \BibitemOpen
  \bibfield  {author} {\bibinfo {author} {\bibfnamefont {V.}~\bibnamefont
  {{See}}}, \bibinfo {author} {\bibfnamefont {M.}~\bibnamefont {{Jardine}}},
  \bibinfo {author} {\bibfnamefont {A.~A.}\ \bibnamefont {{Vidotto}}}, \bibinfo
  {author} {\bibfnamefont {P.}~\bibnamefont {{Petit}}}, \bibinfo {author}
  {\bibfnamefont {S.~C.}\ \bibnamefont {{Marsden}}}, \bibinfo {author}
  {\bibfnamefont {S.~V.}\ \bibnamefont {{Jeffers}}}, \ and\ \bibinfo {author}
  {\bibfnamefont {J.~D.}\ \bibnamefont {{do Nascimento}}},\ }\href {\doibase
  10.1051/0004-6361/201424323} {\bibfield  {journal} {\bibinfo  {journal}
  {Astron. Astrophys.}\ }\textbf {\bibinfo {volume} {570}},\ \bibinfo {eid}
  {A99} (\bibinfo {year} {2014})}\BibitemShut {NoStop}%
\bibitem [{\citenamefont {{Grie{\ss}meier}}\ \emph {et~al.}(2005)\citenamefont
  {{Grie{\ss}meier}}, \citenamefont {{Stadelmann}}, \citenamefont
  {{Motschmann}}, \citenamefont {{Belisheva}}, \citenamefont {{Lammer}},\ and\
  \citenamefont {{Biernat}}}]{GSM05}%
  \BibitemOpen
  \bibfield  {author} {\bibinfo {author} {\bibfnamefont {J.-M.}\ \bibnamefont
  {{Grie{\ss}meier}}}, \bibinfo {author} {\bibfnamefont {A.}~\bibnamefont
  {{Stadelmann}}}, \bibinfo {author} {\bibfnamefont {U.}~\bibnamefont
  {{Motschmann}}}, \bibinfo {author} {\bibfnamefont {N.~K.}\ \bibnamefont
  {{Belisheva}}}, \bibinfo {author} {\bibfnamefont {H.}~\bibnamefont
  {{Lammer}}}, \ and\ \bibinfo {author} {\bibfnamefont {H.~K.}\ \bibnamefont
  {{Biernat}}},\ }\href {\doibase 10.1089/ast.2005.5.587} {\bibfield  {journal}
  {\bibinfo  {journal} {Astrobiology}\ }\textbf {\bibinfo {volume} {5}},\
  \bibinfo {pages} {587} (\bibinfo {year} {2005})}\BibitemShut {NoStop}%
\bibitem [{\citenamefont {{L{\'o}pez-Morales}}\ \emph
  {et~al.}(2011)\citenamefont {{L{\'o}pez-Morales}}, \citenamefont
  {{G{\'o}mez-P{\'e}rez}},\ and\ \citenamefont {{Ruedas}}}]{LG11}%
  \BibitemOpen
  \bibfield  {author} {\bibinfo {author} {\bibfnamefont {M.}~\bibnamefont
  {{L{\'o}pez-Morales}}}, \bibinfo {author} {\bibfnamefont {N.}~\bibnamefont
  {{G{\'o}mez-P{\'e}rez}}}, \ and\ \bibinfo {author} {\bibfnamefont
  {T.}~\bibnamefont {{Ruedas}}},\ }\href {\doibase 10.1007/s11084-012-9263-8}
  {\bibfield  {journal} {\bibinfo  {journal} {Orig. Life Evol. Biosph.}\
  }\textbf {\bibinfo {volume} {41}},\ \bibinfo {pages} {533} (\bibinfo {year}
  {2011})}\BibitemShut {NoStop}%
\bibitem [{\citenamefont {{Christensen}}(2010)}]{Ch10}%
  \BibitemOpen
  \bibfield  {author} {\bibinfo {author} {\bibfnamefont {U.~R.}\ \bibnamefont
  {{Christensen}}},\ }\href {\doibase 10.1007/s11214-009-9553-2} {\bibfield
  {journal} {\bibinfo  {journal} {Space Sci. Rev.}\ }\textbf {\bibinfo {volume}
  {152}},\ \bibinfo {pages} {565} (\bibinfo {year} {2010})}\BibitemShut
  {NoStop}%
\bibitem [{\citenamefont {{Grie{\ss}meier}}(2015)}]{Grie15}%
  \BibitemOpen
  \bibfield  {author} {\bibinfo {author} {\bibfnamefont {J.-M.}\ \bibnamefont
  {{Grie{\ss}meier}}},\ }in\ \href {\doibase 10.1007/978-3-319-09749-7_11}
  {\emph {\bibinfo {booktitle} {Characterizing Stellar and Exoplanetary
  Environments}}},\ \bibinfo {series} {Astrophysics and Space Science Library},
  Vol.\ \bibinfo {volume} {411},\ \bibinfo {editor} {edited by\ \bibinfo
  {editor} {\bibfnamefont {H.}~\bibnamefont {{Lammer}}}\ and\ \bibinfo {editor}
  {\bibfnamefont {M.}~\bibnamefont {{Khodachenko}}}}\ (\bibinfo {year} {2015})\
  pp.\ \bibinfo {pages} {213--237}\BibitemShut {NoStop}%
\bibitem [{\citenamefont {{Leconte}}\ \emph {et~al.}(2015)\citenamefont
  {{Leconte}}, \citenamefont {{Wu}}, \citenamefont {{Menou}},\ and\
  \citenamefont {{Murray}}}]{LWMM15}%
  \BibitemOpen
  \bibfield  {author} {\bibinfo {author} {\bibfnamefont {J.}~\bibnamefont
  {{Leconte}}}, \bibinfo {author} {\bibfnamefont {H.}~\bibnamefont {{Wu}}},
  \bibinfo {author} {\bibfnamefont {K.}~\bibnamefont {{Menou}}}, \ and\
  \bibinfo {author} {\bibfnamefont {N.}~\bibnamefont {{Murray}}},\ }\href
  {\doibase 10.1126/science.1258686} {\bibfield  {journal} {\bibinfo  {journal}
  {Science}\ }\textbf {\bibinfo {volume} {347}},\ \bibinfo {pages} {632}
  (\bibinfo {year} {2015})}\BibitemShut {NoStop}%
\bibitem [{\citenamefont {{Makarov}}(2015)}]{Mak15}%
  \BibitemOpen
  \bibfield  {author} {\bibinfo {author} {\bibfnamefont {V.~V.}\ \bibnamefont
  {{Makarov}}},\ }\href {\doibase 10.1088/0004-637X/810/1/12} {\bibfield
  {journal} {\bibinfo  {journal} {Astrophys. J.}\ }\textbf {\bibinfo {volume}
  {810}},\ \bibinfo {eid} {12} (\bibinfo {year} {2015})}\BibitemShut {NoStop}%
\bibitem [{\citenamefont {{Vinson}}\ and\ \citenamefont
  {{Hansen}}(2017)}]{VH17}%
  \BibitemOpen
  \bibfield  {author} {\bibinfo {author} {\bibfnamefont {A.~M.}\ \bibnamefont
  {{Vinson}}}\ and\ \bibinfo {author} {\bibfnamefont {B.~M.~S.}\ \bibnamefont
  {{Hansen}}},\ }\href {\doibase 10.1093/mnras/stx2100} {\bibfield  {journal}
  {\bibinfo  {journal} {Mon. Not. R. Astron. Soc.}\ }\textbf {\bibinfo {volume}
  {472}},\ \bibinfo {pages} {3217} (\bibinfo {year} {2017})}\BibitemShut
  {NoStop}%
\bibitem [{\citenamefont {{Bolmont}}\ \emph {et~al.}(2011)\citenamefont
  {{Bolmont}}, \citenamefont {{Raymond}},\ and\ \citenamefont
  {{Leconte}}}]{BRL11}%
  \BibitemOpen
  \bibfield  {author} {\bibinfo {author} {\bibfnamefont {E.}~\bibnamefont
  {{Bolmont}}}, \bibinfo {author} {\bibfnamefont {S.~N.}\ \bibnamefont
  {{Raymond}}}, \ and\ \bibinfo {author} {\bibfnamefont {J.}~\bibnamefont
  {{Leconte}}},\ }\href {\doibase 10.1051/0004-6361/201117734} {\bibfield
  {journal} {\bibinfo  {journal} {Astron. Astrophys.}\ }\textbf {\bibinfo
  {volume} {535}},\ \bibinfo {eid} {A94} (\bibinfo {year} {2011})}\BibitemShut
  {NoStop}%
\bibitem [{\citenamefont {{Barnes}}(2017)}]{Bar17}%
  \BibitemOpen
  \bibfield  {author} {\bibinfo {author} {\bibfnamefont {R.}~\bibnamefont
  {{Barnes}}},\ }\href {\doibase 10.1007/s10569-017-9783-7} {\bibfield
  {journal} {\bibinfo  {journal} {Celest. Mech. Dyn. Astron.}\ }\textbf
  {\bibinfo {volume} {129}},\ \bibinfo {pages} {509} (\bibinfo {year}
  {2017})}\BibitemShut {NoStop}%
\bibitem [{\citenamefont {{Zuluaga}}\ \emph {et~al.}(2013)\citenamefont
  {{Zuluaga}}, \citenamefont {{Bustamante}}, \citenamefont {{Cuartas}},\ and\
  \citenamefont {{Hoyos}}}]{ZBC13}%
  \BibitemOpen
  \bibfield  {author} {\bibinfo {author} {\bibfnamefont {J.~I.}\ \bibnamefont
  {{Zuluaga}}}, \bibinfo {author} {\bibfnamefont {S.}~\bibnamefont
  {{Bustamante}}}, \bibinfo {author} {\bibfnamefont {P.~A.}\ \bibnamefont
  {{Cuartas}}}, \ and\ \bibinfo {author} {\bibfnamefont {J.~H.}\ \bibnamefont
  {{Hoyos}}},\ }\href {\doibase 10.1088/0004-637X/770/1/23} {\bibfield
  {journal} {\bibinfo  {journal} {Astrophys. J.}\ }\textbf {\bibinfo {volume}
  {770}},\ \bibinfo {eid} {23} (\bibinfo {year} {2013})}\BibitemShut {NoStop}%
\bibitem [{\citenamefont {{Dong}}\ \emph
  {et~al.}(2018{\natexlab{c}})\citenamefont {{Dong}}, \citenamefont
  {{Luhmann}}, \citenamefont {{Ma}}, \citenamefont {{Lingam}}, \citenamefont
  {{Bhattacharjee}}, \citenamefont {{Lee}}, \citenamefont {{Bougher}},
  \citenamefont {{Wang}}, \citenamefont {{Glocer}}, \citenamefont
  {{Strangeway}}, \citenamefont {{Curry}}, \citenamefont {{Fang}},
  \citenamefont {{Toth}}, \citenamefont {{Nagy}}, \citenamefont {{Lillis}},
  \citenamefont {{Mitchell}}, \citenamefont {{Brain}},\ and\ \citenamefont
  {{Jakosky}}}]{DLML18}%
  \BibitemOpen
  \bibfield  {author} {\bibinfo {author} {\bibfnamefont {C.}~\bibnamefont
  {{Dong}}}, \bibinfo {author} {\bibfnamefont {J.~G.}\ \bibnamefont
  {{Luhmann}}}, \bibinfo {author} {\bibfnamefont {Y.}~\bibnamefont {{Ma}}},
  \bibinfo {author} {\bibfnamefont {M.}~\bibnamefont {{Lingam}}}, \bibinfo
  {author} {\bibfnamefont {A.}~\bibnamefont {{Bhattacharjee}}}, \bibinfo
  {author} {\bibfnamefont {Y.}~\bibnamefont {{Lee}}}, \bibinfo {author}
  {\bibfnamefont {S.~W.}\ \bibnamefont {{Bougher}}}, \bibinfo {author}
  {\bibfnamefont {L.}~\bibnamefont {{Wang}}}, \bibinfo {author} {\bibfnamefont
  {A.}~\bibnamefont {{Glocer}}}, \bibinfo {author} {\bibfnamefont {R.~J.}\
  \bibnamefont {{Strangeway}}}, \bibinfo {author} {\bibfnamefont
  {S.}~\bibnamefont {{Curry}}}, \bibinfo {author} {\bibfnamefont
  {X.}~\bibnamefont {{Fang}}}, \bibinfo {author} {\bibfnamefont
  {G.}~\bibnamefont {{Toth}}}, \bibinfo {author} {\bibfnamefont {A.~F.}\
  \bibnamefont {{Nagy}}}, \bibinfo {author} {\bibfnamefont {R.~J.}\
  \bibnamefont {{Lillis}}}, \bibinfo {author} {\bibfnamefont {D.~L.}\
  \bibnamefont {{Mitchell}}}, \bibinfo {author} {\bibfnamefont
  {D.}~\bibnamefont {{Brain}}}, \ and\ \bibinfo {author} {\bibfnamefont
  {B.~M.}\ \bibnamefont {{Jakosky}}},\ }\href@noop {} {\bibfield  {journal}
  {\bibinfo  {journal} {AGU Fall Meeting Abstracts}\ }\textbf {\bibinfo
  {volume} {P31C-3736}} (\bibinfo {year} {2018}{\natexlab{c}})}\BibitemShut
  {NoStop}%
\bibitem [{\citenamefont {{Blackman}}\ and\ \citenamefont
  {{Tarduno}}(2018)}]{BT18}%
  \BibitemOpen
  \bibfield  {author} {\bibinfo {author} {\bibfnamefont {E.~G.}\ \bibnamefont
  {{Blackman}}}\ and\ \bibinfo {author} {\bibfnamefont {J.~A.}\ \bibnamefont
  {{Tarduno}}},\ }\href {\doibase 10.1093/mnras/sty2640} {\bibfield  {journal}
  {\bibinfo  {journal} {Mon. Not. R. Astron. Soc.}\ }\textbf {\bibinfo {volume}
  {481}},\ \bibinfo {pages} {5146} (\bibinfo {year} {2018})}\BibitemShut
  {NoStop}%
\bibitem [{\citenamefont {{Gunell}}\ \emph {et~al.}(2018)\citenamefont
  {{Gunell}}, \citenamefont {{Maggiolo}}, \citenamefont {{Nilsson}},
  \citenamefont {{Stenberg Wieser}}, \citenamefont {{Slapak}}, \citenamefont
  {{Lindkvist}}, \citenamefont {{Hamrin}},\ and\ \citenamefont {{De
  Keyser}}}]{GMN18}%
  \BibitemOpen
  \bibfield  {author} {\bibinfo {author} {\bibfnamefont {H.}~\bibnamefont
  {{Gunell}}}, \bibinfo {author} {\bibfnamefont {R.}~\bibnamefont
  {{Maggiolo}}}, \bibinfo {author} {\bibfnamefont {H.}~\bibnamefont
  {{Nilsson}}}, \bibinfo {author} {\bibfnamefont {G.}~\bibnamefont {{Stenberg
  Wieser}}}, \bibinfo {author} {\bibfnamefont {R.}~\bibnamefont {{Slapak}}},
  \bibinfo {author} {\bibfnamefont {J.}~\bibnamefont {{Lindkvist}}}, \bibinfo
  {author} {\bibfnamefont {M.}~\bibnamefont {{Hamrin}}}, \ and\ \bibinfo
  {author} {\bibfnamefont {J.}~\bibnamefont {{De Keyser}}},\ }\href {\doibase
  10.1051/0004-6361/201832934} {\bibfield  {journal} {\bibinfo  {journal}
  {Astron. Astrophys.}\ }\textbf {\bibinfo {volume} {614}},\ \bibinfo {eid}
  {L3} (\bibinfo {year} {2018})}\BibitemShut {NoStop}%
\bibitem [{\citenamefont {{Lingam}}(2019)}]{Ling19}%
  \BibitemOpen
  \bibfield  {author} {\bibinfo {author} {\bibfnamefont {M.}~\bibnamefont
  {{Lingam}}},\ }\href {\doibase 10.3847/2041-8213/ab12eb} {\bibfield
  {journal} {\bibinfo  {journal} {Astrophys. J. Lett.}\ }\textbf {\bibinfo
  {volume} {874}},\ \bibinfo {eid} {L28} (\bibinfo {year} {2019})}\BibitemShut
  {NoStop}%
\bibitem [{\citenamefont {{Axford}}(1968)}]{Ax68}%
  \BibitemOpen
  \bibfield  {author} {\bibinfo {author} {\bibfnamefont {W.~I.}\ \bibnamefont
  {{Axford}}},\ }\href {\doibase 10.1029/JA073i021p06855} {\bibfield  {journal}
  {\bibinfo  {journal} {J. Geophys. Res.}\ }\textbf {\bibinfo {volume} {73}},\
  \bibinfo {pages} {6855} (\bibinfo {year} {1968})}\BibitemShut {NoStop}%
\bibitem [{\citenamefont {{Schunk}}\ and\ \citenamefont
  {{Nagy}}(2009)}]{schunk2009}%
  \BibitemOpen
  \bibfield  {author} {\bibinfo {author} {\bibfnamefont {R.}~\bibnamefont
  {{Schunk}}}\ and\ \bibinfo {author} {\bibfnamefont {A.}~\bibnamefont
  {{Nagy}}},\ }\href {\doibase 10.1017/CBO9780511635342} {\emph {\bibinfo
  {title} {{Ionospheres: Physics, Plasma Physics, and Chemistry}}}},\ Cambridge
  Atmospheric and Space Science Series\ (\bibinfo  {publisher} {Cambridge Univ.
  Press},\ \bibinfo {year} {2009})\BibitemShut {NoStop}%
\bibitem [{\citenamefont {{Sakai}}\ \emph {et~al.}(2018)\citenamefont
  {{Sakai}}, \citenamefont {{Seki}}, \citenamefont {{Terada}}, \citenamefont
  {{Shinagawa}}, \citenamefont {{Tanaka}},\ and\ \citenamefont
  {{Ebihara}}}]{SS18}%
  \BibitemOpen
  \bibfield  {author} {\bibinfo {author} {\bibfnamefont {S.}~\bibnamefont
  {{Sakai}}}, \bibinfo {author} {\bibfnamefont {K.}~\bibnamefont {{Seki}}},
  \bibinfo {author} {\bibfnamefont {N.}~\bibnamefont {{Terada}}}, \bibinfo
  {author} {\bibfnamefont {H.}~\bibnamefont {{Shinagawa}}}, \bibinfo {author}
  {\bibfnamefont {T.}~\bibnamefont {{Tanaka}}}, \ and\ \bibinfo {author}
  {\bibfnamefont {Y.}~\bibnamefont {{Ebihara}}},\ }\href {\doibase
  10.1029/2018GL079972} {\bibfield  {journal} {\bibinfo  {journal} {Geophys.
  Res. Lett.}\ }\textbf {\bibinfo {volume} {45}},\ \bibinfo {pages} {9336}
  (\bibinfo {year} {2018})}\BibitemShut {NoStop}%
\bibitem [{\citenamefont {{Dartnell}}(2011)}]{Dart11}%
  \BibitemOpen
  \bibfield  {author} {\bibinfo {author} {\bibfnamefont {L.~R.}\ \bibnamefont
  {{Dartnell}}},\ }\href {\doibase 10.1089/ast.2010.0528} {\bibfield  {journal}
  {\bibinfo  {journal} {Astrobiology}\ }\textbf {\bibinfo {volume} {11}},\
  \bibinfo {pages} {551} (\bibinfo {year} {2011})}\BibitemShut {NoStop}%
\bibitem [{\citenamefont {{Horvath}}\ and\ \citenamefont
  {{Galante}}(2012)}]{HG12}%
  \BibitemOpen
  \bibfield  {author} {\bibinfo {author} {\bibfnamefont {J.~E.}\ \bibnamefont
  {{Horvath}}}\ and\ \bibinfo {author} {\bibfnamefont {D.}~\bibnamefont
  {{Galante}}},\ }\href {\doibase 10.1017/S1473550412000304} {\bibfield
  {journal} {\bibinfo  {journal} {Int. J. Astrobiol.}\ }\textbf {\bibinfo
  {volume} {11}},\ \bibinfo {pages} {279} (\bibinfo {year} {2012})}\BibitemShut
  {NoStop}%
\bibitem [{\citenamefont {{Grie{\ss}meier}}\ \emph {et~al.}(2009)\citenamefont
  {{Grie{\ss}meier}}, \citenamefont {{Stadelmann}}, \citenamefont {{Grenfell}},
  \citenamefont {{Lammer}},\ and\ \citenamefont {{Motschmann}}}]{GSG09}%
  \BibitemOpen
  \bibfield  {author} {\bibinfo {author} {\bibfnamefont {J.-M.}\ \bibnamefont
  {{Grie{\ss}meier}}}, \bibinfo {author} {\bibfnamefont {A.}~\bibnamefont
  {{Stadelmann}}}, \bibinfo {author} {\bibfnamefont {J.~L.}\ \bibnamefont
  {{Grenfell}}}, \bibinfo {author} {\bibfnamefont {H.}~\bibnamefont
  {{Lammer}}}, \ and\ \bibinfo {author} {\bibfnamefont {U.}~\bibnamefont
  {{Motschmann}}},\ }\href {\doibase 10.1016/j.icarus.2008.09.015} {\bibfield
  {journal} {\bibinfo  {journal} {Icarus}\ }\textbf {\bibinfo {volume} {199}},\
  \bibinfo {pages} {526} (\bibinfo {year} {2009})}\BibitemShut {NoStop}%
\bibitem [{\citenamefont {{Atri}}\ \emph {et~al.}(2013)\citenamefont {{Atri}},
  \citenamefont {{Hariharan}},\ and\ \citenamefont {{Grie{\ss}meier}}}]{AHG13}%
  \BibitemOpen
  \bibfield  {author} {\bibinfo {author} {\bibfnamefont {D.}~\bibnamefont
  {{Atri}}}, \bibinfo {author} {\bibfnamefont {B.}~\bibnamefont {{Hariharan}}},
  \ and\ \bibinfo {author} {\bibfnamefont {J.-M.}\ \bibnamefont
  {{Grie{\ss}meier}}},\ }\href {\doibase 10.1089/ast.2013.1052} {\bibfield
  {journal} {\bibinfo  {journal} {Astrobiology}\ }\textbf {\bibinfo {volume}
  {13}},\ \bibinfo {pages} {910} (\bibinfo {year} {2013})}\BibitemShut
  {NoStop}%
\bibitem [{\citenamefont {{Grie{\ss}meier}}\ \emph {et~al.}(2016)\citenamefont
  {{Grie{\ss}meier}}, \citenamefont {{Tabataba-Vakili}}, \citenamefont
  {{Stadelmann}}, \citenamefont {{Grenfell}},\ and\ \citenamefont
  {{Atri}}}]{GTV16}%
  \BibitemOpen
  \bibfield  {author} {\bibinfo {author} {\bibfnamefont {J.-M.}\ \bibnamefont
  {{Grie{\ss}meier}}}, \bibinfo {author} {\bibfnamefont {F.}~\bibnamefont
  {{Tabataba-Vakili}}}, \bibinfo {author} {\bibfnamefont {A.}~\bibnamefont
  {{Stadelmann}}}, \bibinfo {author} {\bibfnamefont {J.~L.}\ \bibnamefont
  {{Grenfell}}}, \ and\ \bibinfo {author} {\bibfnamefont {D.}~\bibnamefont
  {{Atri}}},\ }\href {\doibase 10.1051/0004-6361/201425452} {\bibfield
  {journal} {\bibinfo  {journal} {Astron. Astrophys.}\ }\textbf {\bibinfo
  {volume} {587}},\ \bibinfo {eid} {A159} (\bibinfo {year} {2016})}\BibitemShut
  {NoStop}%
\bibitem [{\citenamefont {{Grenfell}}(2017)}]{Gren17}%
  \BibitemOpen
  \bibfield  {author} {\bibinfo {author} {\bibfnamefont {J.~L.}\ \bibnamefont
  {{Grenfell}}},\ }\href {\doibase 10.1016/j.physrep.2017.08.003} {\bibfield
  {journal} {\bibinfo  {journal} {Phys. Rep.}\ }\textbf {\bibinfo {volume}
  {713}},\ \bibinfo {pages} {1} (\bibinfo {year} {2017})}\BibitemShut {NoStop}%
\bibitem [{\citenamefont {{Tian}}(2015{\natexlab{a}})}]{Ti15}%
  \BibitemOpen
  \bibfield  {author} {\bibinfo {author} {\bibfnamefont {F.}~\bibnamefont
  {{Tian}}},\ }\href {\doibase 10.1146/annurev-earth-060313-054834} {\bibfield
  {journal} {\bibinfo  {journal} {Annu. Rev. Earth Planet. Sci.}\ }\textbf
  {\bibinfo {volume} {43}},\ \bibinfo {pages} {459} (\bibinfo {year}
  {2015}{\natexlab{a}})}\BibitemShut {NoStop}%
\bibitem [{\citenamefont {{Lammer}}\ \emph {et~al.}(2008)\citenamefont
  {{Lammer}}, \citenamefont {{Kasting}}, \citenamefont {{Chassefi{\`e}re}},
  \citenamefont {{Johnson}}, \citenamefont {{Kulikov}},\ and\ \citenamefont
  {{Tian}}}]{LK08}%
  \BibitemOpen
  \bibfield  {author} {\bibinfo {author} {\bibfnamefont {H.}~\bibnamefont
  {{Lammer}}}, \bibinfo {author} {\bibfnamefont {J.~F.}\ \bibnamefont
  {{Kasting}}}, \bibinfo {author} {\bibfnamefont {E.}~\bibnamefont
  {{Chassefi{\`e}re}}}, \bibinfo {author} {\bibfnamefont {R.~E.}\ \bibnamefont
  {{Johnson}}}, \bibinfo {author} {\bibfnamefont {Y.~N.}\ \bibnamefont
  {{Kulikov}}}, \ and\ \bibinfo {author} {\bibfnamefont {F.}~\bibnamefont
  {{Tian}}},\ }\href {\doibase 10.1007/s11214-008-9413-5} {\bibfield  {journal}
  {\bibinfo  {journal} {Space Sci. Rev.}\ }\textbf {\bibinfo {volume} {139}},\
  \bibinfo {pages} {399} (\bibinfo {year} {2008})}\BibitemShut {NoStop}%
\bibitem [{\citenamefont {{Brain}}\ \emph {et~al.}(2016)\citenamefont
  {{Brain}}, \citenamefont {{Bagenal}}, \citenamefont {{Ma}}, \citenamefont
  {{Nilsson}},\ and\ \citenamefont {{Stenberg Wieser}}}]{BB16}%
  \BibitemOpen
  \bibfield  {author} {\bibinfo {author} {\bibfnamefont {D.~A.}\ \bibnamefont
  {{Brain}}}, \bibinfo {author} {\bibfnamefont {F.}~\bibnamefont {{Bagenal}}},
  \bibinfo {author} {\bibfnamefont {Y.-J.}\ \bibnamefont {{Ma}}}, \bibinfo
  {author} {\bibfnamefont {H.}~\bibnamefont {{Nilsson}}}, \ and\ \bibinfo
  {author} {\bibfnamefont {G.}~\bibnamefont {{Stenberg Wieser}}},\ }\href
  {\doibase 10.1002/2016JE005162} {\bibfield  {journal} {\bibinfo  {journal}
  {J. Geophys. Res. E}\ }\textbf {\bibinfo {volume} {121}},\ \bibinfo {pages}
  {2364} (\bibinfo {year} {2016})}\BibitemShut {NoStop}%
\bibitem [{\citenamefont {{Freidberg}}(2014)}]{Fri14}%
  \BibitemOpen
  \bibfield  {author} {\bibinfo {author} {\bibfnamefont {J.~P.}\ \bibnamefont
  {{Freidberg}}},\ }\href@noop {} {\emph {\bibinfo {title} {{Ideal MHD}}}}\
  (\bibinfo  {publisher} {Cambridge University Press},\ \bibinfo {year}
  {2014})\BibitemShut {NoStop}%
\bibitem [{\citenamefont {{Lingam}}\ \emph {et~al.}(2016)\citenamefont
  {{Lingam}}, \citenamefont {{Miloshevich}},\ and\ \citenamefont
  {{Morrison}}}]{LMM16}%
  \BibitemOpen
  \bibfield  {author} {\bibinfo {author} {\bibfnamefont {M.}~\bibnamefont
  {{Lingam}}}, \bibinfo {author} {\bibfnamefont {G.}~\bibnamefont
  {{Miloshevich}}}, \ and\ \bibinfo {author} {\bibfnamefont {P.~J.}\
  \bibnamefont {{Morrison}}},\ }\href {\doibase 10.1016/j.physleta.2016.05.024}
  {\bibfield  {journal} {\bibinfo  {journal} {Phys. Lett. A}\ }\textbf
  {\bibinfo {volume} {380}},\ \bibinfo {pages} {2400} (\bibinfo {year}
  {2016})}\BibitemShut {NoStop}%
\bibitem [{\citenamefont {{Lingam}}\ \emph {et~al.}(2017)\citenamefont
  {{Lingam}}, \citenamefont {{Hirvijoki}}, \citenamefont {{Pfefferl{\'e}}},
  \citenamefont {{Comisso}},\ and\ \citenamefont {{Bhattacharjee}}}]{LH17}%
  \BibitemOpen
  \bibfield  {author} {\bibinfo {author} {\bibfnamefont {M.}~\bibnamefont
  {{Lingam}}}, \bibinfo {author} {\bibfnamefont {E.}~\bibnamefont
  {{Hirvijoki}}}, \bibinfo {author} {\bibfnamefont {D.}~\bibnamefont
  {{Pfefferl{\'e}}}}, \bibinfo {author} {\bibfnamefont {L.}~\bibnamefont
  {{Comisso}}}, \ and\ \bibinfo {author} {\bibfnamefont {A.}~\bibnamefont
  {{Bhattacharjee}}},\ }\href {\doibase 10.1063/1.4980838} {\bibfield
  {journal} {\bibinfo  {journal} {Phys. Plasmas}\ }\textbf {\bibinfo {volume}
  {24}},\ \bibinfo {eid} {042120} (\bibinfo {year} {2017})}\BibitemShut
  {NoStop}%
\bibitem [{\citenamefont {{Lammer}}(2013)}]{Lam13}%
  \BibitemOpen
  \bibfield  {author} {\bibinfo {author} {\bibfnamefont {H.}~\bibnamefont
  {{Lammer}}},\ }\href {\doibase 10.1007/978-3-642-32087-3} {\emph {\bibinfo
  {title} {{Origin and Evolution of Planetary Atmospheres: Implications for
  Habitability}}}},\ Springer Briefs in Astronomy\ (\bibinfo  {publisher}
  {Springer},\ \bibinfo {year} {2013})\BibitemShut {NoStop}%
\bibitem [{\citenamefont {{Acuna}}\ \emph {et~al.}(1998)\citenamefont
  {{Acuna}}, \citenamefont {{Connerney}}, \citenamefont {{Wasilewski}},
  \citenamefont {{Lin}}, \citenamefont {{Anderson}}, \citenamefont {{Carlson}},
  \citenamefont {{McFadden}}, \citenamefont {{Curtis}}, \citenamefont
  {{Mitchell}}, \citenamefont {{Reme}}, \citenamefont {{Mazelle}},
  \citenamefont {{Sauvaud}}, \citenamefont {{D'Uston}}, \citenamefont {{Cros}},
  \citenamefont {{Medale}}, \citenamefont {{Bauer}}, \citenamefont
  {{Cloutier}}, \citenamefont {{Mayhew}}, \citenamefont {{Winterhalter}},\ and\
  \citenamefont {{Ness}}}]{AC98}%
  \BibitemOpen
  \bibfield  {author} {\bibinfo {author} {\bibfnamefont {M.~H.}\ \bibnamefont
  {{Acuna}}}, \bibinfo {author} {\bibfnamefont {J.~E.~P.}\ \bibnamefont
  {{Connerney}}}, \bibinfo {author} {\bibfnamefont {P.}~\bibnamefont
  {{Wasilewski}}}, \bibinfo {author} {\bibfnamefont {R.~P.}\ \bibnamefont
  {{Lin}}}, \bibinfo {author} {\bibfnamefont {K.~A.}\ \bibnamefont
  {{Anderson}}}, \bibinfo {author} {\bibfnamefont {C.~W.}\ \bibnamefont
  {{Carlson}}}, \bibinfo {author} {\bibfnamefont {J.}~\bibnamefont
  {{McFadden}}}, \bibinfo {author} {\bibfnamefont {D.~W.}\ \bibnamefont
  {{Curtis}}}, \bibinfo {author} {\bibfnamefont {D.}~\bibnamefont
  {{Mitchell}}}, \bibinfo {author} {\bibfnamefont {H.}~\bibnamefont {{Reme}}},
  \bibinfo {author} {\bibfnamefont {C.}~\bibnamefont {{Mazelle}}}, \bibinfo
  {author} {\bibfnamefont {J.~A.}\ \bibnamefont {{Sauvaud}}}, \bibinfo {author}
  {\bibfnamefont {C.}~\bibnamefont {{D'Uston}}}, \bibinfo {author}
  {\bibfnamefont {A.}~\bibnamefont {{Cros}}}, \bibinfo {author} {\bibfnamefont
  {J.~L.}\ \bibnamefont {{Medale}}}, \bibinfo {author} {\bibfnamefont {S.~J.}\
  \bibnamefont {{Bauer}}}, \bibinfo {author} {\bibfnamefont {P.}~\bibnamefont
  {{Cloutier}}}, \bibinfo {author} {\bibfnamefont {M.}~\bibnamefont
  {{Mayhew}}}, \bibinfo {author} {\bibfnamefont {D.}~\bibnamefont
  {{Winterhalter}}}, \ and\ \bibinfo {author} {\bibfnamefont {N.~F.}\
  \bibnamefont {{Ness}}},\ }\href {\doibase 10.1126/science.279.5357.1676}
  {\bibfield  {journal} {\bibinfo  {journal} {Science}\ }\textbf {\bibinfo
  {volume} {279}},\ \bibinfo {pages} {1676} (\bibinfo {year}
  {1998})}\BibitemShut {NoStop}%
\bibitem [{\citenamefont {{Zendejas}}\ \emph {et~al.}(2010)\citenamefont
  {{Zendejas}}, \citenamefont {{Segura}},\ and\ \citenamefont
  {{Raga}}}]{ZSR10}%
  \BibitemOpen
  \bibfield  {author} {\bibinfo {author} {\bibfnamefont {J.}~\bibnamefont
  {{Zendejas}}}, \bibinfo {author} {\bibfnamefont {A.}~\bibnamefont
  {{Segura}}}, \ and\ \bibinfo {author} {\bibfnamefont {A.~C.}\ \bibnamefont
  {{Raga}}},\ }\href {\doibase 10.1016/j.icarus.2010.07.013} {\bibfield
  {journal} {\bibinfo  {journal} {Icarus}\ }\textbf {\bibinfo {volume} {210}},\
  \bibinfo {pages} {539} (\bibinfo {year} {2010})}\BibitemShut {NoStop}%
\bibitem [{\citenamefont {{Zahnle}}\ and\ \citenamefont
  {{Catling}}(2017)}]{ZC17}%
  \BibitemOpen
  \bibfield  {author} {\bibinfo {author} {\bibfnamefont {K.~J.}\ \bibnamefont
  {{Zahnle}}}\ and\ \bibinfo {author} {\bibfnamefont {D.~C.}\ \bibnamefont
  {{Catling}}},\ }\href {\doibase 10.3847/1538-4357/aa7846} {\bibfield
  {journal} {\bibinfo  {journal} {Astrophys. J.}\ }\textbf {\bibinfo {volume}
  {843}},\ \bibinfo {eid} {122} (\bibinfo {year} {2017})}\BibitemShut {NoStop}%
\bibitem [{\citenamefont {{Lingam}}\ and\ \citenamefont
  {{Loeb}}(2018{\natexlab{b}})}]{LL18b}%
  \BibitemOpen
  \bibfield  {author} {\bibinfo {author} {\bibfnamefont {M.}~\bibnamefont
  {{Lingam}}}\ and\ \bibinfo {author} {\bibfnamefont {A.}~\bibnamefont
  {{Loeb}}},\ }\href {\doibase 10.1017/S1473550417000179} {\bibfield  {journal}
  {\bibinfo  {journal} {Int. J. Astrobiol.}\ }\textbf {\bibinfo {volume}
  {17}},\ \bibinfo {pages} {116} (\bibinfo {year}
  {2018}{\natexlab{b}})}\BibitemShut {NoStop}%
\bibitem [{\citenamefont {{Wood}}\ \emph {et~al.}(2005)\citenamefont {{Wood}},
  \citenamefont {{M{\"u}ller}}, \citenamefont {{Zank}}, \citenamefont
  {{Linsky}},\ and\ \citenamefont {{Redfield}}}]{WM05}%
  \BibitemOpen
  \bibfield  {author} {\bibinfo {author} {\bibfnamefont {B.~E.}\ \bibnamefont
  {{Wood}}}, \bibinfo {author} {\bibfnamefont {H.-R.}\ \bibnamefont
  {{M{\"u}ller}}}, \bibinfo {author} {\bibfnamefont {G.~P.}\ \bibnamefont
  {{Zank}}}, \bibinfo {author} {\bibfnamefont {J.~L.}\ \bibnamefont
  {{Linsky}}}, \ and\ \bibinfo {author} {\bibfnamefont {S.}~\bibnamefont
  {{Redfield}}},\ }\href {\doibase 10.1086/432716} {\bibfield  {journal}
  {\bibinfo  {journal} {Astrophys. J. Lett.}\ }\textbf {\bibinfo {volume}
  {628}},\ \bibinfo {pages} {L143} (\bibinfo {year} {2005})}\BibitemShut
  {NoStop}%
\bibitem [{\citenamefont {{Cranmer}}\ and\ \citenamefont
  {{Saar}}(2011)}]{CS11}%
  \BibitemOpen
  \bibfield  {author} {\bibinfo {author} {\bibfnamefont {S.~R.}\ \bibnamefont
  {{Cranmer}}}\ and\ \bibinfo {author} {\bibfnamefont {S.~H.}\ \bibnamefont
  {{Saar}}},\ }\href {\doibase 10.1088/0004-637X/741/1/54} {\bibfield
  {journal} {\bibinfo  {journal} {Astrophys. J.}\ }\textbf {\bibinfo {volume}
  {741}},\ \bibinfo {eid} {54} (\bibinfo {year} {2011})}\BibitemShut {NoStop}%
\bibitem [{\citenamefont {{Johnstone}}\ \emph {et~al.}(2015)\citenamefont
  {{Johnstone}}, \citenamefont {{G{\"u}del}}, \citenamefont {{Brott}},\ and\
  \citenamefont {{L{\"u}ftinger}}}]{JG15}%
  \BibitemOpen
  \bibfield  {author} {\bibinfo {author} {\bibfnamefont {C.~P.}\ \bibnamefont
  {{Johnstone}}}, \bibinfo {author} {\bibfnamefont {M.}~\bibnamefont
  {{G{\"u}del}}}, \bibinfo {author} {\bibfnamefont {I.}~\bibnamefont
  {{Brott}}}, \ and\ \bibinfo {author} {\bibfnamefont {T.}~\bibnamefont
  {{L{\"u}ftinger}}},\ }\href {\doibase 10.1051/0004-6361/201425301} {\bibfield
   {journal} {\bibinfo  {journal} {Astron. Astrophys.}\ }\textbf {\bibinfo
  {volume} {577}},\ \bibinfo {eid} {A28} (\bibinfo {year} {2015})}\BibitemShut
  {NoStop}%
\bibitem [{\citenamefont {{Valencia}}\ \emph {et~al.}(2006)\citenamefont
  {{Valencia}}, \citenamefont {{O'Connell}},\ and\ \citenamefont
  {{Sasselov}}}]{VOCS06}%
  \BibitemOpen
  \bibfield  {author} {\bibinfo {author} {\bibfnamefont {D.}~\bibnamefont
  {{Valencia}}}, \bibinfo {author} {\bibfnamefont {R.~J.}\ \bibnamefont
  {{O'Connell}}}, \ and\ \bibinfo {author} {\bibfnamefont {D.}~\bibnamefont
  {{Sasselov}}},\ }\href {\doibase 10.1016/j.icarus.2005.11.021} {\bibfield
  {journal} {\bibinfo  {journal} {Icarus}\ }\textbf {\bibinfo {volume} {181}},\
  \bibinfo {pages} {545} (\bibinfo {year} {2006})}\BibitemShut {NoStop}%
\bibitem [{\citenamefont {{Zeng}}\ \emph {et~al.}(2016)\citenamefont {{Zeng}},
  \citenamefont {{Sasselov}},\ and\ \citenamefont {{Jacobsen}}}]{ZSJ16}%
  \BibitemOpen
  \bibfield  {author} {\bibinfo {author} {\bibfnamefont {L.}~\bibnamefont
  {{Zeng}}}, \bibinfo {author} {\bibfnamefont {D.~D.}\ \bibnamefont
  {{Sasselov}}}, \ and\ \bibinfo {author} {\bibfnamefont {S.~B.}\ \bibnamefont
  {{Jacobsen}}},\ }\href {\doibase 10.3847/0004-637X/819/2/127} {\bibfield
  {journal} {\bibinfo  {journal} {Astrophys. J.}\ }\textbf {\bibinfo {volume}
  {819}},\ \bibinfo {eid} {127} (\bibinfo {year} {2016})}\BibitemShut {NoStop}%
\bibitem [{\citenamefont {{Airapetian}}\ \emph
  {et~al.}(2017{\natexlab{a}})\citenamefont {{Airapetian}}, \citenamefont
  {{Glocer}}, \citenamefont {{Khazanov}}, \citenamefont {{Loyd}}, \citenamefont
  {{France}}, \citenamefont {{Sojka}}, \citenamefont {{Danchi}},\ and\
  \citenamefont {{Liemohn}}}]{AGK17}%
  \BibitemOpen
  \bibfield  {author} {\bibinfo {author} {\bibfnamefont {V.~S.}\ \bibnamefont
  {{Airapetian}}}, \bibinfo {author} {\bibfnamefont {A.}~\bibnamefont
  {{Glocer}}}, \bibinfo {author} {\bibfnamefont {G.~V.}\ \bibnamefont
  {{Khazanov}}}, \bibinfo {author} {\bibfnamefont {R.~O.~P.}\ \bibnamefont
  {{Loyd}}}, \bibinfo {author} {\bibfnamefont {K.}~\bibnamefont {{France}}},
  \bibinfo {author} {\bibfnamefont {J.}~\bibnamefont {{Sojka}}}, \bibinfo
  {author} {\bibfnamefont {W.~C.}\ \bibnamefont {{Danchi}}}, \ and\ \bibinfo
  {author} {\bibfnamefont {M.~W.}\ \bibnamefont {{Liemohn}}},\ }\href {\doibase
  10.3847/2041-8213/836/1/L3} {\bibfield  {journal} {\bibinfo  {journal}
  {Astrophys. J. Lett.}\ }\textbf {\bibinfo {volume} {836}},\ \bibinfo {eid}
  {L3} (\bibinfo {year} {2017}{\natexlab{a}})}\BibitemShut {NoStop}%
\bibitem [{\citenamefont {{Garcia-Sage}}\ \emph {et~al.}(2017)\citenamefont
  {{Garcia-Sage}}, \citenamefont {{Glocer}}, \citenamefont {{Drake}},
  \citenamefont {{Gronoff}},\ and\ \citenamefont {{Cohen}}}]{GG17}%
  \BibitemOpen
  \bibfield  {author} {\bibinfo {author} {\bibfnamefont {K.}~\bibnamefont
  {{Garcia-Sage}}}, \bibinfo {author} {\bibfnamefont {A.}~\bibnamefont
  {{Glocer}}}, \bibinfo {author} {\bibfnamefont {J.~J.}\ \bibnamefont
  {{Drake}}}, \bibinfo {author} {\bibfnamefont {G.}~\bibnamefont {{Gronoff}}},
  \ and\ \bibinfo {author} {\bibfnamefont {O.}~\bibnamefont {{Cohen}}},\ }\href
  {\doibase 10.3847/2041-8213/aa7eca} {\bibfield  {journal} {\bibinfo
  {journal} {Astrophys. J. Lett.}\ }\textbf {\bibinfo {volume} {844}},\
  \bibinfo {eid} {L13} (\bibinfo {year} {2017})}\BibitemShut {NoStop}%
\bibitem [{\citenamefont {{Kislyakova}}\ \emph {et~al.}(2014)\citenamefont
  {{Kislyakova}}, \citenamefont {{Johnstone}}, \citenamefont {{Odert}},
  \citenamefont {{Erkaev}}, \citenamefont {{Lammer}}, \citenamefont
  {{L{\"u}ftinger}}, \citenamefont {{Holmstr{\"o}m}}, \citenamefont
  {{Khodachenko}},\ and\ \citenamefont {{G{\"u}del}}}]{KJO14}%
  \BibitemOpen
  \bibfield  {author} {\bibinfo {author} {\bibfnamefont {K.~G.}\ \bibnamefont
  {{Kislyakova}}}, \bibinfo {author} {\bibfnamefont {C.~P.}\ \bibnamefont
  {{Johnstone}}}, \bibinfo {author} {\bibfnamefont {P.}~\bibnamefont
  {{Odert}}}, \bibinfo {author} {\bibfnamefont {N.~V.}\ \bibnamefont
  {{Erkaev}}}, \bibinfo {author} {\bibfnamefont {H.}~\bibnamefont {{Lammer}}},
  \bibinfo {author} {\bibfnamefont {T.}~\bibnamefont {{L{\"u}ftinger}}},
  \bibinfo {author} {\bibfnamefont {M.}~\bibnamefont {{Holmstr{\"o}m}}},
  \bibinfo {author} {\bibfnamefont {M.~L.}\ \bibnamefont {{Khodachenko}}}, \
  and\ \bibinfo {author} {\bibfnamefont {M.}~\bibnamefont {{G{\"u}del}}},\
  }\href {\doibase 10.1051/0004-6361/201322933} {\bibfield  {journal} {\bibinfo
   {journal} {Astron. Astrophys.}\ }\textbf {\bibinfo {volume} {562}},\
  \bibinfo {eid} {A116} (\bibinfo {year} {2014})}\BibitemShut {NoStop}%
\bibitem [{\citenamefont {{Cohen}}\ \emph {et~al.}(2015)\citenamefont
  {{Cohen}}, \citenamefont {{Ma}}, \citenamefont {{Drake}}, \citenamefont
  {{Glocer}}, \citenamefont {{Garraffo}}, \citenamefont {{Bell}},\ and\
  \citenamefont {{Gombosi}}}]{CMD}%
  \BibitemOpen
  \bibfield  {author} {\bibinfo {author} {\bibfnamefont {O.}~\bibnamefont
  {{Cohen}}}, \bibinfo {author} {\bibfnamefont {Y.}~\bibnamefont {{Ma}}},
  \bibinfo {author} {\bibfnamefont {J.~J.}\ \bibnamefont {{Drake}}}, \bibinfo
  {author} {\bibfnamefont {A.}~\bibnamefont {{Glocer}}}, \bibinfo {author}
  {\bibfnamefont {C.}~\bibnamefont {{Garraffo}}}, \bibinfo {author}
  {\bibfnamefont {J.~M.}\ \bibnamefont {{Bell}}}, \ and\ \bibinfo {author}
  {\bibfnamefont {T.~I.}\ \bibnamefont {{Gombosi}}},\ }\href {\doibase
  10.1088/0004-637X/806/1/41} {\bibfield  {journal} {\bibinfo  {journal}
  {Astrophys. J.}\ }\textbf {\bibinfo {volume} {806}},\ \bibinfo {eid} {41}
  (\bibinfo {year} {2015})}\BibitemShut {NoStop}%
\bibitem [{\citenamefont {{Seki}}\ \emph {et~al.}(2001)\citenamefont {{Seki}},
  \citenamefont {{Elphic}}, \citenamefont {{Hirahara}}, \citenamefont
  {{Terasawa}},\ and\ \citenamefont {{Mukai}}}]{SEH01}%
  \BibitemOpen
  \bibfield  {author} {\bibinfo {author} {\bibfnamefont {K.}~\bibnamefont
  {{Seki}}}, \bibinfo {author} {\bibfnamefont {R.~C.}\ \bibnamefont
  {{Elphic}}}, \bibinfo {author} {\bibfnamefont {M.}~\bibnamefont
  {{Hirahara}}}, \bibinfo {author} {\bibfnamefont {T.}~\bibnamefont
  {{Terasawa}}}, \ and\ \bibinfo {author} {\bibfnamefont {T.}~\bibnamefont
  {{Mukai}}},\ }\href {\doibase 10.1126/science.1058913} {\bibfield  {journal}
  {\bibinfo  {journal} {Science}\ }\textbf {\bibinfo {volume} {291}},\ \bibinfo
  {pages} {1939} (\bibinfo {year} {2001})}\BibitemShut {NoStop}%
\bibitem [{\citenamefont {{Dong}}\ \emph
  {et~al.}(2017{\natexlab{b}})\citenamefont {{Dong}}, \citenamefont {{Huang}},
  \citenamefont {{Lingam}}, \citenamefont {{T{\'o}th}}, \citenamefont
  {{Gombosi}},\ and\ \citenamefont {{Bhattacharjee}}}]{DH17}%
  \BibitemOpen
  \bibfield  {author} {\bibinfo {author} {\bibfnamefont {C.}~\bibnamefont
  {{Dong}}}, \bibinfo {author} {\bibfnamefont {Z.}~\bibnamefont {{Huang}}},
  \bibinfo {author} {\bibfnamefont {M.}~\bibnamefont {{Lingam}}}, \bibinfo
  {author} {\bibfnamefont {G.}~\bibnamefont {{T{\'o}th}}}, \bibinfo {author}
  {\bibfnamefont {T.}~\bibnamefont {{Gombosi}}}, \ and\ \bibinfo {author}
  {\bibfnamefont {A.}~\bibnamefont {{Bhattacharjee}}},\ }\href {\doibase
  10.3847/2041-8213/aa8a60} {\bibfield  {journal} {\bibinfo  {journal}
  {Astrophys. J. Lett.}\ }\textbf {\bibinfo {volume} {847}},\ \bibinfo {eid}
  {L4} (\bibinfo {year} {2017}{\natexlab{b}})}\BibitemShut {NoStop}%
\bibitem [{\citenamefont {{Pearce}}\ \emph {et~al.}(2018)\citenamefont
  {{Pearce}}, \citenamefont {{Tupper}}, \citenamefont {{Pudritz}},\ and\
  \citenamefont {{Higgs}}}]{PT18}%
  \BibitemOpen
  \bibfield  {author} {\bibinfo {author} {\bibfnamefont {B.~K.~D.}\
  \bibnamefont {{Pearce}}}, \bibinfo {author} {\bibfnamefont {A.~S.}\
  \bibnamefont {{Tupper}}}, \bibinfo {author} {\bibfnamefont {R.~E.}\
  \bibnamefont {{Pudritz}}}, \ and\ \bibinfo {author} {\bibfnamefont {P.~G.}\
  \bibnamefont {{Higgs}}},\ }\href {\doibase 10.1089/ast.2017.1674} {\bibfield
  {journal} {\bibinfo  {journal} {Astrobiology}\ }\textbf {\bibinfo {volume}
  {18}},\ \bibinfo {pages} {343} (\bibinfo {year} {2018})}\BibitemShut
  {NoStop}%
\bibitem [{\citenamefont {{Spiegel}}\ and\ \citenamefont
  {{Turner}}(2012)}]{ST12}%
  \BibitemOpen
  \bibfield  {author} {\bibinfo {author} {\bibfnamefont {D.~S.}\ \bibnamefont
  {{Spiegel}}}\ and\ \bibinfo {author} {\bibfnamefont {E.~L.}\ \bibnamefont
  {{Turner}}},\ }\href {\doibase 10.1073/pnas.1111694108} {\bibfield  {journal}
  {\bibinfo  {journal} {Proc. Natl. Acad. Sci. USA}\ }\textbf {\bibinfo
  {volume} {109}},\ \bibinfo {pages} {395} (\bibinfo {year}
  {2012})}\BibitemShut {NoStop}%
\bibitem [{\citenamefont {{Rushby}}\ \emph {et~al.}(2013)\citenamefont
  {{Rushby}}, \citenamefont {{Claire}}, \citenamefont {{Osborn}},\ and\
  \citenamefont {{Watson}}}]{RC13}%
  \BibitemOpen
  \bibfield  {author} {\bibinfo {author} {\bibfnamefont {A.~J.}\ \bibnamefont
  {{Rushby}}}, \bibinfo {author} {\bibfnamefont {M.~W.}\ \bibnamefont
  {{Claire}}}, \bibinfo {author} {\bibfnamefont {H.}~\bibnamefont {{Osborn}}},
  \ and\ \bibinfo {author} {\bibfnamefont {A.~J.}\ \bibnamefont {{Watson}}},\
  }\href {\doibase 10.1089/ast.2012.0938} {\bibfield  {journal} {\bibinfo
  {journal} {Astrobiology}\ }\textbf {\bibinfo {volume} {13}},\ \bibinfo
  {pages} {833} (\bibinfo {year} {2013})}\BibitemShut {NoStop}%
\bibitem [{\citenamefont {{Lingam}}\ and\ \citenamefont
  {{Loeb}}(2019{\natexlab{a}})}]{LL18c}%
  \BibitemOpen
  \bibfield  {author} {\bibinfo {author} {\bibfnamefont {M.}~\bibnamefont
  {{Lingam}}}\ and\ \bibinfo {author} {\bibfnamefont {A.}~\bibnamefont
  {{Loeb}}},\ }\href {\doibase 10.1017/S1473550419000016} {\bibfield  {journal}
  {\bibinfo  {journal} {Int. J. Astrobiol.}\ } (\bibinfo {year}
  {2019}{\natexlab{a}}),\ 10.1017/S1473550419000016},\ \Eprint
  {http://arxiv.org/abs/1804.02271} {arXiv:1804.02271 [physics.pop-ph]}
  \BibitemShut {NoStop}%
\bibitem [{\citenamefont {{Russell}}(1983)}]{Russ83}%
  \BibitemOpen
  \bibfield  {author} {\bibinfo {author} {\bibfnamefont {D.~A.}\ \bibnamefont
  {{Russell}}},\ }\href {\doibase 10.1016/0273-1177(83)90045-5} {\bibfield
  {journal} {\bibinfo  {journal} {Adv. Space Res.}\ }\textbf {\bibinfo {volume}
  {3}},\ \bibinfo {pages} {95} (\bibinfo {year} {1983})}\BibitemShut {NoStop}%
\bibitem [{\citenamefont {{Purvis}}\ and\ \citenamefont
  {{Hector}}(2000)}]{PuHe00}%
  \BibitemOpen
  \bibfield  {author} {\bibinfo {author} {\bibfnamefont {A.}~\bibnamefont
  {{Purvis}}}\ and\ \bibinfo {author} {\bibfnamefont {A.}~\bibnamefont
  {{Hector}}},\ }\href {\doibase 10.1038/35012221} {\bibfield  {journal}
  {\bibinfo  {journal} {Nature}\ }\textbf {\bibinfo {volume} {405}},\ \bibinfo
  {pages} {212} (\bibinfo {year} {2000})}\BibitemShut {NoStop}%
\bibitem [{\citenamefont {{Lingam}}\ and\ \citenamefont
  {{Loeb}}(2017{\natexlab{a}})}]{LL17a}%
  \BibitemOpen
  \bibfield  {author} {\bibinfo {author} {\bibfnamefont {M.}~\bibnamefont
  {{Lingam}}}\ and\ \bibinfo {author} {\bibfnamefont {A.}~\bibnamefont
  {{Loeb}}},\ }\href {\doibase 10.3847/2041-8213/aa8860} {\bibfield  {journal}
  {\bibinfo  {journal} {Astrophys. J. Lett.}\ }\textbf {\bibinfo {volume}
  {846}},\ \bibinfo {eid} {L21} (\bibinfo {year}
  {2017}{\natexlab{a}})}\BibitemShut {NoStop}%
\bibitem [{\citenamefont {{Locey}}\ and\ \citenamefont
  {{Lennon}}(2016)}]{LL16}%
  \BibitemOpen
  \bibfield  {author} {\bibinfo {author} {\bibfnamefont {K.~J.}\ \bibnamefont
  {{Locey}}}\ and\ \bibinfo {author} {\bibfnamefont {J.~T.}\ \bibnamefont
  {{Lennon}}},\ }\href {\doibase 10.1073/pnas.1521291113} {\bibfield  {journal}
  {\bibinfo  {journal} {Proc. Natl. Acad. Sci. USA}\ }\textbf {\bibinfo
  {volume} {113}},\ \bibinfo {pages} {5970} (\bibinfo {year}
  {2016})}\BibitemShut {NoStop}%
\bibitem [{\citenamefont {{Lingam}}\ and\ \citenamefont
  {{Loeb}}(2018{\natexlab{c}})}]{LL18g}%
  \BibitemOpen
  \bibfield  {author} {\bibinfo {author} {\bibfnamefont {M.}~\bibnamefont
  {{Lingam}}}\ and\ \bibinfo {author} {\bibfnamefont {A.}~\bibnamefont
  {{Loeb}}},\ }\href {\doibase 10.1088/1475-7516/2018/05/020} {\bibfield
  {journal} {\bibinfo  {journal} {J. Cosmol. Astropart. Phys.}\ }\textbf
  {\bibinfo {volume} {5}},\ \bibinfo {eid} {020} (\bibinfo {year}
  {2018}{\natexlab{c}})}\BibitemShut {NoStop}%
\bibitem [{\citenamefont {{Erkaev}}\ \emph {et~al.}(2007)\citenamefont
  {{Erkaev}}, \citenamefont {{Kulikov}}, \citenamefont {{Lammer}},
  \citenamefont {{Selsis}}, \citenamefont {{Langmayr}}, \citenamefont
  {{Jaritz}},\ and\ \citenamefont {{Biernat}}}]{EK07}%
  \BibitemOpen
  \bibfield  {author} {\bibinfo {author} {\bibfnamefont {N.~V.}\ \bibnamefont
  {{Erkaev}}}, \bibinfo {author} {\bibfnamefont {Y.~N.}\ \bibnamefont
  {{Kulikov}}}, \bibinfo {author} {\bibfnamefont {H.}~\bibnamefont {{Lammer}}},
  \bibinfo {author} {\bibfnamefont {F.}~\bibnamefont {{Selsis}}}, \bibinfo
  {author} {\bibfnamefont {D.}~\bibnamefont {{Langmayr}}}, \bibinfo {author}
  {\bibfnamefont {G.~F.}\ \bibnamefont {{Jaritz}}}, \ and\ \bibinfo {author}
  {\bibfnamefont {H.~K.}\ \bibnamefont {{Biernat}}},\ }\href {\doibase
  10.1051/0004-6361:20066929} {\bibfield  {journal} {\bibinfo  {journal}
  {Astron. Astrophys.}\ }\textbf {\bibinfo {volume} {472}},\ \bibinfo {pages}
  {329} (\bibinfo {year} {2007})}\BibitemShut {NoStop}%
\bibitem [{\citenamefont {{Ribas}}\ \emph {et~al.}(2016)\citenamefont
  {{Ribas}}, \citenamefont {{Bolmont}}, \citenamefont {{Selsis}}, \citenamefont
  {{Reiners}}, \citenamefont {{Leconte}}, \citenamefont {{Raymond}},
  \citenamefont {{Engle}}, \citenamefont {{Guinan}}, \citenamefont {{Morin}},
  \citenamefont {{Turbet}}, \citenamefont {{Forget}},\ and\ \citenamefont
  {{Anglada-Escud{\'e}}}}]{RB16}%
  \BibitemOpen
  \bibfield  {author} {\bibinfo {author} {\bibfnamefont {I.}~\bibnamefont
  {{Ribas}}}, \bibinfo {author} {\bibfnamefont {E.}~\bibnamefont {{Bolmont}}},
  \bibinfo {author} {\bibfnamefont {F.}~\bibnamefont {{Selsis}}}, \bibinfo
  {author} {\bibfnamefont {A.}~\bibnamefont {{Reiners}}}, \bibinfo {author}
  {\bibfnamefont {J.}~\bibnamefont {{Leconte}}}, \bibinfo {author}
  {\bibfnamefont {S.~N.}\ \bibnamefont {{Raymond}}}, \bibinfo {author}
  {\bibfnamefont {S.~G.}\ \bibnamefont {{Engle}}}, \bibinfo {author}
  {\bibfnamefont {E.~F.}\ \bibnamefont {{Guinan}}}, \bibinfo {author}
  {\bibfnamefont {J.}~\bibnamefont {{Morin}}}, \bibinfo {author} {\bibfnamefont
  {M.}~\bibnamefont {{Turbet}}}, \bibinfo {author} {\bibfnamefont
  {F.}~\bibnamefont {{Forget}}}, \ and\ \bibinfo {author} {\bibfnamefont
  {G.}~\bibnamefont {{Anglada-Escud{\'e}}}},\ }\href {\doibase
  10.1051/0004-6361/201629576} {\bibfield  {journal} {\bibinfo  {journal}
  {Astron. Astrophys.}\ }\textbf {\bibinfo {volume} {596}},\ \bibinfo {eid}
  {A111} (\bibinfo {year} {2016})}\BibitemShut {NoStop}%
\bibitem [{\citenamefont {{Owen}}(2019)}]{Ow18}%
  \BibitemOpen
  \bibfield  {author} {\bibinfo {author} {\bibfnamefont {J.~E.}\ \bibnamefont
  {{Owen}}},\ }\href {\doibase 10.1146/annurev-earth-053018-060246} {\bibfield
  {journal} {\bibinfo  {journal} {Annu. Rev. Earth Planet. Sci.}\ }\textbf
  {\bibinfo {volume} {47}},\ \bibinfo {pages} {67} (\bibinfo {year}
  {2019})}\BibitemShut {NoStop}%
\bibitem [{\citenamefont {{Owen}}\ and\ \citenamefont
  {{Alvarez}}(2016)}]{OA16}%
  \BibitemOpen
  \bibfield  {author} {\bibinfo {author} {\bibfnamefont {J.~E.}\ \bibnamefont
  {{Owen}}}\ and\ \bibinfo {author} {\bibfnamefont {M.~A.}\ \bibnamefont
  {{Alvarez}}},\ }\href {\doibase 10.3847/0004-637X/816/1/34} {\bibfield
  {journal} {\bibinfo  {journal} {Astrophys. J.}\ }\textbf {\bibinfo {volume}
  {816}},\ \bibinfo {eid} {34} (\bibinfo {year} {2016})}\BibitemShut {NoStop}%
\bibitem [{\citenamefont {{McKee}}\ and\ \citenamefont
  {{Ostriker}}(2007)}]{MO07}%
  \BibitemOpen
  \bibfield  {author} {\bibinfo {author} {\bibfnamefont {C.~F.}\ \bibnamefont
  {{McKee}}}\ and\ \bibinfo {author} {\bibfnamefont {E.~C.}\ \bibnamefont
  {{Ostriker}}},\ }\href {\doibase 10.1146/annurev.astro.45.051806.110602}
  {\bibfield  {journal} {\bibinfo  {journal} {Annu. Rev. Astron. Astrophys.}\
  }\textbf {\bibinfo {volume} {45}},\ \bibinfo {pages} {565} (\bibinfo {year}
  {2007})}\BibitemShut {NoStop}%
\bibitem [{\citenamefont {{Baraffe}}\ \emph {et~al.}(2002)\citenamefont
  {{Baraffe}}, \citenamefont {{Chabrier}}, \citenamefont {{Allard}},\ and\
  \citenamefont {{Hauschildt}}}]{BC02}%
  \BibitemOpen
  \bibfield  {author} {\bibinfo {author} {\bibfnamefont {I.}~\bibnamefont
  {{Baraffe}}}, \bibinfo {author} {\bibfnamefont {G.}~\bibnamefont
  {{Chabrier}}}, \bibinfo {author} {\bibfnamefont {F.}~\bibnamefont
  {{Allard}}}, \ and\ \bibinfo {author} {\bibfnamefont {P.~H.}\ \bibnamefont
  {{Hauschildt}}},\ }\href {\doibase 10.1051/0004-6361:20011638} {\bibfield
  {journal} {\bibinfo  {journal} {Astron. Astrophys.}\ }\textbf {\bibinfo
  {volume} {382}},\ \bibinfo {pages} {563} (\bibinfo {year}
  {2002})}\BibitemShut {NoStop}%
\bibitem [{\citenamefont {{Ramirez}}\ and\ \citenamefont
  {{Kaltenegger}}(2014)}]{RK14}%
  \BibitemOpen
  \bibfield  {author} {\bibinfo {author} {\bibfnamefont {R.~M.}\ \bibnamefont
  {{Ramirez}}}\ and\ \bibinfo {author} {\bibfnamefont {L.}~\bibnamefont
  {{Kaltenegger}}},\ }\href {\doibase 10.1088/2041-8205/797/2/L25} {\bibfield
  {journal} {\bibinfo  {journal} {Astrophys. J. Lett.}\ }\textbf {\bibinfo
  {volume} {797}},\ \bibinfo {eid} {L25} (\bibinfo {year} {2014})}\BibitemShut
  {NoStop}%
\bibitem [{\citenamefont {{Luger}}\ and\ \citenamefont
  {{Barnes}}(2015)}]{LB15}%
  \BibitemOpen
  \bibfield  {author} {\bibinfo {author} {\bibfnamefont {R.}~\bibnamefont
  {{Luger}}}\ and\ \bibinfo {author} {\bibfnamefont {R.}~\bibnamefont
  {{Barnes}}},\ }\href {\doibase 10.1089/ast.2014.1231} {\bibfield  {journal}
  {\bibinfo  {journal} {Astrobiology}\ }\textbf {\bibinfo {volume} {15}},\
  \bibinfo {pages} {119} (\bibinfo {year} {2015})}\BibitemShut {NoStop}%
\bibitem [{\citenamefont {{Bolmont}}\ \emph {et~al.}(2017)\citenamefont
  {{Bolmont}}, \citenamefont {{Selsis}}, \citenamefont {{Owen}}, \citenamefont
  {{Ribas}}, \citenamefont {{Raymond}}, \citenamefont {{Leconte}},\ and\
  \citenamefont {{Gillon}}}]{BS17}%
  \BibitemOpen
  \bibfield  {author} {\bibinfo {author} {\bibfnamefont {E.}~\bibnamefont
  {{Bolmont}}}, \bibinfo {author} {\bibfnamefont {F.}~\bibnamefont {{Selsis}}},
  \bibinfo {author} {\bibfnamefont {J.~E.}\ \bibnamefont {{Owen}}}, \bibinfo
  {author} {\bibfnamefont {I.}~\bibnamefont {{Ribas}}}, \bibinfo {author}
  {\bibfnamefont {S.~N.}\ \bibnamefont {{Raymond}}}, \bibinfo {author}
  {\bibfnamefont {J.}~\bibnamefont {{Leconte}}}, \ and\ \bibinfo {author}
  {\bibfnamefont {M.}~\bibnamefont {{Gillon}}},\ }\href {\doibase
  10.1093/mnras/stw2578} {\bibfield  {journal} {\bibinfo  {journal} {Mon. Not.
  R. Astron. Soc.}\ }\textbf {\bibinfo {volume} {464}},\ \bibinfo {pages}
  {3728} (\bibinfo {year} {2017})}\BibitemShut {NoStop}%
\bibitem [{\citenamefont {{Tian}}\ \emph {et~al.}(2018)\citenamefont {{Tian}},
  \citenamefont {{G{\"u}del}}, \citenamefont {{Johnstone}}, \citenamefont
  {{Lammer}}, \citenamefont {{Luger}},\ and\ \citenamefont {{Odert}}}]{TGJ18}%
  \BibitemOpen
  \bibfield  {author} {\bibinfo {author} {\bibfnamefont {F.}~\bibnamefont
  {{Tian}}}, \bibinfo {author} {\bibfnamefont {M.}~\bibnamefont {{G{\"u}del}}},
  \bibinfo {author} {\bibfnamefont {C.~P.}\ \bibnamefont {{Johnstone}}},
  \bibinfo {author} {\bibfnamefont {H.}~\bibnamefont {{Lammer}}}, \bibinfo
  {author} {\bibfnamefont {R.}~\bibnamefont {{Luger}}}, \ and\ \bibinfo
  {author} {\bibfnamefont {P.}~\bibnamefont {{Odert}}},\ }\href {\doibase
  10.1007/s11214-018-0490-9} {\bibfield  {journal} {\bibinfo  {journal} {Space
  Sci. Rev.}\ }\textbf {\bibinfo {volume} {214}},\ \bibinfo {eid} {65}
  (\bibinfo {year} {2018})}\BibitemShut {NoStop}%
\bibitem [{\citenamefont {{Tian}}(2015{\natexlab{b}})}]{Tian}%
  \BibitemOpen
  \bibfield  {author} {\bibinfo {author} {\bibfnamefont {F.}~\bibnamefont
  {{Tian}}},\ }\href {\doibase 10.1016/j.epsl.2015.09.051} {\bibfield
  {journal} {\bibinfo  {journal} {Earth Planet. Sci. Lett.}\ }\textbf {\bibinfo
  {volume} {432}},\ \bibinfo {pages} {126} (\bibinfo {year}
  {2015}{\natexlab{b}})}\BibitemShut {NoStop}%
\bibitem [{\citenamefont {{Holland}}(2002)}]{Holl02}%
  \BibitemOpen
  \bibfield  {author} {\bibinfo {author} {\bibfnamefont {H.~D.}\ \bibnamefont
  {{Holland}}},\ }\href {\doibase 10.1016/S0016-7037(02)00950-X} {\bibfield
  {journal} {\bibinfo  {journal} {Geochim. Cosmochim. Acta}\ }\textbf {\bibinfo
  {volume} {66}},\ \bibinfo {pages} {3811} (\bibinfo {year}
  {2002})}\BibitemShut {NoStop}%
\bibitem [{\citenamefont {{Segura}}\ \emph {et~al.}(2007)\citenamefont
  {{Segura}}, \citenamefont {{Meadows}}, \citenamefont {{Kasting}},
  \citenamefont {{Crisp}},\ and\ \citenamefont {{Cohen}}}]{SM07}%
  \BibitemOpen
  \bibfield  {author} {\bibinfo {author} {\bibfnamefont {A.}~\bibnamefont
  {{Segura}}}, \bibinfo {author} {\bibfnamefont {V.~S.}\ \bibnamefont
  {{Meadows}}}, \bibinfo {author} {\bibfnamefont {J.~F.}\ \bibnamefont
  {{Kasting}}}, \bibinfo {author} {\bibfnamefont {D.}~\bibnamefont {{Crisp}}},
  \ and\ \bibinfo {author} {\bibfnamefont {M.}~\bibnamefont {{Cohen}}},\ }\href
  {\doibase 10.1051/0004-6361:20066663} {\bibfield  {journal} {\bibinfo
  {journal} {Astron. Astrophys.}\ }\textbf {\bibinfo {volume} {472}},\ \bibinfo
  {pages} {665} (\bibinfo {year} {2007})}\BibitemShut {NoStop}%
\bibitem [{\citenamefont {{Wordsworth}}\ and\ \citenamefont
  {{Pierrehumbert}}(2013)}]{WP13}%
  \BibitemOpen
  \bibfield  {author} {\bibinfo {author} {\bibfnamefont {R.~D.}\ \bibnamefont
  {{Wordsworth}}}\ and\ \bibinfo {author} {\bibfnamefont {R.~T.}\ \bibnamefont
  {{Pierrehumbert}}},\ }\href {\doibase 10.1088/0004-637X/778/2/154} {\bibfield
   {journal} {\bibinfo  {journal} {Astrophys. J.}\ }\textbf {\bibinfo {volume}
  {778}},\ \bibinfo {eid} {154} (\bibinfo {year} {2013})}\BibitemShut {NoStop}%
\bibitem [{\citenamefont {{Wordsworth}}\ and\ \citenamefont
  {{Pierrehumbert}}(2014)}]{WP14}%
  \BibitemOpen
  \bibfield  {author} {\bibinfo {author} {\bibfnamefont {R.}~\bibnamefont
  {{Wordsworth}}}\ and\ \bibinfo {author} {\bibfnamefont {R.}~\bibnamefont
  {{Pierrehumbert}}},\ }\href {\doibase 10.1088/2041-8205/785/2/L20} {\bibfield
   {journal} {\bibinfo  {journal} {Astrophys. J. Lett.}\ }\textbf {\bibinfo
  {volume} {785}},\ \bibinfo {eid} {L20} (\bibinfo {year} {2014})}\BibitemShut
  {NoStop}%
\bibitem [{\citenamefont {{Tian}}\ \emph {et~al.}(2014)\citenamefont {{Tian}},
  \citenamefont {{France}}, \citenamefont {{Linsky}}, \citenamefont {{Mauas}},\
  and\ \citenamefont {{Vieytes}}}]{TF14}%
  \BibitemOpen
  \bibfield  {author} {\bibinfo {author} {\bibfnamefont {F.}~\bibnamefont
  {{Tian}}}, \bibinfo {author} {\bibfnamefont {K.}~\bibnamefont {{France}}},
  \bibinfo {author} {\bibfnamefont {J.~L.}\ \bibnamefont {{Linsky}}}, \bibinfo
  {author} {\bibfnamefont {P.~J.~D.}\ \bibnamefont {{Mauas}}}, \ and\ \bibinfo
  {author} {\bibfnamefont {M.~C.}\ \bibnamefont {{Vieytes}}},\ }\href {\doibase
  10.1016/j.epsl.2013.10.024} {\bibfield  {journal} {\bibinfo  {journal} {Earth
  Planet. Sci. Lett.}\ }\textbf {\bibinfo {volume} {385}},\ \bibinfo {pages}
  {22} (\bibinfo {year} {2014})}\BibitemShut {NoStop}%
\bibitem [{\citenamefont {{Harman}}\ \emph {et~al.}(2015)\citenamefont
  {{Harman}}, \citenamefont {{Schwieterman}}, \citenamefont {{Schottelkotte}},\
  and\ \citenamefont {{Kasting}}}]{HS15}%
  \BibitemOpen
  \bibfield  {author} {\bibinfo {author} {\bibfnamefont {C.~E.}\ \bibnamefont
  {{Harman}}}, \bibinfo {author} {\bibfnamefont {E.~W.}\ \bibnamefont
  {{Schwieterman}}}, \bibinfo {author} {\bibfnamefont {J.~C.}\ \bibnamefont
  {{Schottelkotte}}}, \ and\ \bibinfo {author} {\bibfnamefont {J.~F.}\
  \bibnamefont {{Kasting}}},\ }\href {\doibase 10.1088/0004-637X/812/2/137}
  {\bibfield  {journal} {\bibinfo  {journal} {Astrophys. J.}\ }\textbf
  {\bibinfo {volume} {812}},\ \bibinfo {eid} {137} (\bibinfo {year}
  {2015})}\BibitemShut {NoStop}%
\bibitem [{\citenamefont {{Narita}}\ \emph {et~al.}(2015)\citenamefont
  {{Narita}}, \citenamefont {{Enomoto}}, \citenamefont {{Masaoka}},\ and\
  \citenamefont {{Kusakabe}}}]{NEM15}%
  \BibitemOpen
  \bibfield  {author} {\bibinfo {author} {\bibfnamefont {N.}~\bibnamefont
  {{Narita}}}, \bibinfo {author} {\bibfnamefont {T.}~\bibnamefont {{Enomoto}}},
  \bibinfo {author} {\bibfnamefont {S.}~\bibnamefont {{Masaoka}}}, \ and\
  \bibinfo {author} {\bibfnamefont {N.}~\bibnamefont {{Kusakabe}}},\ }\href
  {\doibase 10.1038/srep13977} {\bibfield  {journal} {\bibinfo  {journal} {Sci.
  Rep.}\ }\textbf {\bibinfo {volume} {5}},\ \bibinfo {eid} {13977} (\bibinfo
  {year} {2015})}\BibitemShut {NoStop}%
\bibitem [{\citenamefont {{Gao}}\ \emph {et~al.}(2015)\citenamefont {{Gao}},
  \citenamefont {{Hu}}, \citenamefont {{Robinson}}, \citenamefont {{Li}},\ and\
  \citenamefont {{Yung}}}]{GHR15}%
  \BibitemOpen
  \bibfield  {author} {\bibinfo {author} {\bibfnamefont {P.}~\bibnamefont
  {{Gao}}}, \bibinfo {author} {\bibfnamefont {R.}~\bibnamefont {{Hu}}},
  \bibinfo {author} {\bibfnamefont {T.~D.}\ \bibnamefont {{Robinson}}},
  \bibinfo {author} {\bibfnamefont {C.}~\bibnamefont {{Li}}}, \ and\ \bibinfo
  {author} {\bibfnamefont {Y.~L.}\ \bibnamefont {{Yung}}},\ }\href {\doibase
  10.1088/0004-637X/806/2/249} {\bibfield  {journal} {\bibinfo  {journal}
  {Astrophys. J.}\ }\textbf {\bibinfo {volume} {806}},\ \bibinfo {eid} {249}
  (\bibinfo {year} {2015})}\BibitemShut {NoStop}%
\bibitem [{\citenamefont {{Meadows}}(2017)}]{Mea17}%
  \BibitemOpen
  \bibfield  {author} {\bibinfo {author} {\bibfnamefont {V.~S.}\ \bibnamefont
  {{Meadows}}},\ }\href {\doibase 10.1089/ast.2016.1578} {\bibfield  {journal}
  {\bibinfo  {journal} {Astrobiology}\ }\textbf {\bibinfo {volume} {17}},\
  \bibinfo {pages} {1022} (\bibinfo {year} {2017})}\BibitemShut {NoStop}%
\bibitem [{\citenamefont {{Meadows}}\ \emph
  {et~al.}(2018{\natexlab{b}})\citenamefont {{Meadows}}, \citenamefont
  {{Reinhard}}, \citenamefont {{Arney}}, \citenamefont {{Parenteau}},
  \citenamefont {{Schwieterman}}, \citenamefont {{Domagal-Goldman}},
  \citenamefont {{Lincowski}}, \citenamefont {{Stapelfeldt}}, \citenamefont
  {{Rauer}}, \citenamefont {{DasSarma}}, \citenamefont {{Hegde}}, \citenamefont
  {{Narita}}, \citenamefont {{Deitrick}}, \citenamefont {{Lustig-Yaeger}},
  \citenamefont {{Lyons}}, \citenamefont {{Siegler}},\ and\ \citenamefont
  {{Grenfell}}}]{MR18}%
  \BibitemOpen
  \bibfield  {author} {\bibinfo {author} {\bibfnamefont {V.~S.}\ \bibnamefont
  {{Meadows}}}, \bibinfo {author} {\bibfnamefont {C.~T.}\ \bibnamefont
  {{Reinhard}}}, \bibinfo {author} {\bibfnamefont {G.~N.}\ \bibnamefont
  {{Arney}}}, \bibinfo {author} {\bibfnamefont {M.~N.}\ \bibnamefont
  {{Parenteau}}}, \bibinfo {author} {\bibfnamefont {E.~W.}\ \bibnamefont
  {{Schwieterman}}}, \bibinfo {author} {\bibfnamefont {S.~D.}\ \bibnamefont
  {{Domagal-Goldman}}}, \bibinfo {author} {\bibfnamefont {A.~P.}\ \bibnamefont
  {{Lincowski}}}, \bibinfo {author} {\bibfnamefont {K.~R.}\ \bibnamefont
  {{Stapelfeldt}}}, \bibinfo {author} {\bibfnamefont {H.}~\bibnamefont
  {{Rauer}}}, \bibinfo {author} {\bibfnamefont {S.}~\bibnamefont {{DasSarma}}},
  \bibinfo {author} {\bibfnamefont {S.}~\bibnamefont {{Hegde}}}, \bibinfo
  {author} {\bibfnamefont {N.}~\bibnamefont {{Narita}}}, \bibinfo {author}
  {\bibfnamefont {R.}~\bibnamefont {{Deitrick}}}, \bibinfo {author}
  {\bibfnamefont {J.}~\bibnamefont {{Lustig-Yaeger}}}, \bibinfo {author}
  {\bibfnamefont {T.~W.}\ \bibnamefont {{Lyons}}}, \bibinfo {author}
  {\bibfnamefont {N.}~\bibnamefont {{Siegler}}}, \ and\ \bibinfo {author}
  {\bibfnamefont {J.~L.}\ \bibnamefont {{Grenfell}}},\ }\href {\doibase
  10.1089/ast.2017.1727} {\bibfield  {journal} {\bibinfo  {journal}
  {Astrobiology}\ }\textbf {\bibinfo {volume} {18}},\ \bibinfo {pages} {630}
  (\bibinfo {year} {2018}{\natexlab{b}})}\BibitemShut {NoStop}%
\bibitem [{\citenamefont {{Bourrier}}\ \emph {et~al.}(2017)\citenamefont
  {{Bourrier}}, \citenamefont {{Ehrenreich}}, \citenamefont {{Wheatley}},
  \citenamefont {{Bolmont}}, \citenamefont {{Gillon}}, \citenamefont {{de
  Wit}}, \citenamefont {{Burgasser}}, \citenamefont {{Jehin}}, \citenamefont
  {{Queloz}},\ and\ \citenamefont {{Triaud}}}]{BE17}%
  \BibitemOpen
  \bibfield  {author} {\bibinfo {author} {\bibfnamefont {V.}~\bibnamefont
  {{Bourrier}}}, \bibinfo {author} {\bibfnamefont {D.}~\bibnamefont
  {{Ehrenreich}}}, \bibinfo {author} {\bibfnamefont {P.~J.}\ \bibnamefont
  {{Wheatley}}}, \bibinfo {author} {\bibfnamefont {E.}~\bibnamefont
  {{Bolmont}}}, \bibinfo {author} {\bibfnamefont {M.}~\bibnamefont {{Gillon}}},
  \bibinfo {author} {\bibfnamefont {J.}~\bibnamefont {{de Wit}}}, \bibinfo
  {author} {\bibfnamefont {A.~J.}\ \bibnamefont {{Burgasser}}}, \bibinfo
  {author} {\bibfnamefont {E.}~\bibnamefont {{Jehin}}}, \bibinfo {author}
  {\bibfnamefont {D.}~\bibnamefont {{Queloz}}}, \ and\ \bibinfo {author}
  {\bibfnamefont {A.~H.~M.~J.}\ \bibnamefont {{Triaud}}},\ }\href {\doibase
  10.1051/0004-6361/201630238} {\bibfield  {journal} {\bibinfo  {journal}
  {Astron. Astrophys.}\ }\textbf {\bibinfo {volume} {599}},\ \bibinfo {eid}
  {L3} (\bibinfo {year} {2017})}\BibitemShut {NoStop}%
\bibitem [{\citenamefont {{O'Brien}}\ \emph {et~al.}(2018)\citenamefont
  {{O'Brien}}, \citenamefont {{Izidoro}}, \citenamefont {{Jacobson}},
  \citenamefont {{Raymond}},\ and\ \citenamefont {{Rubie}}}]{OI18}%
  \BibitemOpen
  \bibfield  {author} {\bibinfo {author} {\bibfnamefont {D.~P.}\ \bibnamefont
  {{O'Brien}}}, \bibinfo {author} {\bibfnamefont {A.}~\bibnamefont
  {{Izidoro}}}, \bibinfo {author} {\bibfnamefont {S.~A.}\ \bibnamefont
  {{Jacobson}}}, \bibinfo {author} {\bibfnamefont {S.~N.}\ \bibnamefont
  {{Raymond}}}, \ and\ \bibinfo {author} {\bibfnamefont {D.~C.}\ \bibnamefont
  {{Rubie}}},\ }\href {\doibase 10.1007/s11214-018-0475-8} {\bibfield
  {journal} {\bibinfo  {journal} {Space Sci. Rev.}\ }\textbf {\bibinfo {volume}
  {214}},\ \bibinfo {eid} {47} (\bibinfo {year} {2018})}\BibitemShut {NoStop}%
\bibitem [{\citenamefont {{Raymond}}\ \emph {et~al.}(2007)\citenamefont
  {{Raymond}}, \citenamefont {{Scalo}},\ and\ \citenamefont
  {{Meadows}}}]{RSM07}%
  \BibitemOpen
  \bibfield  {author} {\bibinfo {author} {\bibfnamefont {S.~N.}\ \bibnamefont
  {{Raymond}}}, \bibinfo {author} {\bibfnamefont {J.}~\bibnamefont {{Scalo}}},
  \ and\ \bibinfo {author} {\bibfnamefont {V.~S.}\ \bibnamefont {{Meadows}}},\
  }\href {\doibase 10.1086/521587} {\bibfield  {journal} {\bibinfo  {journal}
  {Astrophys. J.}\ }\textbf {\bibinfo {volume} {669}},\ \bibinfo {pages} {606}
  (\bibinfo {year} {2007})}\BibitemShut {NoStop}%
\bibitem [{\citenamefont {{Mulders}}\ \emph
  {et~al.}(2015{\natexlab{b}})\citenamefont {{Mulders}}, \citenamefont
  {{Ciesla}}, \citenamefont {{Min}},\ and\ \citenamefont {{Pascucci}}}]{MC15}%
  \BibitemOpen
  \bibfield  {author} {\bibinfo {author} {\bibfnamefont {G.~D.}\ \bibnamefont
  {{Mulders}}}, \bibinfo {author} {\bibfnamefont {F.~J.}\ \bibnamefont
  {{Ciesla}}}, \bibinfo {author} {\bibfnamefont {M.}~\bibnamefont {{Min}}}, \
  and\ \bibinfo {author} {\bibfnamefont {I.}~\bibnamefont {{Pascucci}}},\
  }\href {\doibase 10.1088/0004-637X/807/1/9} {\bibfield  {journal} {\bibinfo
  {journal} {Astrophys. J.}\ }\textbf {\bibinfo {volume} {807}},\ \bibinfo
  {eid} {9} (\bibinfo {year} {2015}{\natexlab{b}})}\BibitemShut {NoStop}%
\bibitem [{\citenamefont {{Raymond}}\ and\ \citenamefont
  {{Izidoro}}(2017)}]{RI17}%
  \BibitemOpen
  \bibfield  {author} {\bibinfo {author} {\bibfnamefont {S.~N.}\ \bibnamefont
  {{Raymond}}}\ and\ \bibinfo {author} {\bibfnamefont {A.}~\bibnamefont
  {{Izidoro}}},\ }\href {\doibase 10.1016/j.icarus.2017.06.030} {\bibfield
  {journal} {\bibinfo  {journal} {Icarus}\ }\textbf {\bibinfo {volume} {297}},\
  \bibinfo {pages} {134} (\bibinfo {year} {2017})}\BibitemShut {NoStop}%
\bibitem [{\citenamefont {{Unterborn}}\ \emph
  {et~al.}(2018{\natexlab{a}})\citenamefont {{Unterborn}}, \citenamefont
  {{Desch}}, \citenamefont {{Hinkel}},\ and\ \citenamefont {{Lorenzo}}}]{UD18}%
  \BibitemOpen
  \bibfield  {author} {\bibinfo {author} {\bibfnamefont {C.~T.}\ \bibnamefont
  {{Unterborn}}}, \bibinfo {author} {\bibfnamefont {S.~J.}\ \bibnamefont
  {{Desch}}}, \bibinfo {author} {\bibfnamefont {N.~R.}\ \bibnamefont
  {{Hinkel}}}, \ and\ \bibinfo {author} {\bibfnamefont {A.}~\bibnamefont
  {{Lorenzo}}},\ }\href {\doibase 10.1038/s41550-018-0411-6} {\bibfield
  {journal} {\bibinfo  {journal} {Nat. Astron.}\ }\textbf {\bibinfo {volume}
  {2}},\ \bibinfo {pages} {297} (\bibinfo {year}
  {2018}{\natexlab{a}})}\BibitemShut {NoStop}%
\bibitem [{\citenamefont {{Unterborn}}\ \emph
  {et~al.}(2018{\natexlab{b}})\citenamefont {{Unterborn}}, \citenamefont
  {{Hinkel}},\ and\ \citenamefont {{Desch}}}]{UHD18}%
  \BibitemOpen
  \bibfield  {author} {\bibinfo {author} {\bibfnamefont {C.~T.}\ \bibnamefont
  {{Unterborn}}}, \bibinfo {author} {\bibfnamefont {N.~R.}\ \bibnamefont
  {{Hinkel}}}, \ and\ \bibinfo {author} {\bibfnamefont {S.~J.}\ \bibnamefont
  {{Desch}}},\ }\href {\doibase 10.3847/2515-5172/aacf43} {\bibfield  {journal}
  {\bibinfo  {journal} {Res. Notes AAS}\ }\textbf {\bibinfo {volume} {2}},\
  \bibinfo {eid} {116} (\bibinfo {year} {2018}{\natexlab{b}})}\BibitemShut
  {NoStop}%
\bibitem [{\citenamefont {{Dorn}}\ \emph {et~al.}(2018)\citenamefont {{Dorn}},
  \citenamefont {{Mosegaard}}, \citenamefont {{Grimm}},\ and\ \citenamefont
  {{Alibert}}}]{DMG18}%
  \BibitemOpen
  \bibfield  {author} {\bibinfo {author} {\bibfnamefont {C.}~\bibnamefont
  {{Dorn}}}, \bibinfo {author} {\bibfnamefont {K.}~\bibnamefont {{Mosegaard}}},
  \bibinfo {author} {\bibfnamefont {S.~L.}\ \bibnamefont {{Grimm}}}, \ and\
  \bibinfo {author} {\bibfnamefont {Y.}~\bibnamefont {{Alibert}}},\ }\href
  {\doibase 10.3847/1538-4357/aad95d} {\bibfield  {journal} {\bibinfo
  {journal} {Astrophys. J.}\ }\textbf {\bibinfo {volume} {865}},\ \bibinfo
  {eid} {20} (\bibinfo {year} {2018})}\BibitemShut {NoStop}%
\bibitem [{\citenamefont {{Rogers}}(2015)}]{Rog15}%
  \BibitemOpen
  \bibfield  {author} {\bibinfo {author} {\bibfnamefont {L.~A.}\ \bibnamefont
  {{Rogers}}},\ }\href {\doibase 10.1088/0004-637X/801/1/41} {\bibfield
  {journal} {\bibinfo  {journal} {Astrophys. J.}\ }\textbf {\bibinfo {volume}
  {801}},\ \bibinfo {eid} {41} (\bibinfo {year} {2015})}\BibitemShut {NoStop}%
\bibitem [{\citenamefont {{Wolfgang}}\ and\ \citenamefont
  {{Lopez}}(2015)}]{WL15}%
  \BibitemOpen
  \bibfield  {author} {\bibinfo {author} {\bibfnamefont {A.}~\bibnamefont
  {{Wolfgang}}}\ and\ \bibinfo {author} {\bibfnamefont {E.}~\bibnamefont
  {{Lopez}}},\ }\href {\doibase 10.1088/0004-637X/806/2/183} {\bibfield
  {journal} {\bibinfo  {journal} {Astrophys. J.}\ }\textbf {\bibinfo {volume}
  {806}},\ \bibinfo {eid} {183} (\bibinfo {year} {2015})}\BibitemShut {NoStop}%
\bibitem [{\citenamefont {{Chen}}\ and\ \citenamefont
  {{Kipping}}(2017)}]{ChKi17}%
  \BibitemOpen
  \bibfield  {author} {\bibinfo {author} {\bibfnamefont {J.}~\bibnamefont
  {{Chen}}}\ and\ \bibinfo {author} {\bibfnamefont {D.}~\bibnamefont
  {{Kipping}}},\ }\href {\doibase 10.3847/1538-4357/834/1/17} {\bibfield
  {journal} {\bibinfo  {journal} {Astrophys. J.}\ }\textbf {\bibinfo {volume}
  {834}},\ \bibinfo {eid} {17} (\bibinfo {year} {2017})}\BibitemShut {NoStop}%
\bibitem [{\citenamefont {{Jin}}\ and\ \citenamefont
  {{Mordasini}}(2018)}]{JM18}%
  \BibitemOpen
  \bibfield  {author} {\bibinfo {author} {\bibfnamefont {S.}~\bibnamefont
  {{Jin}}}\ and\ \bibinfo {author} {\bibfnamefont {C.}~\bibnamefont
  {{Mordasini}}},\ }\href {\doibase 10.3847/1538-4357/aa9f1e} {\bibfield
  {journal} {\bibinfo  {journal} {Astrophys. J.}\ }\textbf {\bibinfo {volume}
  {853}},\ \bibinfo {eid} {163} (\bibinfo {year} {2018})}\BibitemShut {NoStop}%
\bibitem [{\citenamefont {{Lozovsky}}\ \emph {et~al.}(2018)\citenamefont
  {{Lozovsky}}, \citenamefont {{Helled}}, \citenamefont {{Dorn}},\ and\
  \citenamefont {{Venturini}}}]{LHD18}%
  \BibitemOpen
  \bibfield  {author} {\bibinfo {author} {\bibfnamefont {M.}~\bibnamefont
  {{Lozovsky}}}, \bibinfo {author} {\bibfnamefont {R.}~\bibnamefont
  {{Helled}}}, \bibinfo {author} {\bibfnamefont {C.}~\bibnamefont {{Dorn}}}, \
  and\ \bibinfo {author} {\bibfnamefont {J.}~\bibnamefont {{Venturini}}},\
  }\href {\doibase 10.3847/1538-4357/aadd09} {\bibfield  {journal} {\bibinfo
  {journal} {Astrophys. J.}\ }\textbf {\bibinfo {volume} {866}},\ \bibinfo
  {eid} {49} (\bibinfo {year} {2018})}\BibitemShut {NoStop}%
\bibitem [{\citenamefont {{Alibert}}\ and\ \citenamefont
  {{Benz}}(2017)}]{AB17}%
  \BibitemOpen
  \bibfield  {author} {\bibinfo {author} {\bibfnamefont {Y.}~\bibnamefont
  {{Alibert}}}\ and\ \bibinfo {author} {\bibfnamefont {W.}~\bibnamefont
  {{Benz}}},\ }\href {\doibase 10.1051/0004-6361/201629671} {\bibfield
  {journal} {\bibinfo  {journal} {Astron. Astrophys.}\ }\textbf {\bibinfo
  {volume} {598}},\ \bibinfo {eid} {L5} (\bibinfo {year} {2017})}\BibitemShut
  {NoStop}%
\bibitem [{\citenamefont {{Simpson}}(2017)}]{Sim17}%
  \BibitemOpen
  \bibfield  {author} {\bibinfo {author} {\bibfnamefont {F.}~\bibnamefont
  {{Simpson}}},\ }\href {\doibase 10.1093/mnras/stx516} {\bibfield  {journal}
  {\bibinfo  {journal} {Mon. Not. R. Astron. Soc.}\ }\textbf {\bibinfo {volume}
  {468}},\ \bibinfo {pages} {2803} (\bibinfo {year} {2017})}\BibitemShut
  {NoStop}%
\bibitem [{\citenamefont {{Kuchner}}(2003)}]{Kuc03}%
  \BibitemOpen
  \bibfield  {author} {\bibinfo {author} {\bibfnamefont {M.~J.}\ \bibnamefont
  {{Kuchner}}},\ }\href {\doibase 10.1086/378397} {\bibfield  {journal}
  {\bibinfo  {journal} {Astrophys. J. Lett.}\ }\textbf {\bibinfo {volume}
  {596}},\ \bibinfo {pages} {L105} (\bibinfo {year} {2003})}\BibitemShut
  {NoStop}%
\bibitem [{\citenamefont {{L{\'e}ger}}\ \emph {et~al.}(2004)\citenamefont
  {{L{\'e}ger}}, \citenamefont {{Selsis}}, \citenamefont {{Sotin}},
  \citenamefont {{Guillot}}, \citenamefont {{Despois}}, \citenamefont
  {{Mawet}}, \citenamefont {{Ollivier}}, \citenamefont {{Lab{\`e}que}},
  \citenamefont {{Valette}}, \citenamefont {{Brachet}}, \citenamefont
  {{Chazelas}},\ and\ \citenamefont {{Lammer}}}]{LS04}%
  \BibitemOpen
  \bibfield  {author} {\bibinfo {author} {\bibfnamefont {A.}~\bibnamefont
  {{L{\'e}ger}}}, \bibinfo {author} {\bibfnamefont {F.}~\bibnamefont
  {{Selsis}}}, \bibinfo {author} {\bibfnamefont {C.}~\bibnamefont {{Sotin}}},
  \bibinfo {author} {\bibfnamefont {T.}~\bibnamefont {{Guillot}}}, \bibinfo
  {author} {\bibfnamefont {D.}~\bibnamefont {{Despois}}}, \bibinfo {author}
  {\bibfnamefont {D.}~\bibnamefont {{Mawet}}}, \bibinfo {author} {\bibfnamefont
  {M.}~\bibnamefont {{Ollivier}}}, \bibinfo {author} {\bibfnamefont
  {A.}~\bibnamefont {{Lab{\`e}que}}}, \bibinfo {author} {\bibfnamefont
  {C.}~\bibnamefont {{Valette}}}, \bibinfo {author} {\bibfnamefont
  {F.}~\bibnamefont {{Brachet}}}, \bibinfo {author} {\bibfnamefont
  {B.}~\bibnamefont {{Chazelas}}}, \ and\ \bibinfo {author} {\bibfnamefont
  {H.}~\bibnamefont {{Lammer}}},\ }\href {\doibase
  10.1016/j.icarus.2004.01.001} {\bibfield  {journal} {\bibinfo  {journal}
  {Icarus}\ }\textbf {\bibinfo {volume} {169}},\ \bibinfo {pages} {499}
  (\bibinfo {year} {2004})}\BibitemShut {NoStop}%
\bibitem [{\citenamefont {{Noack}}\ \emph {et~al.}(2017)\citenamefont
  {{Noack}}, \citenamefont {{Snellen}},\ and\ \citenamefont {{Rauer}}}]{NSR17}%
  \BibitemOpen
  \bibfield  {author} {\bibinfo {author} {\bibfnamefont {L.}~\bibnamefont
  {{Noack}}}, \bibinfo {author} {\bibfnamefont {I.}~\bibnamefont {{Snellen}}},
  \ and\ \bibinfo {author} {\bibfnamefont {H.}~\bibnamefont {{Rauer}}},\ }\href
  {\doibase 10.1007/s11214-017-0413-1} {\bibfield  {journal} {\bibinfo
  {journal} {Space Sci. Rev.}\ }\textbf {\bibinfo {volume} {212}},\ \bibinfo
  {pages} {877} (\bibinfo {year} {2017})}\BibitemShut {NoStop}%
\bibitem [{\citenamefont {{Kite}}\ and\ \citenamefont {{Ford}}(2018)}]{KF18}%
  \BibitemOpen
  \bibfield  {author} {\bibinfo {author} {\bibfnamefont {E.~S.}\ \bibnamefont
  {{Kite}}}\ and\ \bibinfo {author} {\bibfnamefont {E.~B.}\ \bibnamefont
  {{Ford}}},\ }\href {\doibase 10.3847/1538-4357/aad6e0} {\bibfield  {journal}
  {\bibinfo  {journal} {Astrophys. J.}\ }\textbf {\bibinfo {volume} {864}},\
  \bibinfo {eid} {75} (\bibinfo {year} {2018})}\BibitemShut {NoStop}%
\bibitem [{\citenamefont {{Ramirez}}\ and\ \citenamefont
  {{Levi}}(2018)}]{RL18}%
  \BibitemOpen
  \bibfield  {author} {\bibinfo {author} {\bibfnamefont {R.~M.}\ \bibnamefont
  {{Ramirez}}}\ and\ \bibinfo {author} {\bibfnamefont {A.}~\bibnamefont
  {{Levi}}},\ }\href {\doibase 10.1093/mnras/sty761} {\bibfield  {journal}
  {\bibinfo  {journal} {Mon. Not. R. Astron. Soc.}\ }\textbf {\bibinfo {volume}
  {477}},\ \bibinfo {pages} {4627} (\bibinfo {year} {2018})}\BibitemShut
  {NoStop}%
\bibitem [{\citenamefont {{Hirschmann}}(2006)}]{Hir06}%
  \BibitemOpen
  \bibfield  {author} {\bibinfo {author} {\bibfnamefont {M.~M.}\ \bibnamefont
  {{Hirschmann}}},\ }\href {\doibase 10.1146/annurev.earth.34.031405.125211}
  {\bibfield  {journal} {\bibinfo  {journal} {Annu. Rev. Earth Planet. Sci.}\
  }\textbf {\bibinfo {volume} {34}},\ \bibinfo {pages} {629} (\bibinfo {year}
  {2006})}\BibitemShut {NoStop}%
\bibitem [{\citenamefont {{Korenaga}}(2008)}]{Kor08}%
  \BibitemOpen
  \bibfield  {author} {\bibinfo {author} {\bibfnamefont {J.}~\bibnamefont
  {{Korenaga}}},\ }\href {\doibase 10.1111/j.1365-3121.2008.00843.x} {\bibfield
   {journal} {\bibinfo  {journal} {Terra Nova}\ }\textbf {\bibinfo {volume}
  {20}},\ \bibinfo {pages} {419} (\bibinfo {year} {2008})}\BibitemShut
  {NoStop}%
\bibitem [{\citenamefont {{Ni}}\ \emph {et~al.}(2017)\citenamefont {{Ni}},
  \citenamefont {{Zheng}}, \citenamefont {{Mao}}, \citenamefont {{Wang}},
  \citenamefont {{Chen}},\ and\ \citenamefont {{Zhang}}}]{NZM17}%
  \BibitemOpen
  \bibfield  {author} {\bibinfo {author} {\bibfnamefont {H.}~\bibnamefont
  {{Ni}}}, \bibinfo {author} {\bibfnamefont {Y.-F.}\ \bibnamefont {{Zheng}}},
  \bibinfo {author} {\bibfnamefont {Z.}~\bibnamefont {{Mao}}}, \bibinfo
  {author} {\bibfnamefont {Q.}~\bibnamefont {{Wang}}}, \bibinfo {author}
  {\bibfnamefont {R.-X.}\ \bibnamefont {{Chen}}}, \ and\ \bibinfo {author}
  {\bibfnamefont {L.}~\bibnamefont {{Zhang}}},\ }\href {\doibase
  10.1093/nsr/nwx130} {\bibfield  {journal} {\bibinfo  {journal} {Natl. Sci.
  Rev.}\ }\textbf {\bibinfo {volume} {4}},\ \bibinfo {pages} {879} (\bibinfo
  {year} {2017})}\BibitemShut {NoStop}%
\bibitem [{\citenamefont {{Abe}}\ \emph {et~al.}(2011)\citenamefont {{Abe}},
  \citenamefont {{Abe-Ouchi}}, \citenamefont {{Sleep}},\ and\ \citenamefont
  {{Zahnle}}}]{AA11}%
  \BibitemOpen
  \bibfield  {author} {\bibinfo {author} {\bibfnamefont {Y.}~\bibnamefont
  {{Abe}}}, \bibinfo {author} {\bibfnamefont {A.}~\bibnamefont {{Abe-Ouchi}}},
  \bibinfo {author} {\bibfnamefont {N.~H.}\ \bibnamefont {{Sleep}}}, \ and\
  \bibinfo {author} {\bibfnamefont {K.~J.}\ \bibnamefont {{Zahnle}}},\ }\href
  {\doibase 10.1089/ast.2010.0545} {\bibfield  {journal} {\bibinfo  {journal}
  {Astrobiology}\ }\textbf {\bibinfo {volume} {11}},\ \bibinfo {pages} {443}
  (\bibinfo {year} {2011})}\BibitemShut {NoStop}%
\bibitem [{\citenamefont {{Zsom}}\ \emph {et~al.}(2013)\citenamefont {{Zsom}},
  \citenamefont {{Seager}}, \citenamefont {{de Wit}},\ and\ \citenamefont
  {{Stamenkovi{\'c}}}}]{ZS13}%
  \BibitemOpen
  \bibfield  {author} {\bibinfo {author} {\bibfnamefont {A.}~\bibnamefont
  {{Zsom}}}, \bibinfo {author} {\bibfnamefont {S.}~\bibnamefont {{Seager}}},
  \bibinfo {author} {\bibfnamefont {J.}~\bibnamefont {{de Wit}}}, \ and\
  \bibinfo {author} {\bibfnamefont {V.}~\bibnamefont {{Stamenkovi{\'c}}}},\
  }\href {\doibase 10.1088/0004-637X/778/2/109} {\bibfield  {journal} {\bibinfo
   {journal} {Astrophys. J.}\ }\textbf {\bibinfo {volume} {778}},\ \bibinfo
  {eid} {109} (\bibinfo {year} {2013})}\BibitemShut {NoStop}%
\bibitem [{\citenamefont {{Tian}}\ and\ \citenamefont {{Ida}}(2015)}]{TiId}%
  \BibitemOpen
  \bibfield  {author} {\bibinfo {author} {\bibfnamefont {F.}~\bibnamefont
  {{Tian}}}\ and\ \bibinfo {author} {\bibfnamefont {S.}~\bibnamefont {{Ida}}},\
  }\href {\doibase 10.1038/ngeo2372} {\bibfield  {journal} {\bibinfo  {journal}
  {Nat. Geosci.}\ }\textbf {\bibinfo {volume} {8}},\ \bibinfo {pages} {177}
  (\bibinfo {year} {2015})}\BibitemShut {NoStop}%
\bibitem [{\citenamefont {{Zain}}\ \emph {et~al.}(2018)\citenamefont {{Zain}},
  \citenamefont {{de El{\'{\i}}a}}, \citenamefont {{Ronco}},\ and\
  \citenamefont {{Guilera}}}]{ZDR18}%
  \BibitemOpen
  \bibfield  {author} {\bibinfo {author} {\bibfnamefont {P.~S.}\ \bibnamefont
  {{Zain}}}, \bibinfo {author} {\bibfnamefont {G.~C.}\ \bibnamefont {{de
  El{\'{\i}}a}}}, \bibinfo {author} {\bibfnamefont {M.~P.}\ \bibnamefont
  {{Ronco}}}, \ and\ \bibinfo {author} {\bibfnamefont {O.~M.}\ \bibnamefont
  {{Guilera}}},\ }\href {\doibase 10.1051/0004-6361/201730848} {\bibfield
  {journal} {\bibinfo  {journal} {Astron. Astrophys.}\ }\textbf {\bibinfo
  {volume} {609}},\ \bibinfo {eid} {A76} (\bibinfo {year} {2018})}\BibitemShut
  {NoStop}%
\bibitem [{\citenamefont {{Hadley}}\ and\ \citenamefont
  {{Szarek}}(1981)}]{HS81}%
  \BibitemOpen
  \bibfield  {author} {\bibinfo {author} {\bibfnamefont {N.~F.}\ \bibnamefont
  {{Hadley}}}\ and\ \bibinfo {author} {\bibfnamefont {S.~R.}\ \bibnamefont
  {{Szarek}}},\ }\href {\doibase 10.2307/1308782} {\bibfield  {journal}
  {\bibinfo  {journal} {BioScience}\ }\textbf {\bibinfo {volume} {31}},\
  \bibinfo {pages} {747} (\bibinfo {year} {1981})}\BibitemShut {NoStop}%
\bibitem [{\citenamefont {{Tyrrell}}(1999)}]{Tyrr99}%
  \BibitemOpen
  \bibfield  {author} {\bibinfo {author} {\bibfnamefont {T.}~\bibnamefont
  {{Tyrrell}}},\ }\href {\doibase 10.1038/22941} {\bibfield  {journal}
  {\bibinfo  {journal} {Nature}\ }\textbf {\bibinfo {volume} {400}},\ \bibinfo
  {pages} {525} (\bibinfo {year} {1999})}\BibitemShut {NoStop}%
\bibitem [{\citenamefont {{Paytan}}\ and\ \citenamefont
  {{McLaughlin}}(2007)}]{PM07}%
  \BibitemOpen
  \bibfield  {author} {\bibinfo {author} {\bibfnamefont {A.}~\bibnamefont
  {{Paytan}}}\ and\ \bibinfo {author} {\bibfnamefont {K.}~\bibnamefont
  {{McLaughlin}}},\ }\href {\doibase 10.1021/cr0503613} {\bibfield  {journal}
  {\bibinfo  {journal} {Chem. Rev.}\ }\textbf {\bibinfo {volume} {107}},\
  \bibinfo {pages} {563} (\bibinfo {year} {2007})}\BibitemShut {NoStop}%
\bibitem [{\citenamefont {{Lingam}}\ and\ \citenamefont
  {{Loeb}}(2019{\natexlab{b}})}]{LL17c}%
  \BibitemOpen
  \bibfield  {author} {\bibinfo {author} {\bibfnamefont {M.}~\bibnamefont
  {{Lingam}}}\ and\ \bibinfo {author} {\bibfnamefont {A.}~\bibnamefont
  {{Loeb}}},\ }\href {\doibase 10.1017/S1473550418000083} {\bibfield  {journal}
  {\bibinfo  {journal} {Int. J. Astrobiol.}\ }\textbf {\bibinfo {volume}
  {18}},\ \bibinfo {pages} {112} (\bibinfo {year}
  {2019}{\natexlab{b}})}\BibitemShut {NoStop}%
\bibitem [{\citenamefont {{Lingam}}\ and\ \citenamefont
  {{Loeb}}(2018{\natexlab{d}})}]{LL18d}%
  \BibitemOpen
  \bibfield  {author} {\bibinfo {author} {\bibfnamefont {M.}~\bibnamefont
  {{Lingam}}}\ and\ \bibinfo {author} {\bibfnamefont {A.}~\bibnamefont
  {{Loeb}}},\ }\href {\doibase 10.3847/1538-3881/aada02} {\bibfield  {journal}
  {\bibinfo  {journal} {Astron. J.}\ }\textbf {\bibinfo {volume} {156}},\
  \bibinfo {eid} {151} (\bibinfo {year} {2018}{\natexlab{d}})}\BibitemShut
  {NoStop}%
\bibitem [{\citenamefont {{Lingam}}\ and\ \citenamefont
  {{Loeb}}(2019{\natexlab{c}})}]{LL18e}%
  \BibitemOpen
  \bibfield  {author} {\bibinfo {author} {\bibfnamefont {M.}~\bibnamefont
  {{Lingam}}}\ and\ \bibinfo {author} {\bibfnamefont {A.}~\bibnamefont
  {{Loeb}}},\ }\href {\doibase 10.3847/1538-3881/aaf420} {\bibfield  {journal}
  {\bibinfo  {journal} {Astron. J.}\ }\textbf {\bibinfo {volume} {157}},\
  \bibinfo {eid} {25} (\bibinfo {year} {2019}{\natexlab{c}})}\BibitemShut
  {NoStop}%
\bibitem [{\citenamefont {{McCollom}}(2013)}]{McCo13}%
  \BibitemOpen
  \bibfield  {author} {\bibinfo {author} {\bibfnamefont {T.~M.}\ \bibnamefont
  {{McCollom}}},\ }\href {\doibase 10.1146/annurev-earth-040610-133457}
  {\bibfield  {journal} {\bibinfo  {journal} {Annu. Rev. Earth Planet. Sci.}\
  }\textbf {\bibinfo {volume} {41}},\ \bibinfo {pages} {207} (\bibinfo {year}
  {2013})}\BibitemShut {NoStop}%
\bibitem [{\citenamefont {{Luisi}}(2016)}]{Lu16}%
  \BibitemOpen
  \bibfield  {author} {\bibinfo {author} {\bibfnamefont {P.~L.}\ \bibnamefont
  {{Luisi}}},\ }\href@noop {} {\emph {\bibinfo {title} {The Emergence of Life:
  From Chemical Origins to Synthetic Biology}}}\ (\bibinfo  {publisher}
  {Cambridge Univ. Press},\ \bibinfo {year} {2016})\BibitemShut {NoStop}%
\bibitem [{\citenamefont {{Sutherland}}(2016)}]{Suth16}%
  \BibitemOpen
  \bibfield  {author} {\bibinfo {author} {\bibfnamefont {J.~D.}\ \bibnamefont
  {{Sutherland}}},\ }\href {\doibase 10.1002/anie.201506585} {\bibfield
  {journal} {\bibinfo  {journal} {Angew. Chem. Int. Ed.}\ }\textbf {\bibinfo
  {volume} {55}},\ \bibinfo {pages} {104} (\bibinfo {year} {2016})}\BibitemShut
  {NoStop}%
\bibitem [{\citenamefont {{Gilbert}}(1986)}]{Gil86}%
  \BibitemOpen
  \bibfield  {author} {\bibinfo {author} {\bibfnamefont {W.}~\bibnamefont
  {{Gilbert}}},\ }\href {\doibase 10.1038/319618a0} {\bibfield  {journal}
  {\bibinfo  {journal} {Nature}\ }\textbf {\bibinfo {volume} {319}},\ \bibinfo
  {pages} {618} (\bibinfo {year} {1986})}\BibitemShut {NoStop}%
\bibitem [{\citenamefont {{Joyce}}(2002)}]{Joy02}%
  \BibitemOpen
  \bibfield  {author} {\bibinfo {author} {\bibfnamefont {G.~F.}\ \bibnamefont
  {{Joyce}}},\ }\href {\doibase 10.1038/418214a} {\bibfield  {journal}
  {\bibinfo  {journal} {Nature}\ }\textbf {\bibinfo {volume} {418}},\ \bibinfo
  {pages} {214} (\bibinfo {year} {2002})}\BibitemShut {NoStop}%
\bibitem [{\citenamefont {{Orgel}}(2004)}]{Org04}%
  \BibitemOpen
  \bibfield  {author} {\bibinfo {author} {\bibfnamefont {L.~E.}\ \bibnamefont
  {{Orgel}}},\ }\href {\doibase 10.1080/10409230490460765} {\bibfield
  {journal} {\bibinfo  {journal} {Crit. Rev. Biochem. Mol. Biol.}\ }\textbf
  {\bibinfo {volume} {39}},\ \bibinfo {pages} {99} (\bibinfo {year}
  {2004})}\BibitemShut {NoStop}%
\bibitem [{\citenamefont {{Neveu}}\ \emph {et~al.}(2013)\citenamefont
  {{Neveu}}, \citenamefont {{Kim}},\ and\ \citenamefont {{Benner}}}]{NKB13}%
  \BibitemOpen
  \bibfield  {author} {\bibinfo {author} {\bibfnamefont {M.}~\bibnamefont
  {{Neveu}}}, \bibinfo {author} {\bibfnamefont {H.-J.}\ \bibnamefont {{Kim}}},
  \ and\ \bibinfo {author} {\bibfnamefont {S.~A.}\ \bibnamefont {{Benner}}},\
  }\href {\doibase 10.1089/ast.2012.0868} {\bibfield  {journal} {\bibinfo
  {journal} {Astrobiology}\ }\textbf {\bibinfo {volume} {13}},\ \bibinfo
  {pages} {391} (\bibinfo {year} {2013})}\BibitemShut {NoStop}%
\bibitem [{\citenamefont {{Higgs}}\ and\ \citenamefont
  {{Lehman}}(2015)}]{HL15}%
  \BibitemOpen
  \bibfield  {author} {\bibinfo {author} {\bibfnamefont {P.~G.}\ \bibnamefont
  {{Higgs}}}\ and\ \bibinfo {author} {\bibfnamefont {N.}~\bibnamefont
  {{Lehman}}},\ }\href {\doibase 10.1038/nrg3841} {\bibfield  {journal}
  {\bibinfo  {journal} {Nat. Rev. Genet.}\ }\textbf {\bibinfo {volume} {16}},\
  \bibinfo {pages} {7} (\bibinfo {year} {2015})}\BibitemShut {NoStop}%
\bibitem [{\citenamefont {{W{\"a}chtersh{\"a}user}}(2007)}]{Wac07}%
  \BibitemOpen
  \bibfield  {author} {\bibinfo {author} {\bibfnamefont {G.}~\bibnamefont
  {{W{\"a}chtersh{\"a}user}}},\ }\href {\doibase 10.1002/cbdv.200790052}
  {\bibfield  {journal} {\bibinfo  {journal} {Chem. Biodivers.}\ }\textbf
  {\bibinfo {volume} {4}},\ \bibinfo {pages} {584} (\bibinfo {year}
  {2007})}\BibitemShut {NoStop}%
\bibitem [{\citenamefont {{Ruiz-Mirazo}}\ \emph {et~al.}(2014)\citenamefont
  {{Ruiz-Mirazo}}, \citenamefont {{Briones}},\ and\ \citenamefont {{de la
  Escosura}}}]{RBD14}%
  \BibitemOpen
  \bibfield  {author} {\bibinfo {author} {\bibfnamefont {K.}~\bibnamefont
  {{Ruiz-Mirazo}}}, \bibinfo {author} {\bibfnamefont {C.}~\bibnamefont
  {{Briones}}}, \ and\ \bibinfo {author} {\bibfnamefont {A.}~\bibnamefont {{de
  la Escosura}}},\ }\href {\doibase 10.1021/cr2004844} {\bibfield  {journal}
  {\bibinfo  {journal} {Chem. Rev.}\ }\textbf {\bibinfo {volume} {114}},\
  \bibinfo {pages} {285} (\bibinfo {year} {2014})}\BibitemShut {NoStop}%
\bibitem [{\citenamefont {{Goldford}}\ and\ \citenamefont
  {{Segr{\`e}}}(2018)}]{GS18}%
  \BibitemOpen
  \bibfield  {author} {\bibinfo {author} {\bibfnamefont {J.~E.}\ \bibnamefont
  {{Goldford}}}\ and\ \bibinfo {author} {\bibfnamefont {D.}~\bibnamefont
  {{Segr{\`e}}}},\ }\href {\doibase 10.1016/j.coisb.2018.01.004} {\bibfield
  {journal} {\bibinfo  {journal} {Curr. Opin. Syst. Biol.}\ }\textbf {\bibinfo
  {volume} {4}},\ \bibinfo {pages} {117} (\bibinfo {year} {2018})}\BibitemShut
  {NoStop}%
\bibitem [{\citenamefont {{St{\"u}eken}}\ \emph {et~al.}(2013)\citenamefont
  {{St{\"u}eken}}, \citenamefont {{Anderson}}, \citenamefont {{Bowman}},
  \citenamefont {{Brazelton}}, \citenamefont {{Colangelo-Lillis}},
  \citenamefont {{Goldman}}, \citenamefont {{Som}},\ and\ \citenamefont
  {{Baross}}}]{SAB13}%
  \BibitemOpen
  \bibfield  {author} {\bibinfo {author} {\bibfnamefont {E.~E.}\ \bibnamefont
  {{St{\"u}eken}}}, \bibinfo {author} {\bibfnamefont {R.~E.}\ \bibnamefont
  {{Anderson}}}, \bibinfo {author} {\bibfnamefont {J.~S.}\ \bibnamefont
  {{Bowman}}}, \bibinfo {author} {\bibfnamefont {W.~J.}\ \bibnamefont
  {{Brazelton}}}, \bibinfo {author} {\bibfnamefont {J.}~\bibnamefont
  {{Colangelo-Lillis}}}, \bibinfo {author} {\bibfnamefont {A.~D.}\ \bibnamefont
  {{Goldman}}}, \bibinfo {author} {\bibfnamefont {S.~M.}\ \bibnamefont
  {{Som}}}, \ and\ \bibinfo {author} {\bibfnamefont {J.~A.}\ \bibnamefont
  {{Baross}}},\ }\href {\doibase 10.1111/gbi.12025} {\bibfield  {journal}
  {\bibinfo  {journal} {Geobiology}\ }\textbf {\bibinfo {volume} {11}},\
  \bibinfo {pages} {101} (\bibinfo {year} {2013})}\BibitemShut {NoStop}%
\bibitem [{\citenamefont {{Kitadai}}\ and\ \citenamefont
  {{Maruyama}}(2018)}]{KM18}%
  \BibitemOpen
  \bibfield  {author} {\bibinfo {author} {\bibfnamefont {N.}~\bibnamefont
  {{Kitadai}}}\ and\ \bibinfo {author} {\bibfnamefont {S.}~\bibnamefont
  {{Maruyama}}},\ }\href {\doibase 10.1016/j.gsf.2017.07.007} {\bibfield
  {journal} {\bibinfo  {journal} {Geosci. Front.}\ }\textbf {\bibinfo {volume}
  {9}},\ \bibinfo {pages} {1117} (\bibinfo {year} {2018})}\BibitemShut
  {NoStop}%
\bibitem [{\citenamefont {{Romer}}(1933)}]{Rom33}%
  \BibitemOpen
  \bibfield  {author} {\bibinfo {author} {\bibfnamefont {A.~S.}\ \bibnamefont
  {{Romer}}},\ }\href@noop {} {\emph {\bibinfo {title} {{Man and the
  vertebrates}}}}\ (\bibinfo  {publisher} {The University of Chicago Press.},\
  \bibinfo {year} {1933})\BibitemShut {NoStop}%
\bibitem [{\citenamefont {{Lathe}}(2004)}]{Lat04}%
  \BibitemOpen
  \bibfield  {author} {\bibinfo {author} {\bibfnamefont {R.}~\bibnamefont
  {{Lathe}}},\ }\href {\doibase 10.1016/j.icarus.2003.10.018} {\bibfield
  {journal} {\bibinfo  {journal} {Icarus}\ }\textbf {\bibinfo {volume} {168}},\
  \bibinfo {pages} {18} (\bibinfo {year} {2004})}\BibitemShut {NoStop}%
\bibitem [{\citenamefont {{Balbus}}(2014)}]{Bal14}%
  \BibitemOpen
  \bibfield  {author} {\bibinfo {author} {\bibfnamefont {S.~A.}\ \bibnamefont
  {{Balbus}}},\ }\href {\doibase 10.1098/rspa.2014.0263} {\bibfield  {journal}
  {\bibinfo  {journal} {Proc. R. Soc. London Ser. A}\ }\textbf {\bibinfo
  {volume} {470}},\ \bibinfo {pages} {20140263} (\bibinfo {year}
  {2014})}\BibitemShut {NoStop}%
\bibitem [{\citenamefont {{Lingam}}\ and\ \citenamefont
  {{Loeb}}(2018{\natexlab{e}})}]{LL18f}%
  \BibitemOpen
  \bibfield  {author} {\bibinfo {author} {\bibfnamefont {M.}~\bibnamefont
  {{Lingam}}}\ and\ \bibinfo {author} {\bibfnamefont {A.}~\bibnamefont
  {{Loeb}}},\ }\href {\doibase 10.1089/ast.2017.1718} {\bibfield  {journal}
  {\bibinfo  {journal} {Astrobiology}\ }\textbf {\bibinfo {volume} {18}},\
  \bibinfo {pages} {967} (\bibinfo {year} {2018}{\natexlab{e}})}\BibitemShut
  {NoStop}%
\bibitem [{\citenamefont {{Benner}}\ \emph {et~al.}(2012)\citenamefont
  {{Benner}}, \citenamefont {{Kim}},\ and\ \citenamefont {{Carrigan}}}]{BKC12}%
  \BibitemOpen
  \bibfield  {author} {\bibinfo {author} {\bibfnamefont {S.~A.}\ \bibnamefont
  {{Benner}}}, \bibinfo {author} {\bibfnamefont {H.-J.}\ \bibnamefont {{Kim}}},
  \ and\ \bibinfo {author} {\bibfnamefont {M.~A.}\ \bibnamefont {{Carrigan}}},\
  }\href {\doibase 10.1021/ar200332w} {\bibfield  {journal} {\bibinfo
  {journal} {Acc. Chem. Res.}\ }\textbf {\bibinfo {volume} {45}},\ \bibinfo
  {pages} {2025} (\bibinfo {year} {2012})}\BibitemShut {NoStop}%
\bibitem [{\citenamefont {{Mulkidjanian}}\ \emph {et~al.}(2012)\citenamefont
  {{Mulkidjanian}}, \citenamefont {{Bychkov}}, \citenamefont {{Dibrova}},
  \citenamefont {{Galperin}},\ and\ \citenamefont {{Koonin}}}]{MB12}%
  \BibitemOpen
  \bibfield  {author} {\bibinfo {author} {\bibfnamefont {A.~Y.}\ \bibnamefont
  {{Mulkidjanian}}}, \bibinfo {author} {\bibfnamefont {A.~Y.}\ \bibnamefont
  {{Bychkov}}}, \bibinfo {author} {\bibfnamefont {D.~V.}\ \bibnamefont
  {{Dibrova}}}, \bibinfo {author} {\bibfnamefont {M.~Y.}\ \bibnamefont
  {{Galperin}}}, \ and\ \bibinfo {author} {\bibfnamefont {E.~V.}\ \bibnamefont
  {{Koonin}}},\ }\href {\doibase 10.1073/pnas.1117774109} {\bibfield  {journal}
  {\bibinfo  {journal} {Proc. Natl. Acad. Sci. USA}\ }\textbf {\bibinfo
  {volume} {109}},\ \bibinfo {pages} {E821} (\bibinfo {year}
  {2012})}\BibitemShut {NoStop}%
\bibitem [{\citenamefont {{Martin}}\ \emph {et~al.}(2008)\citenamefont
  {{Martin}}, \citenamefont {{Baross}}, \citenamefont {{Kelley}},\ and\
  \citenamefont {{Russell}}}]{MB08}%
  \BibitemOpen
  \bibfield  {author} {\bibinfo {author} {\bibfnamefont {W.}~\bibnamefont
  {{Martin}}}, \bibinfo {author} {\bibfnamefont {J.}~\bibnamefont {{Baross}}},
  \bibinfo {author} {\bibfnamefont {D.}~\bibnamefont {{Kelley}}}, \ and\
  \bibinfo {author} {\bibfnamefont {M.~J.}\ \bibnamefont {{Russell}}},\ }\href
  {\doibase 10.1038/nrmicro1991} {\bibfield  {journal} {\bibinfo  {journal}
  {Nat. Rev. Microbiol.}\ }\textbf {\bibinfo {volume} {6}},\ \bibinfo {pages}
  {805} (\bibinfo {year} {2008})}\BibitemShut {NoStop}%
\bibitem [{\citenamefont {{Sojo}}\ \emph {et~al.}(2016)\citenamefont {{Sojo}},
  \citenamefont {{Herschy}}, \citenamefont {{Whicher}}, \citenamefont
  {{Camprub{\'{\i}}}},\ and\ \citenamefont {{Lane}}}]{SHW16}%
  \BibitemOpen
  \bibfield  {author} {\bibinfo {author} {\bibfnamefont {V.}~\bibnamefont
  {{Sojo}}}, \bibinfo {author} {\bibfnamefont {B.}~\bibnamefont {{Herschy}}},
  \bibinfo {author} {\bibfnamefont {A.}~\bibnamefont {{Whicher}}}, \bibinfo
  {author} {\bibfnamefont {E.}~\bibnamefont {{Camprub{\'{\i}}}}}, \ and\
  \bibinfo {author} {\bibfnamefont {N.}~\bibnamefont {{Lane}}},\ }\href
  {\doibase 10.1089/ast.2015.1406} {\bibfield  {journal} {\bibinfo  {journal}
  {Astrobiology}\ }\textbf {\bibinfo {volume} {16}},\ \bibinfo {pages} {181}
  (\bibinfo {year} {2016})}\BibitemShut {NoStop}%
\bibitem [{\citenamefont {{Cleaves}}\ \emph {et~al.}(2012)\citenamefont
  {{Cleaves}}, \citenamefont {{Scott}}, \citenamefont {{Hill}}, \citenamefont
  {{Leszczynski}}, \citenamefont {{Sahai}},\ and\ \citenamefont
  {{Hazen}}}]{CSH12}%
  \BibitemOpen
  \bibfield  {author} {\bibinfo {author} {\bibfnamefont {H.~J.}\ \bibnamefont
  {{Cleaves}}}, \bibinfo {author} {\bibfnamefont {A.~M.}\ \bibnamefont
  {{Scott}}}, \bibinfo {author} {\bibfnamefont {F.~C.}\ \bibnamefont {{Hill}}},
  \bibinfo {author} {\bibfnamefont {J.}~\bibnamefont {{Leszczynski}}}, \bibinfo
  {author} {\bibfnamefont {N.}~\bibnamefont {{Sahai}}}, \ and\ \bibinfo
  {author} {\bibfnamefont {R.}~\bibnamefont {{Hazen}}},\ }\href {\doibase
  10.1039/C2CS35112A} {\bibfield  {journal} {\bibinfo  {journal} {Chem. Soc.
  Rev.}\ }\textbf {\bibinfo {volume} {41}},\ \bibinfo {pages} {5502} (\bibinfo
  {year} {2012})}\BibitemShut {NoStop}%
\bibitem [{\citenamefont {{Hazen}}(2017)}]{Haz17}%
  \BibitemOpen
  \bibfield  {author} {\bibinfo {author} {\bibfnamefont {R.~M.}\ \bibnamefont
  {{Hazen}}},\ }\href {\doibase 10.1098/rsta.2016.0353} {\bibfield  {journal}
  {\bibinfo  {journal} {Phil. Trans. R. Soc. A}\ }\textbf {\bibinfo {volume}
  {375}},\ \bibinfo {pages} {20160353} (\bibinfo {year} {2017})}\BibitemShut
  {NoStop}%
\bibitem [{\citenamefont {{Noack}}\ \emph {et~al.}(2016)\citenamefont
  {{Noack}}, \citenamefont {{H{\"o}ning}}, \citenamefont {{Rivoldini}},
  \citenamefont {{Heistracher}}, \citenamefont {{Zimov}}, \citenamefont
  {{Journaux}}, \citenamefont {{Lammer}}, \citenamefont {{Van Hoolst}},\ and\
  \citenamefont {{Bredeh{\"o}ft}}}]{NH16}%
  \BibitemOpen
  \bibfield  {author} {\bibinfo {author} {\bibfnamefont {L.}~\bibnamefont
  {{Noack}}}, \bibinfo {author} {\bibfnamefont {D.}~\bibnamefont
  {{H{\"o}ning}}}, \bibinfo {author} {\bibfnamefont {A.}~\bibnamefont
  {{Rivoldini}}}, \bibinfo {author} {\bibfnamefont {C.}~\bibnamefont
  {{Heistracher}}}, \bibinfo {author} {\bibfnamefont {N.}~\bibnamefont
  {{Zimov}}}, \bibinfo {author} {\bibfnamefont {B.}~\bibnamefont {{Journaux}}},
  \bibinfo {author} {\bibfnamefont {H.}~\bibnamefont {{Lammer}}}, \bibinfo
  {author} {\bibfnamefont {T.}~\bibnamefont {{Van Hoolst}}}, \ and\ \bibinfo
  {author} {\bibfnamefont {J.~H.}\ \bibnamefont {{Bredeh{\"o}ft}}},\ }\href
  {\doibase 10.1016/j.icarus.2016.05.009} {\bibfield  {journal} {\bibinfo
  {journal} {Icarus}\ }\textbf {\bibinfo {volume} {277}},\ \bibinfo {pages}
  {215} (\bibinfo {year} {2016})}\BibitemShut {NoStop}%
\bibitem [{\citenamefont {{Sagan}}\ and\ \citenamefont {{Khare}}(1971)}]{SK71}%
  \BibitemOpen
  \bibfield  {author} {\bibinfo {author} {\bibfnamefont {C.}~\bibnamefont
  {{Sagan}}}\ and\ \bibinfo {author} {\bibfnamefont {B.~N.}\ \bibnamefont
  {{Khare}}},\ }\href {\doibase 10.1126/science.173.3995.417} {\bibfield
  {journal} {\bibinfo  {journal} {Science}\ }\textbf {\bibinfo {volume}
  {173}},\ \bibinfo {pages} {417} (\bibinfo {year} {1971})}\BibitemShut
  {NoStop}%
\bibitem [{\citenamefont {{Sutherland}}(2017)}]{Suth17}%
  \BibitemOpen
  \bibfield  {author} {\bibinfo {author} {\bibfnamefont {J.~D.}\ \bibnamefont
  {{Sutherland}}},\ }\href {\doibase 10.1038/s41570-016-0012} {\bibfield
  {journal} {\bibinfo  {journal} {Nat. Rev. Chem.}\ }\textbf {\bibinfo {volume}
  {1}},\ \bibinfo {pages} {0012} (\bibinfo {year} {2017})}\BibitemShut
  {NoStop}%
\bibitem [{\citenamefont {{Gustavsson}}\ \emph {et~al.}(2010)\citenamefont
  {{Gustavsson}}, \citenamefont {{Improta}},\ and\ \citenamefont
  {{Markovitsi}}}]{GIM10}%
  \BibitemOpen
  \bibfield  {author} {\bibinfo {author} {\bibfnamefont {T.}~\bibnamefont
  {{Gustavsson}}}, \bibinfo {author} {\bibfnamefont {R.}~\bibnamefont
  {{Improta}}}, \ and\ \bibinfo {author} {\bibfnamefont {D.}~\bibnamefont
  {{Markovitsi}}},\ }\href {\doibase 10.1021/jz1004973} {\bibfield  {journal}
  {\bibinfo  {journal} {J. Phys. Chem. Lett.}\ }\textbf {\bibinfo {volume}
  {1}},\ \bibinfo {pages} {2025} (\bibinfo {year} {2010})}\BibitemShut
  {NoStop}%
\bibitem [{\citenamefont {{{\v S}poner}}\ \emph {et~al.}(2016)\citenamefont
  {{{\v S}poner}}, \citenamefont {{Szabla}}, \citenamefont {{G{\'o}ra}},
  \citenamefont {{Saitta}}, \citenamefont {{Pietrucci}}, \citenamefont
  {{Saija}}, \citenamefont {{Di Mauro}}, \citenamefont {{Saladino}},
  \citenamefont {{Ferus}}, \citenamefont {{Civi{\v s}}},\ and\ \citenamefont
  {{{\v S}poner}}}]{SSG16}%
  \BibitemOpen
  \bibfield  {author} {\bibinfo {author} {\bibfnamefont {J.~E.}\ \bibnamefont
  {{{\v S}poner}}}, \bibinfo {author} {\bibfnamefont {R.}~\bibnamefont
  {{Szabla}}}, \bibinfo {author} {\bibfnamefont {R.~W.}\ \bibnamefont
  {{G{\'o}ra}}}, \bibinfo {author} {\bibfnamefont {A.~M.}\ \bibnamefont
  {{Saitta}}}, \bibinfo {author} {\bibfnamefont {F.}~\bibnamefont
  {{Pietrucci}}}, \bibinfo {author} {\bibfnamefont {F.}~\bibnamefont
  {{Saija}}}, \bibinfo {author} {\bibfnamefont {E.}~\bibnamefont {{Di Mauro}}},
  \bibinfo {author} {\bibfnamefont {R.}~\bibnamefont {{Saladino}}}, \bibinfo
  {author} {\bibfnamefont {M.}~\bibnamefont {{Ferus}}}, \bibinfo {author}
  {\bibfnamefont {S.}~\bibnamefont {{Civi{\v s}}}}, \ and\ \bibinfo {author}
  {\bibfnamefont {J.}~\bibnamefont {{{\v S}poner}}},\ }\href {\doibase
  10.1039/C6CP00670A} {\bibfield  {journal} {\bibinfo  {journal} {Phys. Chem.
  Chem. Phys.}\ }\textbf {\bibinfo {volume} {18}},\ \bibinfo {pages} {20047}
  (\bibinfo {year} {2016})}\BibitemShut {NoStop}%
\bibitem [{\citenamefont {{Dibrova}}\ \emph {et~al.}(2012)\citenamefont
  {{Dibrova}}, \citenamefont {{Chudetsky}}, \citenamefont {{Galperin}},
  \citenamefont {{Koonin}},\ and\ \citenamefont {{Mulkidjanian}}}]{DCG12}%
  \BibitemOpen
  \bibfield  {author} {\bibinfo {author} {\bibfnamefont {D.~V.}\ \bibnamefont
  {{Dibrova}}}, \bibinfo {author} {\bibfnamefont {M.~Y.}\ \bibnamefont
  {{Chudetsky}}}, \bibinfo {author} {\bibfnamefont {M.~Y.}\ \bibnamefont
  {{Galperin}}}, \bibinfo {author} {\bibfnamefont {E.~V.}\ \bibnamefont
  {{Koonin}}}, \ and\ \bibinfo {author} {\bibfnamefont {A.~Y.}\ \bibnamefont
  {{Mulkidjanian}}},\ }\href {\doibase 10.1007/s11084-012-9308-z} {\bibfield
  {journal} {\bibinfo  {journal} {Orig. Life Evol. Biosph.}\ }\textbf {\bibinfo
  {volume} {42}},\ \bibinfo {pages} {459} (\bibinfo {year} {2012})}\BibitemShut
  {NoStop}%
\bibitem [{\citenamefont {{Powner}}\ \emph {et~al.}(2009)\citenamefont
  {{Powner}}, \citenamefont {{Gerland}},\ and\ \citenamefont
  {{Sutherland}}}]{PGS09}%
  \BibitemOpen
  \bibfield  {author} {\bibinfo {author} {\bibfnamefont {M.~W.}\ \bibnamefont
  {{Powner}}}, \bibinfo {author} {\bibfnamefont {B.}~\bibnamefont {{Gerland}}},
  \ and\ \bibinfo {author} {\bibfnamefont {J.~D.}\ \bibnamefont
  {{Sutherland}}},\ }\href {\doibase 10.1038/nature08013} {\bibfield  {journal}
  {\bibinfo  {journal} {Nature}\ }\textbf {\bibinfo {volume} {459}},\ \bibinfo
  {pages} {239} (\bibinfo {year} {2009})}\BibitemShut {NoStop}%
\bibitem [{\citenamefont {{Islam}}\ and\ \citenamefont
  {{Powner}}(2017)}]{IP17}%
  \BibitemOpen
  \bibfield  {author} {\bibinfo {author} {\bibfnamefont {S.}~\bibnamefont
  {{Islam}}}\ and\ \bibinfo {author} {\bibfnamefont {M.~W.}\ \bibnamefont
  {{Powner}}},\ }\href {\doibase 10.1016/j.chempr.2017.03.001} {\bibfield
  {journal} {\bibinfo  {journal} {Chem}\ }\textbf {\bibinfo {volume} {2}},\
  \bibinfo {pages} {470} (\bibinfo {year} {2017})}\BibitemShut {NoStop}%
\bibitem [{\citenamefont {{Ritson}}\ and\ \citenamefont
  {{Sutherland}}(2012)}]{RS12}%
  \BibitemOpen
  \bibfield  {author} {\bibinfo {author} {\bibfnamefont {D.}~\bibnamefont
  {{Ritson}}}\ and\ \bibinfo {author} {\bibfnamefont {J.~D.}\ \bibnamefont
  {{Sutherland}}},\ }\href {\doibase 10.1038/nchem.1467} {\bibfield  {journal}
  {\bibinfo  {journal} {Nat. Chem.}\ }\textbf {\bibinfo {volume} {4}},\
  \bibinfo {pages} {895} (\bibinfo {year} {2012})}\BibitemShut {NoStop}%
\bibitem [{\citenamefont {{Ritson}}\ and\ \citenamefont
  {{Sutherland}}(2013)}]{RS13}%
  \BibitemOpen
  \bibfield  {author} {\bibinfo {author} {\bibfnamefont {D.~J.}\ \bibnamefont
  {{Ritson}}}\ and\ \bibinfo {author} {\bibfnamefont {J.~D.}\ \bibnamefont
  {{Sutherland}}},\ }\href {\doibase 10.1002/anie.201300321} {\bibfield
  {journal} {\bibinfo  {journal} {Angew. Chem. Int. Ed.}\ }\textbf {\bibinfo
  {volume} {52}},\ \bibinfo {pages} {5845} (\bibinfo {year}
  {2013})}\BibitemShut {NoStop}%
\bibitem [{\citenamefont {{Todd}}\ \emph {et~al.}(2018)\citenamefont {{Todd}},
  \citenamefont {{Fahrenbach}}, \citenamefont {{Magnani}}, \citenamefont
  {{Ranjan}}, \citenamefont {{Bj{\"o}rkbom}}, \citenamefont {{Szostak}},\ and\
  \citenamefont {{Sasselov}}}]{TFM18}%
  \BibitemOpen
  \bibfield  {author} {\bibinfo {author} {\bibfnamefont {Z.~R.}\ \bibnamefont
  {{Todd}}}, \bibinfo {author} {\bibfnamefont {A.~C.}\ \bibnamefont
  {{Fahrenbach}}}, \bibinfo {author} {\bibfnamefont {C.~J.}\ \bibnamefont
  {{Magnani}}}, \bibinfo {author} {\bibfnamefont {S.}~\bibnamefont {{Ranjan}}},
  \bibinfo {author} {\bibfnamefont {A.}~\bibnamefont {{Bj{\"o}rkbom}}},
  \bibinfo {author} {\bibfnamefont {J.~W.}\ \bibnamefont {{Szostak}}}, \ and\
  \bibinfo {author} {\bibfnamefont {D.~D.}\ \bibnamefont {{Sasselov}}},\ }\href
  {\doibase 10.1039/c7cc07748c} {\bibfield  {journal} {\bibinfo  {journal}
  {Chem. Commun.}\ }\textbf {\bibinfo {volume} {54}},\ \bibinfo {pages} {1121}
  (\bibinfo {year} {2018})}\BibitemShut {NoStop}%
\bibitem [{\citenamefont {{Patel}}\ \emph {et~al.}(2015)\citenamefont
  {{Patel}}, \citenamefont {{Percivalle}}, \citenamefont {{Ritson}},
  \citenamefont {{Duffy}},\ and\ \citenamefont {{Sutherland}}}]{Pat15}%
  \BibitemOpen
  \bibfield  {author} {\bibinfo {author} {\bibfnamefont {B.~H.}\ \bibnamefont
  {{Patel}}}, \bibinfo {author} {\bibfnamefont {C.}~\bibnamefont
  {{Percivalle}}}, \bibinfo {author} {\bibfnamefont {D.~J.}\ \bibnamefont
  {{Ritson}}}, \bibinfo {author} {\bibfnamefont {C.~D.}\ \bibnamefont
  {{Duffy}}}, \ and\ \bibinfo {author} {\bibfnamefont {J.~D.}\ \bibnamefont
  {{Sutherland}}},\ }\href {\doibase 10.1038/nchem.2202} {\bibfield  {journal}
  {\bibinfo  {journal} {Nat. Chem.}\ }\textbf {\bibinfo {volume} {7}},\
  \bibinfo {pages} {301} (\bibinfo {year} {2015})}\BibitemShut {NoStop}%
\bibitem [{\citenamefont {{Xu}}\ \emph {et~al.}(2018)\citenamefont {{Xu}},
  \citenamefont {{Ritson}}, \citenamefont {{Ranjan}}, \citenamefont {{Todd}},
  \citenamefont {{Sasselov}},\ and\ \citenamefont {{Sutherland}}}]{XRR18}%
  \BibitemOpen
  \bibfield  {author} {\bibinfo {author} {\bibfnamefont {J.}~\bibnamefont
  {{Xu}}}, \bibinfo {author} {\bibfnamefont {D.~J.}\ \bibnamefont {{Ritson}}},
  \bibinfo {author} {\bibfnamefont {S.}~\bibnamefont {{Ranjan}}}, \bibinfo
  {author} {\bibfnamefont {Z.~R.}\ \bibnamefont {{Todd}}}, \bibinfo {author}
  {\bibfnamefont {D.~D.}\ \bibnamefont {{Sasselov}}}, \ and\ \bibinfo {author}
  {\bibfnamefont {J.~D.}\ \bibnamefont {{Sutherland}}},\ }\href {\doibase
  10.1039/c8cc01499j} {\bibfield  {journal} {\bibinfo  {journal} {Chem.
  Commun.}\ }\textbf {\bibinfo {volume} {54}},\ \bibinfo {pages} {5566}
  (\bibinfo {year} {2018})}\BibitemShut {NoStop}%
\bibitem [{\citenamefont {{Bonfio}}\ \emph {et~al.}(2017)\citenamefont
  {{Bonfio}}, \citenamefont {{Valer}}, \citenamefont {{Scintilla}},
  \citenamefont {{Shah}}, \citenamefont {{Evans}}, \citenamefont {{Jin}},
  \citenamefont {{Szostak}}, \citenamefont {{Sasselov}}, \citenamefont
  {{Sutherland}},\ and\ \citenamefont {{Mansy}}}]{BVS17}%
  \BibitemOpen
  \bibfield  {author} {\bibinfo {author} {\bibfnamefont {C.}~\bibnamefont
  {{Bonfio}}}, \bibinfo {author} {\bibfnamefont {L.}~\bibnamefont {{Valer}}},
  \bibinfo {author} {\bibfnamefont {S.}~\bibnamefont {{Scintilla}}}, \bibinfo
  {author} {\bibfnamefont {S.}~\bibnamefont {{Shah}}}, \bibinfo {author}
  {\bibfnamefont {D.~J.}\ \bibnamefont {{Evans}}}, \bibinfo {author}
  {\bibfnamefont {L.}~\bibnamefont {{Jin}}}, \bibinfo {author} {\bibfnamefont
  {J.~W.}\ \bibnamefont {{Szostak}}}, \bibinfo {author} {\bibfnamefont {D.~D.}\
  \bibnamefont {{Sasselov}}}, \bibinfo {author} {\bibfnamefont {J.~D.}\
  \bibnamefont {{Sutherland}}}, \ and\ \bibinfo {author} {\bibfnamefont
  {S.~S.}\ \bibnamefont {{Mansy}}},\ }\href {\doibase 10.1038/nchem.2817}
  {\bibfield  {journal} {\bibinfo  {journal} {Nat. Chem.}\ }\textbf {\bibinfo
  {volume} {9}},\ \bibinfo {pages} {1229} (\bibinfo {year} {2017})}\BibitemShut
  {NoStop}%
\bibitem [{\citenamefont {{Rios}}\ and\ \citenamefont {{Tor}}(2013)}]{RT13}%
  \BibitemOpen
  \bibfield  {author} {\bibinfo {author} {\bibfnamefont {A.~C.}\ \bibnamefont
  {{Rios}}}\ and\ \bibinfo {author} {\bibfnamefont {Y.}~\bibnamefont {{Tor}}},\
  }\href {\doibase 10.1002/ijch.201300009} {\bibfield  {journal} {\bibinfo
  {journal} {Isr. J. Chem.}\ }\textbf {\bibinfo {volume} {53}},\ \bibinfo
  {pages} {469} (\bibinfo {year} {2013})}\BibitemShut {NoStop}%
\bibitem [{\citenamefont {{Beckstead}}\ \emph {et~al.}(2016)\citenamefont
  {{Beckstead}}, \citenamefont {{Zhang}}, \citenamefont {{de Vries}},\ and\
  \citenamefont {{Kohler}}}]{Beck16}%
  \BibitemOpen
  \bibfield  {author} {\bibinfo {author} {\bibfnamefont {A.~A.}\ \bibnamefont
  {{Beckstead}}}, \bibinfo {author} {\bibfnamefont {Y.}~\bibnamefont
  {{Zhang}}}, \bibinfo {author} {\bibfnamefont {M.~S.}\ \bibnamefont {{de
  Vries}}}, \ and\ \bibinfo {author} {\bibfnamefont {B.}~\bibnamefont
  {{Kohler}}},\ }\href {\doibase 10.1039/C6CP04230A} {\bibfield  {journal}
  {\bibinfo  {journal} {Phys. Chem. Chem. Phys.}\ }\textbf {\bibinfo {volume}
  {18}},\ \bibinfo {pages} {24228} (\bibinfo {year} {2016})}\BibitemShut
  {NoStop}%
\bibitem [{\citenamefont {{Ranjan}}\ and\ \citenamefont
  {{Sasselov}}(2016)}]{RS16}%
  \BibitemOpen
  \bibfield  {author} {\bibinfo {author} {\bibfnamefont {S.}~\bibnamefont
  {{Ranjan}}}\ and\ \bibinfo {author} {\bibfnamefont {D.~D.}\ \bibnamefont
  {{Sasselov}}},\ }\href {\doibase 10.1089/ast.2015.1359} {\bibfield  {journal}
  {\bibinfo  {journal} {Astrobiology}\ }\textbf {\bibinfo {volume} {16}},\
  \bibinfo {pages} {68} (\bibinfo {year} {2016})}\BibitemShut {NoStop}%
\bibitem [{\citenamefont {{Rugheimer}}\ \emph
  {et~al.}(2015{\natexlab{a}})\citenamefont {{Rugheimer}}, \citenamefont
  {{Segura}}, \citenamefont {{Kaltenegger}},\ and\ \citenamefont
  {{Sasselov}}}]{RSK15}%
  \BibitemOpen
  \bibfield  {author} {\bibinfo {author} {\bibfnamefont {S.}~\bibnamefont
  {{Rugheimer}}}, \bibinfo {author} {\bibfnamefont {A.}~\bibnamefont
  {{Segura}}}, \bibinfo {author} {\bibfnamefont {L.}~\bibnamefont
  {{Kaltenegger}}}, \ and\ \bibinfo {author} {\bibfnamefont {D.}~\bibnamefont
  {{Sasselov}}},\ }\href {\doibase 10.1088/0004-637X/806/1/137} {\bibfield
  {journal} {\bibinfo  {journal} {Astrophys. J.}\ }\textbf {\bibinfo {volume}
  {806}},\ \bibinfo {eid} {137} (\bibinfo {year}
  {2015}{\natexlab{a}})}\BibitemShut {NoStop}%
\bibitem [{\citenamefont {{Ranjan}}\ \emph {et~al.}(2017)\citenamefont
  {{Ranjan}}, \citenamefont {{Wordsworth}},\ and\ \citenamefont
  {{Sasselov}}}]{RWS17}%
  \BibitemOpen
  \bibfield  {author} {\bibinfo {author} {\bibfnamefont {S.}~\bibnamefont
  {{Ranjan}}}, \bibinfo {author} {\bibfnamefont {R.}~\bibnamefont
  {{Wordsworth}}}, \ and\ \bibinfo {author} {\bibfnamefont {D.~D.}\
  \bibnamefont {{Sasselov}}},\ }\href {\doibase 10.3847/1538-4357/aa773e}
  {\bibfield  {journal} {\bibinfo  {journal} {Astrophys. J.}\ }\textbf
  {\bibinfo {volume} {843}},\ \bibinfo {eid} {110} (\bibinfo {year}
  {2017})}\BibitemShut {NoStop}%
\bibitem [{\citenamefont {{Buccino}}\ \emph {et~al.}(2007)\citenamefont
  {{Buccino}}, \citenamefont {{Lemarchand}},\ and\ \citenamefont
  {{Mauas}}}]{BLM07}%
  \BibitemOpen
  \bibfield  {author} {\bibinfo {author} {\bibfnamefont {A.~P.}\ \bibnamefont
  {{Buccino}}}, \bibinfo {author} {\bibfnamefont {G.~A.}\ \bibnamefont
  {{Lemarchand}}}, \ and\ \bibinfo {author} {\bibfnamefont {P.~J.~D.}\
  \bibnamefont {{Mauas}}},\ }\href {\doibase 10.1016/j.icarus.2007.08.012}
  {\bibfield  {journal} {\bibinfo  {journal} {Icarus}\ }\textbf {\bibinfo
  {volume} {192}},\ \bibinfo {pages} {582} (\bibinfo {year}
  {2007})}\BibitemShut {NoStop}%
\bibitem [{\citenamefont {{Guo}}\ \emph {et~al.}(2010)\citenamefont {{Guo}},
  \citenamefont {{Zhang}}, \citenamefont {{Zhang}},\ and\ \citenamefont
  {{Han}}}]{GZZ10}%
  \BibitemOpen
  \bibfield  {author} {\bibinfo {author} {\bibfnamefont {J.}~\bibnamefont
  {{Guo}}}, \bibinfo {author} {\bibfnamefont {F.}~\bibnamefont {{Zhang}}},
  \bibinfo {author} {\bibfnamefont {X.}~\bibnamefont {{Zhang}}}, \ and\
  \bibinfo {author} {\bibfnamefont {Z.}~\bibnamefont {{Han}}},\ }\href
  {\doibase 10.1007/s10509-009-0173-9} {\bibfield  {journal} {\bibinfo
  {journal} {Astrophys. Space Sci.}\ }\textbf {\bibinfo {volume} {325}},\
  \bibinfo {pages} {25} (\bibinfo {year} {2010})}\BibitemShut {NoStop}%
\bibitem [{\citenamefont {{Rimmer}}\ \emph {et~al.}(2018)\citenamefont
  {{Rimmer}}, \citenamefont {{Xu}}, \citenamefont {{Thompson}}, \citenamefont
  {{Gillen}}, \citenamefont {{Sutherland}},\ and\ \citenamefont
  {{Queloz}}}]{RXT18}%
  \BibitemOpen
  \bibfield  {author} {\bibinfo {author} {\bibfnamefont {P.~B.}\ \bibnamefont
  {{Rimmer}}}, \bibinfo {author} {\bibfnamefont {J.}~\bibnamefont {{Xu}}},
  \bibinfo {author} {\bibfnamefont {S.~J.}\ \bibnamefont {{Thompson}}},
  \bibinfo {author} {\bibfnamefont {E.}~\bibnamefont {{Gillen}}}, \bibinfo
  {author} {\bibfnamefont {J.~D.}\ \bibnamefont {{Sutherland}}}, \ and\
  \bibinfo {author} {\bibfnamefont {D.}~\bibnamefont {{Queloz}}},\ }\href
  {\doibase 10.1126/sciadv.aar3302} {\bibfield  {journal} {\bibinfo  {journal}
  {Sci. Adv.}\ }\textbf {\bibinfo {volume} {4}},\ \bibinfo {pages} {eaar3302}
  (\bibinfo {year} {2018})}\BibitemShut {NoStop}%
\bibitem [{\citenamefont {{Oishi}}\ and\ \citenamefont
  {{Kamaya}}(2016)}]{OK16}%
  \BibitemOpen
  \bibfield  {author} {\bibinfo {author} {\bibfnamefont {M.}~\bibnamefont
  {{Oishi}}}\ and\ \bibinfo {author} {\bibfnamefont {H.}~\bibnamefont
  {{Kamaya}}},\ }\href {\doibase 10.3847/1538-4357/833/2/293} {\bibfield
  {journal} {\bibinfo  {journal} {Astrophys. J.}\ }\textbf {\bibinfo {volume}
  {833}},\ \bibinfo {eid} {293} (\bibinfo {year} {2016})}\BibitemShut {NoStop}%
\bibitem [{\citenamefont {{Voet}}\ \emph {et~al.}(1963)\citenamefont {{Voet}},
  \citenamefont {{Gratzer}}, \citenamefont {{Cox}},\ and\ \citenamefont
  {{Doty}}}]{VGCD}%
  \BibitemOpen
  \bibfield  {author} {\bibinfo {author} {\bibfnamefont {D.}~\bibnamefont
  {{Voet}}}, \bibinfo {author} {\bibfnamefont {W.~B.}\ \bibnamefont
  {{Gratzer}}}, \bibinfo {author} {\bibfnamefont {R.~A.}\ \bibnamefont
  {{Cox}}}, \ and\ \bibinfo {author} {\bibfnamefont {P.}~\bibnamefont
  {{Doty}}},\ }\href {\doibase 10.1002/bip.360010302} {\bibfield  {journal}
  {\bibinfo  {journal} {Biopolymers}\ }\textbf {\bibinfo {volume} {1}},\
  \bibinfo {pages} {193} (\bibinfo {year} {1963})}\BibitemShut {NoStop}%
\bibitem [{\citenamefont {{Sagan}}(1973)}]{Sag73}%
  \BibitemOpen
  \bibfield  {author} {\bibinfo {author} {\bibfnamefont {C.}~\bibnamefont
  {{Sagan}}},\ }\href {\doibase 10.1016/0022-5193(73)90216-6} {\bibfield
  {journal} {\bibinfo  {journal} {J. Theor. Biol.}\ }\textbf {\bibinfo {volume}
  {39}},\ \bibinfo {pages} {195} (\bibinfo {year} {1973})}\BibitemShut
  {NoStop}%
\bibitem [{\citenamefont {{Teramura}}\ and\ \citenamefont
  {{Sullivan}}(1994)}]{TS94}%
  \BibitemOpen
  \bibfield  {author} {\bibinfo {author} {\bibfnamefont {A.~H.}\ \bibnamefont
  {{Teramura}}}\ and\ \bibinfo {author} {\bibfnamefont {J.~H.}\ \bibnamefont
  {{Sullivan}}},\ }\href {\doibase 10.1007/BF00014599} {\bibfield  {journal}
  {\bibinfo  {journal} {Photosynth. Res.}\ }\textbf {\bibinfo {volume} {39}},\
  \bibinfo {pages} {463} (\bibinfo {year} {1994})}\BibitemShut {NoStop}%
\bibitem [{\citenamefont {{Cadet}}\ \emph {et~al.}(2005)\citenamefont
  {{Cadet}}, \citenamefont {{Sage}},\ and\ \citenamefont {{Douki}}}]{CSD05}%
  \BibitemOpen
  \bibfield  {author} {\bibinfo {author} {\bibfnamefont {J.}~\bibnamefont
  {{Cadet}}}, \bibinfo {author} {\bibfnamefont {E.}~\bibnamefont {{Sage}}}, \
  and\ \bibinfo {author} {\bibfnamefont {T.}~\bibnamefont {{Douki}}},\ }\href
  {\doibase 10.1016/j.mrfmmm.2004.09.012} {\bibfield  {journal} {\bibinfo
  {journal} {Mutat. Res.}\ }\textbf {\bibinfo {volume} {571}},\ \bibinfo
  {pages} {3} (\bibinfo {year} {2005})}\BibitemShut {NoStop}%
\bibitem [{\citenamefont {{Rothschild}}(1999)}]{Roth99}%
  \BibitemOpen
  \bibfield  {author} {\bibinfo {author} {\bibfnamefont {L.~J.}\ \bibnamefont
  {{Rothschild}}},\ }\href {\doibase 10.1111/j.1550-7408.1999.tb06074.x}
  {\bibfield  {journal} {\bibinfo  {journal} {J. Eukaryot. Microbiol.}\
  }\textbf {\bibinfo {volume} {46}},\ \bibinfo {pages} {548} (\bibinfo {year}
  {1999})}\BibitemShut {NoStop}%
\bibitem [{\citenamefont {{Evans}}\ and\ \citenamefont
  {{Gaston}}(2005)}]{EG05}%
  \BibitemOpen
  \bibfield  {author} {\bibinfo {author} {\bibfnamefont {K.~L.}\ \bibnamefont
  {{Evans}}}\ and\ \bibinfo {author} {\bibfnamefont {K.~J.}\ \bibnamefont
  {{Gaston}}},\ }\href {\doibase 10.1111/j.1365-2435.2005.01046.x} {\bibfield
  {journal} {\bibinfo  {journal} {Funct. Ecol.}\ }\textbf {\bibinfo {volume}
  {19}},\ \bibinfo {pages} {899} (\bibinfo {year} {2005})}\BibitemShut
  {NoStop}%
\bibitem [{\citenamefont {{Cleaves}}\ and\ \citenamefont
  {{Miller}}(1998)}]{CM98}%
  \BibitemOpen
  \bibfield  {author} {\bibinfo {author} {\bibfnamefont {H.~J.}\ \bibnamefont
  {{Cleaves}}}\ and\ \bibinfo {author} {\bibfnamefont {S.~L.}\ \bibnamefont
  {{Miller}}},\ }\href {\doibase 10.1073/pnas.95.13.7260} {\bibfield  {journal}
  {\bibinfo  {journal} {Proc. Natl. Acad. Sci. USA}\ }\textbf {\bibinfo
  {volume} {95}},\ \bibinfo {pages} {7260} (\bibinfo {year}
  {1998})}\BibitemShut {NoStop}%
\bibitem [{\citenamefont {{Arney}}\ \emph {et~al.}(2016)\citenamefont
  {{Arney}}, \citenamefont {{Domagal-Goldman}}, \citenamefont {{Meadows}},
  \citenamefont {{Wolf}}, \citenamefont {{Schwieterman}}, \citenamefont
  {{Charnay}}, \citenamefont {{Claire}}, \citenamefont {{H{\'e}brard}},\ and\
  \citenamefont {{Trainer}}}]{ADG16}%
  \BibitemOpen
  \bibfield  {author} {\bibinfo {author} {\bibfnamefont {G.}~\bibnamefont
  {{Arney}}}, \bibinfo {author} {\bibfnamefont {S.~D.}\ \bibnamefont
  {{Domagal-Goldman}}}, \bibinfo {author} {\bibfnamefont {V.~S.}\ \bibnamefont
  {{Meadows}}}, \bibinfo {author} {\bibfnamefont {E.~T.}\ \bibnamefont
  {{Wolf}}}, \bibinfo {author} {\bibfnamefont {E.}~\bibnamefont
  {{Schwieterman}}}, \bibinfo {author} {\bibfnamefont {B.}~\bibnamefont
  {{Charnay}}}, \bibinfo {author} {\bibfnamefont {M.}~\bibnamefont {{Claire}}},
  \bibinfo {author} {\bibfnamefont {E.}~\bibnamefont {{H{\'e}brard}}}, \ and\
  \bibinfo {author} {\bibfnamefont {M.~G.}\ \bibnamefont {{Trainer}}},\ }\href
  {\doibase 10.1089/ast.2015.1422} {\bibfield  {journal} {\bibinfo  {journal}
  {Astrobiology}\ }\textbf {\bibinfo {volume} {16}},\ \bibinfo {pages} {873}
  (\bibinfo {year} {2016})}\BibitemShut {NoStop}%
\bibitem [{\citenamefont {{Cockell}}\ and\ \citenamefont
  {{Knowland}}(1999)}]{CK99}%
  \BibitemOpen
  \bibfield  {author} {\bibinfo {author} {\bibfnamefont {C.~S.}\ \bibnamefont
  {{Cockell}}}\ and\ \bibinfo {author} {\bibfnamefont {J.}~\bibnamefont
  {{Knowland}}},\ }\href {\doibase 10.1017/S0006323199005356} {\bibfield
  {journal} {\bibinfo  {journal} {Biol. Rev.}\ }\textbf {\bibinfo {volume}
  {74}},\ \bibinfo {pages} {311} (\bibinfo {year} {1999})}\BibitemShut
  {NoStop}%
\bibitem [{\citenamefont {{Gao}}\ and\ \citenamefont
  {{Garcia-Pichel}}(2011)}]{GG11}%
  \BibitemOpen
  \bibfield  {author} {\bibinfo {author} {\bibfnamefont {Q.}~\bibnamefont
  {{Gao}}}\ and\ \bibinfo {author} {\bibfnamefont {F.}~\bibnamefont
  {{Garcia-Pichel}}},\ }\href {\doibase 10.1038/nrmicro2649} {\bibfield
  {journal} {\bibinfo  {journal} {Nat. Rev. Microbiol.}\ }\textbf {\bibinfo
  {volume} {9}},\ \bibinfo {pages} {791} (\bibinfo {year} {2011})}\BibitemShut
  {NoStop}%
\bibitem [{\citenamefont {{Gabani}}\ and\ \citenamefont
  {{Singh}}(2013)}]{GS13}%
  \BibitemOpen
  \bibfield  {author} {\bibinfo {author} {\bibfnamefont {P.}~\bibnamefont
  {{Gabani}}}\ and\ \bibinfo {author} {\bibfnamefont {O.~V.}\ \bibnamefont
  {{Singh}}},\ }\href {\doibase 10.1007/s00253-012-4642-7} {\bibfield
  {journal} {\bibinfo  {journal} {Appl. Microbiol. Biotechnol.}\ }\textbf
  {\bibinfo {volume} {97}},\ \bibinfo {pages} {993} (\bibinfo {year}
  {2013})}\BibitemShut {NoStop}%
\bibitem [{\citenamefont {{Pacelli}}\ \emph {et~al.}(2017)\citenamefont
  {{Pacelli}}, \citenamefont {{Bryan}}, \citenamefont {{Onofri}}, \citenamefont
  {{Selbmann}}, \citenamefont {{Shuryak}},\ and\ \citenamefont
  {{Dadachova}}}]{PBO17}%
  \BibitemOpen
  \bibfield  {author} {\bibinfo {author} {\bibfnamefont {C.}~\bibnamefont
  {{Pacelli}}}, \bibinfo {author} {\bibfnamefont {R.~A.}\ \bibnamefont
  {{Bryan}}}, \bibinfo {author} {\bibfnamefont {S.}~\bibnamefont {{Onofri}}},
  \bibinfo {author} {\bibfnamefont {L.}~\bibnamefont {{Selbmann}}}, \bibinfo
  {author} {\bibfnamefont {I.}~\bibnamefont {{Shuryak}}}, \ and\ \bibinfo
  {author} {\bibfnamefont {E.}~\bibnamefont {{Dadachova}}},\ }\href {\doibase
  10.1111/1462-2920.13681} {\bibfield  {journal} {\bibinfo  {journal} {Environ.
  Microbiol.}\ }\textbf {\bibinfo {volume} {19}},\ \bibinfo {pages} {1612}
  (\bibinfo {year} {2017})}\BibitemShut {NoStop}%
\bibitem [{\citenamefont {{Jung}}\ \emph {et~al.}(2017)\citenamefont {{Jung}},
  \citenamefont {{Lim}},\ and\ \citenamefont {{Bahn}}}]{JLB17}%
  \BibitemOpen
  \bibfield  {author} {\bibinfo {author} {\bibfnamefont {K.-W.}\ \bibnamefont
  {{Jung}}}, \bibinfo {author} {\bibfnamefont {S.}~\bibnamefont {{Lim}}}, \
  and\ \bibinfo {author} {\bibfnamefont {Y.-S.}\ \bibnamefont {{Bahn}}},\
  }\href {\doibase 10.1007/s12275-017-7242-5} {\bibfield  {journal} {\bibinfo
  {journal} {J. Microbiol.}\ }\textbf {\bibinfo {volume} {55}},\ \bibinfo
  {pages} {499} (\bibinfo {year} {2017})}\BibitemShut {NoStop}%
\bibitem [{\citenamefont {{O'Malley-James}}\ and\ \citenamefont
  {{Kaltenegger}}(2017)}]{OMJ17}%
  \BibitemOpen
  \bibfield  {author} {\bibinfo {author} {\bibfnamefont {J.~T.}\ \bibnamefont
  {{O'Malley-James}}}\ and\ \bibinfo {author} {\bibfnamefont {L.}~\bibnamefont
  {{Kaltenegger}}},\ }\href {\doibase 10.1093/mnrasl/slx047} {\bibfield
  {journal} {\bibinfo  {journal} {Mon. Not. R. Astron. Soc. Lett.}\ }\textbf
  {\bibinfo {volume} {469}},\ \bibinfo {pages} {L26} (\bibinfo {year}
  {2017})}\BibitemShut {NoStop}%
\bibitem [{\citenamefont {{O'Malley-James}}\ and\ \citenamefont
  {{Kaltenegger}}(2019)}]{OMJK19}%
  \BibitemOpen
  \bibfield  {author} {\bibinfo {author} {\bibfnamefont {J.~T.}\ \bibnamefont
  {{O'Malley-James}}}\ and\ \bibinfo {author} {\bibfnamefont {L.}~\bibnamefont
  {{Kaltenegger}}},\ }\href {\doibase 10.1093/mnras/stz724} {\bibfield
  {journal} {\bibinfo  {journal} {Mon. Not. R. Astron. Soc.}\ }\textbf
  {\bibinfo {volume} {485}},\ \bibinfo {pages} {5598–5603} (\bibinfo {year}
  {2019})}\BibitemShut {NoStop}%
\bibitem [{\citenamefont {{Hohmann-Marriott}}\ and\ \citenamefont
  {{Blankenship}}(2011)}]{HB11}%
  \BibitemOpen
  \bibfield  {author} {\bibinfo {author} {\bibfnamefont {M.~F.}\ \bibnamefont
  {{Hohmann-Marriott}}}\ and\ \bibinfo {author} {\bibfnamefont {R.~E.}\
  \bibnamefont {{Blankenship}}},\ }\href {\doibase
  10.1146/annurev-arplant-042110-103811} {\bibfield  {journal} {\bibinfo
  {journal} {Annu. Rev. Plant Biol.}\ }\textbf {\bibinfo {volume} {62}},\
  \bibinfo {pages} {515} (\bibinfo {year} {2011})}\BibitemShut {NoStop}%
\bibitem [{\citenamefont {{Kiang}}\ \emph
  {et~al.}(2007{\natexlab{a}})\citenamefont {{Kiang}}, \citenamefont
  {{Siefert}}, \citenamefont {{Govindjee}},\ and\ \citenamefont
  {{Blankenship}}}]{KS07a}%
  \BibitemOpen
  \bibfield  {author} {\bibinfo {author} {\bibfnamefont {N.~Y.}\ \bibnamefont
  {{Kiang}}}, \bibinfo {author} {\bibfnamefont {J.}~\bibnamefont {{Siefert}}},
  \bibinfo {author} {\bibnamefont {{Govindjee}}}, \ and\ \bibinfo {author}
  {\bibfnamefont {R.~E.}\ \bibnamefont {{Blankenship}}},\ }\href {\doibase
  10.1089/ast.2006.0105} {\bibfield  {journal} {\bibinfo  {journal}
  {Astrobiology}\ }\textbf {\bibinfo {volume} {7}},\ \bibinfo {pages} {222}
  (\bibinfo {year} {2007}{\natexlab{a}})}\BibitemShut {NoStop}%
\bibitem [{\citenamefont {{Bains}}\ \emph {et~al.}(2014)\citenamefont
  {{Bains}}, \citenamefont {{Seager}},\ and\ \citenamefont {{Zsom}}}]{BSZ14}%
  \BibitemOpen
  \bibfield  {author} {\bibinfo {author} {\bibfnamefont {W.}~\bibnamefont
  {{Bains}}}, \bibinfo {author} {\bibfnamefont {S.}~\bibnamefont {{Seager}}}, \
  and\ \bibinfo {author} {\bibfnamefont {A.}~\bibnamefont {{Zsom}}},\ }\href
  {\doibase 10.3390/life4040716} {\bibfield  {journal} {\bibinfo  {journal}
  {Life}\ }\textbf {\bibinfo {volume} {4}},\ \bibinfo {pages} {716} (\bibinfo
  {year} {2014})}\BibitemShut {NoStop}%
\bibitem [{\citenamefont {{Seager}}\ \emph {et~al.}(2005)\citenamefont
  {{Seager}}, \citenamefont {{Turner}}, \citenamefont {{Schafer}},\ and\
  \citenamefont {{Ford}}}]{STS05}%
  \BibitemOpen
  \bibfield  {author} {\bibinfo {author} {\bibfnamefont {S.}~\bibnamefont
  {{Seager}}}, \bibinfo {author} {\bibfnamefont {E.~L.}\ \bibnamefont
  {{Turner}}}, \bibinfo {author} {\bibfnamefont {J.}~\bibnamefont {{Schafer}}},
  \ and\ \bibinfo {author} {\bibfnamefont {E.~B.}\ \bibnamefont {{Ford}}},\
  }\href {\doibase 10.1089/ast.2005.5.372} {\bibfield  {journal} {\bibinfo
  {journal} {Astrobiology}\ }\textbf {\bibinfo {volume} {5}},\ \bibinfo {pages}
  {372} (\bibinfo {year} {2005})}\BibitemShut {NoStop}%
\bibitem [{\citenamefont {{Kaltenegger}}(2017)}]{Kal17}%
  \BibitemOpen
  \bibfield  {author} {\bibinfo {author} {\bibfnamefont {L.}~\bibnamefont
  {{Kaltenegger}}},\ }\href {\doibase 10.1146/annurev-astro-082214-122238}
  {\bibfield  {journal} {\bibinfo  {journal} {Annu. Rev. Astron. Astrophys.}\
  }\textbf {\bibinfo {volume} {55}},\ \bibinfo {pages} {433} (\bibinfo {year}
  {2017})}\BibitemShut {NoStop}%
\bibitem [{\citenamefont {{Kiang}}\ \emph
  {et~al.}(2007{\natexlab{b}})\citenamefont {{Kiang}}, \citenamefont
  {{Segura}}, \citenamefont {{Tinetti}}, \citenamefont {{Govindjee}},
  \citenamefont {{Blankenship}}, \citenamefont {{Cohen}}, \citenamefont
  {{Siefert}}, \citenamefont {{Crisp}},\ and\ \citenamefont
  {{Meadows}}}]{KS07b}%
  \BibitemOpen
  \bibfield  {author} {\bibinfo {author} {\bibfnamefont {N.~Y.}\ \bibnamefont
  {{Kiang}}}, \bibinfo {author} {\bibfnamefont {A.}~\bibnamefont {{Segura}}},
  \bibinfo {author} {\bibfnamefont {G.}~\bibnamefont {{Tinetti}}}, \bibinfo
  {author} {\bibnamefont {{Govindjee}}}, \bibinfo {author} {\bibfnamefont
  {R.~E.}\ \bibnamefont {{Blankenship}}}, \bibinfo {author} {\bibfnamefont
  {M.}~\bibnamefont {{Cohen}}}, \bibinfo {author} {\bibfnamefont
  {J.}~\bibnamefont {{Siefert}}}, \bibinfo {author} {\bibfnamefont
  {D.}~\bibnamefont {{Crisp}}}, \ and\ \bibinfo {author} {\bibfnamefont
  {V.~S.}\ \bibnamefont {{Meadows}}},\ }\href {\doibase 10.1089/ast.2006.0108}
  {\bibfield  {journal} {\bibinfo  {journal} {Astrobiology}\ }\textbf {\bibinfo
  {volume} {7}},\ \bibinfo {pages} {252} (\bibinfo {year}
  {2007}{\natexlab{b}})}\BibitemShut {NoStop}%
\bibitem [{\citenamefont {{Takizawa}}\ \emph {et~al.}(2017)\citenamefont
  {{Takizawa}}, \citenamefont {{Minagawa}}, \citenamefont {{Tamura}},
  \citenamefont {{Kusakabe}},\ and\ \citenamefont {{Narita}}}]{TMT17}%
  \BibitemOpen
  \bibfield  {author} {\bibinfo {author} {\bibfnamefont {K.}~\bibnamefont
  {{Takizawa}}}, \bibinfo {author} {\bibfnamefont {J.}~\bibnamefont
  {{Minagawa}}}, \bibinfo {author} {\bibfnamefont {M.}~\bibnamefont
  {{Tamura}}}, \bibinfo {author} {\bibfnamefont {N.}~\bibnamefont
  {{Kusakabe}}}, \ and\ \bibinfo {author} {\bibfnamefont {N.}~\bibnamefont
  {{Narita}}},\ }\href {\doibase 10.1038/s41598-017-07948-5} {\bibfield
  {journal} {\bibinfo  {journal} {Sci. Rep.}\ }\textbf {\bibinfo {volume}
  {7}},\ \bibinfo {eid} {7561} (\bibinfo {year} {2017})}\BibitemShut {NoStop}%
\bibitem [{\citenamefont {{Sparks}}\ \emph {et~al.}(2009)\citenamefont
  {{Sparks}}, \citenamefont {{Hough}}, \citenamefont {{Germer}}, \citenamefont
  {{Chen}}, \citenamefont {{DasSarma}}, \citenamefont {{DasSarma}},
  \citenamefont {{Robb}}, \citenamefont {{Manset}}, \citenamefont
  {{Kolokolova}}, \citenamefont {{Reid}}, \citenamefont {{Macchetto}},\ and\
  \citenamefont {{Martin}}}]{SHG09}%
  \BibitemOpen
  \bibfield  {author} {\bibinfo {author} {\bibfnamefont {W.~B.}\ \bibnamefont
  {{Sparks}}}, \bibinfo {author} {\bibfnamefont {J.}~\bibnamefont {{Hough}}},
  \bibinfo {author} {\bibfnamefont {T.~A.}\ \bibnamefont {{Germer}}}, \bibinfo
  {author} {\bibfnamefont {F.}~\bibnamefont {{Chen}}}, \bibinfo {author}
  {\bibfnamefont {S.}~\bibnamefont {{DasSarma}}}, \bibinfo {author}
  {\bibfnamefont {P.}~\bibnamefont {{DasSarma}}}, \bibinfo {author}
  {\bibfnamefont {F.~T.}\ \bibnamefont {{Robb}}}, \bibinfo {author}
  {\bibfnamefont {N.}~\bibnamefont {{Manset}}}, \bibinfo {author}
  {\bibfnamefont {L.}~\bibnamefont {{Kolokolova}}}, \bibinfo {author}
  {\bibfnamefont {N.}~\bibnamefont {{Reid}}}, \bibinfo {author} {\bibfnamefont
  {F.~D.}\ \bibnamefont {{Macchetto}}}, \ and\ \bibinfo {author} {\bibfnamefont
  {W.}~\bibnamefont {{Martin}}},\ }\href {\doibase 10.1073/pnas.0810215106}
  {\bibfield  {journal} {\bibinfo  {journal} {Proc. Natl. Acad. Sci. USA}\
  }\textbf {\bibinfo {volume} {106}},\ \bibinfo {pages} {7816} (\bibinfo {year}
  {2009})}\BibitemShut {NoStop}%
\bibitem [{\citenamefont {{Berdyugina}}\ \emph {et~al.}(2016)\citenamefont
  {{Berdyugina}}, \citenamefont {{Kuhn}}, \citenamefont {{Harrington}},
  \citenamefont {{{\v S}antl-Temkiv}},\ and\ \citenamefont
  {{Messersmith}}}]{BKH16}%
  \BibitemOpen
  \bibfield  {author} {\bibinfo {author} {\bibfnamefont {S.~V.}\ \bibnamefont
  {{Berdyugina}}}, \bibinfo {author} {\bibfnamefont {J.~R.}\ \bibnamefont
  {{Kuhn}}}, \bibinfo {author} {\bibfnamefont {D.~M.}\ \bibnamefont
  {{Harrington}}}, \bibinfo {author} {\bibfnamefont {T.}~\bibnamefont {{{\v
  S}antl-Temkiv}}}, \ and\ \bibinfo {author} {\bibfnamefont {E.~J.}\
  \bibnamefont {{Messersmith}}},\ }\href {\doibase 10.1017/S1473550415000129}
  {\bibfield  {journal} {\bibinfo  {journal} {Int. J. Astrobiol.}\ }\textbf
  {\bibinfo {volume} {15}},\ \bibinfo {pages} {45} (\bibinfo {year}
  {2016})}\BibitemShut {NoStop}%
\bibitem [{\citenamefont {{Doughty}}\ and\ \citenamefont
  {{Wolf}}(2010)}]{DW10}%
  \BibitemOpen
  \bibfield  {author} {\bibinfo {author} {\bibfnamefont {C.~E.}\ \bibnamefont
  {{Doughty}}}\ and\ \bibinfo {author} {\bibfnamefont {A.}~\bibnamefont
  {{Wolf}}},\ }\href {\doibase 10.1089/ast.2010.0495} {\bibfield  {journal}
  {\bibinfo  {journal} {Astrobiology}\ }\textbf {\bibinfo {volume} {10}},\
  \bibinfo {pages} {869} (\bibinfo {year} {2010})}\BibitemShut {NoStop}%
\bibitem [{\citenamefont {{Fischer}}\ \emph {et~al.}(2016)\citenamefont
  {{Fischer}}, \citenamefont {{Hemp}},\ and\ \citenamefont
  {{Johnson}}}]{FHJ16}%
  \BibitemOpen
  \bibfield  {author} {\bibinfo {author} {\bibfnamefont {W.~W.}\ \bibnamefont
  {{Fischer}}}, \bibinfo {author} {\bibfnamefont {J.}~\bibnamefont {{Hemp}}}, \
  and\ \bibinfo {author} {\bibfnamefont {J.~E.}\ \bibnamefont {{Johnson}}},\
  }\href {\doibase 10.1146/annurev-earth-060313-054810} {\bibfield  {journal}
  {\bibinfo  {journal} {Annu. Rev. Earth Planet. Sci.}\ }\textbf {\bibinfo
  {volume} {44}},\ \bibinfo {pages} {647} (\bibinfo {year} {2016})}\BibitemShut
  {NoStop}%
\bibitem [{\citenamefont {{Gale}}\ and\ \citenamefont {{Wandel}}(2017)}]{GW17}%
  \BibitemOpen
  \bibfield  {author} {\bibinfo {author} {\bibfnamefont {J.}~\bibnamefont
  {{Gale}}}\ and\ \bibinfo {author} {\bibfnamefont {A.}~\bibnamefont
  {{Wandel}}},\ }\href {\doibase 10.1017/S1473550415000440} {\bibfield
  {journal} {\bibinfo  {journal} {Int. J. Astrobiol.}\ }\textbf {\bibinfo
  {volume} {16}},\ \bibinfo {pages} {1} (\bibinfo {year} {2017})}\BibitemShut
  {NoStop}%
\bibitem [{\citenamefont {{Ritchie}}\ \emph {et~al.}(2018)\citenamefont
  {{Ritchie}}, \citenamefont {{Larkum}},\ and\ \citenamefont
  {{Ribas}}}]{RLR18}%
  \BibitemOpen
  \bibfield  {author} {\bibinfo {author} {\bibfnamefont {R.~J.}\ \bibnamefont
  {{Ritchie}}}, \bibinfo {author} {\bibfnamefont {A.~W.~D.}\ \bibnamefont
  {{Larkum}}}, \ and\ \bibinfo {author} {\bibfnamefont {I.}~\bibnamefont
  {{Ribas}}},\ }\href {\doibase 10.1017/S1473550417000167} {\bibfield
  {journal} {\bibinfo  {journal} {Int. J. Astrobiol.}\ }\textbf {\bibinfo
  {volume} {17}},\ \bibinfo {pages} {147} (\bibinfo {year} {2018})}\BibitemShut
  {NoStop}%
\bibitem [{\citenamefont {{Lehmer}}\ \emph {et~al.}(2018)\citenamefont
  {{Lehmer}}, \citenamefont {{Catling}}, \citenamefont {{Parenteau}},\ and\
  \citenamefont {{Hoehler}}}]{LCP18}%
  \BibitemOpen
  \bibfield  {author} {\bibinfo {author} {\bibfnamefont {O.~R.}\ \bibnamefont
  {{Lehmer}}}, \bibinfo {author} {\bibfnamefont {D.~C.}\ \bibnamefont
  {{Catling}}}, \bibinfo {author} {\bibfnamefont {M.~N.}\ \bibnamefont
  {{Parenteau}}}, \ and\ \bibinfo {author} {\bibfnamefont {T.~M.}\ \bibnamefont
  {{Hoehler}}},\ }\href {\doibase 10.3847/1538-4357/aac104} {\bibfield
  {journal} {\bibinfo  {journal} {Astrophys. J.}\ }\textbf {\bibinfo {volume}
  {859}},\ \bibinfo {eid} {171} (\bibinfo {year} {2018})}\BibitemShut {NoStop}%
\bibitem [{\citenamefont {{Lingam}}\ and\ \citenamefont
  {{Loeb}}(2019{\natexlab{d}})}]{LL19}%
  \BibitemOpen
  \bibfield  {author} {\bibinfo {author} {\bibfnamefont {M.}~\bibnamefont
  {{Lingam}}}\ and\ \bibinfo {author} {\bibfnamefont {A.}~\bibnamefont
  {{Loeb}}},\ }\href {\doibase 10.1093/mnras/stz847} {\bibfield  {journal}
  {\bibinfo  {journal} {Mon. Not. R. Astron. Soc.}\ }\textbf {\bibinfo {volume}
  {485}},\ \bibinfo {pages} {5924} (\bibinfo {year}
  {2019}{\natexlab{d}})}\BibitemShut {NoStop}%
\bibitem [{\citenamefont {{Knoll}}(1985)}]{Kno85}%
  \BibitemOpen
  \bibfield  {author} {\bibinfo {author} {\bibfnamefont {A.~H.}\ \bibnamefont
  {{Knoll}}},\ }in\ \href {\doibase 10.1007/978-94-009-5462-5_29} {\emph
  {\bibinfo {booktitle} {The Search for Extraterrestrial Life: Recent
  Developments}}},\ \bibinfo {series} {IAU Symposium}, Vol.\ \bibinfo {volume}
  {112},\ \bibinfo {editor} {edited by\ \bibinfo {editor} {\bibfnamefont
  {M.~D.}\ \bibnamefont {{Papagiannis}}}}\ (\bibinfo  {publisher}
  {International Astronomical Union},\ \bibinfo {year} {1985})\ pp.\ \bibinfo
  {pages} {201--211}\BibitemShut {NoStop}%
\bibitem [{\citenamefont {{O'Malley}}\ and\ \citenamefont
  {{Powell}}(2016)}]{OMP16}%
  \BibitemOpen
  \bibfield  {author} {\bibinfo {author} {\bibfnamefont {M.~A.}\ \bibnamefont
  {{O'Malley}}}\ and\ \bibinfo {author} {\bibfnamefont {R.}~\bibnamefont
  {{Powell}}},\ }\href {\doibase 10.1007/s10539-015-9513-z} {\bibfield
  {journal} {\bibinfo  {journal} {Biol. Philos.}\ }\textbf {\bibinfo {volume}
  {31}},\ \bibinfo {pages} {159} (\bibinfo {year} {2016})}\BibitemShut
  {NoStop}%
\bibitem [{\citenamefont {{Judson}}(2017)}]{Jud17}%
  \BibitemOpen
  \bibfield  {author} {\bibinfo {author} {\bibfnamefont {O.~P.}\ \bibnamefont
  {{Judson}}},\ }\href {\doibase 10.1038/s41559-017-0138} {\bibfield  {journal}
  {\bibinfo  {journal} {Nat. Ecol. Evol.}\ }\textbf {\bibinfo {volume} {1}},\
  \bibinfo {pages} {0138} (\bibinfo {year} {2017})}\BibitemShut {NoStop}%
\bibitem [{\citenamefont {{Koch}}\ and\ \citenamefont
  {{Britton}}(2008)}]{KB08}%
  \BibitemOpen
  \bibfield  {author} {\bibinfo {author} {\bibfnamefont {L.~G.}\ \bibnamefont
  {{Koch}}}\ and\ \bibinfo {author} {\bibfnamefont {S.~L.}\ \bibnamefont
  {{Britton}}},\ }\href {\doibase 10.1113/jphysiol.2007.144709} {\bibfield
  {journal} {\bibinfo  {journal} {J. Physiol.}\ }\textbf {\bibinfo {volume}
  {586}},\ \bibinfo {pages} {83} (\bibinfo {year} {2008})}\BibitemShut
  {NoStop}%
\bibitem [{\citenamefont {{Gumsley}}\ \emph {et~al.}(2017)\citenamefont
  {{Gumsley}}, \citenamefont {{Chamberlain}}, \citenamefont {{Bleeker}},
  \citenamefont {{S{\"o}derlund}}, \citenamefont {{de Kock}}, \citenamefont
  {{Larsson}},\ and\ \citenamefont {{Bekker}}}]{GCBS17}%
  \BibitemOpen
  \bibfield  {author} {\bibinfo {author} {\bibfnamefont {A.~P.}\ \bibnamefont
  {{Gumsley}}}, \bibinfo {author} {\bibfnamefont {K.~R.}\ \bibnamefont
  {{Chamberlain}}}, \bibinfo {author} {\bibfnamefont {W.}~\bibnamefont
  {{Bleeker}}}, \bibinfo {author} {\bibfnamefont {U.}~\bibnamefont
  {{S{\"o}derlund}}}, \bibinfo {author} {\bibfnamefont {M.~O.}\ \bibnamefont
  {{de Kock}}}, \bibinfo {author} {\bibfnamefont {E.~R.}\ \bibnamefont
  {{Larsson}}}, \ and\ \bibinfo {author} {\bibfnamefont {A.}~\bibnamefont
  {{Bekker}}},\ }\href {\doibase 10.1073/pnas.1608824114} {\bibfield  {journal}
  {\bibinfo  {journal} {Proc. Natl. Acad. Sci. USA}\ }\textbf {\bibinfo
  {volume} {114}},\ \bibinfo {pages} {1811} (\bibinfo {year}
  {2017})}\BibitemShut {NoStop}%
\bibitem [{\citenamefont {{Knoll}}\ and\ \citenamefont {{Nowak}}(2017)}]{KN17}%
  \BibitemOpen
  \bibfield  {author} {\bibinfo {author} {\bibfnamefont {A.~H.}\ \bibnamefont
  {{Knoll}}}\ and\ \bibinfo {author} {\bibfnamefont {M.~A.}\ \bibnamefont
  {{Nowak}}},\ }\href {\doibase 10.1126/sciadv.1603076} {\bibfield  {journal}
  {\bibinfo  {journal} {Sci. Adv.}\ }\textbf {\bibinfo {volume} {3}},\ \bibinfo
  {pages} {e1603076} (\bibinfo {year} {2017})}\BibitemShut {NoStop}%
\bibitem [{\citenamefont {{Kasting}}(2013)}]{Kast13}%
  \BibitemOpen
  \bibfield  {author} {\bibinfo {author} {\bibfnamefont {J.~F.}\ \bibnamefont
  {{Kasting}}},\ }\href {\doibase 10.1016/j.chemgeo.2013.05.039} {\bibfield
  {journal} {\bibinfo  {journal} {Chem. Geol.}\ }\textbf {\bibinfo {volume}
  {362}},\ \bibinfo {pages} {13} (\bibinfo {year} {2013})}\BibitemShut
  {NoStop}%
\bibitem [{\citenamefont {{Lyons}}\ \emph {et~al.}(2014)\citenamefont
  {{Lyons}}, \citenamefont {{Reinhard}},\ and\ \citenamefont
  {{Planavsky}}}]{LRP14}%
  \BibitemOpen
  \bibfield  {author} {\bibinfo {author} {\bibfnamefont {T.~W.}\ \bibnamefont
  {{Lyons}}}, \bibinfo {author} {\bibfnamefont {C.~T.}\ \bibnamefont
  {{Reinhard}}}, \ and\ \bibinfo {author} {\bibfnamefont {N.~J.}\ \bibnamefont
  {{Planavsky}}},\ }\href {\doibase 10.1038/nature13068} {\bibfield  {journal}
  {\bibinfo  {journal} {Nature}\ }\textbf {\bibinfo {volume} {506}},\ \bibinfo
  {pages} {307} (\bibinfo {year} {2014})}\BibitemShut {NoStop}%
\bibitem [{\citenamefont {{Catling}}\ \emph {et~al.}(2001)\citenamefont
  {{Catling}}, \citenamefont {{Zahnle}},\ and\ \citenamefont
  {{McKay}}}]{CZM01}%
  \BibitemOpen
  \bibfield  {author} {\bibinfo {author} {\bibfnamefont {D.~C.}\ \bibnamefont
  {{Catling}}}, \bibinfo {author} {\bibfnamefont {K.~J.}\ \bibnamefont
  {{Zahnle}}}, \ and\ \bibinfo {author} {\bibfnamefont {C.~P.}\ \bibnamefont
  {{McKay}}},\ }\href {\doibase 10.1126/science.1061976} {\bibfield  {journal}
  {\bibinfo  {journal} {Science}\ }\textbf {\bibinfo {volume} {293}},\ \bibinfo
  {pages} {839} (\bibinfo {year} {2001})}\BibitemShut {NoStop}%
\bibitem [{\citenamefont {{Catling}}\ and\ \citenamefont
  {{Kasting}}(2017{\natexlab{b}})}]{CaKa17}%
  \BibitemOpen
  \bibfield  {author} {\bibinfo {author} {\bibfnamefont {D.~C.}\ \bibnamefont
  {{Catling}}}\ and\ \bibinfo {author} {\bibfnamefont {J.~F.}\ \bibnamefont
  {{Kasting}}},\ }\href@noop {} {\emph {\bibinfo {title} {{Atmospheric
  Evolution on Inhabited and Lifeless Worlds}}}}\ (\bibinfo  {publisher}
  {Cambridge University Press},\ \bibinfo {year} {2017})\BibitemShut {NoStop}%
\bibitem [{\citenamefont {{Linsky}}\ \emph {et~al.}(2013)\citenamefont
  {{Linsky}}, \citenamefont {{France}},\ and\ \citenamefont {{Ayres}}}]{LFA13}%
  \BibitemOpen
  \bibfield  {author} {\bibinfo {author} {\bibfnamefont {J.~L.}\ \bibnamefont
  {{Linsky}}}, \bibinfo {author} {\bibfnamefont {K.}~\bibnamefont {{France}}},
  \ and\ \bibinfo {author} {\bibfnamefont {T.}~\bibnamefont {{Ayres}}},\ }\href
  {\doibase 10.1088/0004-637X/766/2/69} {\bibfield  {journal} {\bibinfo
  {journal} {Astrophys. J.}\ }\textbf {\bibinfo {volume} {766}},\ \bibinfo
  {eid} {69} (\bibinfo {year} {2013})}\BibitemShut {NoStop}%
\bibitem [{\citenamefont {{Livio}}(1999)}]{Liv99}%
  \BibitemOpen
  \bibfield  {author} {\bibinfo {author} {\bibfnamefont {M.}~\bibnamefont
  {{Livio}}},\ }\href {\doibase 10.1086/306668} {\bibfield  {journal} {\bibinfo
   {journal} {Astrophys. J.}\ }\textbf {\bibinfo {volume} {511}},\ \bibinfo
  {pages} {429} (\bibinfo {year} {1999})}\BibitemShut {NoStop}%
\bibitem [{\citenamefont {{Benz}}(2017)}]{Benz17}%
  \BibitemOpen
  \bibfield  {author} {\bibinfo {author} {\bibfnamefont {A.~O.}\ \bibnamefont
  {{Benz}}},\ }\href {\doibase 10.1007/s41116-016-0004-3} {\bibfield  {journal}
  {\bibinfo  {journal} {Living Rev. Sol. Phys.}\ }\textbf {\bibinfo {volume}
  {14}},\ \bibinfo {eid} {2} (\bibinfo {year} {2017})}\BibitemShut {NoStop}%
\bibitem [{\citenamefont {{Priest}}\ and\ \citenamefont
  {{Forbes}}(2002)}]{PF02}%
  \BibitemOpen
  \bibfield  {author} {\bibinfo {author} {\bibfnamefont {E.~R.}\ \bibnamefont
  {{Priest}}}\ and\ \bibinfo {author} {\bibfnamefont {T.~G.}\ \bibnamefont
  {{Forbes}}},\ }\href {\doibase 10.1007/s001590100013} {\bibfield  {journal}
  {\bibinfo  {journal} {Astron. Astrophys. Rev.}\ }\textbf {\bibinfo {volume}
  {10}},\ \bibinfo {pages} {313} (\bibinfo {year} {2002})}\BibitemShut
  {NoStop}%
\bibitem [{\citenamefont {{Biskamp}}(2000)}]{Bis00}%
  \BibitemOpen
  \bibfield  {author} {\bibinfo {author} {\bibfnamefont {D.}~\bibnamefont
  {{Biskamp}}},\ }\href {\doibase 10.1017/CBO9780511599958} {\emph {\bibinfo
  {title} {{Magnetic Reconnection in Plasmas}}}},\ \bibinfo {series} {Cambridge
  Monographs on Plasma Physics}, Vol.~\bibinfo {volume} {3}\ (\bibinfo
  {publisher} {Cambridge University Press},\ \bibinfo {year}
  {2000})\BibitemShut {NoStop}%
\bibitem [{\citenamefont {{Shibata}}\ and\ \citenamefont
  {{Magara}}(2011)}]{ShiMa11}%
  \BibitemOpen
  \bibfield  {author} {\bibinfo {author} {\bibfnamefont {K.}~\bibnamefont
  {{Shibata}}}\ and\ \bibinfo {author} {\bibfnamefont {T.}~\bibnamefont
  {{Magara}}},\ }\href {\doibase 10.12942/lrsp-2011-6} {\bibfield  {journal}
  {\bibinfo  {journal} {Living Rev. Sol. Phys.}\ }\textbf {\bibinfo {volume}
  {14}},\ \bibinfo {eid} {6} (\bibinfo {year} {2011})}\BibitemShut {NoStop}%
\bibitem [{\citenamefont {{Comisso}}\ \emph {et~al.}(2016)\citenamefont
  {{Comisso}}, \citenamefont {{Lingam}}, \citenamefont {{Huang}},\ and\
  \citenamefont {{Bhattacharjee}}}]{CLHB16}%
  \BibitemOpen
  \bibfield  {author} {\bibinfo {author} {\bibfnamefont {L.}~\bibnamefont
  {{Comisso}}}, \bibinfo {author} {\bibfnamefont {M.}~\bibnamefont {{Lingam}}},
  \bibinfo {author} {\bibfnamefont {Y.-M.}\ \bibnamefont {{Huang}}}, \ and\
  \bibinfo {author} {\bibfnamefont {A.}~\bibnamefont {{Bhattacharjee}}},\
  }\href {\doibase 10.1063/1.4964481} {\bibfield  {journal} {\bibinfo
  {journal} {Phys. Plasmas}\ }\textbf {\bibinfo {volume} {23}},\ \bibinfo {eid}
  {100702} (\bibinfo {year} {2016})}\BibitemShut {NoStop}%
\bibitem [{\citenamefont {{Carrington}}(1859)}]{Carr59}%
  \BibitemOpen
  \bibfield  {author} {\bibinfo {author} {\bibfnamefont {R.~C.}\ \bibnamefont
  {{Carrington}}},\ }\href {\doibase 10.1093/mnras/20.1.13} {\bibfield
  {journal} {\bibinfo  {journal} {Mon. Not. R. Astron. Soc.}\ }\textbf
  {\bibinfo {volume} {20}},\ \bibinfo {pages} {13} (\bibinfo {year}
  {1859})}\BibitemShut {NoStop}%
\bibitem [{\citenamefont {{Cliver}}\ and\ \citenamefont
  {{Dietrich}}(2013)}]{CD13}%
  \BibitemOpen
  \bibfield  {author} {\bibinfo {author} {\bibfnamefont {E.~W.}\ \bibnamefont
  {{Cliver}}}\ and\ \bibinfo {author} {\bibfnamefont {W.~F.}\ \bibnamefont
  {{Dietrich}}},\ }\href {\doibase 10.1051/swsc/2013053} {\bibfield  {journal}
  {\bibinfo  {journal} {J. Space Weather Space Clim.}\ }\textbf {\bibinfo
  {volume} {3}},\ \bibinfo {eid} {A31} (\bibinfo {year} {2013})}\BibitemShut
  {NoStop}%
\bibitem [{\citenamefont {{Maehara}}\ \emph {et~al.}(2012)\citenamefont
  {{Maehara}}, \citenamefont {{Shibayama}}, \citenamefont {{Notsu}},
  \citenamefont {{Notsu}}, \citenamefont {{Nagao}}, \citenamefont {{Kusaba}},
  \citenamefont {{Honda}}, \citenamefont {{Nogami}},\ and\ \citenamefont
  {{Shibata}}}]{Mae12}%
  \BibitemOpen
  \bibfield  {author} {\bibinfo {author} {\bibfnamefont {H.}~\bibnamefont
  {{Maehara}}}, \bibinfo {author} {\bibfnamefont {T.}~\bibnamefont
  {{Shibayama}}}, \bibinfo {author} {\bibfnamefont {S.}~\bibnamefont
  {{Notsu}}}, \bibinfo {author} {\bibfnamefont {Y.}~\bibnamefont {{Notsu}}},
  \bibinfo {author} {\bibfnamefont {T.}~\bibnamefont {{Nagao}}}, \bibinfo
  {author} {\bibfnamefont {S.}~\bibnamefont {{Kusaba}}}, \bibinfo {author}
  {\bibfnamefont {S.}~\bibnamefont {{Honda}}}, \bibinfo {author} {\bibfnamefont
  {D.}~\bibnamefont {{Nogami}}}, \ and\ \bibinfo {author} {\bibfnamefont
  {K.}~\bibnamefont {{Shibata}}},\ }\href {\doibase 10.1038/nature11063}
  {\bibfield  {journal} {\bibinfo  {journal} {Nature}\ }\textbf {\bibinfo
  {volume} {485}},\ \bibinfo {pages} {478} (\bibinfo {year}
  {2012})}\BibitemShut {NoStop}%
\bibitem [{\citenamefont {{Shibayama}}\ \emph {et~al.}(2013)\citenamefont
  {{Shibayama}}, \citenamefont {{Maehara}}, \citenamefont {{Notsu}},
  \citenamefont {{Notsu}}, \citenamefont {{Nagao}}, \citenamefont {{Honda}},
  \citenamefont {{Ishii}}, \citenamefont {{Nogami}},\ and\ \citenamefont
  {{Shibata}}}]{Shi13}%
  \BibitemOpen
  \bibfield  {author} {\bibinfo {author} {\bibfnamefont {T.}~\bibnamefont
  {{Shibayama}}}, \bibinfo {author} {\bibfnamefont {H.}~\bibnamefont
  {{Maehara}}}, \bibinfo {author} {\bibfnamefont {S.}~\bibnamefont {{Notsu}}},
  \bibinfo {author} {\bibfnamefont {Y.}~\bibnamefont {{Notsu}}}, \bibinfo
  {author} {\bibfnamefont {T.}~\bibnamefont {{Nagao}}}, \bibinfo {author}
  {\bibfnamefont {S.}~\bibnamefont {{Honda}}}, \bibinfo {author} {\bibfnamefont
  {T.~T.}\ \bibnamefont {{Ishii}}}, \bibinfo {author} {\bibfnamefont
  {D.}~\bibnamefont {{Nogami}}}, \ and\ \bibinfo {author} {\bibfnamefont
  {K.}~\bibnamefont {{Shibata}}},\ }\href {\doibase 10.1088/0067-0049/209/1/5}
  {\bibfield  {journal} {\bibinfo  {journal} {Astrophys. J. Suppl.}\ }\textbf
  {\bibinfo {volume} {209}},\ \bibinfo {eid} {5} (\bibinfo {year}
  {2013})}\BibitemShut {NoStop}%
\bibitem [{\citenamefont {{Davenport}}(2016)}]{Dav16}%
  \BibitemOpen
  \bibfield  {author} {\bibinfo {author} {\bibfnamefont {J.~R.~A.}\
  \bibnamefont {{Davenport}}},\ }\href {\doibase 10.3847/0004-637X/829/1/23}
  {\bibfield  {journal} {\bibinfo  {journal} {Astrophys. J.}\ }\textbf
  {\bibinfo {volume} {829}},\ \bibinfo {eid} {23} (\bibinfo {year}
  {2016})}\BibitemShut {NoStop}%
\bibitem [{\citenamefont {{Namekata}}\ \emph {et~al.}(2017)\citenamefont
  {{Namekata}}, \citenamefont {{Sakaue}}, \citenamefont {{Watanabe}},
  \citenamefont {{Asai}}, \citenamefont {{Maehara}}, \citenamefont {{Notsu}},
  \citenamefont {{Notsu}}, \citenamefont {{Honda}}, \citenamefont {{Ishii}},
  \citenamefont {{Ikuta}}, \citenamefont {{Nogami}},\ and\ \citenamefont
  {{Shibata}}}]{NSW17}%
  \BibitemOpen
  \bibfield  {author} {\bibinfo {author} {\bibfnamefont {K.}~\bibnamefont
  {{Namekata}}}, \bibinfo {author} {\bibfnamefont {T.}~\bibnamefont
  {{Sakaue}}}, \bibinfo {author} {\bibfnamefont {K.}~\bibnamefont
  {{Watanabe}}}, \bibinfo {author} {\bibfnamefont {A.}~\bibnamefont {{Asai}}},
  \bibinfo {author} {\bibfnamefont {H.}~\bibnamefont {{Maehara}}}, \bibinfo
  {author} {\bibfnamefont {Y.}~\bibnamefont {{Notsu}}}, \bibinfo {author}
  {\bibfnamefont {S.}~\bibnamefont {{Notsu}}}, \bibinfo {author} {\bibfnamefont
  {S.}~\bibnamefont {{Honda}}}, \bibinfo {author} {\bibfnamefont {T.~T.}\
  \bibnamefont {{Ishii}}}, \bibinfo {author} {\bibfnamefont {K.}~\bibnamefont
  {{Ikuta}}}, \bibinfo {author} {\bibfnamefont {D.}~\bibnamefont {{Nogami}}}, \
  and\ \bibinfo {author} {\bibfnamefont {K.}~\bibnamefont {{Shibata}}},\ }\href
  {\doibase 10.3847/1538-4357/aa9b34} {\bibfield  {journal} {\bibinfo
  {journal} {Astrophys. J.}\ }\textbf {\bibinfo {volume} {851}},\ \bibinfo
  {eid} {91} (\bibinfo {year} {2017})}\BibitemShut {NoStop}%
\bibitem [{\citenamefont {{Notsu}}\ \emph {et~al.}(2019)\citenamefont
  {{Notsu}}, \citenamefont {{Maehara}}, \citenamefont {{Honda}}, \citenamefont
  {{Hawley}}, \citenamefont {{Davenport}}, \citenamefont {{Namekata}},
  \citenamefont {{Notsu}}, \citenamefont {{Ikuta}}, \citenamefont {{Nogami}},\
  and\ \citenamefont {{Shibata}}}]{NMH19}%
  \BibitemOpen
  \bibfield  {author} {\bibinfo {author} {\bibfnamefont {Y.}~\bibnamefont
  {{Notsu}}}, \bibinfo {author} {\bibfnamefont {H.}~\bibnamefont {{Maehara}}},
  \bibinfo {author} {\bibfnamefont {S.}~\bibnamefont {{Honda}}}, \bibinfo
  {author} {\bibfnamefont {S.~L.}\ \bibnamefont {{Hawley}}}, \bibinfo {author}
  {\bibfnamefont {J.~R.~A.}\ \bibnamefont {{Davenport}}}, \bibinfo {author}
  {\bibfnamefont {K.}~\bibnamefont {{Namekata}}}, \bibinfo {author}
  {\bibfnamefont {S.}~\bibnamefont {{Notsu}}}, \bibinfo {author} {\bibfnamefont
  {K.}~\bibnamefont {{Ikuta}}}, \bibinfo {author} {\bibfnamefont
  {D.}~\bibnamefont {{Nogami}}}, \ and\ \bibinfo {author} {\bibfnamefont
  {K.}~\bibnamefont {{Shibata}}},\ }\href {\doibase 10.3847/1538-4357/ab14e6}
  {\bibfield  {journal} {\bibinfo  {journal} {Astrophys. J.}\ }\textbf
  {\bibinfo {volume} {876}},\ \bibinfo {eid} {58} (\bibinfo {year}
  {2019})}\BibitemShut {NoStop}%
\bibitem [{\citenamefont {{Maehara}}\ \emph {et~al.}(2015)\citenamefont
  {{Maehara}}, \citenamefont {{Shibayama}}, \citenamefont {{Notsu}},
  \citenamefont {{Notsu}}, \citenamefont {{Honda}}, \citenamefont {{Nogami}},\
  and\ \citenamefont {{Shibata}}}]{Mae15}%
  \BibitemOpen
  \bibfield  {author} {\bibinfo {author} {\bibfnamefont {H.}~\bibnamefont
  {{Maehara}}}, \bibinfo {author} {\bibfnamefont {T.}~\bibnamefont
  {{Shibayama}}}, \bibinfo {author} {\bibfnamefont {Y.}~\bibnamefont
  {{Notsu}}}, \bibinfo {author} {\bibfnamefont {S.}~\bibnamefont {{Notsu}}},
  \bibinfo {author} {\bibfnamefont {S.}~\bibnamefont {{Honda}}}, \bibinfo
  {author} {\bibfnamefont {D.}~\bibnamefont {{Nogami}}}, \ and\ \bibinfo
  {author} {\bibfnamefont {K.}~\bibnamefont {{Shibata}}},\ }\href {\doibase
  10.1186/s40623-015-0217-z} {\bibfield  {journal} {\bibinfo  {journal} {Earth,
  Planets, and Space}\ }\textbf {\bibinfo {volume} {67}},\ \bibinfo {eid} {59}
  (\bibinfo {year} {2015})}\BibitemShut {NoStop}%
\bibitem [{\citenamefont {{G{\"u}nther}}\ \emph {et~al.}(2019)\citenamefont
  {{G{\"u}nther}}, \citenamefont {{Zhan}}, \citenamefont {{Seager}},
  \citenamefont {{Rimmer}}, \citenamefont {{Ranjan}}, \citenamefont
  {{Stassun}}, \citenamefont {{Oelkers}}, \citenamefont {{Daylan}},
  \citenamefont {{Newton}}, \citenamefont {{Gillen}}, \citenamefont
  {{Rappaport}}, \citenamefont {{Ricker}}, \citenamefont {{Latham}},
  \citenamefont {{Winn}}, \citenamefont {{Jenkins}}, \citenamefont {{Glidden}},
  \citenamefont {{Fausnaugh}}, \citenamefont {{Levine}}, \citenamefont
  {{Dittmann}}, \citenamefont {{Quinn}}, \citenamefont {{Krishnamurthy}},\ and\
  \citenamefont {{Ting}}}]{GZS19}%
  \BibitemOpen
  \bibfield  {author} {\bibinfo {author} {\bibfnamefont {M.~N.}\ \bibnamefont
  {{G{\"u}nther}}}, \bibinfo {author} {\bibfnamefont {Z.}~\bibnamefont
  {{Zhan}}}, \bibinfo {author} {\bibfnamefont {S.}~\bibnamefont {{Seager}}},
  \bibinfo {author} {\bibfnamefont {P.~B.}\ \bibnamefont {{Rimmer}}}, \bibinfo
  {author} {\bibfnamefont {S.}~\bibnamefont {{Ranjan}}}, \bibinfo {author}
  {\bibfnamefont {K.~G.}\ \bibnamefont {{Stassun}}}, \bibinfo {author}
  {\bibfnamefont {R.~J.}\ \bibnamefont {{Oelkers}}}, \bibinfo {author}
  {\bibfnamefont {T.}~\bibnamefont {{Daylan}}}, \bibinfo {author}
  {\bibfnamefont {E.}~\bibnamefont {{Newton}}}, \bibinfo {author}
  {\bibfnamefont {E.}~\bibnamefont {{Gillen}}}, \bibinfo {author}
  {\bibfnamefont {S.}~\bibnamefont {{Rappaport}}}, \bibinfo {author}
  {\bibfnamefont {G.~R.}\ \bibnamefont {{Ricker}}}, \bibinfo {author}
  {\bibfnamefont {D.~W.}\ \bibnamefont {{Latham}}}, \bibinfo {author}
  {\bibfnamefont {J.~N.}\ \bibnamefont {{Winn}}}, \bibinfo {author}
  {\bibfnamefont {J.~M.}\ \bibnamefont {{Jenkins}}}, \bibinfo {author}
  {\bibfnamefont {A.}~\bibnamefont {{Glidden}}}, \bibinfo {author}
  {\bibfnamefont {M.}~\bibnamefont {{Fausnaugh}}}, \bibinfo {author}
  {\bibfnamefont {A.~M.}\ \bibnamefont {{Levine}}}, \bibinfo {author}
  {\bibfnamefont {J.~A.}\ \bibnamefont {{Dittmann}}}, \bibinfo {author}
  {\bibfnamefont {S.~N.}\ \bibnamefont {{Quinn}}}, \bibinfo {author}
  {\bibfnamefont {A.}~\bibnamefont {{Krishnamurthy}}}, \ and\ \bibinfo {author}
  {\bibfnamefont {E.~B.}\ \bibnamefont {{Ting}}},\ }\href@noop {} {\bibfield
  {journal} {\bibinfo  {journal} {Astrophys. J.}\ } (\bibinfo {year} {2019})},\
  \Eprint {http://arxiv.org/abs/1901.00443} {arXiv:1901.00443 [astro-ph.EP]}
  \BibitemShut {NoStop}%
\bibitem [{\citenamefont {{Hannah}}\ \emph {et~al.}(2011)\citenamefont
  {{Hannah}}, \citenamefont {{Hudson}}, \citenamefont {{Battaglia}},
  \citenamefont {{Christe}}, \citenamefont {{Ka{\v s}parov{\'a}}},
  \citenamefont {{Krucker}}, \citenamefont {{Kundu}},\ and\ \citenamefont
  {{Veronig}}}]{Han11}%
  \BibitemOpen
  \bibfield  {author} {\bibinfo {author} {\bibfnamefont {I.~G.}\ \bibnamefont
  {{Hannah}}}, \bibinfo {author} {\bibfnamefont {H.~S.}\ \bibnamefont
  {{Hudson}}}, \bibinfo {author} {\bibfnamefont {M.}~\bibnamefont
  {{Battaglia}}}, \bibinfo {author} {\bibfnamefont {S.}~\bibnamefont
  {{Christe}}}, \bibinfo {author} {\bibfnamefont {J.}~\bibnamefont {{Ka{\v
  s}parov{\'a}}}}, \bibinfo {author} {\bibfnamefont {S.}~\bibnamefont
  {{Krucker}}}, \bibinfo {author} {\bibfnamefont {M.~R.}\ \bibnamefont
  {{Kundu}}}, \ and\ \bibinfo {author} {\bibfnamefont {A.}~\bibnamefont
  {{Veronig}}},\ }\href {\doibase 10.1007/s11214-010-9705-4} {\bibfield
  {journal} {\bibinfo  {journal} {Space Sci. Rev.}\ }\textbf {\bibinfo {volume}
  {159}},\ \bibinfo {pages} {263} (\bibinfo {year} {2011})}\BibitemShut
  {NoStop}%
\bibitem [{\citenamefont {{Shibata}}\ \emph {et~al.}(2013)\citenamefont
  {{Shibata}}, \citenamefont {{Isobe}}, \citenamefont {{Hillier}},
  \citenamefont {{Choudhuri}}, \citenamefont {{Maehara}}, \citenamefont
  {{Ishii}}, \citenamefont {{Shibayama}}, \citenamefont {{Notsu}},
  \citenamefont {{Notsu}}, \citenamefont {{Nagao}}, \citenamefont {{Honda}},\
  and\ \citenamefont {{Nogami}}}]{KS13}%
  \BibitemOpen
  \bibfield  {author} {\bibinfo {author} {\bibfnamefont {K.}~\bibnamefont
  {{Shibata}}}, \bibinfo {author} {\bibfnamefont {H.}~\bibnamefont {{Isobe}}},
  \bibinfo {author} {\bibfnamefont {A.}~\bibnamefont {{Hillier}}}, \bibinfo
  {author} {\bibfnamefont {A.~R.}\ \bibnamefont {{Choudhuri}}}, \bibinfo
  {author} {\bibfnamefont {H.}~\bibnamefont {{Maehara}}}, \bibinfo {author}
  {\bibfnamefont {T.~T.}\ \bibnamefont {{Ishii}}}, \bibinfo {author}
  {\bibfnamefont {T.}~\bibnamefont {{Shibayama}}}, \bibinfo {author}
  {\bibfnamefont {S.}~\bibnamefont {{Notsu}}}, \bibinfo {author} {\bibfnamefont
  {Y.}~\bibnamefont {{Notsu}}}, \bibinfo {author} {\bibfnamefont
  {T.}~\bibnamefont {{Nagao}}}, \bibinfo {author} {\bibfnamefont
  {S.}~\bibnamefont {{Honda}}}, \ and\ \bibinfo {author} {\bibfnamefont
  {D.}~\bibnamefont {{Nogami}}},\ }\href {\doibase 10.1093/pasj/65.3.49}
  {\bibfield  {journal} {\bibinfo  {journal} {Publ. Astron. Soc. Jpn}\ }\textbf
  {\bibinfo {volume} {65}},\ \bibinfo {eid} {49} (\bibinfo {year}
  {2013})}\BibitemShut {NoStop}%
\bibitem [{\citenamefont {{Miyake}}\ \emph {et~al.}(2012)\citenamefont
  {{Miyake}}, \citenamefont {{Nagaya}}, \citenamefont {{Masuda}},\ and\
  \citenamefont {{Nakamura}}}]{MNMN}%
  \BibitemOpen
  \bibfield  {author} {\bibinfo {author} {\bibfnamefont {F.}~\bibnamefont
  {{Miyake}}}, \bibinfo {author} {\bibfnamefont {K.}~\bibnamefont {{Nagaya}}},
  \bibinfo {author} {\bibfnamefont {K.}~\bibnamefont {{Masuda}}}, \ and\
  \bibinfo {author} {\bibfnamefont {T.}~\bibnamefont {{Nakamura}}},\ }\href
  {\doibase 10.1038/nature11123} {\bibfield  {journal} {\bibinfo  {journal}
  {Nature}\ }\textbf {\bibinfo {volume} {486}},\ \bibinfo {pages} {240}
  (\bibinfo {year} {2012})}\BibitemShut {NoStop}%
\bibitem [{\citenamefont {{Miyake}}\ \emph {et~al.}(2013)\citenamefont
  {{Miyake}}, \citenamefont {{Masuda}},\ and\ \citenamefont
  {{Nakamura}}}]{MMN13}%
  \BibitemOpen
  \bibfield  {author} {\bibinfo {author} {\bibfnamefont {F.}~\bibnamefont
  {{Miyake}}}, \bibinfo {author} {\bibfnamefont {K.}~\bibnamefont {{Masuda}}},
  \ and\ \bibinfo {author} {\bibfnamefont {T.}~\bibnamefont {{Nakamura}}},\
  }\href {\doibase 10.1038/ncomms2783} {\bibfield  {journal} {\bibinfo
  {journal} {Nat. Commun.}\ }\textbf {\bibinfo {volume} {4}},\ \bibinfo {eid}
  {1748} (\bibinfo {year} {2013})}\BibitemShut {NoStop}%
\bibitem [{\citenamefont {{Schrijver}}\ \emph {et~al.}(2012)\citenamefont
  {{Schrijver}}, \citenamefont {{Beer}}, \citenamefont {{Baltensperger}},
  \citenamefont {{Cliver}}, \citenamefont {{G{\"u}del}}, \citenamefont
  {{Hudson}}, \citenamefont {{McCracken}}, \citenamefont {{Osten}},
  \citenamefont {{Peter}}, \citenamefont {{Soderblom}}, \citenamefont
  {{Usoskin}},\ and\ \citenamefont {{Wolff}}}]{Sch12}%
  \BibitemOpen
  \bibfield  {author} {\bibinfo {author} {\bibfnamefont {C.~J.}\ \bibnamefont
  {{Schrijver}}}, \bibinfo {author} {\bibfnamefont {J.}~\bibnamefont {{Beer}}},
  \bibinfo {author} {\bibfnamefont {U.}~\bibnamefont {{Baltensperger}}},
  \bibinfo {author} {\bibfnamefont {E.~W.}\ \bibnamefont {{Cliver}}}, \bibinfo
  {author} {\bibfnamefont {M.}~\bibnamefont {{G{\"u}del}}}, \bibinfo {author}
  {\bibfnamefont {H.~S.}\ \bibnamefont {{Hudson}}}, \bibinfo {author}
  {\bibfnamefont {K.~G.}\ \bibnamefont {{McCracken}}}, \bibinfo {author}
  {\bibfnamefont {R.~A.}\ \bibnamefont {{Osten}}}, \bibinfo {author}
  {\bibfnamefont {T.}~\bibnamefont {{Peter}}}, \bibinfo {author} {\bibfnamefont
  {D.~R.}\ \bibnamefont {{Soderblom}}}, \bibinfo {author} {\bibfnamefont
  {I.~G.}\ \bibnamefont {{Usoskin}}}, \ and\ \bibinfo {author} {\bibfnamefont
  {E.~W.}\ \bibnamefont {{Wolff}}},\ }\href {\doibase 10.1029/2012JA017706}
  {\bibfield  {journal} {\bibinfo  {journal} {J. Geophys. Res. A}\ }\textbf
  {\bibinfo {volume} {117}},\ \bibinfo {eid} {A08103} (\bibinfo {year}
  {2012})}\BibitemShut {NoStop}%
\bibitem [{\citenamefont {{Lingam}}\ and\ \citenamefont
  {{Loeb}}(2017{\natexlab{b}})}]{LL17b}%
  \BibitemOpen
  \bibfield  {author} {\bibinfo {author} {\bibfnamefont {M.}~\bibnamefont
  {{Lingam}}}\ and\ \bibinfo {author} {\bibfnamefont {A.}~\bibnamefont
  {{Loeb}}},\ }\href {\doibase 10.3847/1538-4357/aa8e96} {\bibfield  {journal}
  {\bibinfo  {journal} {Astrophys. J.}\ }\textbf {\bibinfo {volume} {848}},\
  \bibinfo {eid} {41} (\bibinfo {year} {2017}{\natexlab{b}})}\BibitemShut
  {NoStop}%
\bibitem [{\citenamefont {{Davenport}}\ \emph {et~al.}(2016)\citenamefont
  {{Davenport}}, \citenamefont {{Kipping}}, \citenamefont {{Sasselov}},
  \citenamefont {{Matthews}},\ and\ \citenamefont {{Cameron}}}]{DKS16}%
  \BibitemOpen
  \bibfield  {author} {\bibinfo {author} {\bibfnamefont {J.~R.~A.}\
  \bibnamefont {{Davenport}}}, \bibinfo {author} {\bibfnamefont {D.~M.}\
  \bibnamefont {{Kipping}}}, \bibinfo {author} {\bibfnamefont {D.}~\bibnamefont
  {{Sasselov}}}, \bibinfo {author} {\bibfnamefont {J.~M.}\ \bibnamefont
  {{Matthews}}}, \ and\ \bibinfo {author} {\bibfnamefont {C.}~\bibnamefont
  {{Cameron}}},\ }\href {\doibase 10.3847/2041-8205/829/2/L31} {\bibfield
  {journal} {\bibinfo  {journal} {Astrophys. J. Lett.}\ }\textbf {\bibinfo
  {volume} {829}},\ \bibinfo {eid} {L31} (\bibinfo {year} {2016})}\BibitemShut
  {NoStop}%
\bibitem [{\citenamefont {{Vida}}\ \emph {et~al.}(2017)\citenamefont {{Vida}},
  \citenamefont {{K{\H o}v{\'a}ri}}, \citenamefont {{P{\'a}l}}, \citenamefont
  {{Ol{\'a}h}},\ and\ \citenamefont {{Kriskovics}}}]{VK17}%
  \BibitemOpen
  \bibfield  {author} {\bibinfo {author} {\bibfnamefont {K.}~\bibnamefont
  {{Vida}}}, \bibinfo {author} {\bibfnamefont {Z.}~\bibnamefont {{K{\H
  o}v{\'a}ri}}}, \bibinfo {author} {\bibfnamefont {A.}~\bibnamefont
  {{P{\'a}l}}}, \bibinfo {author} {\bibfnamefont {K.}~\bibnamefont
  {{Ol{\'a}h}}}, \ and\ \bibinfo {author} {\bibfnamefont {L.}~\bibnamefont
  {{Kriskovics}}},\ }\href {\doibase 10.3847/1538-4357/aa6f05} {\bibfield
  {journal} {\bibinfo  {journal} {Astrophys. J.}\ }\textbf {\bibinfo {volume}
  {841}},\ \bibinfo {eid} {124} (\bibinfo {year} {2017})}\BibitemShut {NoStop}%
\bibitem [{\citenamefont {{Howard}}\ \emph {et~al.}(2018)\citenamefont
  {{Howard}}, \citenamefont {{Tilley}}, \citenamefont {{Corbett}},
  \citenamefont {{Youngblood}}, \citenamefont {{Loyd}}, \citenamefont
  {{Ratzloff}}, \citenamefont {{Law}}, \citenamefont {{Fors}}, \citenamefont
  {{del Ser}}, \citenamefont {{Shkolnik}}, \citenamefont {{Ziegler}},
  \citenamefont {{Goeke}}, \citenamefont {{Pietraallo}},\ and\ \citenamefont
  {{Haislip}}}]{HT18}%
  \BibitemOpen
  \bibfield  {author} {\bibinfo {author} {\bibfnamefont {W.~S.}\ \bibnamefont
  {{Howard}}}, \bibinfo {author} {\bibfnamefont {M.~A.}\ \bibnamefont
  {{Tilley}}}, \bibinfo {author} {\bibfnamefont {H.}~\bibnamefont {{Corbett}}},
  \bibinfo {author} {\bibfnamefont {A.}~\bibnamefont {{Youngblood}}}, \bibinfo
  {author} {\bibfnamefont {R.~O.~P.}\ \bibnamefont {{Loyd}}}, \bibinfo {author}
  {\bibfnamefont {J.~K.}\ \bibnamefont {{Ratzloff}}}, \bibinfo {author}
  {\bibfnamefont {N.~M.}\ \bibnamefont {{Law}}}, \bibinfo {author}
  {\bibfnamefont {O.}~\bibnamefont {{Fors}}}, \bibinfo {author} {\bibfnamefont
  {D.}~\bibnamefont {{del Ser}}}, \bibinfo {author} {\bibfnamefont {E.~L.}\
  \bibnamefont {{Shkolnik}}}, \bibinfo {author} {\bibfnamefont
  {C.}~\bibnamefont {{Ziegler}}}, \bibinfo {author} {\bibfnamefont {E.~E.}\
  \bibnamefont {{Goeke}}}, \bibinfo {author} {\bibfnamefont {A.~D.}\
  \bibnamefont {{Pietraallo}}}, \ and\ \bibinfo {author} {\bibfnamefont
  {J.}~\bibnamefont {{Haislip}}},\ }\href {\doibase 10.3847/2041-8213/aacaf3}
  {\bibfield  {journal} {\bibinfo  {journal} {Astrophys. J. Lett.}\ }\textbf
  {\bibinfo {volume} {860}},\ \bibinfo {eid} {L30} (\bibinfo {year}
  {2018})}\BibitemShut {NoStop}%
\bibitem [{\citenamefont {{MacGregor}}\ \emph {et~al.}(2018)\citenamefont
  {{MacGregor}}, \citenamefont {{Weinberger}}, \citenamefont {{Wilner}},
  \citenamefont {{Kowalski}},\ and\ \citenamefont {{Cranmer}}}]{MW18}%
  \BibitemOpen
  \bibfield  {author} {\bibinfo {author} {\bibfnamefont {M.~A.}\ \bibnamefont
  {{MacGregor}}}, \bibinfo {author} {\bibfnamefont {A.~J.}\ \bibnamefont
  {{Weinberger}}}, \bibinfo {author} {\bibfnamefont {D.~J.}\ \bibnamefont
  {{Wilner}}}, \bibinfo {author} {\bibfnamefont {A.~F.}\ \bibnamefont
  {{Kowalski}}}, \ and\ \bibinfo {author} {\bibfnamefont {S.~R.}\ \bibnamefont
  {{Cranmer}}},\ }\href {\doibase 10.3847/2041-8213/aaad6b} {\bibfield
  {journal} {\bibinfo  {journal} {Astrophys. J. Lett.}\ }\textbf {\bibinfo
  {volume} {855}},\ \bibinfo {eid} {L2} (\bibinfo {year} {2018})}\BibitemShut
  {NoStop}%
\bibitem [{\citenamefont {{Heath}}\ \emph {et~al.}(1999)\citenamefont
  {{Heath}}, \citenamefont {{Doyle}}, \citenamefont {{Joshi}},\ and\
  \citenamefont {{Haberle}}}]{HDJ99}%
  \BibitemOpen
  \bibfield  {author} {\bibinfo {author} {\bibfnamefont {M.~J.}\ \bibnamefont
  {{Heath}}}, \bibinfo {author} {\bibfnamefont {L.~R.}\ \bibnamefont
  {{Doyle}}}, \bibinfo {author} {\bibfnamefont {M.~M.}\ \bibnamefont
  {{Joshi}}}, \ and\ \bibinfo {author} {\bibfnamefont {R.~M.}\ \bibnamefont
  {{Haberle}}},\ }\href {\doibase 10.1023/A:1006596718708} {\bibfield
  {journal} {\bibinfo  {journal} {Orig. Life Evol. Biosph.}\ }\textbf {\bibinfo
  {volume} {29}},\ \bibinfo {pages} {405} (\bibinfo {year} {1999})}\BibitemShut
  {NoStop}%
\bibitem [{\citenamefont {{Smith}}\ \emph
  {et~al.}(2004{\natexlab{a}})\citenamefont {{Smith}}, \citenamefont
  {{Scalo}},\ and\ \citenamefont {{Wheeler}}}]{SS04}%
  \BibitemOpen
  \bibfield  {author} {\bibinfo {author} {\bibfnamefont {D.~S.}\ \bibnamefont
  {{Smith}}}, \bibinfo {author} {\bibfnamefont {J.}~\bibnamefont {{Scalo}}}, \
  and\ \bibinfo {author} {\bibfnamefont {J.~C.}\ \bibnamefont {{Wheeler}}},\
  }\href {\doibase 10.1016/j.icarus.2004.04.009} {\bibfield  {journal}
  {\bibinfo  {journal} {Icarus}\ }\textbf {\bibinfo {volume} {171}},\ \bibinfo
  {pages} {229} (\bibinfo {year} {2004}{\natexlab{a}})}\BibitemShut {NoStop}%
\bibitem [{\citenamefont {{Segura}}\ \emph {et~al.}(2010)\citenamefont
  {{Segura}}, \citenamefont {{Walkowicz}}, \citenamefont {{Meadows}},
  \citenamefont {{Kasting}},\ and\ \citenamefont {{Hawley}}}]{SWM10}%
  \BibitemOpen
  \bibfield  {author} {\bibinfo {author} {\bibfnamefont {A.}~\bibnamefont
  {{Segura}}}, \bibinfo {author} {\bibfnamefont {L.~M.}\ \bibnamefont
  {{Walkowicz}}}, \bibinfo {author} {\bibfnamefont {V.}~\bibnamefont
  {{Meadows}}}, \bibinfo {author} {\bibfnamefont {J.}~\bibnamefont
  {{Kasting}}}, \ and\ \bibinfo {author} {\bibfnamefont {S.}~\bibnamefont
  {{Hawley}}},\ }\href {\doibase 10.1089/ast.2009.0376} {\bibfield  {journal}
  {\bibinfo  {journal} {Astrobiology}\ }\textbf {\bibinfo {volume} {10}},\
  \bibinfo {pages} {751} (\bibinfo {year} {2010})}\BibitemShut {NoStop}%
\bibitem [{\citenamefont {{Estrela}}\ and\ \citenamefont
  {{Valio}}(2018)}]{EV18}%
  \BibitemOpen
  \bibfield  {author} {\bibinfo {author} {\bibfnamefont {R.}~\bibnamefont
  {{Estrela}}}\ and\ \bibinfo {author} {\bibfnamefont {A.}~\bibnamefont
  {{Valio}}},\ }\href {\doibase 10.1089/ast.2017.1724} {\bibfield  {journal}
  {\bibinfo  {journal} {Astrobiology}\ }\textbf {\bibinfo {volume} {18}},\
  \bibinfo {pages} {1414} (\bibinfo {year} {2018})}\BibitemShut {NoStop}%
\bibitem [{\citenamefont {{Mullan}}\ and\ \citenamefont {{Bais}}(2018)}]{MB18}%
  \BibitemOpen
  \bibfield  {author} {\bibinfo {author} {\bibfnamefont {D.~J.}\ \bibnamefont
  {{Mullan}}}\ and\ \bibinfo {author} {\bibfnamefont {H.~P.}\ \bibnamefont
  {{Bais}}},\ }\href {\doibase 10.3847/1538-4357/aadfd1} {\bibfield  {journal}
  {\bibinfo  {journal} {Astron. J.}\ }\textbf {\bibinfo {volume} {865}},\
  \bibinfo {eid} {101} (\bibinfo {year} {2018})}\BibitemShut {NoStop}%
\bibitem [{\citenamefont {{Scharf}}(2019)}]{Sch19}%
  \BibitemOpen
  \bibfield  {author} {\bibinfo {author} {\bibfnamefont {C.}~\bibnamefont
  {{Scharf}}},\ }\href {\doibase 10.3847/1538-4357/ab12ec} {\bibfield
  {journal} {\bibinfo  {journal} {Astrophys. J.}\ }\textbf {\bibinfo {volume}
  {876}},\ \bibinfo {eid} {16} (\bibinfo {year} {2019})}\BibitemShut {NoStop}%
\bibitem [{\citenamefont {{Smith}}\ \emph
  {et~al.}(2004{\natexlab{b}})\citenamefont {{Smith}}, \citenamefont
  {{Scalo}},\ and\ \citenamefont {{Wheeler}}}]{SSW04}%
  \BibitemOpen
  \bibfield  {author} {\bibinfo {author} {\bibfnamefont {D.~S.}\ \bibnamefont
  {{Smith}}}, \bibinfo {author} {\bibfnamefont {J.}~\bibnamefont {{Scalo}}}, \
  and\ \bibinfo {author} {\bibfnamefont {J.~C.}\ \bibnamefont {{Wheeler}}},\
  }\href {\doibase 10.1023/B:ORIG.0000043120.28077.c9} {\bibfield  {journal}
  {\bibinfo  {journal} {Orig. Life Evol. Biosph.}\ }\textbf {\bibinfo {volume}
  {34}},\ \bibinfo {pages} {513} (\bibinfo {year}
  {2004}{\natexlab{b}})}\BibitemShut {NoStop}%
\bibitem [{\citenamefont {{Webb}}\ and\ \citenamefont {{Howard}}(2012)}]{WH12}%
  \BibitemOpen
  \bibfield  {author} {\bibinfo {author} {\bibfnamefont {D.~F.}\ \bibnamefont
  {{Webb}}}\ and\ \bibinfo {author} {\bibfnamefont {T.~A.}\ \bibnamefont
  {{Howard}}},\ }\href {\doibase 10.12942/lrsp-2012-3} {\bibfield  {journal}
  {\bibinfo  {journal} {Living Rev. Sol. Phys.}\ }\textbf {\bibinfo {volume}
  {9}},\ \bibinfo {eid} {3} (\bibinfo {year} {2012})}\BibitemShut {NoStop}%
\bibitem [{\citenamefont {{Kilpua}}\ \emph {et~al.}(2017)\citenamefont
  {{Kilpua}}, \citenamefont {{Koskinen}},\ and\ \citenamefont
  {{Pulkkinen}}}]{KKP17}%
  \BibitemOpen
  \bibfield  {author} {\bibinfo {author} {\bibfnamefont {E.}~\bibnamefont
  {{Kilpua}}}, \bibinfo {author} {\bibfnamefont {H.~E.~J.}\ \bibnamefont
  {{Koskinen}}}, \ and\ \bibinfo {author} {\bibfnamefont {T.~I.}\ \bibnamefont
  {{Pulkkinen}}},\ }\href {\doibase 10.1007/s41116-017-0009-6} {\bibfield
  {journal} {\bibinfo  {journal} {Living Rev. Sol. Phys.}\ }\textbf {\bibinfo
  {volume} {14}},\ \bibinfo {eid} {5} (\bibinfo {year} {2017})}\BibitemShut
  {NoStop}%
\bibitem [{\citenamefont {{Ngwira}}\ \emph {et~al.}(2014)\citenamefont
  {{Ngwira}}, \citenamefont {{Pulkkinen}}, \citenamefont {{Kuznetsova}},\ and\
  \citenamefont {{Glocer}}}]{ng14}%
  \BibitemOpen
  \bibfield  {author} {\bibinfo {author} {\bibfnamefont {C.~M.}\ \bibnamefont
  {{Ngwira}}}, \bibinfo {author} {\bibfnamefont {A.}~\bibnamefont
  {{Pulkkinen}}}, \bibinfo {author} {\bibfnamefont {M.~M.}\ \bibnamefont
  {{Kuznetsova}}}, \ and\ \bibinfo {author} {\bibfnamefont {A.}~\bibnamefont
  {{Glocer}}},\ }\href {\doibase 10.1002/2013JA019661} {\bibfield  {journal}
  {\bibinfo  {journal} {J. Geophys. Res. A}\ }\textbf {\bibinfo {volume}
  {119}},\ \bibinfo {pages} {4456} (\bibinfo {year} {2014})}\BibitemShut
  {NoStop}%
\bibitem [{\citenamefont {{Dong}}\ \emph {et~al.}(2015)\citenamefont {{Dong}},
  \citenamefont {{Ma}}, \citenamefont {{Bougher}}, \citenamefont {{Toth}},
  \citenamefont {{Nagy}}, \citenamefont {{Halekas}}, \citenamefont {{Dong}},
  \citenamefont {{Curry}}, \citenamefont {{Luhmann}}, \citenamefont {{Brain}},
  \citenamefont {{Connerney}}, \citenamefont {{Espley}}, \citenamefont
  {{Mahaffy}}, \citenamefont {{Benna}}, \citenamefont {{McFadden}},
  \citenamefont {{Mitchell}}, \citenamefont {{DiBraccio}}, \citenamefont
  {{Lillis}}, \citenamefont {{Jakosky}},\ and\ \citenamefont
  {{Grebowsky}}}]{dong15a}%
  \BibitemOpen
  \bibfield  {author} {\bibinfo {author} {\bibfnamefont {C.}~\bibnamefont
  {{Dong}}}, \bibinfo {author} {\bibfnamefont {Y.}~\bibnamefont {{Ma}}},
  \bibinfo {author} {\bibfnamefont {S.~W.}\ \bibnamefont {{Bougher}}}, \bibinfo
  {author} {\bibfnamefont {G.}~\bibnamefont {{Toth}}}, \bibinfo {author}
  {\bibfnamefont {A.~F.}\ \bibnamefont {{Nagy}}}, \bibinfo {author}
  {\bibfnamefont {J.~S.}\ \bibnamefont {{Halekas}}}, \bibinfo {author}
  {\bibfnamefont {Y.}~\bibnamefont {{Dong}}}, \bibinfo {author} {\bibfnamefont
  {S.~M.}\ \bibnamefont {{Curry}}}, \bibinfo {author} {\bibfnamefont {J.~G.}\
  \bibnamefont {{Luhmann}}}, \bibinfo {author} {\bibfnamefont {D.}~\bibnamefont
  {{Brain}}}, \bibinfo {author} {\bibfnamefont {J.~E.~P.}\ \bibnamefont
  {{Connerney}}}, \bibinfo {author} {\bibfnamefont {J.}~\bibnamefont
  {{Espley}}}, \bibinfo {author} {\bibfnamefont {P.}~\bibnamefont {{Mahaffy}}},
  \bibinfo {author} {\bibfnamefont {M.}~\bibnamefont {{Benna}}}, \bibinfo
  {author} {\bibfnamefont {J.~P.}\ \bibnamefont {{McFadden}}}, \bibinfo
  {author} {\bibfnamefont {D.~L.}\ \bibnamefont {{Mitchell}}}, \bibinfo
  {author} {\bibfnamefont {G.~A.}\ \bibnamefont {{DiBraccio}}}, \bibinfo
  {author} {\bibfnamefont {R.~J.}\ \bibnamefont {{Lillis}}}, \bibinfo {author}
  {\bibfnamefont {B.~M.}\ \bibnamefont {{Jakosky}}}, \ and\ \bibinfo {author}
  {\bibfnamefont {J.~M.}\ \bibnamefont {{Grebowsky}}},\ }\href {\doibase
  10.1002/2015GL065944} {\bibfield  {journal} {\bibinfo  {journal} {Geophys.
  Res. Lett.}\ }\textbf {\bibinfo {volume} {42}},\ \bibinfo {pages} {9103}
  (\bibinfo {year} {2015})}\BibitemShut {NoStop}%
\bibitem [{\citenamefont {{Jakosky}}\ \emph {et~al.}(2015)\citenamefont
  {{Jakosky}}, \citenamefont {{Grebowsky}}, \citenamefont {{Luhmann}},
  \citenamefont {{Connerney}}, \citenamefont {{Eparvier}}, \citenamefont
  {{Ergun}}, \citenamefont {{Halekas}}, \citenamefont {{Larson}}, \citenamefont
  {{Mahaffy}}, \citenamefont {{McFadden}}, \citenamefont {{Mitchell}},
  \citenamefont {{Schneider}}, \citenamefont {{Zurek}}, \citenamefont
  {{Bougher}}, \citenamefont {{Brain}}, \citenamefont {{Ma}}, \citenamefont
  {{Mazelle}}, \citenamefont {{Andersson}}, \citenamefont {{Andrews}},
  \citenamefont {{Baird}}, \citenamefont {{Baker}}, \citenamefont {{Bell}},
  \citenamefont {{Benna}}, \citenamefont {{Chaffin}}, \citenamefont
  {{Chamberlin}}, \citenamefont {{Chaufray}}, \citenamefont {{Clarke}},
  \citenamefont {{Collinson}}, \citenamefont {{Combi}}, \citenamefont
  {{Crary}}, \citenamefont {{Cravens}}, \citenamefont {{Crismani}},
  \citenamefont {{Curry}}, \citenamefont {{Curtis}}, \citenamefont {{Deighan}},
  \citenamefont {{Delory}}, \citenamefont {{Dewey}}, \citenamefont
  {{DiBraccio}}, \citenamefont {{Dong}}, \citenamefont {{Dong}}, \citenamefont
  {{Dunn}}, \citenamefont {{Elrod}}, \citenamefont {{England}}, \citenamefont
  {{Eriksson}}, \citenamefont {{Espley}}, \citenamefont {{Evans}},
  \citenamefont {{Fang}}, \citenamefont {{Fillingim}}, \citenamefont
  {{Fortier}}, \citenamefont {{Fowler}}, \citenamefont {{Fox}}, \citenamefont
  {{Gr{\"o}ller}}, \citenamefont {{Guzewich}}, \citenamefont {{Hara}},
  \citenamefont {{Harada}}, \citenamefont {{Holsclaw}}, \citenamefont {{Jain}},
  \citenamefont {{Jolitz}}, \citenamefont {{Leblanc}}, \citenamefont {{Lee}},
  \citenamefont {{Lee}}, \citenamefont {{Lefevre}}, \citenamefont {{Lillis}},
  \citenamefont {{Livi}}, \citenamefont {{Lo}}, \citenamefont {{Mayyasi}},
  \citenamefont {{McClintock}}, \citenamefont {{McEnulty}}, \citenamefont
  {{Modolo}}, \citenamefont {{Montmessin}}, \citenamefont {{Morooka}},
  \citenamefont {{Nagy}}, \citenamefont {{Olsen}}, \citenamefont {{Peterson}},
  \citenamefont {{Rahmati}}, \citenamefont {{Ruhunusiri}}, \citenamefont
  {{Russell}}, \citenamefont {{Sakai}}, \citenamefont {{Sauvaud}},
  \citenamefont {{Seki}}, \citenamefont {{Steckiewicz}}, \citenamefont
  {{Stevens}}, \citenamefont {{Stewart}}, \citenamefont {{Stiepen}},
  \citenamefont {{Stone}}, \citenamefont {{Tenishev}}, \citenamefont
  {{Thiemann}}, \citenamefont {{Tolson}}, \citenamefont {{Toublanc}},
  \citenamefont {{Vogt}}, \citenamefont {{Weber}}, \citenamefont {{Withers}},
  \citenamefont {{Woods}},\ and\ \citenamefont {{Yelle}}}]{jak15b}%
  \BibitemOpen
  \bibfield  {author} {\bibinfo {author} {\bibfnamefont {B.~M.}\ \bibnamefont
  {{Jakosky}}}, \bibinfo {author} {\bibfnamefont {J.~M.}\ \bibnamefont
  {{Grebowsky}}}, \bibinfo {author} {\bibfnamefont {J.~G.}\ \bibnamefont
  {{Luhmann}}}, \bibinfo {author} {\bibfnamefont {J.}~\bibnamefont
  {{Connerney}}}, \bibinfo {author} {\bibfnamefont {F.}~\bibnamefont
  {{Eparvier}}}, \bibinfo {author} {\bibfnamefont {R.}~\bibnamefont {{Ergun}}},
  \bibinfo {author} {\bibfnamefont {J.}~\bibnamefont {{Halekas}}}, \bibinfo
  {author} {\bibfnamefont {D.}~\bibnamefont {{Larson}}}, \bibinfo {author}
  {\bibfnamefont {P.}~\bibnamefont {{Mahaffy}}}, \bibinfo {author}
  {\bibfnamefont {J.}~\bibnamefont {{McFadden}}}, \bibinfo {author}
  {\bibfnamefont {D.~F.}\ \bibnamefont {{Mitchell}}}, \bibinfo {author}
  {\bibfnamefont {N.}~\bibnamefont {{Schneider}}}, \bibinfo {author}
  {\bibfnamefont {R.}~\bibnamefont {{Zurek}}}, \bibinfo {author} {\bibfnamefont
  {S.}~\bibnamefont {{Bougher}}}, \bibinfo {author} {\bibfnamefont
  {D.}~\bibnamefont {{Brain}}}, \bibinfo {author} {\bibfnamefont {Y.~J.}\
  \bibnamefont {{Ma}}}, \bibinfo {author} {\bibfnamefont {C.}~\bibnamefont
  {{Mazelle}}}, \bibinfo {author} {\bibfnamefont {L.}~\bibnamefont
  {{Andersson}}}, \bibinfo {author} {\bibfnamefont {D.}~\bibnamefont
  {{Andrews}}}, \bibinfo {author} {\bibfnamefont {D.}~\bibnamefont {{Baird}}},
  \bibinfo {author} {\bibfnamefont {D.}~\bibnamefont {{Baker}}}, \bibinfo
  {author} {\bibfnamefont {J.~M.}\ \bibnamefont {{Bell}}}, \bibinfo {author}
  {\bibfnamefont {M.}~\bibnamefont {{Benna}}}, \bibinfo {author} {\bibfnamefont
  {M.}~\bibnamefont {{Chaffin}}}, \bibinfo {author} {\bibfnamefont
  {P.}~\bibnamefont {{Chamberlin}}}, \bibinfo {author} {\bibfnamefont {Y.-Y.}\
  \bibnamefont {{Chaufray}}}, \bibinfo {author} {\bibfnamefont
  {J.}~\bibnamefont {{Clarke}}}, \bibinfo {author} {\bibfnamefont
  {G.}~\bibnamefont {{Collinson}}}, \bibinfo {author} {\bibfnamefont
  {M.}~\bibnamefont {{Combi}}}, \bibinfo {author} {\bibfnamefont
  {F.}~\bibnamefont {{Crary}}}, \bibinfo {author} {\bibfnamefont
  {T.}~\bibnamefont {{Cravens}}}, \bibinfo {author} {\bibfnamefont
  {M.}~\bibnamefont {{Crismani}}}, \bibinfo {author} {\bibfnamefont
  {S.}~\bibnamefont {{Curry}}}, \bibinfo {author} {\bibfnamefont
  {D.}~\bibnamefont {{Curtis}}}, \bibinfo {author} {\bibfnamefont
  {J.}~\bibnamefont {{Deighan}}}, \bibinfo {author} {\bibfnamefont
  {G.}~\bibnamefont {{Delory}}}, \bibinfo {author} {\bibfnamefont
  {R.}~\bibnamefont {{Dewey}}}, \bibinfo {author} {\bibfnamefont
  {G.}~\bibnamefont {{DiBraccio}}}, \bibinfo {author} {\bibfnamefont
  {C.}~\bibnamefont {{Dong}}}, \bibinfo {author} {\bibfnamefont
  {Y.}~\bibnamefont {{Dong}}}, \bibinfo {author} {\bibfnamefont
  {P.}~\bibnamefont {{Dunn}}}, \bibinfo {author} {\bibfnamefont
  {M.}~\bibnamefont {{Elrod}}}, \bibinfo {author} {\bibfnamefont
  {S.}~\bibnamefont {{England}}}, \bibinfo {author} {\bibfnamefont
  {A.}~\bibnamefont {{Eriksson}}}, \bibinfo {author} {\bibfnamefont
  {J.}~\bibnamefont {{Espley}}}, \bibinfo {author} {\bibfnamefont
  {S.}~\bibnamefont {{Evans}}}, \bibinfo {author} {\bibfnamefont
  {X.}~\bibnamefont {{Fang}}}, \bibinfo {author} {\bibfnamefont
  {M.}~\bibnamefont {{Fillingim}}}, \bibinfo {author} {\bibfnamefont
  {K.}~\bibnamefont {{Fortier}}}, \bibinfo {author} {\bibfnamefont {C.~M.}\
  \bibnamefont {{Fowler}}}, \bibinfo {author} {\bibfnamefont {J.}~\bibnamefont
  {{Fox}}}, \bibinfo {author} {\bibfnamefont {H.}~\bibnamefont
  {{Gr{\"o}ller}}}, \bibinfo {author} {\bibfnamefont {S.}~\bibnamefont
  {{Guzewich}}}, \bibinfo {author} {\bibfnamefont {T.}~\bibnamefont {{Hara}}},
  \bibinfo {author} {\bibfnamefont {Y.}~\bibnamefont {{Harada}}}, \bibinfo
  {author} {\bibfnamefont {G.}~\bibnamefont {{Holsclaw}}}, \bibinfo {author}
  {\bibfnamefont {S.~K.}\ \bibnamefont {{Jain}}}, \bibinfo {author}
  {\bibfnamefont {R.}~\bibnamefont {{Jolitz}}}, \bibinfo {author}
  {\bibfnamefont {F.}~\bibnamefont {{Leblanc}}}, \bibinfo {author}
  {\bibfnamefont {C.~O.}\ \bibnamefont {{Lee}}}, \bibinfo {author}
  {\bibfnamefont {Y.}~\bibnamefont {{Lee}}}, \bibinfo {author} {\bibfnamefont
  {F.}~\bibnamefont {{Lefevre}}}, \bibinfo {author} {\bibfnamefont
  {R.}~\bibnamefont {{Lillis}}}, \bibinfo {author} {\bibfnamefont
  {R.}~\bibnamefont {{Livi}}}, \bibinfo {author} {\bibfnamefont
  {D.}~\bibnamefont {{Lo}}}, \bibinfo {author} {\bibfnamefont {M.}~\bibnamefont
  {{Mayyasi}}}, \bibinfo {author} {\bibfnamefont {W.}~\bibnamefont
  {{McClintock}}}, \bibinfo {author} {\bibfnamefont {T.}~\bibnamefont
  {{McEnulty}}}, \bibinfo {author} {\bibfnamefont {R.}~\bibnamefont
  {{Modolo}}}, \bibinfo {author} {\bibfnamefont {F.}~\bibnamefont
  {{Montmessin}}}, \bibinfo {author} {\bibfnamefont {M.}~\bibnamefont
  {{Morooka}}}, \bibinfo {author} {\bibfnamefont {A.}~\bibnamefont {{Nagy}}},
  \bibinfo {author} {\bibfnamefont {K.}~\bibnamefont {{Olsen}}}, \bibinfo
  {author} {\bibfnamefont {W.}~\bibnamefont {{Peterson}}}, \bibinfo {author}
  {\bibfnamefont {A.}~\bibnamefont {{Rahmati}}}, \bibinfo {author}
  {\bibfnamefont {S.}~\bibnamefont {{Ruhunusiri}}}, \bibinfo {author}
  {\bibfnamefont {C.~T.}\ \bibnamefont {{Russell}}}, \bibinfo {author}
  {\bibfnamefont {S.}~\bibnamefont {{Sakai}}}, \bibinfo {author} {\bibfnamefont
  {J.-A.}\ \bibnamefont {{Sauvaud}}}, \bibinfo {author} {\bibfnamefont
  {K.}~\bibnamefont {{Seki}}}, \bibinfo {author} {\bibfnamefont
  {M.}~\bibnamefont {{Steckiewicz}}}, \bibinfo {author} {\bibfnamefont
  {M.}~\bibnamefont {{Stevens}}}, \bibinfo {author} {\bibfnamefont {A.~I.~F.}\
  \bibnamefont {{Stewart}}}, \bibinfo {author} {\bibfnamefont {A.}~\bibnamefont
  {{Stiepen}}}, \bibinfo {author} {\bibfnamefont {S.}~\bibnamefont {{Stone}}},
  \bibinfo {author} {\bibfnamefont {V.}~\bibnamefont {{Tenishev}}}, \bibinfo
  {author} {\bibfnamefont {E.}~\bibnamefont {{Thiemann}}}, \bibinfo {author}
  {\bibfnamefont {R.}~\bibnamefont {{Tolson}}}, \bibinfo {author}
  {\bibfnamefont {D.}~\bibnamefont {{Toublanc}}}, \bibinfo {author}
  {\bibfnamefont {M.}~\bibnamefont {{Vogt}}}, \bibinfo {author} {\bibfnamefont
  {T.}~\bibnamefont {{Weber}}}, \bibinfo {author} {\bibfnamefont
  {P.}~\bibnamefont {{Withers}}}, \bibinfo {author} {\bibfnamefont
  {T.}~\bibnamefont {{Woods}}}, \ and\ \bibinfo {author} {\bibfnamefont
  {R.}~\bibnamefont {{Yelle}}},\ }\href {\doibase 10.1126/science.aad0210}
  {\bibfield  {journal} {\bibinfo  {journal} {Science}\ }\textbf {\bibinfo
  {volume} {350}},\ \bibinfo {eid} {0210} (\bibinfo {year} {2015})}\BibitemShut
  {NoStop}%
\bibitem [{\citenamefont {{Kay}}\ \emph {et~al.}(2016)\citenamefont {{Kay}},
  \citenamefont {{Opher}},\ and\ \citenamefont {{Kornbleuth}}}]{KOK16}%
  \BibitemOpen
  \bibfield  {author} {\bibinfo {author} {\bibfnamefont {C.}~\bibnamefont
  {{Kay}}}, \bibinfo {author} {\bibfnamefont {M.}~\bibnamefont {{Opher}}}, \
  and\ \bibinfo {author} {\bibfnamefont {M.}~\bibnamefont {{Kornbleuth}}},\
  }\href {\doibase 10.3847/0004-637X/826/2/195} {\bibfield  {journal} {\bibinfo
   {journal} {Astrophys. J.}\ }\textbf {\bibinfo {volume} {826}},\ \bibinfo
  {eid} {195} (\bibinfo {year} {2016})}\BibitemShut {NoStop}%
\bibitem [{\citenamefont {{Khodachenko}}\ \emph {et~al.}(2007)\citenamefont
  {{Khodachenko}}, \citenamefont {{Ribas}}, \citenamefont {{Lammer}},
  \citenamefont {{Grie{\ss}meier}}, \citenamefont {{Leitner}}, \citenamefont
  {{Selsis}}, \citenamefont {{Eiroa}}, \citenamefont {{Hanslmeier}},
  \citenamefont {{Biernat}}, \citenamefont {{Farrugia}},\ and\ \citenamefont
  {{Rucker}}}]{KRL07}%
  \BibitemOpen
  \bibfield  {author} {\bibinfo {author} {\bibfnamefont {M.~L.}\ \bibnamefont
  {{Khodachenko}}}, \bibinfo {author} {\bibfnamefont {I.}~\bibnamefont
  {{Ribas}}}, \bibinfo {author} {\bibfnamefont {H.}~\bibnamefont {{Lammer}}},
  \bibinfo {author} {\bibfnamefont {J.-M.}\ \bibnamefont {{Grie{\ss}meier}}},
  \bibinfo {author} {\bibfnamefont {M.}~\bibnamefont {{Leitner}}}, \bibinfo
  {author} {\bibfnamefont {F.}~\bibnamefont {{Selsis}}}, \bibinfo {author}
  {\bibfnamefont {C.}~\bibnamefont {{Eiroa}}}, \bibinfo {author} {\bibfnamefont
  {A.}~\bibnamefont {{Hanslmeier}}}, \bibinfo {author} {\bibfnamefont {H.~K.}\
  \bibnamefont {{Biernat}}}, \bibinfo {author} {\bibfnamefont {C.~J.}\
  \bibnamefont {{Farrugia}}}, \ and\ \bibinfo {author} {\bibfnamefont {H.~O.}\
  \bibnamefont {{Rucker}}},\ }\href {\doibase 10.1089/ast.2006.0127} {\bibfield
   {journal} {\bibinfo  {journal} {Astrobiology}\ }\textbf {\bibinfo {volume}
  {7}},\ \bibinfo {pages} {167} (\bibinfo {year} {2007})}\BibitemShut {NoStop}%
\bibitem [{\citenamefont {{Lammer}}\ \emph {et~al.}(2007)\citenamefont
  {{Lammer}}, \citenamefont {{Lichtenegger}}, \citenamefont {{Kulikov}},
  \citenamefont {{Grie{\ss}meier}}, \citenamefont {{Terada}}, \citenamefont
  {{Erkaev}}, \citenamefont {{Biernat}}, \citenamefont {{Khodachenko}},
  \citenamefont {{Ribas}}, \citenamefont {{Penz}},\ and\ \citenamefont
  {{Selsis}}}]{LLK07}%
  \BibitemOpen
  \bibfield  {author} {\bibinfo {author} {\bibfnamefont {H.}~\bibnamefont
  {{Lammer}}}, \bibinfo {author} {\bibfnamefont {H.~I.~M.}\ \bibnamefont
  {{Lichtenegger}}}, \bibinfo {author} {\bibfnamefont {Y.~N.}\ \bibnamefont
  {{Kulikov}}}, \bibinfo {author} {\bibfnamefont {J.-M.}\ \bibnamefont
  {{Grie{\ss}meier}}}, \bibinfo {author} {\bibfnamefont {N.}~\bibnamefont
  {{Terada}}}, \bibinfo {author} {\bibfnamefont {N.~V.}\ \bibnamefont
  {{Erkaev}}}, \bibinfo {author} {\bibfnamefont {H.~K.}\ \bibnamefont
  {{Biernat}}}, \bibinfo {author} {\bibfnamefont {M.~L.}\ \bibnamefont
  {{Khodachenko}}}, \bibinfo {author} {\bibfnamefont {I.}~\bibnamefont
  {{Ribas}}}, \bibinfo {author} {\bibfnamefont {T.}~\bibnamefont {{Penz}}}, \
  and\ \bibinfo {author} {\bibfnamefont {F.}~\bibnamefont {{Selsis}}},\ }\href
  {\doibase 10.1089/ast.2006.0128} {\bibfield  {journal} {\bibinfo  {journal}
  {Astrobiology}\ }\textbf {\bibinfo {volume} {7}},\ \bibinfo {pages} {185}
  (\bibinfo {year} {2007})}\BibitemShut {NoStop}%
\bibitem [{\citenamefont {{Odert}}\ \emph {et~al.}(2017)\citenamefont
  {{Odert}}, \citenamefont {{Leitzinger}}, \citenamefont {{Hanslmeier}},\ and\
  \citenamefont {{Lammer}}}]{OLH17}%
  \BibitemOpen
  \bibfield  {author} {\bibinfo {author} {\bibfnamefont {P.}~\bibnamefont
  {{Odert}}}, \bibinfo {author} {\bibfnamefont {M.}~\bibnamefont
  {{Leitzinger}}}, \bibinfo {author} {\bibfnamefont {A.}~\bibnamefont
  {{Hanslmeier}}}, \ and\ \bibinfo {author} {\bibfnamefont {H.}~\bibnamefont
  {{Lammer}}},\ }\href {\doibase 10.1093/mnras/stx1969} {\bibfield  {journal}
  {\bibinfo  {journal} {Mon. Not. R. Astron. Soc.}\ }\textbf {\bibinfo {volume}
  {472}},\ \bibinfo {pages} {876} (\bibinfo {year} {2017})}\BibitemShut
  {NoStop}%
\bibitem [{\citenamefont {{Alvarado-G{\'o}mez}}\ \emph
  {et~al.}(2018)\citenamefont {{Alvarado-G{\'o}mez}}, \citenamefont {{Drake}},
  \citenamefont {{Cohen}}, \citenamefont {{Moschou}},\ and\ \citenamefont
  {{Garraffo}}}]{AGD18}%
  \BibitemOpen
  \bibfield  {author} {\bibinfo {author} {\bibfnamefont {J.~D.}\ \bibnamefont
  {{Alvarado-G{\'o}mez}}}, \bibinfo {author} {\bibfnamefont {J.~J.}\
  \bibnamefont {{Drake}}}, \bibinfo {author} {\bibfnamefont {O.}~\bibnamefont
  {{Cohen}}}, \bibinfo {author} {\bibfnamefont {S.~P.}\ \bibnamefont
  {{Moschou}}}, \ and\ \bibinfo {author} {\bibfnamefont {C.}~\bibnamefont
  {{Garraffo}}},\ }\href {\doibase 10.3847/1538-4357/aacb7f} {\bibfield
  {journal} {\bibinfo  {journal} {Astrophys. J.}\ }\textbf {\bibinfo {volume}
  {862}},\ \bibinfo {eid} {93} (\bibinfo {year} {2018})}\BibitemShut {NoStop}%
\bibitem [{\citenamefont {{Drake}}\ \emph {et~al.}(2013)\citenamefont
  {{Drake}}, \citenamefont {{Cohen}}, \citenamefont {{Yashiro}},\ and\
  \citenamefont {{Gopalswamy}}}]{DCY13}%
  \BibitemOpen
  \bibfield  {author} {\bibinfo {author} {\bibfnamefont {J.~J.}\ \bibnamefont
  {{Drake}}}, \bibinfo {author} {\bibfnamefont {O.}~\bibnamefont {{Cohen}}},
  \bibinfo {author} {\bibfnamefont {S.}~\bibnamefont {{Yashiro}}}, \ and\
  \bibinfo {author} {\bibfnamefont {N.}~\bibnamefont {{Gopalswamy}}},\ }\href
  {\doibase 10.1088/0004-637X/764/2/170} {\bibfield  {journal} {\bibinfo
  {journal} {Astrophys. J.}\ }\textbf {\bibinfo {volume} {764}},\ \bibinfo
  {eid} {170} (\bibinfo {year} {2013})}\BibitemShut {NoStop}%
\bibitem [{\citenamefont {{Cranmer}}(2017)}]{Cran17}%
  \BibitemOpen
  \bibfield  {author} {\bibinfo {author} {\bibfnamefont {S.~R.}\ \bibnamefont
  {{Cranmer}}},\ }\href {\doibase 10.3847/1538-4357/aa6f0e} {\bibfield
  {journal} {\bibinfo  {journal} {Astrophys. J.}\ }\textbf {\bibinfo {volume}
  {840}},\ \bibinfo {eid} {114} (\bibinfo {year} {2017})}\BibitemShut {NoStop}%
\bibitem [{\citenamefont {{Reames}}(2013)}]{Rea13}%
  \BibitemOpen
  \bibfield  {author} {\bibinfo {author} {\bibfnamefont {D.~V.}\ \bibnamefont
  {{Reames}}},\ }\href {\doibase 10.1007/s11214-013-9958-9} {\bibfield
  {journal} {\bibinfo  {journal} {Space Sci. Rev.}\ }\textbf {\bibinfo {volume}
  {175}},\ \bibinfo {pages} {53} (\bibinfo {year} {2013})}\BibitemShut
  {NoStop}%
\bibitem [{\citenamefont {{Desai}}\ and\ \citenamefont
  {{Giacalone}}(2016)}]{DG16}%
  \BibitemOpen
  \bibfield  {author} {\bibinfo {author} {\bibfnamefont {M.}~\bibnamefont
  {{Desai}}}\ and\ \bibinfo {author} {\bibfnamefont {J.}~\bibnamefont
  {{Giacalone}}},\ }\href {\doibase 10.1007/s41116-016-0002-5} {\bibfield
  {journal} {\bibinfo  {journal} {Living Rev. Sol. Phys.}\ }\textbf {\bibinfo
  {volume} {13}},\ \bibinfo {eid} {3} (\bibinfo {year} {2016})}\BibitemShut
  {NoStop}%
\bibitem [{\citenamefont {{Mewaldt}}(2006)}]{Mew06}%
  \BibitemOpen
  \bibfield  {author} {\bibinfo {author} {\bibfnamefont {R.~A.}\ \bibnamefont
  {{Mewaldt}}},\ }\href {\doibase 10.1007/s11214-006-9091-0} {\bibfield
  {journal} {\bibinfo  {journal} {Space Sci. Rev.}\ }\textbf {\bibinfo {volume}
  {124}},\ \bibinfo {pages} {303} (\bibinfo {year} {2006})}\BibitemShut
  {NoStop}%
\bibitem [{\citenamefont {{Takahashi}}\ \emph {et~al.}(2016)\citenamefont
  {{Takahashi}}, \citenamefont {{Mizuno}},\ and\ \citenamefont
  {{Shibata}}}]{TMS16}%
  \BibitemOpen
  \bibfield  {author} {\bibinfo {author} {\bibfnamefont {T.}~\bibnamefont
  {{Takahashi}}}, \bibinfo {author} {\bibfnamefont {Y.}~\bibnamefont
  {{Mizuno}}}, \ and\ \bibinfo {author} {\bibfnamefont {K.}~\bibnamefont
  {{Shibata}}},\ }\href {\doibase 10.3847/2041-8205/833/1/L8} {\bibfield
  {journal} {\bibinfo  {journal} {Astrophys. J. Lett.}\ }\textbf {\bibinfo
  {volume} {833}},\ \bibinfo {eid} {L8} (\bibinfo {year} {2016})}\BibitemShut
  {NoStop}%
\bibitem [{\citenamefont {{Youngblood}}\ \emph {et~al.}(2017)\citenamefont
  {{Youngblood}}, \citenamefont {{France}}, \citenamefont {{Loyd}},
  \citenamefont {{Brown}}, \citenamefont {{Mason}}, \citenamefont
  {{Schneider}}, \citenamefont {{Tilley}}, \citenamefont {{Berta-Thompson}},
  \citenamefont {{Buccino}}, \citenamefont {{Froning}}, \citenamefont
  {{Hawley}}, \citenamefont {{Linsky}}, \citenamefont {{Mauas}}, \citenamefont
  {{Redfield}}, \citenamefont {{Kowalski}}, \citenamefont {{Miguel}},
  \citenamefont {{Newton}}, \citenamefont {{Rugheimer}}, \citenamefont
  {{Segura}}, \citenamefont {{Roberge}},\ and\ \citenamefont
  {{Vieytes}}}]{YFL17}%
  \BibitemOpen
  \bibfield  {author} {\bibinfo {author} {\bibfnamefont {A.}~\bibnamefont
  {{Youngblood}}}, \bibinfo {author} {\bibfnamefont {K.}~\bibnamefont
  {{France}}}, \bibinfo {author} {\bibfnamefont {R.~O.~P.}\ \bibnamefont
  {{Loyd}}}, \bibinfo {author} {\bibfnamefont {A.}~\bibnamefont {{Brown}}},
  \bibinfo {author} {\bibfnamefont {J.~P.}\ \bibnamefont {{Mason}}}, \bibinfo
  {author} {\bibfnamefont {P.~C.}\ \bibnamefont {{Schneider}}}, \bibinfo
  {author} {\bibfnamefont {M.~A.}\ \bibnamefont {{Tilley}}}, \bibinfo {author}
  {\bibfnamefont {Z.~K.}\ \bibnamefont {{Berta-Thompson}}}, \bibinfo {author}
  {\bibfnamefont {A.}~\bibnamefont {{Buccino}}}, \bibinfo {author}
  {\bibfnamefont {C.~S.}\ \bibnamefont {{Froning}}}, \bibinfo {author}
  {\bibfnamefont {S.~L.}\ \bibnamefont {{Hawley}}}, \bibinfo {author}
  {\bibfnamefont {J.}~\bibnamefont {{Linsky}}}, \bibinfo {author}
  {\bibfnamefont {P.~J.~D.}\ \bibnamefont {{Mauas}}}, \bibinfo {author}
  {\bibfnamefont {S.}~\bibnamefont {{Redfield}}}, \bibinfo {author}
  {\bibfnamefont {A.}~\bibnamefont {{Kowalski}}}, \bibinfo {author}
  {\bibfnamefont {Y.}~\bibnamefont {{Miguel}}}, \bibinfo {author}
  {\bibfnamefont {E.~R.}\ \bibnamefont {{Newton}}}, \bibinfo {author}
  {\bibfnamefont {S.}~\bibnamefont {{Rugheimer}}}, \bibinfo {author}
  {\bibfnamefont {A.}~\bibnamefont {{Segura}}}, \bibinfo {author}
  {\bibfnamefont {A.}~\bibnamefont {{Roberge}}}, \ and\ \bibinfo {author}
  {\bibfnamefont {M.}~\bibnamefont {{Vieytes}}},\ }\href {\doibase
  10.3847/1538-4357/aa76dd} {\bibfield  {journal} {\bibinfo  {journal}
  {Astrophys. J.}\ }\textbf {\bibinfo {volume} {843}},\ \bibinfo {eid} {31}
  (\bibinfo {year} {2017})}\BibitemShut {NoStop}%
\bibitem [{\citenamefont {{Hudson}}(2015)}]{Hud15}%
  \BibitemOpen
  \bibfield  {author} {\bibinfo {author} {\bibfnamefont {H.~S.}\ \bibnamefont
  {{Hudson}}},\ }\href {\doibase 10.1088/1742-6596/632/1/012058} {\bibfield
  {journal} {\bibinfo  {journal} {J. Phys. Conf. Ser.}\ }\textbf {\bibinfo
  {volume} {632}},\ \bibinfo {eid} {012058} (\bibinfo {year}
  {2015})}\BibitemShut {NoStop}%
\bibitem [{\citenamefont {{Usoskin}}(2017)}]{Uso17}%
  \BibitemOpen
  \bibfield  {author} {\bibinfo {author} {\bibfnamefont {I.~G.}\ \bibnamefont
  {{Usoskin}}},\ }\href {\doibase 10.1007/s41116-017-0006-9} {\bibfield
  {journal} {\bibinfo  {journal} {Living Rev. Sol. Phys.}\ }\textbf {\bibinfo
  {volume} {14}},\ \bibinfo {eid} {3} (\bibinfo {year} {2017})}\BibitemShut
  {NoStop}%
\bibitem [{\citenamefont {{Crutzen}}(1979)}]{Cru79}%
  \BibitemOpen
  \bibfield  {author} {\bibinfo {author} {\bibfnamefont {P.~J.}\ \bibnamefont
  {{Crutzen}}},\ }\href {\doibase 10.1146/annurev.ea.07.050179.002303}
  {\bibfield  {journal} {\bibinfo  {journal} {Annu. Rev. Earth Planet. Sci.}\
  }\textbf {\bibinfo {volume} {7}},\ \bibinfo {pages} {443} (\bibinfo {year}
  {1979})}\BibitemShut {NoStop}%
\bibitem [{\citenamefont {{L{\'o}pez-Puertas}}\ \emph
  {et~al.}(2005)\citenamefont {{L{\'o}pez-Puertas}}, \citenamefont {{Funke}},
  \citenamefont {{Gil-L{\'o}pez}}, \citenamefont {{von Clarmann}},
  \citenamefont {{Stiller}}, \citenamefont {{H{\"o}pfner}}, \citenamefont
  {{Kellmann}}, \citenamefont {{Fischer}},\ and\ \citenamefont
  {{Jackman}}}]{LPF05}%
  \BibitemOpen
  \bibfield  {author} {\bibinfo {author} {\bibfnamefont {M.}~\bibnamefont
  {{L{\'o}pez-Puertas}}}, \bibinfo {author} {\bibfnamefont {B.}~\bibnamefont
  {{Funke}}}, \bibinfo {author} {\bibfnamefont {S.}~\bibnamefont
  {{Gil-L{\'o}pez}}}, \bibinfo {author} {\bibfnamefont {T.}~\bibnamefont {{von
  Clarmann}}}, \bibinfo {author} {\bibfnamefont {G.~P.}\ \bibnamefont
  {{Stiller}}}, \bibinfo {author} {\bibfnamefont {M.}~\bibnamefont
  {{H{\"o}pfner}}}, \bibinfo {author} {\bibfnamefont {S.}~\bibnamefont
  {{Kellmann}}}, \bibinfo {author} {\bibfnamefont {H.}~\bibnamefont
  {{Fischer}}}, \ and\ \bibinfo {author} {\bibfnamefont {C.~H.}\ \bibnamefont
  {{Jackman}}},\ }\href {\doibase 10.1029/2005JA011050} {\bibfield  {journal}
  {\bibinfo  {journal} {J. Geophys. Res. A}\ }\textbf {\bibinfo {volume}
  {110}},\ \bibinfo {eid} {A09S43} (\bibinfo {year} {2005})}\BibitemShut
  {NoStop}%
\bibitem [{\citenamefont {{Solomon}}(1999)}]{Sol99}%
  \BibitemOpen
  \bibfield  {author} {\bibinfo {author} {\bibfnamefont {S.}~\bibnamefont
  {{Solomon}}},\ }\href {\doibase 10.1029/1999RG900008} {\bibfield  {journal}
  {\bibinfo  {journal} {Rev. Geophys.}\ }\textbf {\bibinfo {volume} {37}},\
  \bibinfo {pages} {275} (\bibinfo {year} {1999})}\BibitemShut {NoStop}%
\bibitem [{\citenamefont {{Crutzen}}\ \emph {et~al.}(1975)\citenamefont
  {{Crutzen}}, \citenamefont {{Isaksen}},\ and\ \citenamefont
  {{Reid}}}]{CIR75}%
  \BibitemOpen
  \bibfield  {author} {\bibinfo {author} {\bibfnamefont {P.~J.}\ \bibnamefont
  {{Crutzen}}}, \bibinfo {author} {\bibfnamefont {I.~S.~A.}\ \bibnamefont
  {{Isaksen}}}, \ and\ \bibinfo {author} {\bibfnamefont {G.~C.}\ \bibnamefont
  {{Reid}}},\ }\href {\doibase 10.1126/science.189.4201.457} {\bibfield
  {journal} {\bibinfo  {journal} {Science}\ }\textbf {\bibinfo {volume}
  {189}},\ \bibinfo {pages} {457} (\bibinfo {year} {1975})}\BibitemShut
  {NoStop}%
\bibitem [{\citenamefont {{Tilley}}\ \emph {et~al.}(2019)\citenamefont
  {{Tilley}}, \citenamefont {{Segura}}, \citenamefont {{Meadows}},
  \citenamefont {{Hawley}},\ and\ \citenamefont {{Davenport}}}]{TSM18}%
  \BibitemOpen
  \bibfield  {author} {\bibinfo {author} {\bibfnamefont {M.~A.}\ \bibnamefont
  {{Tilley}}}, \bibinfo {author} {\bibfnamefont {A.}~\bibnamefont {{Segura}}},
  \bibinfo {author} {\bibfnamefont {V.~S.}\ \bibnamefont {{Meadows}}}, \bibinfo
  {author} {\bibfnamefont {S.}~\bibnamefont {{Hawley}}}, \ and\ \bibinfo
  {author} {\bibfnamefont {J.}~\bibnamefont {{Davenport}}},\ }\href {\doibase
  10.1089/ast.2017.1794} {\bibfield  {journal} {\bibinfo  {journal}
  {Astrobiology}\ }\textbf {\bibinfo {volume} {19}},\ \bibinfo {pages} {64}
  (\bibinfo {year} {2019})}\BibitemShut {NoStop}%
\bibitem [{\citenamefont {{Rugheimer}}\ \emph {et~al.}(2013)\citenamefont
  {{Rugheimer}}, \citenamefont {{Kaltenegger}}, \citenamefont {{Zsom}},
  \citenamefont {{Segura}},\ and\ \citenamefont {{Sasselov}}}]{RKZ13}%
  \BibitemOpen
  \bibfield  {author} {\bibinfo {author} {\bibfnamefont {S.}~\bibnamefont
  {{Rugheimer}}}, \bibinfo {author} {\bibfnamefont {L.}~\bibnamefont
  {{Kaltenegger}}}, \bibinfo {author} {\bibfnamefont {A.}~\bibnamefont
  {{Zsom}}}, \bibinfo {author} {\bibfnamefont {A.}~\bibnamefont {{Segura}}}, \
  and\ \bibinfo {author} {\bibfnamefont {D.}~\bibnamefont {{Sasselov}}},\
  }\href {\doibase 10.1089/ast.2012.0888} {\bibfield  {journal} {\bibinfo
  {journal} {Astrobiology}\ }\textbf {\bibinfo {volume} {13}},\ \bibinfo
  {pages} {251} (\bibinfo {year} {2013})}\BibitemShut {NoStop}%
\bibitem [{\citenamefont {{Rugheimer}}\ \emph
  {et~al.}(2015{\natexlab{b}})\citenamefont {{Rugheimer}}, \citenamefont
  {{Kaltenegger}}, \citenamefont {{Segura}}, \citenamefont {{Linsky}},\ and\
  \citenamefont {{Mohanty}}}]{RKS15}%
  \BibitemOpen
  \bibfield  {author} {\bibinfo {author} {\bibfnamefont {S.}~\bibnamefont
  {{Rugheimer}}}, \bibinfo {author} {\bibfnamefont {L.}~\bibnamefont
  {{Kaltenegger}}}, \bibinfo {author} {\bibfnamefont {A.}~\bibnamefont
  {{Segura}}}, \bibinfo {author} {\bibfnamefont {J.}~\bibnamefont {{Linsky}}},
  \ and\ \bibinfo {author} {\bibfnamefont {S.}~\bibnamefont {{Mohanty}}},\
  }\href {\doibase 10.1088/0004-637X/809/1/57} {\bibfield  {journal} {\bibinfo
  {journal} {Astrophys. J.}\ }\textbf {\bibinfo {volume} {809}},\ \bibinfo
  {eid} {57} (\bibinfo {year} {2015}{\natexlab{b}})}\BibitemShut {NoStop}%
\bibitem [{\citenamefont {{Tabataba-Vakili}}\ \emph {et~al.}(2016)\citenamefont
  {{Tabataba-Vakili}}, \citenamefont {{Grenfell}}, \citenamefont
  {{Grie{\ss}meier}},\ and\ \citenamefont {{Rauer}}}]{TVG16}%
  \BibitemOpen
  \bibfield  {author} {\bibinfo {author} {\bibfnamefont {F.}~\bibnamefont
  {{Tabataba-Vakili}}}, \bibinfo {author} {\bibfnamefont {J.~L.}\ \bibnamefont
  {{Grenfell}}}, \bibinfo {author} {\bibfnamefont {J.-M.}\ \bibnamefont
  {{Grie{\ss}meier}}}, \ and\ \bibinfo {author} {\bibfnamefont
  {H.}~\bibnamefont {{Rauer}}},\ }\href {\doibase 10.1051/0004-6361/201425602}
  {\bibfield  {journal} {\bibinfo  {journal} {Astron. Astrophys.}\ }\textbf
  {\bibinfo {volume} {585}},\ \bibinfo {eid} {A96} (\bibinfo {year}
  {2016})}\BibitemShut {NoStop}%
\bibitem [{\citenamefont {{B{\"o}hm-Vitense}}(1992)}]{BV92}%
  \BibitemOpen
  \bibfield  {author} {\bibinfo {author} {\bibfnamefont {E.}~\bibnamefont
  {{B{\"o}hm-Vitense}}},\ }\href@noop {} {\emph {\bibinfo {title}
  {{Introduction to stellar astrophysics}}}},\ Vol.~\bibinfo {volume} {3}\
  (\bibinfo  {publisher} {Cambridge University Press},\ \bibinfo {year}
  {1992})\BibitemShut {NoStop}%
\bibitem [{\citenamefont {{Atri}}\ and\ \citenamefont {{Melott}}(2014)}]{AT14}%
  \BibitemOpen
  \bibfield  {author} {\bibinfo {author} {\bibfnamefont {D.}~\bibnamefont
  {{Atri}}}\ and\ \bibinfo {author} {\bibfnamefont {A.~L.}\ \bibnamefont
  {{Melott}}},\ }\href {\doibase 10.1016/j.astropartphys.2013.03.001}
  {\bibfield  {journal} {\bibinfo  {journal} {Astropar. Phys.}\ }\textbf
  {\bibinfo {volume} {53}},\ \bibinfo {pages} {186} (\bibinfo {year}
  {2014})}\BibitemShut {NoStop}%
\bibitem [{\citenamefont {{Atri}}(2017)}]{Atri}%
  \BibitemOpen
  \bibfield  {author} {\bibinfo {author} {\bibfnamefont {D.}~\bibnamefont
  {{Atri}}},\ }\href {\doibase 10.1093/mnrasl/slw199} {\bibfield  {journal}
  {\bibinfo  {journal} {Mon. Not. R. Astron. Soc. Lett.}\ }\textbf {\bibinfo
  {volume} {465}},\ \bibinfo {pages} {L34} (\bibinfo {year}
  {2017})}\BibitemShut {NoStop}%
\bibitem [{\citenamefont {{Miller}}\ and\ \citenamefont {{Urey}}(1959)}]{MU59}%
  \BibitemOpen
  \bibfield  {author} {\bibinfo {author} {\bibfnamefont {S.~L.}\ \bibnamefont
  {{Miller}}}\ and\ \bibinfo {author} {\bibfnamefont {H.~C.}\ \bibnamefont
  {{Urey}}},\ }\href {\doibase 10.1126/science.130.3370.245} {\bibfield
  {journal} {\bibinfo  {journal} {Science}\ }\textbf {\bibinfo {volume}
  {130}},\ \bibinfo {pages} {245} (\bibinfo {year} {1959})}\BibitemShut
  {NoStop}%
\bibitem [{\citenamefont {{Chyba}}\ and\ \citenamefont {{Sagan}}(1992)}]{CS92}%
  \BibitemOpen
  \bibfield  {author} {\bibinfo {author} {\bibfnamefont {C.}~\bibnamefont
  {{Chyba}}}\ and\ \bibinfo {author} {\bibfnamefont {C.}~\bibnamefont
  {{Sagan}}},\ }\href {\doibase 10.1038/355125a0} {\bibfield  {journal}
  {\bibinfo  {journal} {Nature}\ }\textbf {\bibinfo {volume} {355}},\ \bibinfo
  {pages} {125} (\bibinfo {year} {1992})}\BibitemShut {NoStop}%
\bibitem [{\citenamefont {{Deamer}}\ and\ \citenamefont
  {{Weber}}(2010)}]{DeWe10}%
  \BibitemOpen
  \bibfield  {author} {\bibinfo {author} {\bibfnamefont {D.}~\bibnamefont
  {{Deamer}}}\ and\ \bibinfo {author} {\bibfnamefont {A.~L.}\ \bibnamefont
  {{Weber}}},\ }\href {\doibase 10.1101/cshperspect.a004929} {\bibfield
  {journal} {\bibinfo  {journal} {Cold Spring Harb. Perspect. Biol.}\ }\textbf
  {\bibinfo {volume} {2}},\ \bibinfo {pages} {a004929} (\bibinfo {year}
  {2010})}\BibitemShut {NoStop}%
\bibitem [{\citenamefont {{Airapetian}}\ \emph {et~al.}(2016)\citenamefont
  {{Airapetian}}, \citenamefont {{Glocer}}, \citenamefont {{Gronoff}},
  \citenamefont {{H{\'e}brard}},\ and\ \citenamefont {{Danchi}}}]{Aira16}%
  \BibitemOpen
  \bibfield  {author} {\bibinfo {author} {\bibfnamefont {V.~S.}\ \bibnamefont
  {{Airapetian}}}, \bibinfo {author} {\bibfnamefont {A.}~\bibnamefont
  {{Glocer}}}, \bibinfo {author} {\bibfnamefont {G.}~\bibnamefont {{Gronoff}}},
  \bibinfo {author} {\bibfnamefont {E.}~\bibnamefont {{H{\'e}brard}}}, \ and\
  \bibinfo {author} {\bibfnamefont {W.}~\bibnamefont {{Danchi}}},\ }\href
  {\doibase 10.1038/ngeo2719} {\bibfield  {journal} {\bibinfo  {journal} {Nat.
  Geosci.}\ }\textbf {\bibinfo {volume} {9}},\ \bibinfo {pages} {452} (\bibinfo
  {year} {2016})}\BibitemShut {NoStop}%
\bibitem [{\citenamefont {{Wong}}\ \emph {et~al.}(2017)\citenamefont {{Wong}},
  \citenamefont {{Charnay}}, \citenamefont {{Gao}}, \citenamefont {{Yung}},\
  and\ \citenamefont {{Russell}}}]{WCG17}%
  \BibitemOpen
  \bibfield  {author} {\bibinfo {author} {\bibfnamefont {M.~L.}\ \bibnamefont
  {{Wong}}}, \bibinfo {author} {\bibfnamefont {B.~D.}\ \bibnamefont
  {{Charnay}}}, \bibinfo {author} {\bibfnamefont {P.}~\bibnamefont {{Gao}}},
  \bibinfo {author} {\bibfnamefont {Y.~L.}\ \bibnamefont {{Yung}}}, \ and\
  \bibinfo {author} {\bibfnamefont {M.~J.}\ \bibnamefont {{Russell}}},\ }\href
  {\doibase 10.1089/ast.2016.1473} {\bibfield  {journal} {\bibinfo  {journal}
  {Astrobiology}\ }\textbf {\bibinfo {volume} {17}},\ \bibinfo {pages} {975}
  (\bibinfo {year} {2017})}\BibitemShut {NoStop}%
\bibitem [{\citenamefont {{Ranjan}}\ \emph {et~al.}(2019)\citenamefont
  {{Ranjan}}, \citenamefont {{Todd}}, \citenamefont {{Rimmer}}, \citenamefont
  {{Sasselov}},\ and\ \citenamefont {{Babbin}}}]{RTR19}%
  \BibitemOpen
  \bibfield  {author} {\bibinfo {author} {\bibfnamefont {S.}~\bibnamefont
  {{Ranjan}}}, \bibinfo {author} {\bibfnamefont {Z.~R.}\ \bibnamefont
  {{Todd}}}, \bibinfo {author} {\bibfnamefont {P.~B.}\ \bibnamefont
  {{Rimmer}}}, \bibinfo {author} {\bibfnamefont {D.~D.}\ \bibnamefont
  {{Sasselov}}}, \ and\ \bibinfo {author} {\bibfnamefont {A.~R.}\ \bibnamefont
  {{Babbin}}},\ }\href {\doibase 10.1029/2018GC008082} {\bibfield  {journal}
  {\bibinfo  {journal} {Geochem. Geophys. Geosys.}\ }\textbf {\bibinfo {volume}
  {20}},\ \bibinfo {pages} {2021} (\bibinfo {year} {2019})}\BibitemShut
  {NoStop}%
\bibitem [{\citenamefont {{Kobayashi}}\ \emph {et~al.}(1998)\citenamefont
  {{Kobayashi}}, \citenamefont {{Kaneko}}, \citenamefont {{Saito}},\ and\
  \citenamefont {{Oshima}}}]{KKT98}%
  \BibitemOpen
  \bibfield  {author} {\bibinfo {author} {\bibfnamefont {K.}~\bibnamefont
  {{Kobayashi}}}, \bibinfo {author} {\bibfnamefont {T.}~\bibnamefont
  {{Kaneko}}}, \bibinfo {author} {\bibfnamefont {T.}~\bibnamefont {{Saito}}}, \
  and\ \bibinfo {author} {\bibfnamefont {T.}~\bibnamefont {{Oshima}}},\ }\href
  {\doibase 10.1023/A:1006561217063} {\bibfield  {journal} {\bibinfo  {journal}
  {Orig. Life Evol. Biosph.}\ }\textbf {\bibinfo {volume} {28}},\ \bibinfo
  {pages} {155} (\bibinfo {year} {1998})}\BibitemShut {NoStop}%
\bibitem [{\citenamefont {{Miyakawa}}\ \emph {et~al.}(2002)\citenamefont
  {{Miyakawa}}, \citenamefont {{Yamanashi}}, \citenamefont {{Kobayashi}},
  \citenamefont {{Cleaves}},\ and\ \citenamefont {{Miller}}}]{Miy02}%
  \BibitemOpen
  \bibfield  {author} {\bibinfo {author} {\bibfnamefont {S.}~\bibnamefont
  {{Miyakawa}}}, \bibinfo {author} {\bibfnamefont {H.}~\bibnamefont
  {{Yamanashi}}}, \bibinfo {author} {\bibfnamefont {K.}~\bibnamefont
  {{Kobayashi}}}, \bibinfo {author} {\bibfnamefont {H.~J.}\ \bibnamefont
  {{Cleaves}}}, \ and\ \bibinfo {author} {\bibfnamefont {S.~L.}\ \bibnamefont
  {{Miller}}},\ }\href {\doibase 10.1073/pnas.192568299} {\bibfield  {journal}
  {\bibinfo  {journal} {Proc. Natl. Acad. Sci. USA}\ }\textbf {\bibinfo
  {volume} {99}},\ \bibinfo {pages} {14628} (\bibinfo {year}
  {2002})}\BibitemShut {NoStop}%
\bibitem [{\citenamefont {{Lingam}}\ \emph {et~al.}(2018)\citenamefont
  {{Lingam}}, \citenamefont {{Dong}}, \citenamefont {{Fang}}, \citenamefont
  {{Jakosky}},\ and\ \citenamefont {{Loeb}}}]{LDF18}%
  \BibitemOpen
  \bibfield  {author} {\bibinfo {author} {\bibfnamefont {M.}~\bibnamefont
  {{Lingam}}}, \bibinfo {author} {\bibfnamefont {C.}~\bibnamefont {{Dong}}},
  \bibinfo {author} {\bibfnamefont {X.}~\bibnamefont {{Fang}}}, \bibinfo
  {author} {\bibfnamefont {B.~M.}\ \bibnamefont {{Jakosky}}}, \ and\ \bibinfo
  {author} {\bibfnamefont {A.}~\bibnamefont {{Loeb}}},\ }\href {\doibase
  10.3847/1538-4357/aa9fef} {\bibfield  {journal} {\bibinfo  {journal}
  {Astrophys. J.}\ }\textbf {\bibinfo {volume} {853}},\ \bibinfo {eid} {10}
  (\bibinfo {year} {2018})}\BibitemShut {NoStop}%
\bibitem [{\citenamefont {{Fraschetti}}\ \emph {et~al.}(2019)\citenamefont
  {{Fraschetti}}, \citenamefont {{Drake}}, \citenamefont
  {{Alvarado-G{\'o}mez}}, \citenamefont {{Moschou}}, \citenamefont
  {{Garraffo}},\ and\ \citenamefont {{Cohen}}}]{FDA19}%
  \BibitemOpen
  \bibfield  {author} {\bibinfo {author} {\bibfnamefont {F.}~\bibnamefont
  {{Fraschetti}}}, \bibinfo {author} {\bibfnamefont {J.~J.}\ \bibnamefont
  {{Drake}}}, \bibinfo {author} {\bibfnamefont {J.~D.}\ \bibnamefont
  {{Alvarado-G{\'o}mez}}}, \bibinfo {author} {\bibfnamefont {S.~P.}\
  \bibnamefont {{Moschou}}}, \bibinfo {author} {\bibfnamefont {C.}~\bibnamefont
  {{Garraffo}}}, \ and\ \bibinfo {author} {\bibfnamefont {O.}~\bibnamefont
  {{Cohen}}},\ }\href {\doibase 10.3847/1538-4357/ab05e4} {\bibfield  {journal}
  {\bibinfo  {journal} {Astrophys. J.}\ }\textbf {\bibinfo {volume} {874}},\
  \bibinfo {eid} {21} (\bibinfo {year} {2019})}\BibitemShut {NoStop}%
\bibitem [{\citenamefont {{Nava-Sede{\~n}o}}\ \emph {et~al.}(2016)\citenamefont
  {{Nava-Sede{\~n}o}}, \citenamefont {{Ortiz-Cervantes}}, \citenamefont
  {{Segura}},\ and\ \citenamefont {{Domagal-Goldman}}}]{NOSD}%
  \BibitemOpen
  \bibfield  {author} {\bibinfo {author} {\bibfnamefont {J.~M.}\ \bibnamefont
  {{Nava-Sede{\~n}o}}}, \bibinfo {author} {\bibfnamefont {A.}~\bibnamefont
  {{Ortiz-Cervantes}}}, \bibinfo {author} {\bibfnamefont {A.}~\bibnamefont
  {{Segura}}}, \ and\ \bibinfo {author} {\bibfnamefont {S.~D.}\ \bibnamefont
  {{Domagal-Goldman}}},\ }\href {\doibase 10.1089/ast.2015.1435} {\bibfield
  {journal} {\bibinfo  {journal} {Astrobiology}\ }\textbf {\bibinfo {volume}
  {16}},\ \bibinfo {pages} {744} (\bibinfo {year} {2016})}\BibitemShut
  {NoStop}%
\bibitem [{\citenamefont {{Sagan}}\ and\ \citenamefont
  {{Mullen}}(1972)}]{SM72}%
  \BibitemOpen
  \bibfield  {author} {\bibinfo {author} {\bibfnamefont {C.}~\bibnamefont
  {{Sagan}}}\ and\ \bibinfo {author} {\bibfnamefont {G.}~\bibnamefont
  {{Mullen}}},\ }\href {\doibase 10.1126/science.177.4043.52} {\bibfield
  {journal} {\bibinfo  {journal} {Science}\ }\textbf {\bibinfo {volume}
  {177}},\ \bibinfo {pages} {52} (\bibinfo {year} {1972})}\BibitemShut
  {NoStop}%
\bibitem [{\citenamefont {{Feulner}}(2012)}]{Feu12}%
  \BibitemOpen
  \bibfield  {author} {\bibinfo {author} {\bibfnamefont {G.}~\bibnamefont
  {{Feulner}}},\ }\href {\doibase 10.1029/2011RG000375} {\bibfield  {journal}
  {\bibinfo  {journal} {Rev. Geophys.}\ }\textbf {\bibinfo {volume} {50}},\
  \bibinfo {eid} {RG2006} (\bibinfo {year} {2012})}\BibitemShut {NoStop}%
\bibitem [{\citenamefont {{Voigt}}\ \emph {et~al.}(2017)\citenamefont
  {{Voigt}}, \citenamefont {{Marushchak}}, \citenamefont {{Lamprecht}},
  \citenamefont {{Jackowicz-Korczy{\'n}ski}}, \citenamefont {{Lindgren}},
  \citenamefont {{Mastepanov}}, \citenamefont {{Granlund}}, \citenamefont
  {{Christensen}}, \citenamefont {{Tahvanainen}}, \citenamefont
  {{Martikainen}},\ and\ \citenamefont {{Biasi}}}]{VML17}%
  \BibitemOpen
  \bibfield  {author} {\bibinfo {author} {\bibfnamefont {C.}~\bibnamefont
  {{Voigt}}}, \bibinfo {author} {\bibfnamefont {M.~E.}\ \bibnamefont
  {{Marushchak}}}, \bibinfo {author} {\bibfnamefont {R.~E.}\ \bibnamefont
  {{Lamprecht}}}, \bibinfo {author} {\bibfnamefont {M.}~\bibnamefont
  {{Jackowicz-Korczy{\'n}ski}}}, \bibinfo {author} {\bibfnamefont
  {A.}~\bibnamefont {{Lindgren}}}, \bibinfo {author} {\bibfnamefont
  {M.}~\bibnamefont {{Mastepanov}}}, \bibinfo {author} {\bibfnamefont
  {L.}~\bibnamefont {{Granlund}}}, \bibinfo {author} {\bibfnamefont {T.~R.}\
  \bibnamefont {{Christensen}}}, \bibinfo {author} {\bibfnamefont
  {T.}~\bibnamefont {{Tahvanainen}}}, \bibinfo {author} {\bibfnamefont {P.~J.}\
  \bibnamefont {{Martikainen}}}, \ and\ \bibinfo {author} {\bibfnamefont
  {C.}~\bibnamefont {{Biasi}}},\ }\href {\doibase 10.1073/pnas.1702902114}
  {\bibfield  {journal} {\bibinfo  {journal} {Proc. Natl. Acad. Sci. USA}\
  }\textbf {\bibinfo {volume} {114}},\ \bibinfo {pages} {6238} (\bibinfo {year}
  {2017})}\BibitemShut {NoStop}%
\bibitem [{\citenamefont {{Canfield}}\ \emph {et~al.}(2010)\citenamefont
  {{Canfield}}, \citenamefont {{Glazer}},\ and\ \citenamefont
  {{Falkowski}}}]{CGF10}%
  \BibitemOpen
  \bibfield  {author} {\bibinfo {author} {\bibfnamefont {D.~E.}\ \bibnamefont
  {{Canfield}}}, \bibinfo {author} {\bibfnamefont {A.~N.}\ \bibnamefont
  {{Glazer}}}, \ and\ \bibinfo {author} {\bibfnamefont {P.~G.}\ \bibnamefont
  {{Falkowski}}},\ }\href {\doibase 10.1126/science.1186120} {\bibfield
  {journal} {\bibinfo  {journal} {Science}\ }\textbf {\bibinfo {volume}
  {330}},\ \bibinfo {pages} {192} (\bibinfo {year} {2010})}\BibitemShut
  {NoStop}%
\bibitem [{\citenamefont {{Airapetian}}\ \emph
  {et~al.}(2017{\natexlab{b}})\citenamefont {{Airapetian}}, \citenamefont
  {{Jackman}}, \citenamefont {{Mlynczak}}, \citenamefont {{Danchi}},\ and\
  \citenamefont {{Hunt}}}]{AJM17}%
  \BibitemOpen
  \bibfield  {author} {\bibinfo {author} {\bibfnamefont {V.~S.}\ \bibnamefont
  {{Airapetian}}}, \bibinfo {author} {\bibfnamefont {C.~H.}\ \bibnamefont
  {{Jackman}}}, \bibinfo {author} {\bibfnamefont {M.}~\bibnamefont
  {{Mlynczak}}}, \bibinfo {author} {\bibfnamefont {W.}~\bibnamefont
  {{Danchi}}}, \ and\ \bibinfo {author} {\bibfnamefont {L.}~\bibnamefont
  {{Hunt}}},\ }\href {\doibase 10.1038/s41598-017-14192-4} {\bibfield
  {journal} {\bibinfo  {journal} {Sci. Rep.}\ }\textbf {\bibinfo {volume}
  {7}},\ \bibinfo {eid} {14141} (\bibinfo {year}
  {2017}{\natexlab{b}})}\BibitemShut {NoStop}%
\bibitem [{\citenamefont {{Robles}}\ \emph {et~al.}(2008)\citenamefont
  {{Robles}}, \citenamefont {{Lineweaver}}, \citenamefont {{Grether}},
  \citenamefont {{Flynn}}, \citenamefont {{Egan}}, \citenamefont {{Pracy}},
  \citenamefont {{Holmberg}},\ and\ \citenamefont {{Gardner}}}]{RLG08}%
  \BibitemOpen
  \bibfield  {author} {\bibinfo {author} {\bibfnamefont {J.~A.}\ \bibnamefont
  {{Robles}}}, \bibinfo {author} {\bibfnamefont {C.~H.}\ \bibnamefont
  {{Lineweaver}}}, \bibinfo {author} {\bibfnamefont {D.}~\bibnamefont
  {{Grether}}}, \bibinfo {author} {\bibfnamefont {C.}~\bibnamefont {{Flynn}}},
  \bibinfo {author} {\bibfnamefont {C.~A.}\ \bibnamefont {{Egan}}}, \bibinfo
  {author} {\bibfnamefont {M.~B.}\ \bibnamefont {{Pracy}}}, \bibinfo {author}
  {\bibfnamefont {J.}~\bibnamefont {{Holmberg}}}, \ and\ \bibinfo {author}
  {\bibfnamefont {E.}~\bibnamefont {{Gardner}}},\ }\href {\doibase
  10.1086/589985} {\bibfield  {journal} {\bibinfo  {journal} {Astrophys. J.}\
  }\textbf {\bibinfo {volume} {684}},\ \bibinfo {pages} {691} (\bibinfo {year}
  {2008})}\BibitemShut {NoStop}%
\bibitem [{\citenamefont {{Loeb}}\ \emph {et~al.}(2016)\citenamefont {{Loeb}},
  \citenamefont {{Batista}},\ and\ \citenamefont {{Sloan}}}]{LBS16}%
  \BibitemOpen
  \bibfield  {author} {\bibinfo {author} {\bibfnamefont {A.}~\bibnamefont
  {{Loeb}}}, \bibinfo {author} {\bibfnamefont {R.~A.}\ \bibnamefont
  {{Batista}}}, \ and\ \bibinfo {author} {\bibfnamefont {D.}~\bibnamefont
  {{Sloan}}},\ }\href {\doibase 10.1088/1475-7516/2016/08/040} {\bibfield
  {journal} {\bibinfo  {journal} {J. Cosmol. Astropart. Phys.}\ }\textbf
  {\bibinfo {volume} {8}},\ \bibinfo {eid} {040} (\bibinfo {year}
  {2016})}\BibitemShut {NoStop}%
\bibitem [{\citenamefont {{Luger}}\ \emph {et~al.}(2015)\citenamefont
  {{Luger}}, \citenamefont {{Barnes}}, \citenamefont {{Lopez}}, \citenamefont
  {{Fortney}}, \citenamefont {{Jackson}},\ and\ \citenamefont
  {{Meadows}}}]{LBL15}%
  \BibitemOpen
  \bibfield  {author} {\bibinfo {author} {\bibfnamefont {R.}~\bibnamefont
  {{Luger}}}, \bibinfo {author} {\bibfnamefont {R.}~\bibnamefont {{Barnes}}},
  \bibinfo {author} {\bibfnamefont {E.}~\bibnamefont {{Lopez}}}, \bibinfo
  {author} {\bibfnamefont {J.}~\bibnamefont {{Fortney}}}, \bibinfo {author}
  {\bibfnamefont {B.}~\bibnamefont {{Jackson}}}, \ and\ \bibinfo {author}
  {\bibfnamefont {V.}~\bibnamefont {{Meadows}}},\ }\href {\doibase
  10.1089/ast.2014.1215} {\bibfield  {journal} {\bibinfo  {journal}
  {Astrobiology}\ }\textbf {\bibinfo {volume} {15}},\ \bibinfo {pages} {57}
  (\bibinfo {year} {2015})}\BibitemShut {NoStop}%
\bibitem [{\citenamefont {{Chen}}\ \emph {et~al.}(2018)\citenamefont {{Chen}},
  \citenamefont {{Forbes}},\ and\ \citenamefont {{Loeb}}}]{CFL18}%
  \BibitemOpen
  \bibfield  {author} {\bibinfo {author} {\bibfnamefont {H.}~\bibnamefont
  {{Chen}}}, \bibinfo {author} {\bibfnamefont {J.~C.}\ \bibnamefont
  {{Forbes}}}, \ and\ \bibinfo {author} {\bibfnamefont {A.}~\bibnamefont
  {{Loeb}}},\ }\href {\doibase 10.3847/2041-8213/aaab46} {\bibfield  {journal}
  {\bibinfo  {journal} {Astrophys. J. Lett.}\ }\textbf {\bibinfo {volume}
  {855}},\ \bibinfo {eid} {L1} (\bibinfo {year} {2018})}\BibitemShut {NoStop}%
\bibitem [{\citenamefont {{Tamayo}}\ \emph {et~al.}(2017)\citenamefont
  {{Tamayo}}, \citenamefont {{Rein}}, \citenamefont {{Petrovich}},\ and\
  \citenamefont {{Murray}}}]{TRP17}%
  \BibitemOpen
  \bibfield  {author} {\bibinfo {author} {\bibfnamefont {D.}~\bibnamefont
  {{Tamayo}}}, \bibinfo {author} {\bibfnamefont {H.}~\bibnamefont {{Rein}}},
  \bibinfo {author} {\bibfnamefont {C.}~\bibnamefont {{Petrovich}}}, \ and\
  \bibinfo {author} {\bibfnamefont {N.}~\bibnamefont {{Murray}}},\ }\href
  {\doibase 10.3847/2041-8213/aa70ea} {\bibfield  {journal} {\bibinfo
  {journal} {Astrophys. J. Lett.}\ }\textbf {\bibinfo {volume} {840}},\
  \bibinfo {eid} {L19} (\bibinfo {year} {2017})}\BibitemShut {NoStop}%
\bibitem [{\citenamefont {{Ormel}}\ \emph {et~al.}(2017)\citenamefont
  {{Ormel}}, \citenamefont {{Liu}},\ and\ \citenamefont
  {{Schoonenberg}}}]{OLS17}%
  \BibitemOpen
  \bibfield  {author} {\bibinfo {author} {\bibfnamefont {C.~W.}\ \bibnamefont
  {{Ormel}}}, \bibinfo {author} {\bibfnamefont {B.}~\bibnamefont {{Liu}}}, \
  and\ \bibinfo {author} {\bibfnamefont {D.}~\bibnamefont {{Schoonenberg}}},\
  }\href {\doibase 10.1051/0004-6361/201730826} {\bibfield  {journal} {\bibinfo
   {journal} {Astron. Astrophys.}\ }\textbf {\bibinfo {volume} {604}},\
  \bibinfo {eid} {A1} (\bibinfo {year} {2017})}\BibitemShut {NoStop}%
\bibitem [{\citenamefont {{Steffen}}\ and\ \citenamefont {{Li}}(2016)}]{SL16}%
  \BibitemOpen
  \bibfield  {author} {\bibinfo {author} {\bibfnamefont {J.~H.}\ \bibnamefont
  {{Steffen}}}\ and\ \bibinfo {author} {\bibfnamefont {G.}~\bibnamefont
  {{Li}}},\ }\href {\doibase 10.3847/0004-637X/816/2/97} {\bibfield  {journal}
  {\bibinfo  {journal} {Astrophys. J.}\ }\textbf {\bibinfo {volume} {816}},\
  \bibinfo {eid} {97} (\bibinfo {year} {2016})}\BibitemShut {NoStop}%
\bibitem [{\citenamefont {{Lingam}}\ and\ \citenamefont
  {{Loeb}}(2017{\natexlab{c}})}]{LL17d}%
  \BibitemOpen
  \bibfield  {author} {\bibinfo {author} {\bibfnamefont {M.}~\bibnamefont
  {{Lingam}}}\ and\ \bibinfo {author} {\bibfnamefont {A.}~\bibnamefont
  {{Loeb}}},\ }\href {\doibase 10.1073/pnas.1703517114} {\bibfield  {journal}
  {\bibinfo  {journal} {Proc. Natl. Acad. Sci. USA}\ }\textbf {\bibinfo
  {volume} {114}},\ \bibinfo {pages} {6689} (\bibinfo {year}
  {2017}{\natexlab{c}})}\BibitemShut {NoStop}%
\bibitem [{\citenamefont {{Krijt}}\ \emph {et~al.}(2017)\citenamefont
  {{Krijt}}, \citenamefont {{Bowling}}, \citenamefont {{Lyons}},\ and\
  \citenamefont {{Ciesla}}}]{KBL17}%
  \BibitemOpen
  \bibfield  {author} {\bibinfo {author} {\bibfnamefont {S.}~\bibnamefont
  {{Krijt}}}, \bibinfo {author} {\bibfnamefont {T.~J.}\ \bibnamefont
  {{Bowling}}}, \bibinfo {author} {\bibfnamefont {R.~J.}\ \bibnamefont
  {{Lyons}}}, \ and\ \bibinfo {author} {\bibfnamefont {F.~J.}\ \bibnamefont
  {{Ciesla}}},\ }\href {\doibase 10.3847/2041-8213/aa6b9f} {\bibfield
  {journal} {\bibinfo  {journal} {Astrophys. J. Lett.}\ }\textbf {\bibinfo
  {volume} {839}},\ \bibinfo {eid} {L21} (\bibinfo {year} {2017})}\BibitemShut
  {NoStop}%
\bibitem [{\citenamefont {{Rodler}}\ and\ \citenamefont
  {{L{\'o}pez-Morales}}(2014)}]{RLM14}%
  \BibitemOpen
  \bibfield  {author} {\bibinfo {author} {\bibfnamefont {F.}~\bibnamefont
  {{Rodler}}}\ and\ \bibinfo {author} {\bibfnamefont {M.}~\bibnamefont
  {{L{\'o}pez-Morales}}},\ }\href {\doibase 10.1088/0004-637X/781/1/54}
  {\bibfield  {journal} {\bibinfo  {journal} {Astrophys. J.}\ }\textbf
  {\bibinfo {volume} {781}},\ \bibinfo {eid} {54} (\bibinfo {year}
  {2014})}\BibitemShut {NoStop}%
\bibitem [{\citenamefont {{Snellen}}\ \emph {et~al.}(2015)\citenamefont
  {{Snellen}}, \citenamefont {{de Kok}}, \citenamefont {{Birkby}},
  \citenamefont {{Brandl}}, \citenamefont {{Brogi}}, \citenamefont {{Keller}},
  \citenamefont {{Kenworthy}}, \citenamefont {{Schwarz}},\ and\ \citenamefont
  {{Stuik}}}]{SDB15}%
  \BibitemOpen
  \bibfield  {author} {\bibinfo {author} {\bibfnamefont {I.}~\bibnamefont
  {{Snellen}}}, \bibinfo {author} {\bibfnamefont {R.}~\bibnamefont {{de Kok}}},
  \bibinfo {author} {\bibfnamefont {J.~L.}\ \bibnamefont {{Birkby}}}, \bibinfo
  {author} {\bibfnamefont {B.}~\bibnamefont {{Brandl}}}, \bibinfo {author}
  {\bibfnamefont {M.}~\bibnamefont {{Brogi}}}, \bibinfo {author} {\bibfnamefont
  {C.}~\bibnamefont {{Keller}}}, \bibinfo {author} {\bibfnamefont
  {M.}~\bibnamefont {{Kenworthy}}}, \bibinfo {author} {\bibfnamefont
  {H.}~\bibnamefont {{Schwarz}}}, \ and\ \bibinfo {author} {\bibfnamefont
  {R.}~\bibnamefont {{Stuik}}},\ }\href {\doibase 10.1051/0004-6361/201425018}
  {\bibfield  {journal} {\bibinfo  {journal} {Astron. Astrophys.}\ }\textbf
  {\bibinfo {volume} {576}},\ \bibinfo {eid} {A59} (\bibinfo {year}
  {2015})}\BibitemShut {NoStop}%
\bibitem [{\citenamefont {{Barstow}}\ and\ \citenamefont
  {{Irwin}}(2016)}]{BI16}%
  \BibitemOpen
  \bibfield  {author} {\bibinfo {author} {\bibfnamefont {J.~K.}\ \bibnamefont
  {{Barstow}}}\ and\ \bibinfo {author} {\bibfnamefont {P.~G.~J.}\ \bibnamefont
  {{Irwin}}},\ }\href {\doibase 10.1093/mnrasl/slw109} {\bibfield  {journal}
  {\bibinfo  {journal} {Mon. Not. R. Astron. Soc. Lett.}\ }\textbf {\bibinfo
  {volume} {461}},\ \bibinfo {pages} {L92} (\bibinfo {year}
  {2016})}\BibitemShut {NoStop}%
\bibitem [{\citenamefont {{Morley}}\ \emph {et~al.}(2017)\citenamefont
  {{Morley}}, \citenamefont {{Kreidberg}}, \citenamefont {{Rustamkulov}},
  \citenamefont {{Robinson}},\ and\ \citenamefont {{Fortney}}}]{MKR17}%
  \BibitemOpen
  \bibfield  {author} {\bibinfo {author} {\bibfnamefont {C.~V.}\ \bibnamefont
  {{Morley}}}, \bibinfo {author} {\bibfnamefont {L.}~\bibnamefont
  {{Kreidberg}}}, \bibinfo {author} {\bibfnamefont {Z.}~\bibnamefont
  {{Rustamkulov}}}, \bibinfo {author} {\bibfnamefont {T.}~\bibnamefont
  {{Robinson}}}, \ and\ \bibinfo {author} {\bibfnamefont {J.~J.}\ \bibnamefont
  {{Fortney}}},\ }\href {\doibase 10.3847/1538-4357/aa927b} {\bibfield
  {journal} {\bibinfo  {journal} {Astrophys. J.}\ }\textbf {\bibinfo {volume}
  {850}},\ \bibinfo {eid} {121} (\bibinfo {year} {2017})}\BibitemShut {NoStop}%
\bibitem [{\citenamefont {{Kiang}}\ \emph {et~al.}(2018)\citenamefont
  {{Kiang}}, \citenamefont {{Domagal-Goldman}}, \citenamefont {{Parenteau}},
  \citenamefont {{Catling}}, \citenamefont {{Fujii}}, \citenamefont
  {{Meadows}}, \citenamefont {{Schwieterman}},\ and\ \citenamefont
  {{Walker}}}]{KDG18}%
  \BibitemOpen
  \bibfield  {author} {\bibinfo {author} {\bibfnamefont {N.~Y.}\ \bibnamefont
  {{Kiang}}}, \bibinfo {author} {\bibfnamefont {S.}~\bibnamefont
  {{Domagal-Goldman}}}, \bibinfo {author} {\bibfnamefont {M.~N.}\ \bibnamefont
  {{Parenteau}}}, \bibinfo {author} {\bibfnamefont {D.~C.}\ \bibnamefont
  {{Catling}}}, \bibinfo {author} {\bibfnamefont {Y.}~\bibnamefont {{Fujii}}},
  \bibinfo {author} {\bibfnamefont {V.~S.}\ \bibnamefont {{Meadows}}}, \bibinfo
  {author} {\bibfnamefont {E.~W.}\ \bibnamefont {{Schwieterman}}}, \ and\
  \bibinfo {author} {\bibfnamefont {S.~I.}\ \bibnamefont {{Walker}}},\ }\href
  {\doibase 10.1089/ast.2018.1862} {\bibfield  {journal} {\bibinfo  {journal}
  {Astrobiology}\ }\textbf {\bibinfo {volume} {18}},\ \bibinfo {pages} {619}
  (\bibinfo {year} {2018})}\BibitemShut {NoStop}%
\bibitem [{\citenamefont {{Madhusudhan}}(2019)}]{Mad19}%
  \BibitemOpen
  \bibfield  {author} {\bibinfo {author} {\bibfnamefont {N.}~\bibnamefont
  {{Madhusudhan}}},\ }\href@noop {} {\bibfield  {journal} {\bibinfo  {journal}
  {Annu. Rev. Astron. Astrophys.}\ } (\bibinfo {year} {2019})},\ \Eprint
  {http://arxiv.org/abs/1904.03190} {arXiv:1904.03190 [astro-ph.EP]}
  \BibitemShut {NoStop}%
\end{thebibliography}

%

\end{document}